\documentclass[11pt]{article}
\usepackage{subfiles}
\usepackage{xparse}
\usepackage[normalem]{ulem}

\usepackage{amsfonts}
\usepackage{amsmath}
\usepackage{amssymb, bm}
\usepackage{dsfont}
\usepackage{mathtools}
\usepackage{stmaryrd} 
\usepackage{slashed}
\usepackage{esint}
\usepackage{cancel}

\usepackage{feynmp-auto}

\usepackage{graphicx}
\usepackage{float}
\usepackage{caption}
\usepackage{subcaption}

\usepackage[dvipsnames]{xcolor}
\usepackage{hyperref}
\hypersetup{colorlinks,
    citecolor=black,
    filecolor=black,
    linkcolor=black,
    urlcolor=black,
    hyperfootnotes=true,pdftex}

\newcommand{\ctableofcontents}%
{\begingroup
\hypersetup{linkcolor=black}
\tableofcontents
\endgroup}

\usepackage{csquotes}
\usepackage[backend=bibtex,sorting=none,style=ieee,citestyle=numeric-comp,
           maxbibnames=6,maxcitenames=6]{biblatex}
\newcommand{\Section}[1]{\hyperref[#1]{Sec.~\ref*{#1}}}
\newcommand{\Figure}[1]{\hyperref[#1]{Fig.~\ref*{#1}}}
\newcommand{\Equation}[1]{\hyperref[#1]{Eq.~\eqref{#1}}}
\newcommand{\Appendix}[1]{\hyperref[#1]{App.~\ref*{#1}}}
\newcommand{\diagram}[2]{\hyperref[fig:#1#2]{diagram $#2$}}
\def\S{{\mathcal S}}

\DeclareMathOperator{\Tr}{\mathrm{Tr}}
\renewcommand{\exp}[1]{e^{#1}}

\newcommand{\eps}[2]{%
    \ifthenelse{\equal{#1}{0}}
        {\epsilon^{#2}_{\lambda}(q)}
        {\epsilon^{#2}_{\lambda_{#1}}(p_{#1})}}
\newcommand{\epss}[2]{%
    \ifthenelse{\equal{#1}{0}}
        {\epsilon^{#2}_{\lambda}(q)}
        {\epsilon^{#2}_{\lambda_{#1}}(p_{#1})}}
\newcommand{\epsdown}[2]{
    \ifthenelse{\equal{#1}{0}}
        {\epsilon_{#2}^{\lambda}(q)}
        {\epsilon_{#2}^{\lambda_{#1}}(p_{#1})}}
\newcommand{\epssdown}[2]{
    \ifthenelse{\equal{#1}{0}}
        {(\epsilon_{#2}^{\lambda}(q))^*}
        {(\epsilon_{#2}^{\lambda_{#1}}(p_{#1}))^*}}

\def\rmd{{\rm d}}

\def\rmA{{\rm A}}
\def\rmF{{\rm F}}
\def\rmR{{\rm R}}
\def\eik{{\rm Eik.}}
\def\z{{\bf z}}
\def\k{{\bf k}}
\def\x{{\bf x}}
\def\y{{\bf y}}
\def\M{{\cal M}}
\def\P{{\bf P}}
\def\r{{\bf r}}
\def\b{{\bf b}}
\def\p{{\bf p}}
\def\ux{{\underline x}}
\def\uk{{\underline k}}
\def\C{{\mathcal C}}
\def\A{{\mathcal A}}
\NewDocumentCommand\bp{o}{%
  \IfNoValueTF{#1}{%
    {\mathbf{p}}
  }{%
    {\mathbf{p_{#1}}}
  }%
}
\NewDocumentCommand\bk{o}{%
  \IfNoValueTF{#1}{%
    {\mathbf{k}}
  }{%
    {\mathbf{k_{#1}}}
  }%
}
\def\bw{\mathbf{w}}
\NewDocumentCommand\bx{o}{%
  \IfNoValueTF{#1}{%
    {\mathbf{x}}
  }{%
    {\mathbf{x_{#1}}}
  }%
}
\def\by{\mathbf{y}}
\NewDocumentCommand\bz{o}{%
  \IfNoValueTF{#1}{%
    {\mathbf{z}}
  }{%
    {\mathbf{z_{#1}}}
  }%
}
\def\bP{\mathbf{P}}

\def\bb{\mathbf{b}}
\def\br{\mathbf{r}}
\def\bq{\mathbf{q}}
\def\bzer{\mathbf{0}}

\def\Lpartial{\overset{\leftarrow}{\partial}}
\def\Rpartial{\overset{\rightarrow}{\partial}}

\def\LPartial{\overset{\leftarrow}{\mathcal{D}}}
\def\RPartial{\overset{\rightarrow}{\mathcal{D}}}

\def\sun{\mathfrak{su}(N)}

\def\UA{\mathcal{U}_\rmA}
\def\UF{\mathcal{U}_\rmF}
\def\UR{\mathcal{U}_\rmR}
\def\UAd{\mathcal{U}_\rmA^\dagger}
\def\UFd{\mathcal{U}_\rmF^\dagger}

\def\BA{\big|^{\rm BA}_\eik}
\def\BI{\big|^{\rm BI}_\eik}
\def\IA{\big|^{\rm IA}_\eik}
\def\BAq{\big|^{{\rm BA},q}_\eik}
\def\BIq{\big|^{{\rm BI},q}_\eik}
\def\IAq{\big|^{{\rm IA},q}_\eik}
\def\BAbq{\big|^{{\rm BA},\bar{q}}_\eik}
\def\BIbq{\big|^{{\rm BI},\bar{q}}_\eik}
\def\IAbq{\big|^{{\rm IA},\bar{q}}_\eik}

\newcommand \nn {\nonumber}

\title{Quark TMDs from back-to-back dijet production at forward rapidities in pA collisions beyond eikonal accuracy in the CGC}

\author{Tolga Altinoluk$\, {}^a$, Guillaume Beuf$\, {}^a$, Etienne Blanco${}^a$, Swaleha Mulani${}^a$
\bigskip \\
${}^a$ Theoretical Physics Division, National Centre for Nuclear Research, \\ Pasteura 7, Warsaw 02-093, Poland
}
\date{\today}

\begin{document}
\maketitle
\begin{abstract}
{
We study dijet production in pA collisions at forward rapidities at next-to-eikonal accuracy. We restrict ourselves to the next-to-eikonal corrections that are induced by the quark background field of the target. We consider all possible channels, compute scattering amplitudes both in general kinematics and in the back-to-back limit. By using these results, we compute the back-to-back production cross section and obtain a factorized expression with a quark TMD times associated hard factor for each channel.   
}
\end{abstract}
\newpage
\ctableofcontents

\newpage
\section{Introduction}
\label{sec:intro}

The Color Glass Condensate (CGC) effective theory (see \cite{Gelis:2010nm,Albacete:2014fwa,Blaizot:2016qgz} for recent reviews and references therein) is a powerful framework within quantum chromodynamics (QCD) that describes the high-energy behavior of gluons in hadronic systems, such as protons and heavy nuclei. At very high energies, gluons dominate the internal structure of hadrons, reaching a regime where their density becomes so large that nonlinear interactions among them cannot be neglected. This leads to the phenomenon of gluon saturation, where further increases in energy do not significantly enhance the gluon density. The increase in energy (or equivalently in rapidity) is governed by the famous nonlinear evolution equation known as  Balitsky-Kovchegov / Jalilian-Marian-Iancu-McLerran-Weigert-Leonidov-Kovner (BK-JIMWLK) \cite{Balitsky:1995ub,Kovchegov:1999yj,Kovchegov:1999ua,Jalilian-Marian:1996mkd,Jalilian-Marian:1997qno,Jalilian-Marian:1997jhx,Jalilian-Marian:1997ubg,Kovner:2000pt,Weigert:2000gi,Iancu:2000hn,Iancu:2001ad,Ferreiro:2001qy,} equation.  The CGC provides an effective field theory to model this saturation regime, treating the gluons as a classical field generated by color charges moving at near the speed of light. The theory incorporates quantum corrections and enables predictions of saturation phenomena that might be observed in high-energy collisions, such as those at the Large Hadron Collider (LHC), the Relativistic Heavy Ion Collider (RHIC) and the future Electron-Ion Collider (EIC).  

One of the frequently used observables to test the compatibility of the saturation phenomena with the high-energy proton-nucleus (pA) collision data from LHC and the RHIC is single inclusive hadron/jet production in the forward rapidity region. The state-of-the-art calculation framework for production of single inclusive hadron/jet (as well as production of multi hadron/jet) is known as "hybrid factorization" \cite{Dumitru:2005gt}. In this framework, one computes the partonic cross section in the following way. The projectile parton is assumed to be dilute and it is treated in the spirit of collinear factorization. Perturbative corrections to the wave function of the incoming parton is provided by the Dokshitzer-Gribov-Lipatov-Altarelli-Parisi (DGLAP) equations \cite{Gribov:1972ri,Altarelli:1977zs,Dokshitzer:1977sg}. On the other hand, the target is treated (in the spirit of CGC framework) as a distribution of strong semiclassical color fields which during the scattering event transfers transverse momentum to the propagating partonic configuration of the projectile. Upon computing the partonic cross section, one convolutes it with the corresponding Parton Distribution Function (PDF) in the initial state to get the hadronic cross section of jet production. In the case of hadron production, in addition to convoluting the partonic cross section with the corresponding PDF in the initial state, one also convolutes it with corresponding Fragmentation Function (FF) in the final state to obtain the hadronic cross section.  Over the last decade, we have witnessed an immense effort to compute the next-to-leading order (NLO) corrections in $\alpha_s$ to single inclusive hadron/jet production cross section in pA collisions at forward rapidities.  
Both analytical \cite{Altinoluk:2011qy,Chirilli:2011km,Chirilli:2012jd,Stasto:2013cha,Kang:2014lha,Altinoluk:2014eka,Altinoluk:2015vax,Iancu:2016vyg,Liu:2020mpy,Ducloue:2016shw,Shi:2021hwx,Wang:2022zdu,Altinoluk:2023hfz} and numerical \cite{Stasto:2013cha,Watanabe:2015tja,Ducloue:2017dit,Ducloue:2017mpb,Mantysaari:2023vfh} studies of the NLO corrections to single inclusive hadron/jet production at forward rapidity are now available. 

Another observable that adopts hybrid factorization for computations  is dijets/dihadron production in pA collisions at forward rapidities. This observable, provides an opportunity to probe the high energy limit of leading twist gluon transverse momentum dependent distribution functions (TMDs) from the CGC calculations when one considers a specific regime known as the back-to-back limit (see \cite{Petreska:2018cbf,Boussarie:2023izj} for recent reviews). If one considers the production of two jets with transverse momenta $\p_1$ and $\p_2$, and longitudinal momenta $p_1^+$ and $p_2^+$, at transverse positions $\x_1$ and $\x_2$, relative dijet momentum $\P$ and dijet momentum imbalance $\k$ can be defined as  
 \begin{equation}
    \bP \equiv (1-z)\bp[1] - z\bp[2]
     \, , \quad\quad  \quad\quad 
    \bk = \bp[1] + \bp[2] 
    \label{eq:PK}
\end{equation}
where $z$ is the lightcone momentum fraction carried by the first jet and it is defined as 
 \begin{equation}
 \label{eq:mom_frac}
z \equiv \frac{p_1^+}{p_1^++p_2^+} 
     \, , \quad\quad  \quad\quad 
 (1-z)  \equiv \frac{p_2^+}{p_1^++p_2^+}
\end{equation}
Moreover, one can also define their conjugate variables $\r$ (transverse dijet size) and $\b$ (impact parameter) as 
\begin{equation}
\br \equiv \x_1-\x_2
  \, , \quad\quad  \quad\quad 
\bb \equiv z\x_1+(1-z)\x_2    
    \label{eq:br}
\end{equation}
In the kinematic regime that we are interested in, the produced jets fly almost back-to-back in momentum space and it corresponds to the case where $|\P|\gg |\k|$. The effect of this strong ordering in momenta reflects itself on the conjugate coordinates as $|\r|\ll |\b|$. Therefore, in the back-to-back limit one can perform Taylor expansion of the Wilson lines that appear in the amplitude around $\r\sim 0$ and keep only the first nontrivial contribution in this expansion. At the cross section level, this leads to obtaining TMD distributions from the expanded Wilson line structures \cite{Dominguez:2011wm}. These results indicate an equivalence between the CGC and the standard TMD factorization frameworks 
when appropriate limits are applied to both sides. Namely, the high energy limit of the dijet production cross section calculated in the standard TMD factorization (by constructing hadronic matrix elements of bilocal products of field operators that contain gauge links) coincides with the back-to-back limit of the dijet cross section calculated in the CGC framework. These results triggered many studies including the dijet production with massive quarks \cite{Marquet:2016cgx,Marquet:2017xwy}, photon+dijet production \cite{Altinoluk:2018uax,Altinoluk:2018byz} or photoproduction of trijets \cite{Altinoluk:2020qet} in pA collisions at forward rapidities in the back-to-back limit. Moreover,  forward production of Drell-Yan pair and a jet \cite{Taels:2023czt}, photoproduction of inclusive back-to-back dijets \cite{Taels:2022tza} in pA collisions and inclsuive back-to-back dijet production in DIS \cite{Caucal:2022ulg,Caucal:2023nci,Caucal:2023fsf} have been performed at NLO and it is shown that the relation between the CGC and standard TMD factorization frameworks holds not only at leading order but also at NLO accuracy.  
Moreover, in \cite{Altinoluk:2019fui,Altinoluk:2019wyu,Boussarie:2020vzf,}, the equivalence between the CGC and TMD factorization frameworks has been extended beyond the correlation limit for dijet production by performing a resummation of the small transverse size of the dijets. The results of this study have provided a unique opportunity to study the emergence of saturation by distinguishing between the kinematic power corrections (associated with the conjugate transverse dijet size) and geniune saturation eﬀects \cite{Fujii:2020bkl,Altinoluk:2021ygv,Boussarie:2021ybe,}. It is shown that, the resummation of the kinematic power corrections (that are also referred to as kinematic twist corrections) leads to the so-called small-x improved TMD (iTMD) framework (conjectured and studied previously in \cite{Kotko:2015ura,vanHameren:2016ftb,Bury:2020ndc}) which interpolates between the dilute limit of the CGC (also known as High Energy Factorization) and the TMD limit of the CGC.

One of the key improvements needed for the studies of the high-energy scattering in QCD is to explore the eﬀects of relaxing the standard "kinematic" approximations. The most frequently used approximation in this framework is the eikonal one which amounts to keeping only the leading terms in energy and discarding any contribution that is suppressed in energy. Within the CGC framework high energy can be achieved by boosting the target along $x^-$ direction by a boost parameter $\gamma_t$.  In a high-energy dilute-dense scattering, eikonal approximation amounts to the following three assumptions: (i) the highly boosted background field that describes the target is localized in the longitudinal direction (around $x^+$ = 0) due to Lorentz contraction, (ii) the background field of the target is given by its leading component in $\gamma_t$ whereas the subleading components in terms of $\gamma_t$ are neglected and (iii) the background field of the target is assumed to be independent of the light-cone coordinate $x^-$ due to Lorentz time dilation, which amounts to considering the target fields as static and neglecting the dynamics of the target. Relaxing any of the three aforementioned assumptions give corrections to the eikonal approximation. In order to consider the next-to-eikonal (NEik) corrections, one should take into account the corrections that are of the order $1/\gamma_t$ at the level of the boosted background field.  

All the three approximations listed above are valid in the presence of a gluon background field of the target.  Another source of NEik corrections originates from accounting for the quark background of the target such that projectile parton interacts with the target via a $t$-channel quark exchange. Under a boost of $\gamma_t$ along $x^-$ direction, a current associated with the target scale as 
\begin{align}
\label{eq:current_scaling}
J^-(x)\propto \gamma_t, \, \,  J^j(x)\propto (\gamma_t)^0,  \, \, J^+(x)\propto (\gamma_t)^{-1} .
\end{align}
The projections of quark background field $\Psi(x)$ are defined as 
\begin{align}
\label{eq:projections_Psi}
\Psi^{(-)}(x)&=\frac{\gamma^+\gamma^-}{2} \Psi(x) , \\
\Psi^{(+)}(x) &=\frac{\gamma^-\gamma^+}{2} \Psi(x) ,
\end{align}
which are known to be the good and bad components of $\Psi(x)$. The currents associated with the target are constructed as bilinears of $\Psi(x)$ and its components satisfy
\begin{align}
\overline{\Psi}(x)\gamma^-\Psi(x)&=\overline{\Psi^{(-)}}(x)\gamma^-\Psi^{(-)}(x) , \\
\overline{\Psi}(x)\gamma^j\Psi(x)&=\overline{\Psi^{(-)}}(x)\gamma^j\Psi^{(+)}(x)+\overline{\Psi^{(+)}}(x)\gamma^j\Psi^{(-)}(x) , \\
\overline{\Psi}(x)\gamma^+\Psi(x)&=\overline{\Psi^{(+)}}(x)\gamma^-\Psi^{(+)}(x) .
\end{align}
Since the currents associated with the target have to follow the same scaling behaviour introduced in Eq.~\eqref{eq:current_scaling}, the components of the quark background field scale as
\begin{align}
\Psi^{(-)}(x)&\propto(\gamma_t)^{\frac{1}{2}} , \\
\Psi^{(+)}(x)&\propto(\gamma_t)^{-\frac{1}{2}} ,
\end{align} 
with the boosting parameter $\gamma_t$. Thus, the quark background field does not contribute at eikonal order, the enhanced component $\Psi^{(-)}(x)$ contributes at NEik order and the suppressed component $\Psi^{(+)}(x)$ contribute at next-to-next-to-eikonal (NNEik) order and beyond. 

Over the last decade, there has been a great effort to compute the NEik corrections in the CGC. In \cite{Altinoluk:2014oxa,Altinoluk:2015gia}, the first studies that includes the NEik corrections that stem from the finite longitudinal width are performed. These results then were applied to study particle production and correlations  at NEik accuracy both in dilute-dilute \cite{Altinoluk:2015xuy,Agostini:2019avp,Agostini:2019hkj,} and in dilute-dense \cite{Agostini:2022ctk,Agostini:2022oge} collisions. The NEik corrections to quark and scalar propagators have been computed in \cite{Altinoluk:2020oyd,Altinoluk:2021lvu,Agostini:2023cvc} and the results are applied to DIS dijet production in \cite{Altinoluk:2022jkk, Agostini:2024xqs}. In \cite{Altinoluk:2023qfr}, production of back-to-back quark-gluon dijet in DIS at NEik accuracy due to $t$-channel quark exchange is computed to probe the quark TMDs. In \cite{Altinoluk:2024zom}, back-to-back DIS dijet production in a pure gluon background is computed at NEik accuracy to probe gluon TMDs and the interplay between the NEik and kinematic twist corrections are studied. Finally, in \cite{Altinoluk:2024dba}, the gluon propagator is revisited to include all NEik corrections and the results are applied to various parton-nucleus scattering processes.  In \cite{Kovchegov:2015pbl,Kovchegov:2016zex,Kovchegov:2016weo,Kovchegov:2017jxc,Kovchegov:2017lsr,Kovchegov:2018znm,Kovchegov:2018zeq,Kovchegov:2020kxg,Kovchegov:2020hgb,Kovchegov:2021lvz,Kovchegov:2021iyc,Cougoulic:2022gbk,Kovchegov:2022kyy,Borden:2023ugd,Kovchegov:2024aus,Borden:2024bxa}, quark and gluon helicity evolutions as well as observables such as single and/or double spin asymmetries are computed at NEik accuracy. In \cite{Cougoulic:2019aja,Cougoulic:2020tbc}, helicity dependent extensions of the CGC have been studied at NEik accuracy. In \cite{Balitsky:2015qba,Balitsky:2016dgz,Balitsky:2017flc}, rapidity evolution of gluon TMDs that can interpolate between moderate and low values of momentum fraction x have been studied. A similar idea is pursued to study this interpolation for inclusive DIS \cite{Boussarie:2020fpb,Boussarie:2021wkn} as well as exclusive Compton scattering \cite{Boussarie:2023xun}. The NEik corrections to both quark and gluon propagators have been computed in the high energy operator product expansion (OPE) formalism in \cite{Chirilli:2018kkw,Chirilli:2021lif}. In \cite{Li:2023tlw,Li:2024fdb,Li:2024xra}, subeikonal corrections in the CGC are studied in an effective Hamiltonian approach. Finally, subeikonal corrections are studied in an approach that allows longitudinal momentum exchange between the projectile and the target in \cite{Jalilian-Marian:2017ttv,Jalilian-Marian:2018iui,Jalilian-Marian:2019kaf}. Last but not least, the effects of subeikonal corrections are also studied in the context of orbital angular momentum in \cite{Hatta:2016aoc,Kovchegov:2019rrz,Boussarie:2019icw,Kovchegov:2023yzd,Kovchegov:2024wjs}. 

In this paper, we study dijet production in pA collisions at forward rapidities at NEik accuracy. We restrict ourselves to the case where the suppression with respect to the eikonal order arise only from the interaction of the incoming parton with the quark background field of the target. For simplicity, we only consider the partonic cross section for various channels, hence for fixed quarks flavors. In practice, it is straight forward to convolute the computed partonic cross section with the associated PDF and obtain the dijet production cross section in pA collisions. For various channels, we first compute the scattering amplitude in general kinematics, then its back-to-back limit and then compute the partonic cross section. We also discuss the relation between the color structures that appear in the back-to-back cross section and quark TMDs. The paper is organized as follows. In Secs. \ref{sec:ggq}, \ref{sec:qqbq}, \ref{sec:qgg} and \ref{sec:qq}, we compute the scattering amplitudes both in general kinematics and in the back-to-back limit, and also back-to-back production cross sections for $g\to gq$, $q\to q\bar q$, $q\to gg$ and $q\to qq$ channels respectively.  Sec.~\ref{sec:TMD}  is devoted to the discussion of the relations between the color structures that appear in the back-to-back cross sections and the quark TMDs. Finally, in Sec.~\ref{sec:outlook} we present a brief summary of our results and give an outlook. App.~\ref{app:Tech} is devoted to the technical details that are used in our calculations. In App.~\ref{app:eikprop} we present all the parton propagators at Eikonal accuracy that are needed to compute the scattering amplitudes in each channel. Finally, in App.~\ref{sec:antiquarks} we present the results of scattering amplitudes both in general kinematics and in the back-to-back limit together with the back-to-back production cross sections for the channels that are analogous to the ones presented in main body of the paper but with quark and antiquark exchanged, namely, $g\to g\bar q$, $\bar q \to {\bar q} q$, $\bar q\to gg$ and $\bar q \to \bar q \bar q$ channels.

\section{$g\to gq$ channel } 
\label{sec:ggq}     

\begin{figure}[H]
\centering
\begin{subfigure}{0.49\textwidth}
\centering
\includegraphics[width=\textwidth]{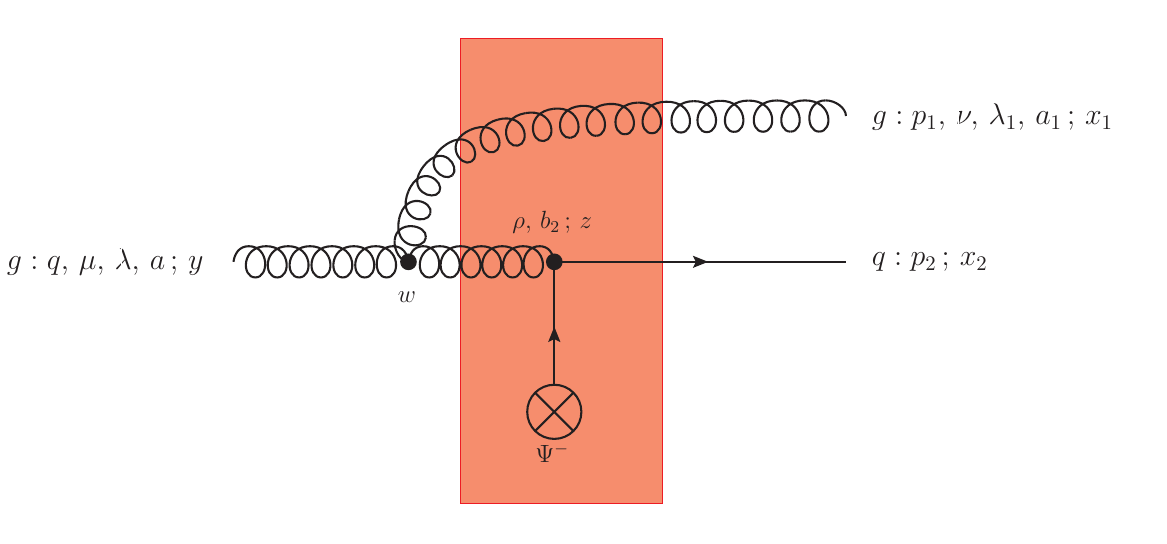}
\caption{Diagram 1}
\label{fig:g-gq1}
\end{subfigure}
\begin{subfigure}{0.49\textwidth}
\centering
\includegraphics[width=\textwidth]{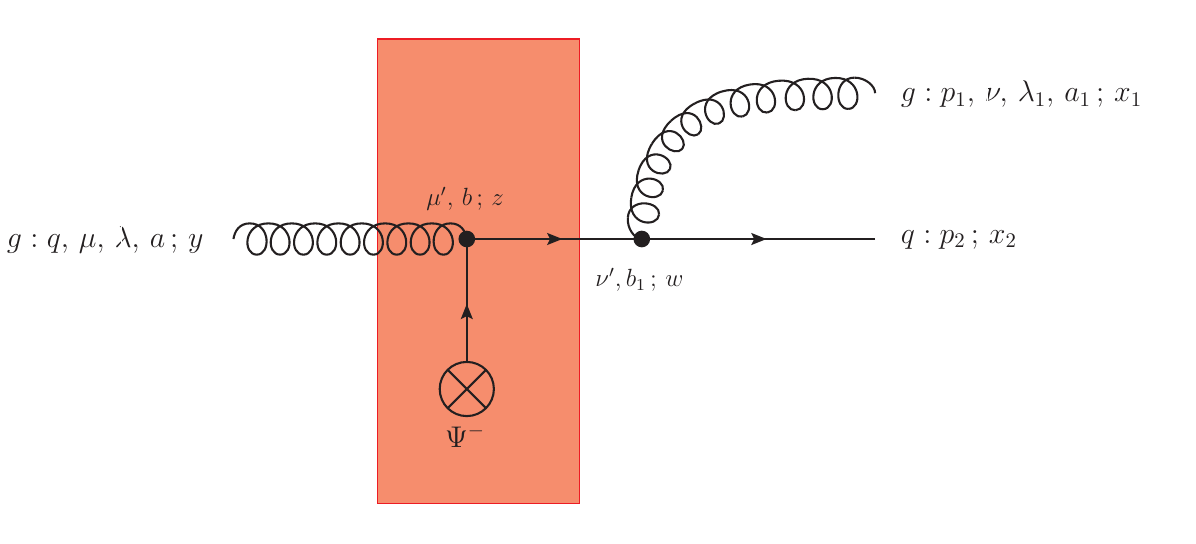}
\caption{Diagram 2}
\label{fig:g-gq2}
\end{subfigure}
\begin{subfigure}{0.5\textwidth}
\centering
\includegraphics[width=\textwidth]{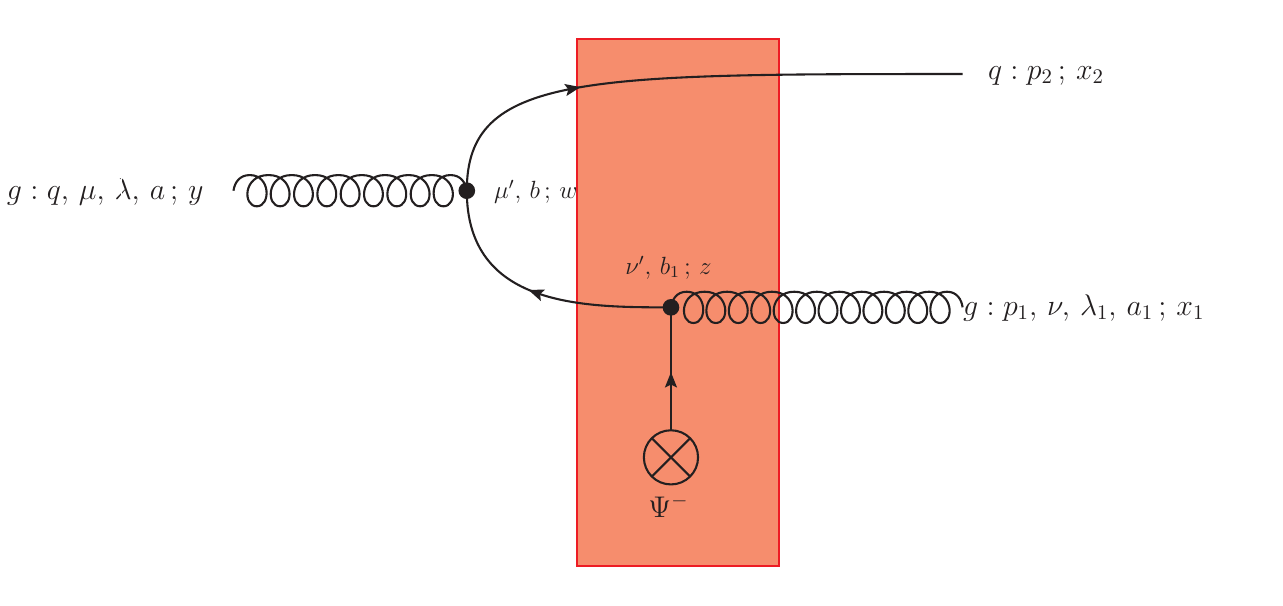}
\caption{Diagram 3}
\label{fig:g-gq3}
\end{subfigure}
\caption{Diagrams contributing to channel $g\to gq$.}
\label{fig:g-gq}
\end{figure}

We start our analysis with the $g\to gq$ channel. In this process, a quark-gluon dijet is produced from an incoming gluon upon interaction with the target. At NEik accuracy, this can be achieved via three different mechanisms. The first mechanism corresponds to the case where an incoming gluon splits into two gluons before the medium, one of the gluons interacts with the target via a $t$-channel quark exchange converting into a quark and one gets a quark-gluon dijet in the final state (see Fig.~\ref{fig:g-gq1}). The second mechanism corresponds to the case where the incoming gluon interacts with the target via a $t$-channel quark exchange converting into a quark and the quark splits into a quark-gluon pair in the final state after interacting with the target (see Fig.~\ref{fig:g-gq2}). The last mechanism corresponds to the case where the incoming gluon splits into a quark-antiquark pair before the interaction with the target, then antiquark interacts with the target via a $t$-channel quark exchange converting into a gluon and finally one gets a quark-gluon dijet in the final state (see Fig.~\ref{fig:g-gq3}). In the following subsections, we study each of these mechanisms in detail. 

\subsection{$g\to gq$ amplitude in general kinematics}
In this channel, we consider the incoming gluon with four momenta $q$, polarization $\lambda$ and color $a$, the produced gluon with four momenta $p_1$, polarization $\lambda_1$ and color $a_1$, and the produced quark with four momenta $p_2$ and helicity $h$. As explained previously, three mechanisms contribute to the scattering amplitude. In that case, one can write the total scattering amplitude for this channel as a sum of the three contributions as 
\begin{align}
\label{eq:M_tot_schm_g_to_gq}
\M_{g\to gq, \, {\rm tot.}} = \M_{g\to gq, \, {\rm 1}}+\M_{g\to gq, \, {\rm 2}}+ \M_{g\to gq, \, {\rm 3}} ,
\end{align}
where each term on the right hand side of Eq.~\eqref{eq:M_tot_schm_g_to_gq} corresponds to the scattering amplitudes computed for each of the three mechanisms described above. 

Let us start our analysis with the first mechanism (see Fig.~\ref{fig:g-gq1}). In that case, the $S$-matrix element can be obtained thanks to the following LSZ-type reduction formula\footnote{
We use the metric signature $(+,-,-,-)$. We use $x^\mu$ for a Minkowski 4-vector. In a light-cone basis we have
$
x^\mu=(x^+,{\bf x},x^-)
$
where $x^\pm=(x^0\pm x^3)/\sqrt{2}$ and ${\bf x}$ denotes a transverse vector with components $x^i$. 
We will also use the notations 
$
\ux = (x^+,{\bf x})
$
and 
$\uk = (k^+,{\bf k})$.}
\footnote{For a given momentum 4-vector $k^{\mu}$, we use the notation $\check{k}^{\mu}$ for its on-shell analog. More precisely, it is defined in such a way that their $+$ and transverse components coincide, $\check{k}^{+}={k}^{+}$ and $\check{\k}=\k$, whereas the $-$ component of $\check{k}^{\mu}$ is adjusted to make it on-shell, i.e. $\check{k}^-=(\k^2+m^2)/(2k^+)$ for a massive quark or $\check{k}^-=\k^2/(2k^+)$ for a gluon.}

\begin{align}
    \mathcal{S}_{g\to gq, \,  {\rm 1}} = &
    \lim_{y^+\to-\infty}\lim_{x_1^+,x_2^+\to\infty}
    \int_{\by,\bx[1],\bx[2]}\int_{y^-,x^-_1,x^-_2}
    \int_{\bw, \bz}
    \int_{w^-, z^-}
    \int_{-\frac{L^+}{2}}^{\frac{L^+}{2}}d z^+
    \int_{-\infty}^{-\frac{L^+}{2}}d w^+ \nn \\
    &\times \exp{ix_1\cdot\Check{p}_1} \ \exp{ix_2\cdot\Check{p}_2}\ \exp{-iy\cdot\Check{q}}
     (-2q^+)\ \eps{0}{\mu} \ (-2p_1^+) \ \eps{1}{\nu} 
    \nn \\
    &\times \left[G_{0,\rmF}^{\mu'\mu}(w,y)\right]_{a'a}
    \left[G_\rmF^{\nu\nu'}(x_1,w)\BA\right]_{a_1b_1}
    \left[G_\rmF^{\rho\rho'}(z,w)\BI\right]_{b_2b}V_{\mu'\nu'\rho'}^{a'b_1b} \nn \\
    &\times \overline{u}(\Check{p}_2, h)\gamma^+\left[S_\rmF(x_2,z)\IAq
    (-ig)t^{b_2}\right]_{\alpha_2\beta}\gamma_{\rho}\Psi_\beta^-(z) ,
\label{eq:ggq1-prop}
\end{align}
where $G_{0,\rmF}^{\mu'\mu}(w,y)$ is the vacuum gluon propagator which in momentum space reads 
\begin{align}
\label{eq:vacuum_gluon_prop}
\tilde{G}^{\mu\nu}_{0,\rmF}(p)=\frac{i}{p^2+i\epsilon}\bigg[ -g^{\mu\nu}+\frac{p^{\mu}\eta^{\nu}+\eta^{\mu}p^{\nu}}{p\cdot \eta}\bigg] ,
\end{align}
with $\eta^{\mu}$ defined as $\eta^{\mu}=g^{\mu +}$, so that $\eta\cdot p=p^+$ and $\eta^2=0$. Since we are in the $\eta \cdot A\equiv A^+=0$ light cone gauge, one has 
\begin{align}
\tilde{G}^{\mu +}_{0,\rmF}(p)=\tilde{G}^{+ \nu}_{0,\rmF}(p) = 0 \, . 
\end{align}
$V_{\mu'\nu'\rho'}^{a'b_1b} $ is the triple gluon vertex which is defined as 
\begin{align}
\label{eq:triple_gluon_vertex}
V_{\lambda_1\lambda_2\lambda_3}^{abc}(p_1,p_2,p_3) =  gf^{abc}
\Big[g_{\lambda_1\lambda_2}(p_1-p_2)_{\lambda_3}
        + g_{\lambda_2\lambda_3}(p_2-p_3)_{\lambda_1}
        + g_{\lambda_3\lambda_1}(p_3-p_1)_{\lambda_2}\Big] ,
\end{align}
%
%
\sloppy with all incoming convention for momenta $p_1$, $p_2$ and $p_3$. Finally, $G_\rmF^{\nu\nu'}(x_1,w)\BA$ and $G_\rmF^{\rho\rho'}(z,w)\BI$ are the before-to-after and before-to-inside gluon propagators whose explicit expressions are given in Eqs.~\eqref{gluon_prop_BA} and \eqref{gluon_prop_BI} respectively at eikonal order. Finally, $S_\rmF(x_2,z)\IAq$ is the inside-to-after quark propagator at eikonal order which is given in Eq.~\eqref{quark_prop_IA}.  After using these explicit expressions and contracting the Dirac indices, the $S$-matrix element for the first mechanism (Fig.~\ref{fig:g-gq1}) can be written as  
\begin{align}
\label{eq:S_1_g_to_gq_1}
\mathcal{S}_{g\to gq, \, 1} = &
\frac{-ig^2}{2p_2^+}f^{ab_1b}
\int_{\bz,\bz[1]}
\int_{-\frac{L^+}{2}}^{\frac{L^+}{2}} dz^+
\int_{-\infty}^{-\frac{L^+}{2}}dw^+
\int\frac{d^3 \underline{k}_1}{(2\pi)^2}
\int\frac{d^3 \underline{k}_2}{(2\pi)^2} \
\delta(k_1^+-p_1^+) \, \delta(k_2^+-p_2^+) \
\nn \\
&
\times 
\exp{-iw^+(\check{q}^--\check{k}_1^--\check{k}_2^-)} \
\exp{-i\bz\cdot(\bp[2]-\bk[2])} \ 
\exp{-i\bz[1]\cdot(\bp[1]-\bk[1])} \ 
(2\pi)^3\delta^{(3)}(\underline{q}-\underline{k}_1-\underline{k}_2)
\nn \\
&
\times 
\overline{u}(\Check{p}_2, h) \ 
\UF(\infty,z^+;\bz) \ t^{b_2} \  \UA(z^+,w^+;\bz)_{b_2b} \ 
\UA(\infty,w^+;\bz[1])_{a_1b_1}
\nn \\
&
\times 
\Big\{
\Big[-\eps{0}{\mu'} + \frac{\eps{0}{\mu}q_\mu}{q^+}\eta^{\mu'}\Big]
\Big[-\eps{1}{\nu'} + \frac{\eps{1}{\nu}k_{1\nu}}{p_1^+}\eta^{\nu'}\Big]
\Big[g^{l\rho'} - \frac{\eta^{\rho'}}{p_2^+}\bk[2]^l\Big] 
\nn \\
&
\times 
\Big[-g_{\mu'\nu'}(q+k_1)_{\rho'} + g_{\nu'\rho'}(k_1-k_2)_{\mu'} + g_{\rho'\mu'}(k_2+q)_{\nu'}\Big]\Big\}
\gamma^l\frac{\gamma^+\gamma^-}{2}\Psi(\underline{z}) .
\end{align}
Note that the second term in the first square brackets vanish by transversality of the gluon polarization vector. Also, we would like to mention that the covariant derivative terms in the before-to-inside gluon propagator given in Eq.~\eqref{gluon_prop_BI} and the inside-to-after quark propagator given in Eq.~\eqref{quark_prop_IA}, come in between two $\gamma^+$ matrices in Eq.~\eqref{eq:S_1_g_to_gq_1} and therefore vanish. The momentum structure appearing in Eq.~\eqref{eq:S_1_g_to_gq_1} can be further simplified by using Eq.~\eqref{eq:polarization_vecs} and it can be written as 
\begin{align}
\label{eq:simpf_1}
&
\Big\{
-\epsilon_{\lambda}^{\mu'}(q)
\Big[-\eps{1}{\nu'} + \frac{\eps{1}{\nu}k_{1\nu}}{p_1^+}\eta^{\nu'}\Big]
\Big[g^{l\rho'} - \frac{\eta^{\rho'}}{p_2^+}\bk[2]^l\Big]
\nn \\
& \hspace{4cm}
\times 
\Big[-g_{\mu'\nu'}(q+k_1)_{\rho'} + g_{\nu'\rho'}(k_1-k_2)_{\mu'} + g_{\rho'\mu'}(k_2+q)_{\nu'}\Big]\Big\}  
\nn \\
&= -\varepsilon_{\lambda}^i \ \varepsilon_{\lambda_1}^{j*} \
\Big\{ \
g^{ij}\Big[\bq^l + \bk[1]^l - \frac{q^++p_1^+}{p_2^+}\bk[2]^l\Big]
-
g^{jl}\Big[\frac{p_2^+-p_1^+}{q^+}\bq^i + \bk[1]^i - \bk[2]^i\Big] 
\nn \\
& \hspace{2.5cm}
-g^{il}\Big[\bq^j - \frac{p_2^++q^+}{p_1^+}\bk[1]^j + \bk[2]^j\Big]\Big\} ,
\end{align}
where we have used the fact that in the light-cone gauge $A^+=0$, the gluon polarization vectors satisfy
\begin{align}
\label{eq:polarization_vecs}
    &\epsilon^+_\lambda(p) = \epsilon^\lambda_-(p) = 0 , \nn \\
    &\epsilon^i_\lambda(p) = -\epsilon^\lambda_i(p) = \varepsilon^i_\lambda , \nn \\
    &\epsilon^-_\lambda(p) = \epsilon^\lambda_+(p) =  \frac{\varepsilon_\lambda\cdot\bp}{p^+} .
\end{align} 
Plugging the simplified momentum structure given in Eq.~\eqref{eq:simpf_1} back into Eq.~\eqref{eq:S_1_g_to_gq_1} and performing the integration over $\k_2$, one can write the $S$-matrix element for the first mechanism as 
\begin{align}
\label{eq:S_1_g_to_gq_2}
\mathcal{S}_{g\to gq, \, 1} = &
\frac{-ig^2}{2p_2^+} \ f^{ab_1b} \ (2\pi)\delta(q^+-p_1^+-p_2^+)
\int_{\bz,\bz[1]}
\int_{-\frac{L^+}{2}}^{\frac{L^+}{2}} dz^+
\int_{-\infty}^{-\frac{L^+}{2}}dw^+
\int\frac{d^2 \bk[1]}{(2\pi)^2} 
\nn \\
&
\times 
\exp{-iw^+\left(\frac{\bq^2}{2q^+}-\frac{\bk[1]^2}{2p_1^+}-\frac{(\bk[1]-\bq)^2
}{2p_2^+}\right)}
\exp{-i\bz\cdot(\bp[2]+\bk[1]-\bq)}
\ \exp{-i\bz[1]\cdot(\bp[1]-\bk[1])} 
\nn \\
&
\times 
\overline{u}(\Check{p}_2,h)
\  \UF(\infty,z^+;\bz) \ t^{b_2} \ \UA(z^+,w^+;\bz)_{b_2b}
\ \UA(\infty,w^+;\bz[1])_{a_1b_1} \ 
\gamma^l\gamma^{+}\gamma^{-}
\ \Psi(\underline{z})  
\nn \\
&
\times 
\varepsilon_\lambda^i\varepsilon_{\lambda_1}^{j*}
\bigg[g^{ij}\Big(\frac{p_1^+}{p_2^+}\bq^l 
    - \frac{q^+}{p_2^+}\bk[1]^l\Big)
    +g^{il}\Big(\bq^j - \frac{q^+}{p_1^+}\bk[1]^j\Big)
    -g^{jl}\Big(\frac{p_1^+}{q^+}\bq^i - \bk[1]^i\Big)\bigg] . 
\end{align}
The scattering amplitude can be obtained from the $S$-matrix element by using the relation 
\begin{equation}
    \mathcal{S}_{g\to gq} = (2q^+) (2\pi) \, \delta\left(p_1^+ + p_2^+ - q^+\right)i\mathcal{M}_{g\to gq} .
\label{eq:amplitude}
\end{equation}
Thus, the scattering amplitude for the first mechanism, where an incoming gluon splits into two gluons before the medium, one of the gluons interact with the via a $t$-channel quark exchange converting into a quark and one gets a quark-gluon dijet in the final state (Fig.~\ref{fig:g-gq1}), can be written by using Eq.~\eqref{eq:amplitude} as 
\begin{align}
\label{eq:M_1_g_to_gq_1}
i\mathcal{M}_{g\to gq, \, 1} = &
\frac{-ig^2}{(2p_2^+)(2q^+)} \ f^{ab_1b} 
\int_{\bz,\bz[1]}
\int_{-\frac{L^+}{2}}^{\frac{L^+}{2}} dz^+
\int_{-\infty}^{-\frac{L^+}{2}}dw^+
\int\frac{d^2 \bk[1]}{(2\pi)^2} 
\nn \\
&
\times 
\exp{-iw^+\left(\frac{\bq^2}{2q^+}-\frac{\bk[1]^2}{2p_1^+}-\frac{(\bk[1]-\bq)^2
}{2p_2^+}\right)}
\exp{-i\bz\cdot(\bp[2]+\bk[1]-\bq)}
\ \exp{-i\bz[1]\cdot(\bp[1]-\bk[1])} 
\nn \\
&
\times 
\overline{u}(\Check{p}_2,h)
\  \UF(\infty,z^+;\bz) \ t^{b_2} \ \UA(z^+,w^+;\bz)_{b_2b}
\ \UA(\infty,w^+;\bz[1])_{a_1b_1} \ 
\gamma^l\gamma^+\gamma^-
\ \Psi(\underline{z})  
\nn \\
&
\times 
\varepsilon_\lambda^i\varepsilon_{\lambda_1}^{j*}
\bigg[g^{ij}\Big(\frac{p_1^+}{p_2^+}\bq^l 
    - \frac{q^+}{p_2^+}\bk[1]^l\Big)
    +g^{il}\Big(\bq^j - \frac{q^+}{p_1^+}\bk[1]^j\Big)
    -g^{jl}\Big(\frac{p_1^+}{q^+}\bq^i - \bk[1]^i\Big)\bigg] . 
\end{align}

We can now consider the second mechanism for the quark-gluon pair production in gluon initiated channel described in Fig.~\ref{fig:g-gq2}. The $S$-matrix element can be again obtained via LSZ-type reduction formula which for the second mechanism reads  
\begin{align}
\label{eq:S_2_g_to_gq_1}
& \hspace{-0.4cm}
\mathcal{S}_{g\to gq, \, 2} = 
\lim_{y^+\to-\infty} \ \lim_{x_1^+,x_2^+\to\infty}
\ \int_{\by,\bx[1],\bx[2]} \ \int_{y^-,x^-_1,x^-_2}
\ \int_{\bw, \bz}
\int_{w^-, z^-}
\int_{-\frac{L^+}{2}}^{\frac{L^+}{2}}d z^+
\int_{\frac{L^+}{2}}^{\infty}d w^+  
\nn \\
& 
\times 
\exp{ix_1\cdot\Check{p}_1} \ \exp{ix_2\cdot\Check{p}_2}
\ \exp{-iy\cdot\Check{q}}
\ (-2q^+)\epsilon_\mu^{\lambda}(q) \ (-2p_1^+) \ \epsilon_\nu^{\lambda_1}(p_1)^* 
\Big[G_\rmF^{\mu'\mu}(z,y)\BI\Big]_{ba}
\ \Big[G_{0,\rmF}^{\nu\nu'}(x_1,w)\Big]_{a_1b_1}
\nn \\
&
\times
\overline{u}(\Check{p}_2, h)\, \gamma^+ 
\Big[S_{0,\rmF}(x_2,w)(-ig)t^{b_1}\gamma_{\nu'}
S_\rmF(w,z)\IAq(-ig)t^{b}\Big]_{\alpha_2\beta}
\gamma_{\mu'}\Psi_\beta^-(z) ,
\end{align}
where the explicit expressions for before-to-inside gluon propagator and inside-to-after quark propagator are given in Eqs.~\eqref{gluon_prop_BI} and \eqref{quark_prop_IA} respectively at eikonal accuracy.  The vacuum gluon propagator is given in Eq.~\eqref{eq:vacuum_gluon_prop} and $S_{0,\rmF}(x_2,w)$ is the vacuum quark propagator which is defined as 
\begin{align}
\label{eq:quark_vacuum_prop}
S_{0,\rmF}(x,y)_{\beta\alpha}=({\bf 1})_{\beta\alpha}\int \frac{d^4k}{(2\pi)^4}\, e^{-ik\cdot(x-y)}\, \frac{i(\slashed{k} +m)}{\big[k^2-m^2+i\epsilon\big]} ,
\end{align}
where $m$ is the quark mass. Using explicit expressions of the propagators and Eq.~\eqref{eq:polarization_vecs} for the polarization vectors, the $S$-matrix element for the second mechanism can be written as 
\begin{align}
\label{eq:S_2_g_to_gq_2}
\mathcal{S}_{g\to gq, \, 2}  & =
\frac{-g^2}{(2q^+)(2p_2^+)}
\int_{\bz}
\int_{-\frac{L^+}{2}}^{\frac{L^+}{2}}d z^+
\int_{\frac{L^+}{2}}^{\infty}d w^+  
\int\frac{d^3 \underline{k}_0}{(2\pi)^2}
\ \delta(k_0^+-q^+) 
\ \exp{-iw^+(\check{k}_0^--\check{p}_1^--\check{p}_2^-)}
\nn \\
&
\times    
 \exp{-i\bz\cdot(\bk[0]-\bq)} \ 
(2\pi)^3\delta^{(3)}(\underline{k}_0-\underline{p}_1-\underline{p}_2)
\varepsilon_{\lambda}^i \ \varepsilon_{\lambda_1}^{j*}
\ \overline{u}(\Check{p}_2,h)
\gamma^+ \ (\slashed{\check{p}}_2+m)
\Big[\gamma^j-\frac{\bp[1]^j}{p_1^+}\gamma^+\Big]
\nn \\
&
\times 
(\slashed{\check{k}}_0+m)\gamma^i \ 
t^{a_1} \ \UF(w^+,z^+;\bz)
\ t^{b} \ \UA(z^+,-\infty;\bz)_{ba}
\ \frac{\gamma^{+}\gamma^{-}}{2}\Psi(\underline{z}) .
\end{align}
The Dirac structure appearing in Eq.~\eqref{eq:S_2_g_to_gq_2} can be seen in other channels in this manuscript. It can be simplified and can be written in a generic manner in the following way
\begin{align}
&
\gamma^+(\slashed{k}_1+m)\left[\gamma^j+a\gamma^+\right](\slashed{k}_2+m)\gamma^+ 
\nn \\
 & 
 \hspace{3.5cm}
 =
 -2\gamma^+\left[k_1^+\bk[2]^l\gamma^j\gamma^l + k_2^+\bk[1]^l\gamma^l\gamma^j + m(k_1^+-k_2^+)\gamma^j - 2ak_1^+k_2^+\right] ,
\label{eq:dirac_simplification}
\end{align}
where $a$ is a scalar and $k_1$ and $k_2$ are arbitrary momenta. Adapting Eq.~\eqref{eq:dirac_simplification} to the case that appears in Eq.~\eqref{eq:S_2_g_to_gq_2}, the $S$-matrix element for the second mechanism reads 
\begin{align}
\label{eq:S_2_g_to_gq_3}
\mathcal{S}_{g\to gq, \, 2}  & =
\frac{g^2}{p_2^+(2q^+)} \
\ (2\pi)\delta(q^+-p_1^+-p_2^+)
\int_{\bz}
\int_{-\frac{L^+}{2}}^{\frac{L^+}{2}}d z^+
\int_{\frac{L^+}{2}}^{\infty}d w^+  
\int\frac{d^2 \bk[0]}{(2\pi)^2}
\nn \\
&
\times 
\exp{-iw^+(\check{k}_0^--\check{p}_1^--\check{p}_2^-)}
\  \exp{-i\bz\cdot(\bk[0]-\bq)}
(2\pi)^2\delta^{(2)}(\bk[0]-\bp[1]-\bp[2]) 
\nn \\
&
\times 
\varepsilon_{\lambda}^i\varepsilon_{\lambda_1}^{j*}
\ \overline{u}(\Check{p}_2, h)
\Big[
p_2^+(\bp[1]^l+\bp[2]^l)\gamma^j\gamma^l
+
q^+\bp[2]^l\gamma^l\gamma^j + mp_1^+\gamma^j
+
 2\frac{q^+p_2^+}{p_1^+}\bp[1]^j
 \Big]
\gamma^i
\nn \\
&
\times 
t^{a_1} \ \UF(w^+,z^+;\bz) \ t^{b} \ \UA(z^+,-\infty;\bz)_{ba} \ \frac{\gamma^{+}\gamma^{-}}{2}\Psi(\underline{z}) .
\end{align}
By using Eq.~\eqref{eq:amplitude}, we can write the scattering amplitude for the second mechanism as 
\begin{align}
\label{eq:M_2_g_to_gq_1}
i\mathcal{M}_{g\to gq, \, 2}  & =
\frac{g^2}{p_2^+(2q^+)^2} \
\int_{\bz}
\int_{-\frac{L^+}{2}}^{\frac{L^+}{2}}d z^+
\int_{\frac{L^+}{2}}^{\infty}d w^+  
\int\frac{d^2 \bk[0]}{(2\pi)^2}
\nn \\
&
\times 
\exp{-iw^+(\check{k}_0^--\check{p}_1^--\check{p}_2^-)}
\  \exp{-i\bz\cdot(\bk[0]-\bq)}
(2\pi)^2\delta^{(2)}(\bk[0]-\bp[1]-\bp[2]) 
\nn \\
&
\times 
\varepsilon_{\lambda}^i\varepsilon_{\lambda_1}^{j*}
\ \overline{u}(\Check{p}_2, h)
\Big[
p_2^+(\bp[1]^l+\bp[2]^l)\gamma^j\gamma^l
+
q^+\bp[2]^l\gamma^l\gamma^j + mp_1^+\gamma^j
+
 2\frac{q^+p_2^+}{p_1^+}\bp[1]^j
 \Big]
\gamma^i
\nn \\
&
\times 
t^{a_1} \ \UF(w^+,z^+;\bz) \ t^{b} \ \UA(z^+,-\infty;\bz)_{ba} \ \frac{\gamma^{+}\gamma^{-}}{2}\Psi(\underline{z}) .
\end{align}

Finally, we can consider the third mechanism for quark-gluon pair production in gluon initiated channel described in Fig.~\ref{fig:g-gq3}. For this mechanism, the LSZ-type reduction formula reads 
\begin{align}
\label{eq:S_3_g_to_gq_1}
\mathcal{S}_{g\to gq, \, 3} & =
\lim_{y^+\to-\infty} 
\ \lim_{x_1^+,x_2^+\to\infty}
\ \int_{\by,\bx[1],\bx[2]}\int_{y^-,x^-_1,x^-_2}
\int_{\bw, \bz}
\int_{w^-, z^-}
\int_{-\frac{L^+}{2}}^{\frac{L^+}{2}}d z^+
\int_{-\infty}^{-\frac{L^+}{2}}d w^+  
\nn \\
& \hspace{-1cm}
\times 
\exp{ix_1\cdot\Check{p}_1} \ \exp{ix_2\cdot\Check{p}_2} \ \exp{-iy\cdot\Check{q}}
\ (-2q^+) \ \epsilon_\mu^{\lambda}(q)^* \ (-2p_1^+) \ \epsilon_\nu^{\lambda_1}(p_1)^* 
\Big[G_{0,\rmF}^{\mu'\mu}(w,y)\Big]_{ba}
\Big[G_\rmF^{\nu\nu'}(x_1,z)\IA\Big]_{a_1b_1}
\nn \\
& \hspace{-1cm}
\times 
\overline{u}(\Check{p}_2, h) \ \gamma^+ 
\Big[
S_\rmF(x_2,w)\BAq(-ig)\ t^{b}\ \gamma_{\mu'}
S_\rmF(w,z)\BIbq(-ig) \ t^{b_1}
\Big]_{\alpha_2\beta}
\gamma_{\nu'}\Psi_\beta^-(z) ,
\end{align}
where the explicit expressions for inside-to-after gluon propagator, before-to-after quark propagator and before-to-inside antiquark propagator are given in Eqs.~\eqref{gluon_prop_IA}, \eqref{quark_prop_BA} and \eqref{antiquark_prop_BI} respectively, at eikonal accuracy. Moreover, the vacuum gluon propagator is given in Eq.~\eqref{eq:vacuum_gluon_prop}. Substituting these expressions into Eq.~\eqref{eq:S_3_g_to_gq_1}, the $S$-matrix element for the third mechanism reads 
\begin{align}
\label{eq:S_3_g_to_gq_2}
\mathcal{S}_{g\to gq, \, 3} & =
\frac{g^2}{(2p_1^+)(2p_2^+)}
\int_{\bz,\bz[1]}
\int_{-\frac{L^+}{2}}^{\frac{L^+}{2}}d z^+
\int_{-\infty}^{-\frac{L^+}{2}}d w^+  
\int\frac{d^3 \underline{k}_1}{(2\pi)^2}
\int\frac{d^3 \underline{k}_2}{(2\pi)^2}
\ \delta(k_1^++p_1^+)
\ \delta(k_2^+-p_2^+) 
\nn \\
&
\times 
\exp{-iw^+(\check{q}^-+\check{k}_1^--\check{k}_2^-)}
\ \exp{-i\bz\cdot(\bp[1]+\bk[1])}
\ \exp{-i\bz[1]\cdot(\bp[2]-\bk[2])}
\ (2\pi)^3\delta^{(3)}(\underline{q}+\underline{k}_1-\underline{k}_2) 
\nn \\
&
\times 
\varepsilon_{\lambda}^i\varepsilon_{\lambda_1}^{j*} \ \overline{u}(\Check{p}_2,h)
\  \gamma^+ \ (\slashed{\check{k}}_2+m)
\Big[-\gamma^i+\frac{\bq^i}{q^+}\gamma^+\Big]
\ (\slashed{\check{k}}_1+m) \ \gamma^j
\nn \\
&
\times 
\UA(\infty,z^+;\bz)_{a_1b_1} \ \UF(\infty,w^+;\bz [1]) \ t^{a} \ \UFd(z^+,w^+;\bz) \ t^{b_1}\
\frac{\gamma^{+}\gamma^{-}}{2}\Psi(\underline{z}) .
\end{align}
One can notice that the Dirac structure that appears in Eq.~\eqref{eq:S_3_g_to_gq_2} has the same form as in Eq.~\eqref{eq:dirac_simplification}, thus upon simplification the $S$-matrix element for the third mechanism reads
\begin{align}
\label{eq:S_3_g_to_gq_3}
\mathcal{S}_{g\to gq, \, 3} &  = 
\frac{g^2}{(2p_1^+)(2p_2^+)} \ (2\pi)\delta\left(p_1^+ + p_2^+ - q^+\right)
\int_{-\frac{L^+}{2}}^{\frac{L^+}{2}}d z^+
\int_{-\infty}^{-\frac{L^+}{2}}d w^+  
\int_{\bz,\bz[1]}\int\frac{d^2\bk[1]}{(2\pi)^2}  
\nn \\
& \hspace{-1cm}
\times 
e^{iw^+\left[\frac{(\bk[1]+\bq)^2+m^2}{2p_2^+} + \frac{\bk[1]^2+m^2}{2p_1^+} - \frac{\bq^2}{2q^+}\right]}\ e^{-i\bz\cdot(\bp[1] + \bk[1])}
\ e^{-i\bz[1]\cdot(\bp[2]-\bk[1]-\bq)} 
\nn \\
& \hspace{-1cm}
\times 
\UA(\infty,z^+;\bz)_{a_1b_1}
\ \UF(\infty,w^+;\bz[1]) \ t^{a} \
\UF^\dagger(z^+,w^+;\bz) \ t^{b_1} 
\nn \\
& \hspace{-1cm}
\times 
\overline{u}(\Check{p}_2,h) \ \varepsilon_{\lambda}^i\varepsilon_{\lambda_1}^{j*}
\Big[
p_2^+\bk[1]^l\gamma^i\gamma^l
- p_1^+(\bk[1]^l+\bq^l)\gamma^l\gamma^i 
- mq^+\gamma^i
- 2\frac{p_1^+p_2^+}{q^+}\bq^i
\Big]
\gamma^j\gamma^+\gamma^-\Psi(\underline{z}) .
\end{align}
Finally, by using Eq.~\eqref{eq:amplitude}, the scattering amplitude for the third mechanism described in Fig.~\ref{fig:g-gq3} can be written as 
\begin{align}
\label{eq:M_3_g_to_gq_1}
&
i\mathcal{M}_{g\to gq, \, 3}   = 
\frac{g^2}{(2p_1^+)(2p_2^+)(2q^+)} 
\int_{-\frac{L^+}{2}}^{\frac{L^+}{2}}d z^+
\int_{-\infty}^{-\frac{L^+}{2}}d w^+  
\int_{\bz,\bz[1]}\int\frac{d^2\bk[1]}{(2\pi)^2}  
\nn \\
& \hspace{0.5cm}
\times 
e^{iw^+\left[\frac{(\bk[1]+\bq)^2+m^2}{2p_2^+} + \frac{\bk[1]^2+m^2}{2p_1^+} - \frac{\bq^2}{2q^+}\right]}\ e^{-i\bz\cdot(\bp[1] + \bk[1])}
\ e^{-i\bz[1]\cdot(\bp[2]-\bk[1]-\bq)} 
\nn \\
& \hspace{0.5cm}
\times 
\UA(\infty,z^+;\bz)_{a_1b_1}
\ \UF(\infty,w^+;\bz[1]) \ t^{a} \
\UF^\dagger(z^+,w^+;\bz) \ t^{b_1} 
\nn \\
& \hspace{0.5cm}
\times 
\overline{u}(\Check{p}_2,h) \ \varepsilon_{\lambda}^i\varepsilon_{\lambda_1}^{j*}
\Big[
p_2^+\bk[1]^l\gamma^i\gamma^l
- p_1^+(\bk[1]^l+\bq^l)\gamma^l\gamma^i 
- mq^+\gamma^i
- 2\frac{p_1^+p_2^+}{q^+}\bq^i
\Big]
\gamma^j\gamma^+\gamma^-\Psi(\underline{z}) .
\end{align}

As stated previously, the total scattering amplitude is given as the sum of scattering amplitudes of the three mechanisms, Eq.~\eqref{eq:M_tot_schm_g_to_gq}, for quark-gluon pair production in gluon initiated channel. In general kinematics, those three contributions to the total scattering amplitude are given by Eqs.~\eqref{eq:M_1_g_to_gq_1}, \eqref{eq:M_2_g_to_gq_1} and \eqref{eq:M_3_g_to_gq_1}.   

\subsection{$g\to gq$ amplitude in the back-to-back limit}
\label{subsec:b2b_g_to_gq_Amp}
We are now ready to study the back-to-back limit of the quark-gluon dijet production in gluon initiated channel.  As in the case of general kinematics, in the back-to-back limit the scattering amplitude receives contributions from the three mechanisms described in Fig.~\ref{fig:g-gq} and the total amplitude can be written as
\begin{align}
\label{eq:M_tot_schm_g_to_gq_b2b}
\M_{g\to gq, \, {\rm tot.}}^{\rm b2b}=\M_{g\to gq, \, 1}^{\rm b2b}+\M_{g\to gq, \, 2}^{\rm b2b}+\M_{g\to gq, \, 3}^{\rm b2b} .
\end{align} 
We start with the scattering amplitude computed in general kinematics for the first mechanism which is given in Eq.~\eqref{eq:M_1_g_to_gq_1}. When written in terms of the relative dijet momentum $\P$ and dijet momentum imbalance $\k$ defined in Eq.~\eqref{eq:PK} and their conjugate variables $\r$ and $\b$ defined in Eq.~\eqref{eq:br}, the amplitude for the first mechanism given in Eq.~\eqref{eq:M_1_g_to_gq_1} can be rewritten as 
%
%
\begin{align}
\label{eq:M_1_g_to_gq_2}
i \mathcal{M}_{g\to gq, \, 1}  & = 
\frac{-i\, g^2}{(1-z)(2q^+)^2}f^{ab_1b}
\int_{\bb,\br}
\int_{-\frac{L^+}{2}}^{\frac{L^+}{2}}d z^+
\int_{-\infty}^{-\frac{L^+}{2}}d w^+
\int\frac{d^2 \bk[1]}{(2\pi)^2} 
\nn \\
&
\times 
\exp{-\frac{iw^+}{2z(1-z)q^+}\big[z(1-z)\bq^2-(1-z)\bk[1]^2-z(\bk[1]-\bq)^2 
\big]}
\exp{-i\bb\cdot(\bk - \bq)} 
\ \exp{-i\br\cdot(z\bq + \bP - \bk[1])}
\nn \\
&
\times
\UA\big(z^+,w^+;\bb
-z\br
\big)_{b_2b}
\ \UA(\infty,w^+;\bb
+(1-z)\br
)_{a_1b_1} 
\nn \\
&
\times 
\overline{u}(\Check{p}_2, h)
\ \UF(\infty,z^+;\bb-z\br) \ t^{b_2} \ \gamma^l\gamma^+\gamma^-\Psi(z^+;\bb-z\br)  
\nn \\
&
\times 
\varepsilon_{\lambda}^i \ \varepsilon_{\lambda_1}^{j*}
\Big[
    g^{ij}\Big(\frac{z}{1-z}\bq^l 
    - \frac{1}{1-z}\bk[1]^l\Big)
    +g^{il}\Big(\bq^j - \frac{1}{z}\bk[1]^j\Big)
    -g^{jl}\big(z \bq^i - \bk[1]^i\big)\Big] ,
\end{align}
where $z$ and $(1-z)$ corresponds to longitudinal momentum fraction carried by the gluon and quark respectively and they are defined in Eq.~\eqref{eq:mom_frac}. 

We can now take the back-to-back limit of $i \mathcal{M}_{g\to gq, \, 1} $. As discussed in Sec.~\ref{sec:intro} this limit  corresponds to the kinematic region where the relative dijet momenta $\P$ is much larger than the dijet momentum imbalance $\k$, defined in Eq.~\eqref{eq:PK}. In coordinate space, this limit is given by $|\r|\ll |\b|$. Therefore, in the back-to-back limit one can perform a Taylor expansion of the Wilson lines and of the quark field around $\r=0$. 
Keeping only the zeroth order term in this expansion (since it is the first non-trivial contribution to the back-to-back limit) one observes that $\r$ dependence in the Wilson lines and quark field disappears and the only $\r$ dependence left is in the phase. This allows one to perform the $\r$ and $\k_1$ integrals trivially. All in all, the back-to-back limit of the first mechanism in the quark-gluon dijet production in gluon initiated channel can be written as 
\begin{align}
i\mathcal{M}_{g\to gq, \, 1}^{\rm b2b} & =
\frac{ig^2}{(1-z)(2q^+)^2}f^{ab_1b}
\int_{-\frac{L^+}{2}}^{\frac{L^+}{2}}d z^+
\int_{-\infty}^{-\frac{L^+}{2}}d w^+ 
e^{i\frac{w^+}{2z(1-z)q^+}
\bP^2
} 
\nn \\
&
\times 
\int_{\bb}e^{-i\bb\cdot(\bk-\bq)}
\ \UA(\infty,w^+;\bb)_{a_1b_1} \ \UA(z^+,w^+;\bb)_{b_2b}
\ \UF(\infty,z^+;\bb) \ t^{b_2} 
\nn \\
&
\times 
\overline{u}(\Check{p}_2, h)\ 
\varepsilon_{\lambda}^i \ \varepsilon_{\lambda_1}^{j*}
\Big[
\frac{1}{1-z}g^{ij}\bP^l + \frac{1}{z}g^{il}\bP^j - g^{jl}\bP^i
\Big]\gamma^l\gamma^+\gamma^-\Psi(z^+;\bb) .
\label{eq:M_1_g_to_gq_b2b}
\end{align}
Note that $w^+$ is before the medium and it is integrated up to the edge of the medium $-L^+/2$. Since the background fields vanish outside the medium one can make the replacement $w^+\mapsto-\infty$ in the Wilson lines. Then, the only dependence of $w^+$ remains in the phase and integration over it can be performed trivially.  Upon integration over $w^+$ one gets a phase with a factor of $L^+$ which can be approximated by one since we are performing the computation at NEik accuracy. 
After all said and done, one gets the back-to-back limit of the scattering amplitude for the first mechanism as    
\begin{align}
\label{eq:M_1_g_to_gq_b2b_m0_1}
i\mathcal{M}_{g\to gq, \, 1}^{{\rm b2b}
}  & =
 g^2   \frac{z}{(2q^+)} \frac{1}{\P^2}
 f^{ab_1b}
\int_{-\frac{L^+}{2}}^{\frac{L^+}{2}}d z^+
\!\! \int_{\bb}e^{-i\bp\cdot(\bk-\bq)}
\nn \\
&
\times
\UA(\infty,-\infty;\bb)_{a_1b_1} \ \UA(z^+,-\infty;\bb)_{b_2b}
\ \UF(\infty,z^+;\bb) \ t^{b_2} 
\nn \\
&
\times 
\overline{u}(\Check{p}_2, h)
\ \varepsilon_{\lambda}^i \ \varepsilon_{\lambda_1}^{j*}
\left[
g^{ij}\frac{\bP^l}{(1-z)} + g^{il}\frac{\bP^j}{z} - g^{jl}\bP^i
\right]
\gamma^l\gamma^+\gamma^-\Psi(z^+;\bb) .
\end{align}
%
%
Note that this contribution to the  amplitude is independent of the quark mass.
One should realize that the color structure that appears in Eq.~\eqref{eq:M_1_g_to_gq_b2b_m0_1} can be further simplified. By using the relations presented in Eq.~\eqref{eq:WL-relations} 
from App.~\ref{app:WL}, one can write the color structure as 
\begin{align}
& \UA(\infty,-\infty;\bb)_{a_1b_1} \UA(z^+,-\infty;\bb)_{b_2b}\left[\UF(\infty,z^+;\bb)t^{b_2}f^{ab_1b}\Psi(z^+;\bb)\right]
\nn \\
& =  
\UA(\infty,-\infty;\bb)_{a_1b_1}\UA(\infty,-\infty;\bb)_{cb}f^{ab_1b} 
\UA(\infty,z^+;\bb)_{cb_2}
    \big[\UF(\infty,z^+;\bb)t^{b_2}\Psi(z^+;\bb)\big] 
\nn \\
& = 
f^{a_1cc'} \ \UA(\infty,-\infty;\bb)_{c'a} \ \left[t^c\UF(\infty,z^+;\bb)\Psi(z^+;\bb)\right] 
\nn \\
& =
-i \ \UA(\infty,-\infty;\bb)_{ca}\Big[\left[t^{c},t^{a_1}\right]
    \UF(\infty,z^+;\bb)\Psi(z^+;\bb)\Big] .
\end{align}
Using this simplification in Eq.~\eqref{eq:M_1_g_to_gq_b2b_m0_1}, one obtains the final expression of the scattering amplitude for the first mechanism in the back-to-back 
limit as 
\begin{align}
\label{eq:M_1_g_to_gq_b2b_m0_2}
i \mathcal{M}_{g\to gq, \, 1}^{{\rm b2b}
}  & =
-  \frac{i\, g^2}{(2q^+)}\frac{1}{\P^2} \int_{-\frac{L}{2}}^{\frac{L}{2}}d z^+
\int_{\b}e^{-i\b\cdot(\bk-\bq)}
\ \UA(\infty,-\infty;\bb)_{ca}\left[t^{c},t^{a_1}\right]
\UF(\infty,z^+;\bb) 
\nn \\
&
\times 
\overline{u}(\Check{p}_2, h)\ \varepsilon_{\lambda}^i\ \varepsilon_{\lambda_1}^{j*}
\left[
    \frac{z}{1-z}g^{ij}\bP^l + g^{il}\bP^j - z g^{jl}\bP^i
    \right]\gamma^l\gamma^+\gamma^-\Psi(z^+;\bb) .
\end{align}

We can continue our discussion of the back-to-back limit for the quark-gluon dijet production in gluon initiated channel with the second mechanism (Fig.~\ref{fig:g-gq2}). As discussed previously, this second mechanism corresponds to the case where the incoming gluon converts into a quark via a $t$-channel quark exchange with the target and it splits into a quark-gluon dijet outside the medium. Since the quark-gluon dijet is produced after the medium, the back-to-back limit of the production amplitude is irrelevant in this mechanism. However, we still write the amplitude in terms of the of the relative dijet momentum $\P$ and dijet momentum imbalance $\k$ defined in Eq.~\eqref{eq:PK} and their conjugate variables $\r$ and $\b$ defined in Eq.~\eqref{eq:br}. Moreover, for the consistency of our final results, we refer to the amplitude as back-to-back amplitude after performing the change of variables. Then, starting from Eq.~\eqref{eq:M_2_g_to_gq_1}, we can write the scattering amplitude as 
\begin{align}
\label{eq:M_2_g_to_gq_b2b}
i\mathcal{M}_{g\to gq, \, 2}^{\rm b2b}  & = 
-\frac{g^2}{(1-z)(2q^+)^2}
\int_{-\frac{L^+}{2}}^{\frac{L^+}{2}}d z^+
\int_{\frac{L^+}{2}}^{\infty}d w^+ 
\exp{\frac{iw^+}{2z(1-z)q^+}\left[\bP^2+z^2m^2\right]}
\int_{\bb} \exp{-i\bb\cdot(\bk - \bq)} 
\nn \\
&
\times 
\overline{u}(\Check{p}_2, h) \ \varepsilon_{\lambda}^i \ \varepsilon_{\lambda_1}^{j*}
\Big[\gamma^l\gamma^j\bP^l - z m\gamma^j - 2\frac{1-z}{z}\bP^j\Big]
\gamma^i
\nn \\
& 
\times
 t^{a_1} \ \UF(w^+,z^+;\bb) \ t^{b} \ \UA(z^+,-\infty;\bb)_{ba} \ \frac{\gamma^{+}\gamma^{-}}{2}\Psi(z^+;\bb) .
\end{align}
Note that for this mechanism $w^+$ is after the medium and its integration region is from $L^+/2$ to $\infty$. Since the background fields vanish outside the medium, one can make the replacement $w^+\mapsto +\infty$ in the Wilson line in Eq.~\eqref{eq:M_2_g_to_gq_b2b}. Then the remaining $w^+$ dependence appears only in the phase and integration over $w^+$ can be performed trivially. Upon integration, one again gets a phase factor with $L^+$ which can be set to one in the accuracy we are performing the computations. Finally, performing the integration over $w^+$  one obtains 
\begin{align}
\label{eq:M_2_g_to_gq_b2b_m0_1_massive}
i\mathcal{M}_{g\to gq, \, 2}^{{\rm b2b}}  & =
-  \frac{ig^2}{(2q^+)}\frac{1}{\left[\P^2\!+\!z^2 m^2\right]} 
\int_{-\frac{L}{2}}^{\frac{L}{2}}d z^+
\int_{\bp}
\exp{-i\bp\cdot(\bk - \bq)} 
\ 
t^{a_1} \ \UF(+\infty,z^+;\bb) \ t^{b} \ \UA(z^+,-\infty;\bb)_{ba} \
\nn \\
&
\times \
\overline{u}(\Check{p}_2, h) \ \varepsilon_{\lambda}^i \ \varepsilon_{\lambda_1}^{j*}
\left[z\gamma^l\gamma^j \P^l - 2(1\!-\!z)\P^j- z^2 m\gamma^j\right]
\gamma^i  \
 \frac{\gamma^{+}\gamma^{-}}{2}\Psi(z^+;\bb) ,
\end{align}
or, in the massless quark limit,  
\begin{align}
\label{eq:M_2_g_to_gq_b2b_m0_1}
i\mathcal{M}_{g\to gq, \, 2}^{{\rm b2b}, \, m=0}  & =
-  \frac{ig^2}{(2q^+)}\frac{1}{\P^2} 
\int_{-\frac{L}{2}}^{\frac{L}{2}}d z^+
\int_{\bp}
\exp{-i\bp\cdot(\bk - \bq)} 
\ \overline{u}(\Check{p}_2, h) \ \varepsilon_{\lambda}^i \ \varepsilon_{\lambda_1}^{j*}
\left[z\gamma^l\gamma^j \P^l - 2(1\!-\!z)\P^j\right]
\nn \\
&
\times 
\gamma^i  \
t^{a_1} \ \UF(+\infty,z^+;\bb) \ t^{b} \ \UA(z^+,-\infty;\bb)_{ba} \ \frac{\gamma^{+}\gamma^{-}}{2}\Psi(z^+;\bb) .
\end{align}
The color structure can be further simplified as 
\begin{align}
& \hspace{-1.5cm}
\UA(z^+,-\infty;\bb)_{ab}
 \left[t^{a_1}\UF(\infty,z^+;\bb)t^{b}\Psi(z^+;\bb)\right] \nn \\
    &= \UA(\infty,-\infty;\bb)_{ca}t^{a_1}\UA(\infty,z^+;\bb)_{cb}
    \left[\UF(\infty,z^+;\bb)t^{b}\Psi(z^+;\bb)\right] \nn \\
    &= \UA(\infty,-\infty;\bb)_{ca}\left[t^{a_1}t^{c}
    \UF(\infty,z^+;\bb)\Psi(z^+;\bb)\right] ,
\end{align}
which leads to the following final expression for the scattering amplitude in the back-to-back and massless quark limits of second mechanism 
\begin{align}
\label{eq:M_2_g_to_gq_b2b_m0_2}
i\mathcal{M}_{g\to gq, \, 2}^{{\rm b2b}, \, m=0} & =
-  \frac{ig^2}{(2q^+)}\frac{1}{\P^2}
\int_{-\frac{L}{2}}^{\frac{L}{2}}d z^+
\int_{\b}
\exp{-i\b\cdot(\bk - \bq)} 
\ \overline{u}(\Check{p}_2, h) \ \varepsilon_{\lambda}^i \ \varepsilon_{\lambda_1}^{j*}
\left[z\gamma^l\gamma^j\bP^l - 2(1-z)\bP^j\right]
\nn \\
&
\times 
\gamma^i
\UA(\infty,-\infty;\bb)_{ca} \ t^{a_1} \ t^{c}\ \UF(\infty,z^+;\bb) \ \frac{\gamma^{+}\gamma^{-}}{2}\Psi(z^+;\bb) .
\end{align}

We can perform a similar analysis for the third mechanism in the quark-gluon dijet production in gluon initiated channel (Fig.~\ref{fig:g-gq3}). This mechanism corresponds to the case where the incoming gluon splits into quark-antiquark pair before the medium, then the pair scatters on the target. The antiquark jet interacts with the target via $t$-channel quark exchange converting into a gluon. Thus, in the final state one gets a quark-gluon dijet. The amplitude for this mechanism is given in Eq.~\eqref{eq:M_3_g_to_gq_1} in general kinematics. In order to consider back-to-back limit, we perform the change of variables given in Eqs.~\eqref{eq:PK} and \eqref{eq:br}, and then perform the small $|\r|$ expansion keeping only the zeroth order term in the expansion. The amplitude then reads 
\begin{align}
\label{eq:M_3_g_to_gq_b2b}
i\mathcal{M}_{g\to gq, \, 3}^{\rm b2b}  & =
-\frac{g^2}{z(1-z)(2q^+)^2}
\int_{-\frac{L^+}{2}}^{\frac{L^+}{2}}d z^+
\int_{-\infty}^{-\frac{L^+}{2}}d w^+ 
\int_{\bb}\exp{-i\bb\cdot(\bk -\bq)}
\ \exp{\frac{iw^+}{2z(1-z)q^+}\left[\bP^2+m^2\right]} 
\nn \\
&
\times
\UA(\infty,z^+;\bb)_{a_1b_1}
\ \UF(\infty,w^+;\bb) \ t^{a}
\ \UF^\dagger(z^+,w^+;\bb) \ t^{b_1} 
\nn \\
&
\times 
\overline{u}(\Check{p}_2, h)\ \varepsilon_{\lambda}^i\varepsilon_{\lambda_1}^{j*}
\left[-z \bP^l\gamma^l\gamma^i
 + (1-z)\bP^l\gamma^i\gamma^l 
 + m\gamma^i\right]
\gamma^j\frac{\gamma^{+}\gamma^{-}}{2}\Psi(z^+;\bb) .
\end{align}
Here, one can again make the replacement  $w^+\mapsto -\infty$ in the Wilson lines since the gauge fields vanish outside the medium. Then again the only $w^+$ depends appears in the phase which can be integrated trivially. The integration gives an $L^+$ dependent phase which can be approximated by one within the accuracy of our computations. Finally, one obtains the scattering amplitude as 
\begin{align}
\label{eq:M_3_g_to_gq_b2b_m0_1_massive}
i\mathcal{M}_{g\to gq, \, 3}^{{\rm b2b}}  & =
 \frac{ig^2}{(2q^+)} \frac{1}{\left[\P^2+m^2\right]}
\int_{-\frac{L^+}{2}}^{\frac{L^+}{2}}d z^+
\int_{\b}\exp{-i\bb\cdot(\bk -\bq)} 
\nn \\
&
\times \
\UA(\infty,z^+;\bb)_{a_1b_1}
\  \UF(\infty,-\infty;\bb) \ t^{a}
\UF^\dagger(z^+,-\infty;\bb) \ t^{b_1}
\nn \\
&
\times  
\overline{u}(\check{q}_2, h) \ \varepsilon_{\lambda}^i \ \varepsilon_{\lambda_1}^{j*}
\left[-z \bP^l\gamma^l\gamma^i 
    +(1-z)\bP^l\gamma^i\gamma^l
    + m\gamma^i\right]
    \gamma^j\frac{\gamma^{+}\gamma^{-}}{2}\Psi(z^+;\bb) ,
\end{align}
or, in the massless quark limit, as
\begin{align}
\label{eq:M_3_g_to_gq_b2b_m0_1}
i\mathcal{M}_{g\to gq, \, 3}^{{\rm b2b}, \, m=0}  & =
 \frac{ig^2}{(2q^+)} \frac{1}{\P^2}
\int_{-\frac{L^+}{2}}^{\frac{L^+}{2}}d z^+
\int_{\b}\exp{-i\bb\cdot(\bk -\bq)} 
\ \UA(\infty,z^+;\bb)_{a_1b_1}
\  \UF(\infty,-\infty;\bb) \ t^{a}
\nn \\
&
\times 
\UF^\dagger(z^+,-\infty;\bb) \ t^{b_1} 
\overline{u}(\check{q}_2, h) \ \varepsilon_{\lambda}^i \ \varepsilon_{\lambda_1}^{j*}
\left[-z \bP^l\gamma^l\gamma^i 
    +(1-z)\bP^l\gamma^i\gamma^l\right]
    \gamma^j\frac{\gamma^{+}\gamma^{-}}{2}\Psi(z^+;\bb) .
\end{align}
The color structure can be further simplified as 
\begin{align}
\label{eq:g_to_gq_Simp_Color_3}
& \hspace{-0.5cm}
\UA(\infty,z^+;\bb)_{a_1b_1}
    \left[\UF(\infty,-\infty;\bb)t^{a}
    \UF^\dagger(z^+,-\infty;\bb)t^{b_1}\Psi(z^+;\bb)\right] \nn \\
    &= \UA(\infty,z^+;\bb)_{a_1b_1}\UA(\infty,-\infty;\bb)_{ca}
    \left[t^{c}\UF(\infty,-\infty;\bb)\UF(-\infty,z^+;\bb)
    t^{b_1}\Psi(z^+;\bb)\right] \nn \\
    &= \UA(\infty,-\infty;\bb)_{ca}\left[t^{c}\UA(\infty,z^+;\bb)_{a_1b_1}
    \UF(\infty,z^+;\bb)t^{b_1}\Psi(z^+;\bb)\right] \nn \\
    &= \UA(\infty,-\infty;\bb)_{ca}
    \left[t^{c}t^{a_1}\UF(\infty,z^+;\bb)\Psi(z^+;\bb)\right] .
\end{align}
Substituting Eq.~\eqref{eq:g_to_gq_Simp_Color_3} into the scattering  amplitude given in Eq.~\eqref{eq:M_3_g_to_gq_b2b_m0_1}, we obtain the back-to-back limit of the scattering amplitude of quark-gluon dijet in gluon initiated channel via the third mechanism as 
\begin{align}
\label{eq:M_3_g_to_gq_b2b_m0_2}
i\mathcal{M}_{g\to gq, \, 3}^{{\rm b2b}, \, m=0}   & = 
 \frac{ig^2}{(2q^+)} \frac{1}{\P^2}
\int_{-\frac{L^+}{2}}^{\frac{L^+}{2}}d z^+
\int_{\b}\exp{-i\bb\cdot(\bk -\bq)} 
\ \UA(\infty,-\infty;\bb)_{ca} \ t^{c} \ t^{a_1} \ \UF(\infty,z^+;\bb) 
\nn \\
&
\times 
\overline{u}(\check{q}_2, h) \ \varepsilon^i_{\lambda} \ \varepsilon_{\lambda_1}^{j*}
 \ \left[-z {\bP^l}\gamma^l\gamma^i 
    + (1-z){\bP^l}\gamma^i\gamma^l\right]
    \gamma^j\frac{\gamma^{+}\gamma^{-}}{2}\Psi(z^+;\bb) .
\end{align}

As stated previously, the total scattering amplitude given in Eq.~\eqref{eq:M_tot_schm_g_to_gq_b2b} receives three contributions that are given in Eqs.~\eqref{eq:M_1_g_to_gq_b2b_m0_2}, \eqref{eq:M_2_g_to_gq_b2b_m0_2} and \eqref{eq:M_3_g_to_gq_b2b_m0_2}, and summing them all we can obtain
\begin{align}
\label{eq:M_tot_g_to_gq_b2b_m0_fin}
i\M_{g\to gq, \, {\rm tot.}}^{{\rm b2b}, \, m=0} & =
 \frac{ig^2}{2q^+} \frac{1}{\P^2}
\int_{-\frac{L^+}{2}}^{\frac{L^+}{2}}d z^+
\int_{\b}e^{-i\bb\cdot\left(\bk - \bq\right)}
\ \UA(\infty,-\infty;\bb)_{ba}
\nn \\
&
\times 
\overline{u}(\check{q}_2, h) \
\left[
t^{a_1}t^b\mathfrak{h}^{(1)}_{g\to gq} 
+t^bt^{a_1}\mathfrak{h}^{(2)}_{g\to gq}
\right] \UF(\infty,z^+;\bb)
\frac{\gamma^{+}\gamma^{-}}{2}\Psi(z^+;\bb) ,
\end{align}
with 
%
%
%
%
%
%
%
%
\begin{align}
\label{eq:HF_g_to_gq_1_raw}
\mathfrak{h}^{(1)}_{g\to gq} &= 
\varepsilon_{\lambda}^i \ \varepsilon_{\lambda_1}^{j*}
\left(
 - z{\bP^l} \ \gamma^l\gamma^j\gamma^i
 - 2z{\bP^j} \ \gamma^i 
 + 2z{\bP^i} \gamma^j
 + 2\frac{z}{1-z}{\bP^l} \ g^{ij}\gamma^l\right) ,
\\
\label{eq:HF_g_to_gq_2_raw}
\mathfrak{h}^{(2)}_{g\to gq} &=
\varepsilon_{\lambda}^i \ \varepsilon_{\lambda_1}^{j*}
\left(
- z{\bP^l} \ \gamma^l\gamma^i\gamma^j
+ (1-z){\bP^l} \ \gamma^i\gamma^l\gamma^j 
+ 2{\bP^j} \gamma^i
- 2z{\bP^i} \gamma^j
- 2\frac{z}{1-z}{\bP^l}g^{ij}\gamma^l\right) .
\end{align}
Anti-commuting the gamma matrices in the expressions \eqref{eq:HF_g_to_gq_1_raw} and \eqref{eq:HF_g_to_gq_2_raw} in order to place them in the order $\gamma^i\gamma^j\gamma^l$ for example, we find
\begin{align}
\label{eq:HF_g_to_gq_1}
\mathfrak{h}^{(1)}_{g\to gq} &= 
z \varepsilon_{\lambda}^i \ \varepsilon_{\lambda_1}^{j*}
\left(
\gamma^i\gamma^j
 + \frac{2z}{1\!-\!z} \ g^{ij}\right) {\bP^l}\gamma^l,
\\
\label{eq:HF_g_to_gq_2}
\mathfrak{h}^{(2)}_{g\to gq} &=
-\varepsilon_{\lambda}^i \ \varepsilon_{\lambda_1}^{j*}
\left(
\gamma^i\gamma^j
 + \frac{2z}{1\!-\!z} \ g^{ij}\right) {\bP^l}\gamma^l\, ,
\end{align}
so that in particular 
\begin{align}
\mathfrak{h}^{(1)}_{g\to gq} &= -z\, \mathfrak{h}^{(2)}_{g\to gq}
\, ,
\end{align}
and thus
\begin{align}
\label{eq:M_tot_g_to_gq_b2b_m0_fin_simple}
i\M_{g\to gq, \, {\rm tot.}}^{{\rm b2b}, \, m=0} & =
 \frac{ig^2}{2q^+} \frac{1}{\P^2}
\int_{-\frac{L^+}{2}}^{\frac{L^+}{2}}d z^+
\int_{\b}e^{-i\bb\cdot\left(\bk - \bq\right)}
\ \UA(\infty,-\infty;\bb)_{ba}
\nn \\
&
\times 
\overline{u}(\check{q}_2, h) \
\left[
-z t^{a_1}t^b 
+t^bt^{a_1}
\right] \UF(\infty,z^+;\bb)
\mathfrak{h}^{(2)}_{g\to gq}\, \frac{\gamma^{+}\gamma^{-}}{2}\Psi(z^+;\bb) .
\end{align}
%
%
%
%
%
\subsection{$g\to gq$ production cross section in the back-to-back limit}
\label{sec:ggqXS} 
\begin{figure}[H]
\centering
\begin{subfigure}{0.49\textwidth}
\centering
\includegraphics[height=5cm]{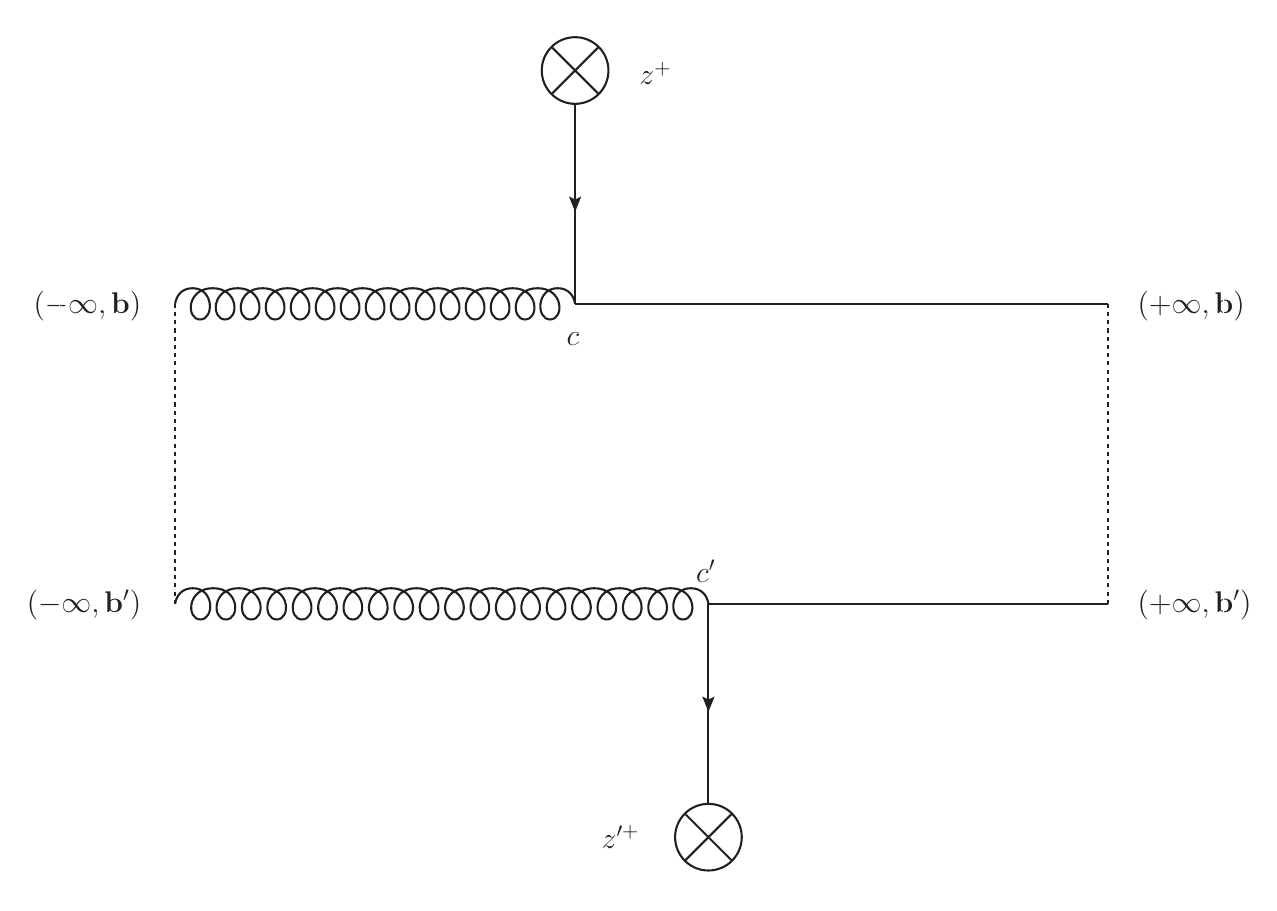}
\caption{$\mathcal{C}^{+g}$}
\label{fig:g-gq:C+g}
\end{subfigure}
\begin{subfigure}{0.49\textwidth}
\centering
\includegraphics[height=5cm]{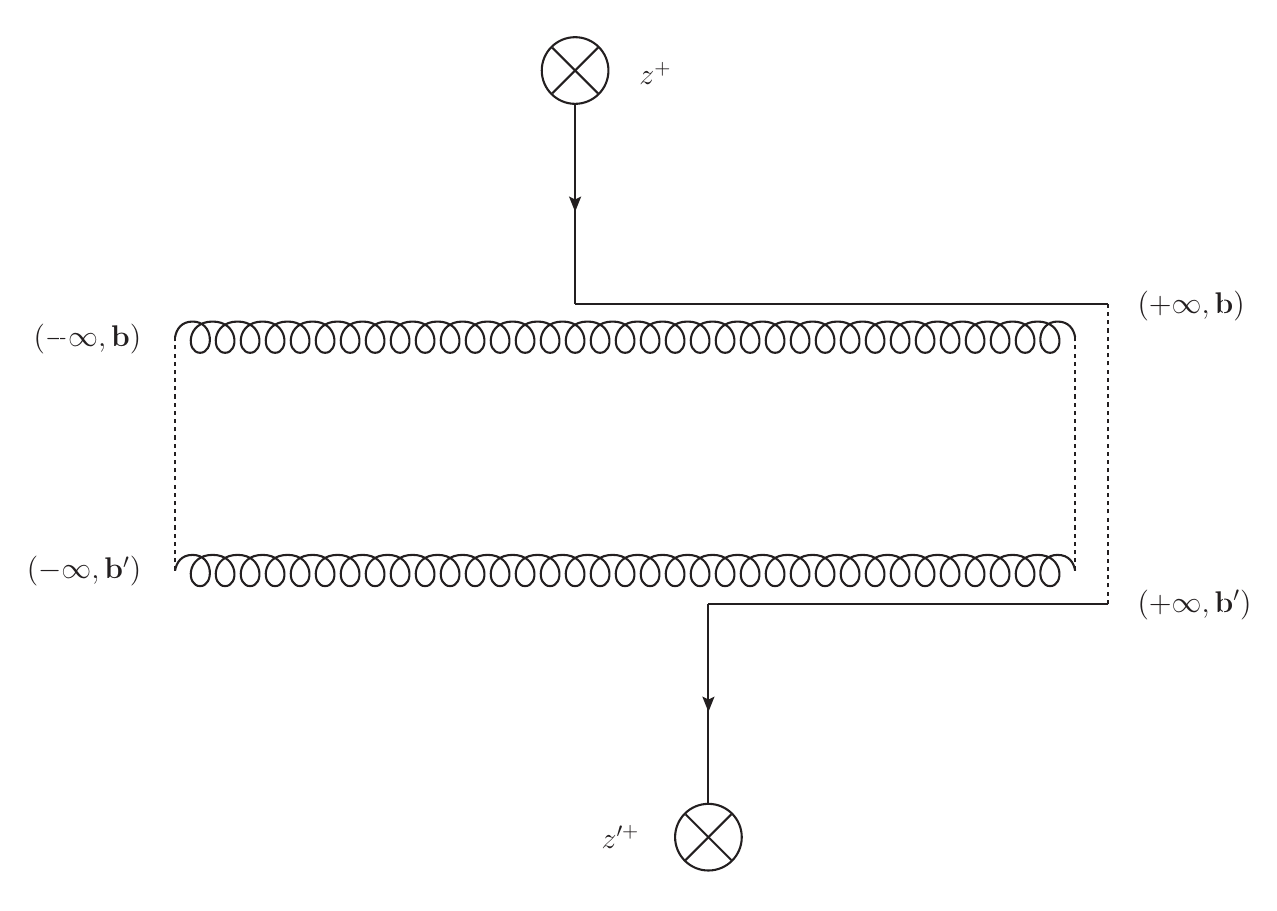}
\caption{$\mathcal{C}^{+\square_g}$}
\label{fig:g-gq:C+lg}
\end{subfigure}
\caption{Color structures appearing in the cross section of $g\to gq$.}
\label{fig:Cg-gq}
\end{figure}

The partonic cross section for the production of quark-gluon dijet in gluon initiated channel in the back-to-back and massless quark limits can be written as 
\begin{equation}
\frac{d\sigma_{g\to gq}^{{\rm b2b}, \, m=0}}{d{\rm P.S.}} =
(2q^+) \ 2\pi  \delta\left(p_1^+ + p_2^+ - q^+\right)
    \frac{1}{2(N_c^2-1)}\sum_{\lambda,\lambda_1}\sum_{h}\sum_{a,a_1}
  \left\langle  \left|i\mathcal{M}^{{\rm b2b}, \, m=0}_{g\to gq, \, {\rm tot.}}\right|^2 \right\rangle ,
\end{equation}
with the total amplitude in the back-to-back and massless limits given in Eq.~\eqref{eq:M_tot_g_to_gq_b2b_m0_fin} together with Eqs.~\eqref{eq:HF_g_to_gq_1} and \eqref{eq:HF_g_to_gq_2}. $\langle \cdots \rangle$ stands for the target averaging within the spirit of CGC formalism. The normalization factors $2$  and $(N_c^2-1)$ arise from the averaging of the helicity and colors respectively. When written in terms of the relative momentum and dijet momentum imbalance, the phase space is defined as 
\begin{align}
\label{def:PS}
d{\rm P.S.} = \frac{d^2\k}{(2\pi)^2}\frac{dk^+}{(2\pi)2k^+} \frac{d^2\P}{(2\pi)^2}\frac{dz}{(2\pi)2z(1-z)} ,
\end{align}
with $k^+=p_1^++p_2^+$ the lightcone momentum of the dijet system. Using the explicit expression of the production amplitude given in Eq.~\eqref{eq:M_tot_g_to_gq_b2b_m0_fin}, the partonic cross section for the $g\to gq$ channel in the back-to-back limit can be written as 
\begin{align}
& \frac{d\sigma_{g\to gq}^{{\rm b2b}, \, m=0}}{d{\rm P.S.}}
  =
 -\frac{g^4(1-z)}{4{\P^4}(N_c^2-1)} \ (2\pi)\ \delta\left(k^+ - q^+\right)
\int_{z^+,{z'}^+}
\int_{\b, \b'}
e^{-i(\bb-\bb')\cdot\left(\bk - \bq\right)} 
\nn \\
&
\times 
\Big\langle \overline{\Psi}({z'}^+;\bb')\ \gamma^- \ \UFd(\infty,{z'}^+;\bb') \ \UA(\infty,-\infty;\bb')_{b'a}
\
\left(\sum_{\lambda,\lambda_1}\left|\mathfrak{h}^{(2)}_{g\to gq}\right|^2\right)
\Bigg[
z^2\, t^{b'}t^{a_1}t^{a_1}t^b 
\nn \\
&\
+t^{a_1}t^{b'}t^bt^{a_1}
-z\, t^{b'}t^{a_1}t^bt^{a_1} 
-z\, t^{a_1}t^{b'}t^{a_1}t^b
\Bigg]\UA(\infty,-\infty;\bb)_{ba} \ \UF(\infty,z^+;\bb) \ \Psi(z^+;\bb)\Big\rangle
\, .
\label{eq:sigma_g_to_qq_i}
\end{align}
The coefficient $\mathfrak{h}^{(2)}_{q\to gq}$ is defined in Eq.\eqref{eq:HF_g_to_gq_2} and its square, summed over the gluon polarizations, has been performed in App.~\ref{app:dirac} and given in Eq.~\eqref{eq:h_2_g_gq} which reads 
\begin{align}
    \sum_{\lambda,\lambda_1}
    \left|\mathfrak{h}^{(2)}_{g\to gq}\right|^2 
    &= -{4\bP^2}\frac{1+z^2}{(1-z)^2} .
\end{align}
Moreover, the color structures appearing in Eq.~\eqref{eq:sigma_g_to_qq_i} can be further simplified. By using the properties of color generators given in App.~\ref{app:sun}, one can easily show that  
\begin{align}
\label{eq:color_iden_1}
t^{a_1}t^{b'}t^bt^{a_1} & = \frac{\delta^{b'b}}{4}\mathds{1}_F - \frac{1}{2N_c}t^{b'}t^b , \\
\label{eq:color_iden_2}
t^{a_1}t^{b'}t^{a_1}t^b & = t^{b'}t^{a_1}t^bt^{a_1} 
= -\frac{1}{2N_c}t^{b'}t^b .
\end{align}
By using Eqs.~\eqref{eq:color_iden_1} and \eqref{eq:color_iden_2}, the color structures in the first, the third and the fourth terms in in the bracket of Eq.~\eqref{eq:sigma_g_to_qq_i} can be written as 
\begin{align} 
& \hspace{-0.8cm}
\Big\langle \overline{\Psi}({z'}^+;\bb') \ \gamma^- \ \UFd(\infty,{z'}^+;\bb') \ \UA(\infty,-\infty;\bb')_{b'a}
\ t^{b'}t^b 
\nn \\
&\hspace{6cm}
\times
\UA(\infty,-\infty;\bb)_{ba} \ \UF(\infty,z^+;\bb) \ \Psi(z^+;\bb)\Big\rangle
\nn \\
&
= \Big\langle\overline{\Psi}({z'}^+;\bb') \ \gamma^- \ \UFd(\infty,{z'}^+;\bb')
\ \UA(\infty,{z'}^+;\bb')_{b'c'} \ \UA({z'}^+,-\infty;\bb')_{c'a} \
t^{b'}t^b 
\nn \\
&\hspace{3.4cm}
\times
\UA(\infty,z^+;\bb)_{bc} \ \UA(z^+,-\infty;\bb)_{ca}
\ \UF(\infty,z^+;\bb) \ \Psi(z^+;\bb)\Big\rangle
\nn \\
&
=
\Big\langle\overline{\Psi}({z'}^+;\bb') \ \gamma^-t^{c'} \ \UFd(\infty,{z'}^+;\bb') \ \UF(\infty,z^+;\bb) \ t^c \ \Psi(z^+;\bb)
\nn \\
& \hspace{6cm}
\times
\UA({z'}^+,-\infty;\bb')_{c'a} \ \UA(z^+,-\infty;\bb)_{ca} \Big\rangle .
\end{align}
On the other hand, the color structure in the second term in the bracket of Eq.~\eqref{eq:sigma_g_to_qq_i} can be simplified, as
\begin{align}
&
\Big\langle\overline{\Psi}({z'}^+;\bb') \ \gamma^- \ \UFd(\infty,{z'}^+;\bb') \ \UA(\infty,-\infty;\bb')_{b'a}
\ t^{a_1}t^{b'}t^bt^{a_1} 
\nn \\
&\hspace{6cm}
\times  \UA(\infty,-\infty;\bb)_{ba} \ \UF(\infty,z^+;\bb) \ \Psi(z^+;\bb)\Big\rangle
\nn \\
&
= \frac{1}{4}\Big\langle\overline{\Psi}({z'}^+;\bb') \ \gamma^- \ \UFd(\infty,{z'}^+;\bb') \ \UF(\infty,z^+;\bb) \ \Psi(z^+;\bb) 
\nn \\
& \hspace{6cm}
\times
\UA(\infty,-\infty;\bb')_{ba} \ \UA(\infty,-\infty;\bb)_{ba} \Big\rangle
\nn \\
&-\frac{1}{2N_c}
\Big\langle\overline{\Psi}({z'}^+;\bb') \ \gamma^- \ \UFd(\infty,{z'}^+;\bb') \ t^{b'}t^b
\ \UF(\infty,z^+;\bb) \ \Psi(z^+;\bb)
\nn \\
& \hspace{6cm}
\times
\UA(\infty,-\infty;\bb')_{b'a} \ \UA(\infty,-\infty;\bb)_{ba} \Big\rangle .
\end{align}
Thus, defining the following two color structures, that are represented in Fig.~\ref{fig:Cg-gq}, as  
\begin{align}
\label{eq:C+g}
\mathcal{C}^{+g} & \equiv 
\Big\langle\overline{\Psi}({z'}^+;\bb') \ \gamma^- \ t^{c'}
\ \UFd(\infty,{z'}^+;\bb') \ \UF(\infty,z^+;\bb) \ t^c \ \Psi({z}^+;\bb) 
\nn \\
& \hspace{6.2cm}
\times \UA({z'}^+,-\infty;\bb')_{c'a} \ \UA(z^+,-\infty;\bb)_{ca} \Big\rangle
\\
\label{eq:C+square_g} 
\mathcal{C}^{+\square_g} & \equiv 
\Big\langle \overline{\Psi}({z'}^+;\bb') \ \gamma^-
\ \UFd(\infty,{z'}^+;\bb') \ \UF(\infty,z^+;\bb) \ \Psi({z}^+;\bb)
\nn \\
& \hspace{6.2cm}
\times \UA(\infty,-\infty;\bb')_{ba} \ \UA(\infty,-\infty;\bb)_{ba} \Big\rangle ,
\end{align}
the production cross section given in Eq.~\eqref{eq:sigma_g_to_qq_i} can be written in the following factorized form 
\begin{align}
\frac{d\sigma_{g\to gq}^{{\rm b2b}, \, m=0}}{d{\rm P.S.}} & =
g^4 \ (2\pi)\delta\left(k^+ - q^+\right)
\int_{-\frac{L^+}{2}}^{\frac{L^+}{2}} dz^+
\int_{-\frac{L^+}{2}}^{\frac{L^+}{2}} d{z'}^+ 
 \nn \\
& \hspace{3.7cm}
\times 
\int_{\bb, \bb'}
    e^{-i(\bb-\bb')\cdot\left(\bk - \bq\right)}
\ \left[\mathcal{H}^{+g}_{g\to gq} \ \mathcal{C}^{+g}
    + \mathcal{H}^{+\square_g}_{g\to gq} \ \mathcal{C}^{+\square_g}\right] ,
\label{eq:sigma_g_to_qq}
\end{align}
where the hard factors are defined as 
\begin{align}
\label{eq:H_g_to_gq_+square_g}
\mathcal{H}^{+\square_g}_{g\to gq} & 
= \frac{1}{4\P^2}\frac{1}{(N_c^2-1)}\frac{(1+z^2)}{(1-z)} , \\
\label{eq:H_g_to_gq_+g}
\mathcal{H}^{+g}_{g\to gq} 
 &= \frac{1}{\P^2}\frac{\left[N_c^2z^2 - (1-z)^2\right]}{2N_c(N_c^2-1)} \frac{(1+z^2)}{(1-z)} .
\end{align}
As a final remark in the $g\to gq$ channel, we would like to mention that the color structures $\mathcal{C}^{+g}$ and $\mathcal{C}^{+\square_g}$ given in Eqs.~\eqref{eq:C+g} and \eqref{eq:C+square_g} respectively, can be rewritten with only fundamental Wilson lines, as discussed in \Appendix{app:fonlycolor}.


\section{$q\to q\Bar{q}$ channel} 
\label{sec:qqbq}                                 
\begin{figure}[H]
\centering
\begin{subfigure}{0.49\textwidth}
\centering
\includegraphics[width=\textwidth]{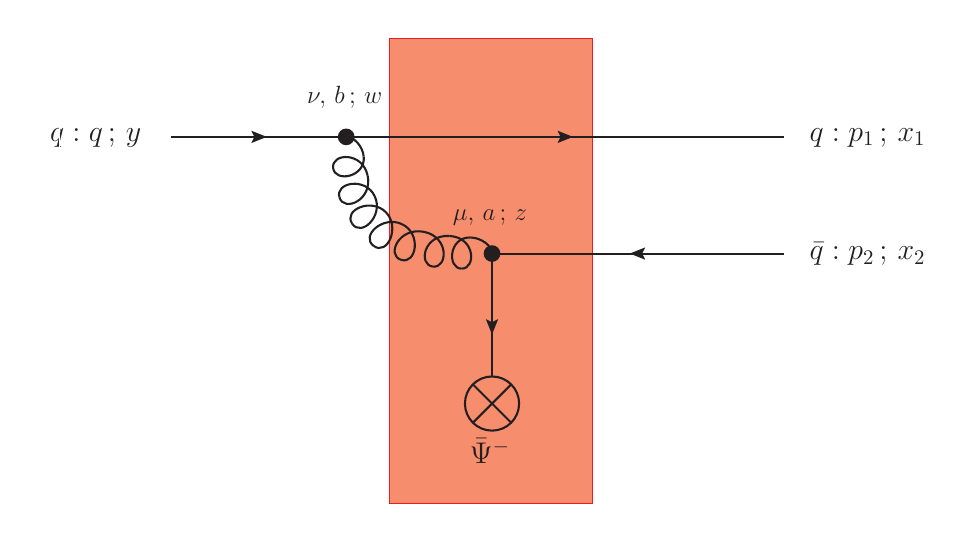}
\caption{Diagram 1}
\label{fig:q-qbq1}
\end{subfigure}
\begin{subfigure}{0.49\textwidth}
\centering
\includegraphics[width=\textwidth]{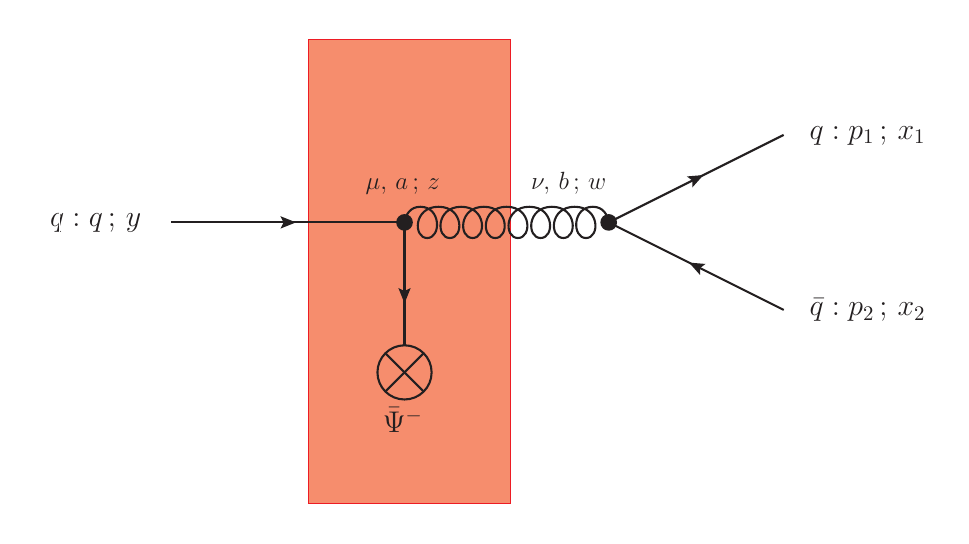}
\caption{Diagram 2}
\label{fig:q-qbq2}
\end{subfigure}
\caption{Diagrams contributing to channel $q \to q\bar{q}$.}
\label{fig:q-qbq}
\end{figure}

The next channel where we present the details of the calculation is the $q\to q\Bar{q}$ channel, first considering that all quarks have the same flavor. At NEik accuracy there are two mechanisms that contribute to the production of a quark-antiquark pair in quark initiated channel. The first mechanism corresponds to the case where the incoming quark splits into a quark-gluon pair before the medium. Then the pair scatters on the target. The quark scatters eikonally while the gluon scatters via a $t$-channel quark exchange and converts into an antiquark. Thus, in the final state a quark-antiquark dijet is produced (see Fig.~\ref{fig:q-qbq1}). In the second mechanism, incoming quark interacts with the target via a $t$-channel quark exchange and converts into a gluon. The gluon then splits into a quark-antiquark pair in the final state (see Fig.~\ref{fig:q-qbq2}).
%
%
%
In the rest of this section, we derive the scattering amplitudes for each mechanism both in general kinematics and in the back-to-back limit. We also compute the quark-antiquark dijet production cross section for quark initiated channel at NEik accuracy in the back-to-back and massless quark limits. 
Finally, we will consider the two other cases where all quarks do not share the same flavor. If the final quarks-antiquark pair has the same flavor, but different than the initial quark one (more explicitly, $q_f\to q_{f'}\bar{q}_{f'}$, with $f\neq f'$), only the second diagram contributes (Fig.~\ref{fig:q-qbq2}).
When the final antiquark has different flavor than the quarks, initial and final (more explicitly, $q_f\to q_{f}q_{f'}$), only the first diagram contributes (Fig.~\ref{fig:q-qbq1}). 

\subsection{$q\to q\Bar{q}$ amplitude in general kinematics }
\label{sub:q-qbq1}

As discussed above, the scattering  amplitude in the $q\to q \bar q$ channel receives two contributions from the aforementioned two mechanisms and it is written as the sum of these two contributions
\begin{align}
\label{eq:M_tot_schm_q_to_qbarq}
\M_{q\to q\bar q, \, {\rm tot.}}= \M_{q\to q\bar q, \, 1} + \M_{q\to q\bar q, \, 2} .
\end{align}
Let us start with the first mechanism which is described in Fig.~\ref{fig:q-qbq1}. In that case, the $S$-matrix element can be obtained from LSZ-type reduction formula which reads 
\begin{align}
\S_{q\to q\bar q, \, 1} &= 
\lim_{y^+\to-\infty}\lim_{x_1^+,x_2^+\to\infty}
\int_{\by,\bx[1],\bx[2]}\int_{y^-,x^-_1,x^-_2}
\int_{\bw, \bz}\int_{w^-, z^-}
\int_{-\frac{L^+}{2}}^{\frac{L^+}{2}}d z^+
\int_{-\infty}^{-\frac{L^+}{2}}d w^+
\nn \\
&
\times 
e^{ix_{1} \cdot \check{p}_{1}} \ e^{ix_{2} \cdot \check{p}_{2}} 
\ e^{-iy \cdot \check{q}} \     
\overline{\Psi}^{-}_{\beta_2}(z)
\Big[ (-igt^{a}\gamma_{\mu}) \ S_{F}(z, x_{2}) \IAbq   \Big]_{\beta_{2}\alpha_{2}} 
  \gamma^{+} v(\check{p}_{2},h_{2}) 
 \nn \\
 &
 \times  
\Big[ G^{\mu\nu}_{F}(z,w) \BI\Big]_{ab}      
\overline{u}(\check{p}_{1},h_{1}) \gamma^{+}     \bigg[  S_{F}(x_{1},w) \BAq (-igt^{b}\gamma_{\nu}) \  S_{0,F}(w,y)\bigg]_{\alpha_1 \alpha} 
\gamma^{+} u(\check{q},h) ,    
\end{align}
where the inside-to-after antiquark propagator is given in Eq.~\eqref{antiquark_prop_IA}, the before-to-inside gluon propagator is given in Eq.~\eqref{gluon_prop_BI}, the before-to-after quark propagator is given in Eq.~\eqref{quark_prop_BA} and the vacuum quark propagator is given in Eq.~\eqref{eq:quark_vacuum_prop}. 
By using these explicit expressions, contracting the $\mu$ and $\nu$ indices, and performing some trivial integrations over $\underline{x_{1}}, \underline{x_{2}}$, and $\underline{y}$, we can take limits over $x_{1}^{-}, x_{2}^{-}$, and $y^{-}$. As a result, we obtain delta functions in + and transverse components of momenta, so performing trivial integrations over delta functions, the $S$-matrix can be written as 
\begin{align}
\S_{q\to q\bar q, \, 1} &= 
-g^2 \int \! \frac{d^{3} \underline{k_{1}}}{(2\pi)^{2}} \int \! \frac{d^{3}\underline{k_{2}}}{(2\pi)^{2}}
\frac{1}{(2k_2^+)(2q^+)} 
\ 2\pi \ \delta(p_{2}^{+} -k_{2}^{+}) \ 2\pi \ \delta(p_{1}^{+} - k_{1}^{+}) 
\nn \\
&
\times 
\int_{\bz} e^{-i\bz \cdot (\p_{2}-\bk_{2})} \int_{\bz_{1}} e^{-i\bz_{1} \cdot(\bp_{1} - \bk_{1})} 
\int_{-\frac{L^+}{2}}^{\frac{L^+}{2}} dz^+
\int_{-\infty}^{-\frac{L^+}{2}} dw^+
e^{iw^{+}\left(\frac{\bk_{1}^{2}+m^{2}}{2p_{1}^{+}} + \frac{\bk_{2}^{2}}{2p_{2}^{+}} - \frac{\bq^{2} + m^{2}}{2q^{+}}\right)}
\nn \\
& 
\times 
(2\pi) \ \delta(k_1^++k_2^+-q^+) \ (2\pi)^{2} \delta^{2}(\bk_{1} + \bk_{2} - \bq)
\ \overline{\Psi}_{\beta_2}(\underline{z}) \frac{\gamma^{-}\gamma^{+}}{2}\gamma^{i}  
\nn \\
& 
\times 
v(\check{p}_{2}, h_{2}) \  \overline{u}(\check{p}_{1}, h_{1}) \ \gamma^{+} \ \frac{(\slashed{\check{k}}_{1} + m)}{2k_{1}^{+}}
\bigg(\gamma^{i} -  \frac{\bk_{2}^{i}\gamma^{+}}{2k_{2}^{+}} \bigg)
(\slashed{\check{q}} + m) \gamma^{+} u(\check{q}, h)  
\nn \\
& 
\times
\Big[ t^{a} \UFd (\infty, z^{+}; \bz)\Big]_{\beta_{2}\alpha_{2}} 
\Big[\UF(\infty, w^{+}; \bz_{1}) t^{b}\Big]_{\alpha_{1} \alpha} 
\UA(z^{+}, w^{+}; \bz)_{ab} .
\end{align}
%
%
%
%
As in the case of $g\to gq$ channel, in the above expression, terms with covariant derivatives vanish as they are multiplied by the projection of quark background field components and as $\gamma^{+}\gamma^{i}\gamma^{+} = -\gamma^{i}\gamma^{+}\gamma^{+} = 0$. Using delta functions to perform the  integrations over $k^{+}_{1}$ and $k_{2}^{+}$, one gets 
\begin{align}
S_{q\to q\bar q, \, 1} &= 
-\frac{g^{2}}{2p_{2}^{+}} \ (2\pi) \ \delta(p_{1}^{+} + p_{2}^{+} - q^{+})
\int \! \frac{d^{2} \bk_{1}}{(2\pi)^{2}} \int \! \frac{d^{2}\bk_{2}}{(2\pi)^{2}} \int_{\bz} e^{-i\bz \cdot (\bP_{2}-\bk_{2})} \int_{\bz_{1}} e^{-i\bz_{1} \cdot(\bp_{1} - \bk_{1})} 
\nn  \\
&
\times 
\int_{-\frac{L^+}{2}}^{\frac{L^+}{2}} dz^+
\int_{-\infty}^{-\frac{L^+}{2}} dw^+
e^{iw^{+}\left(\frac{\bk_{1}^{2}+m^{2}}{2p_{1}^{+}} + \frac{\bk_{2}^{2}}{2p_{2}^{+}} - \frac{\bq^{2} + m^{2}}{2q^{+}}\right)} (2\pi)^{2} \delta^{2}(\bk_{1} + \bk_{2} - \bq)
\ \overline{\Psi}_{\beta_2}(\underline{z}) \frac{\gamma^{-}\gamma^{+}}{2}\gamma^{i}
\nn \\
&
\times 
v(\check{p}_{2}, h_{2})  \ \overline{u}(\check{p}_{1}, h_{1}) \gamma^{+} 
\bigg[\frac{\gamma^{i}\gamma^{l}\bq^{l}}{2q^{+}} + \gamma^{i}\Big(\frac{1}{2q^{+}} - \frac{1}{2p_{1}^{+}}\Big) m + \frac{\gamma^{l} \gamma^{i} \bk_{1}^{l}}{2p_{1}^{+}} + \frac{\bk_{2}^{i}}{p_{2}^{+}}\bigg] 
u(\check{q}, h) 
\nn \\
&
\times 
\Big[ t^{a} \UFd (\infty, z^{+}; \bz)\Big]_{\beta_{2}\alpha_{2}} 
\Big[\UF(\infty, w^{+}; \bz_{1}) t^{b}\Big]_{\alpha_{1} \alpha}  \UA(z^{+}, w^{+}; \bz)_{ab},
\end{align}
where we have used the relation from Eq.~\eqref{eq:dirac_simplification} to simplify the Dirac matrices. Now, the integration over $\bk_{1}$ can be taken trivially and one obtains 
\begin{align}
\S_{q\to q\bar q, \, 1} &= 
-\frac{g^{2}}{2p_{2}^{+}}  \ (2\pi)\  \delta(p_{1}^{+}+p_{2}^{+} - q^{+})
\int\! \frac{d^{2} \bk_{1}}{(2\pi)^{2}} \int_{\bz} \ e^{-i \bz \cdot (\bp_{2}+ \bk_{1}- \bq)} \int_{\bz_{1}}  \ e^{-i\bz_{1} \cdot (\p_{1}-\bk_{1})} 
\nn \\
&
\times
\int_{-\frac{L^+}{2}}^{\frac{L^+}{2}} dz^+
\int_{-\infty}^{-\frac{L^+}{2}} dw^+
e^{iw^{+}\Big(\frac{\bk_{1}^{2}}{2p_{1}^{+}} + \frac{(\bq- \bk_{1})^{2}}{2p_{2}^{+}} -\frac{\bq^{2}}{2q^{+}}+ \frac{m^{2}}{2p^{+}_{1}} -\frac{m^{2}}{2q^{+}}\Big)} \ \overline{\Psi}_{\beta_{2}}(\underline{z}) \frac{\gamma^{-}\gamma^{+}}{2}\gamma^{i}  
\nn \\
&
\times
 v(\check{p}_{2},h_{2}) \ \overline{u}(\check{p}_{1},h_{1}) \gamma^{+}
\bigg[ \frac{\gamma^{i}\gamma^{l} \bq^{l}}{2q^{+}} + \frac{\gamma^{l}\gamma^{i}\bk_{1}^{l}}{2p_{1}^{+}} + \gamma^{i} \Big(\frac{1}{2q^{+}}-\frac{1}{2p_{1}^{+}}\Big)m +\frac{\bq^{i}-\bk_{1}^{i}}{p_{2}^{+}}
\bigg] u(\check{q},h)
\nn \\
&
\times
\Big[ t^{a}\  \UFd (\infty,z^{+};\bz)\ \Big]_{\beta_{2}\alpha_{2}}
\UA(z^{+},w^{+};\bz)_{ab} \ \Big[ \UF(\infty,w^{+};\bz_{1}) \ t^{b}\Big]_{\alpha_{1} \alpha} .
\end{align}
Finally, by using the relation between $S$-matrix and production amplitude, Eq.~\eqref{eq:amplitude}, the scattering amplitude for the first mechanism can be written as 
\begin{align}
\label{eq:M_1_q_to_qbarq}
i\M_{q\to q\bar q, \, 1} 
&= 
-\frac{g^{2}}{(2p_{2}^{+})(2q^+)}  
\int\! \frac{d^{2} \bk_{1}}{(2\pi)^{2}} \int_{\bz} \ e^{-i \bz \cdot (\bp_{2}+ \bk_{1}- \bq)} \int_{\bz_{1}}  \ e^{-i\bz_{1} \cdot (\p_{1}-\bk_{1})} 
\nn \\
&
\times
\int_{-\frac{L^+}{2}}^{\frac{L^+}{2}} dz^+
\int_{-\infty}^{-\frac{L^+}{2}} dw^+
e^{iw^{+}\Big(\frac{\bk_{1}^{2}}{2p_{1}^{+}} + \frac{(\bq- \bk_{1})^{2}}{2p_{2}^{+}} -\frac{\bq^{2}}{2q^{+}}+ \frac{m^{2}}{2p^{+}_{1}} -\frac{m^{2}}{2q^{+}}\Big)} \ \overline{\Psi}_{\beta_{2}}(\underline{z}) \frac{\gamma^{-}\gamma^{+}}{2}\gamma^{i}  
\nn \\
&
\times
v(\check{p}_{2},h_{2}) \ \overline{u}(\check{p}_{1},h_{1}) \gamma^{+}
\bigg[ \frac{\gamma^{i}\gamma^{l} \bq^{l}}{2q^{+}} + \frac{\gamma^{l}\gamma^{i}\bk_{1}^{l}}{2p_{1}^{+}} + \gamma^{i} \Big(\frac{1}{2q^{+}}-\frac{1}{2p_{1}^{+}}\Big)m +\frac{\bq^{i}-\bk_{1}^{i}}{p_{2}^{+}}
\bigg] u(\check{q},h)
\nn \\
&
\times
\Big[ t^{a}\  \UFd (\infty,z^{+};\bz)\ \Big]_{\beta_{2}\alpha_{2}}
\UA(z^{+},w^{+};\bz)_{ab} \ \Big[ \UF(\infty,w^{+};\bz_{1}) \ t^{b}\Big]_{\alpha_{1} \alpha} .
\end{align}
The $S$-matrix element for the second mechanism described in Fig.~\ref{fig:q-qbq2} can be obtained via the following LSZ-type reduction formula:\footnote{The overall minus sign arises when anticommuting quark fields in order to group them pairwise into propagators, following the Wick theorem.}
\begin{align}
S_{q\to q\bar q, \, 2}  &= 
-\lim_{y^+\to-\infty} \ \lim_{x_1^+,x_2^+\to\infty}
\ \int_{\by,\bx[1],\bx[2]} \ \int_{y^-,x^-_1,x^-_2}
\ \int_{\bw, \bz}\int_{w^-, z^-}
\int_{-\frac{L^+}{2}}^{\frac{L^+}{2}}d z^+
\int_{\frac{L^+}{2}}^{\infty}d w^+
\nn \\
&
\times
e^{ix_{1} \cdot \check{p}_{1}} \ e^{ix_{2} \cdot \check{p}_{2}} \ e^{-iy \cdot \check{q}} \
\overline{\Psi}^{-}_{\beta}(z)
\Big[ (-igt^{a}\gamma_{\mu}) \ S_{F}(z,y) \BIq \Big]_{\beta\alpha} 
\gamma^{+} u(\check{q},h) 
\Big[ G^{\nu\mu}_{F}(w,z) \IA \Big]_{ba}  
\nn \\
&
\times
\overline{u}(\check{p}_{1},h_{1}) \ \gamma^{+}  
\Big[ S_{0,F}(x_{1},w) \ (-igt^{b}\gamma_{\nu}) \  S_{0,F}(w,x_{2})\Big]_{\alpha_{1}\alpha_{2}} 
\gamma^{+} v(\check{p}_{2},h_{2}) ,     
\end{align}
where the before-to-inside quark propagator is given in Eq.~\eqref{quark_prop_BI}, the inside-to-after gluon propagator is given in Eq.~\eqref{gluon_prop_IA} and the vacuum quark propagator is given in Eq.~\eqref{eq:quark_vacuum_prop}. Using these expressions for the propagators and performing the trivial integrations as in the case of the first mechanism, the $S$-matrix element can be written as 
\begin{align}
\S_{q\to q\bar q, \, 2} &= 
\frac{g^{2}}{2q^{+}}  \ (2\pi) \ \delta(p_{1}^{+}+p_{2}^{+}-q^{+})
\int_{-\frac{L^+}{2}}^{\frac{L^+}{2}} dz^+
\int_{\frac{L^+}{2}}^{\infty} dw^+
e^{-iw^{+}\Big(\frac{(\bp_{1}+ \bp_{2})^{2}}{2q^{+}}-\frac{\bp_{1}^{2}+m^{2}}{2p_{1}^{+}}-\frac{\bp_{2}^{2}+m^{2}}{2p_{2}^{+}}\big)\Big)} 
\nn \\
&
\times
\int_{\bz}\ e^{-i\bz \cdot (\bp_{1}+\bp_{2}-\bq)}
\ \overline{\Psi}_{\beta}(\underline{z})   
\frac{\gamma^{-}\gamma^{+}}{2}\gamma^{j} \ 
u(\check{q},h) \ \overline{u}(\check{p}_{1},h_{1}) \
\gamma^{+} 
\nn \\
&
\times
\bigg[ 
\frac{\gamma^{j}\gamma^{l}\bp_{2}^{l}}{2p_{2}^{+}} + \frac{\gamma^{l}\gamma^{j}\bp_{1}^{l}}{2p_{1}^{+}} 
- \gamma^{j} \Big(\frac{1}{2p_{1}^{+}}+\frac{1}{2p_{2}^{+}}\Big)m 
+\frac{\bp_{1}^{j}+\bp_{2}^{j}}{q^{+}}
\bigg]
v(\check{p}_{2},h_{2})
\nn \\
&\times
\Big[ t^{a} \UF(z^{+},-\infty;\bz) \ \Big]_{\beta\alpha}
\big(t^{b}\big)_{\alpha_{1}\alpha_{2}} 
\UA(w^{+},z^{+};\bz)_{ba} ,
\end{align}
where we have used Eq.~\eqref{eq:dirac_simplification} to simplify the Dirac structure. Finally, by using Eq.~\eqref{eq:amplitude} we obtain the scattering amplitude for the second mechanism in general kinematics as 
\begin{align}
\label{eq:M_2_q_to_qbarq}
i\M_{q\to q\bar q, \, 2} &= 
\frac{g^{2}}{(2q^{+})^2} 
\int_{-\frac{L^+}{2}}^{\frac{L^+}{2}} dz^+
\int_{\frac{L^+}{2}}^{\infty} dw^+
e^{-iw^{+}\Big(\frac{(\bp_{1}+ \bp_{2})^{2}}{2q^{+}}-\frac{\bp_{1}^{2}+m^{2}}{2p_{1}^{+}}-\frac{\bp_{2}^{2}+m^{2}}{2p_{2}^{+}}\big)\Big)} 
\nn \\
&
\times
\int_{\bz}\ e^{-i\bz \cdot (\bp_{1}+\bp_{2}-\bq)}
\ \overline{\Psi}_{\beta}(\underline{z})   
\frac{\gamma^{-}\gamma^{+}}{2}\gamma^{j} \ 
u(\check{q},h) \ \overline{u}(\check{p}_{1},h_{1}) \
\gamma^{+} 
\nn \\
&
\times
\bigg[ 
\frac{\gamma^{j}\gamma^{l}\bp_{2}^{l}}{2p_{2}^{+}} + \frac{\gamma^{l}\gamma^{j}\bp_{1}^{l}}{2p_{1}^{+}} 
- \gamma^{j} \Big(\frac{1}{2p_{1}^{+}}+\frac{1}{2p_{2}^{+}}\Big)m 
+\frac{\bp_{1}^{j}+\bp_{2}^{j}}{q^{+}}
\bigg]
v(\check{p}_{2},h_{2})
\nn \\
&\times
\Big[ t^{a} \UF(z^{+},-\infty;\bz) \ \Big]_{\beta\alpha}
\big(t^{b}\big)_{\alpha_{1}\alpha_{2}} 
\UA(w^{+},z^{+};\bz)_{ba}.
\end{align}

As stated in Eq.~\eqref{eq:M_tot_schm_q_to_qbarq}, total scattering amplitude in the $q\to q\bar q$ channel in the general kinematics is given by the sum of two production amplitudes, given in Eqs.~\eqref{eq:M_1_q_to_qbarq} and \eqref{eq:M_2_q_to_qbarq}, that correspond to the first and second mechanisms that are illustrated in Fig.~\ref{fig:q-qbq}.

\subsection{$q\to q\Bar{q}$ amplitude in the back-to-back limit}

The back-to-back production limit of the $q \to q\bar q$ dijet can be considered in the same way as discussed in Sec.~\ref{sec:intro} and applied in Sec.~\ref{subsec:b2b_g_to_gq_Amp}. Since there are two mechanisms that contribute to the production of quark-antiquark dijet in quark initiated channel, in the back-to-back limit, the total scattering amplitude can be written as 
\begin{align}
\label{eq:M_tot_schm_q_to_qbarq_b2b}
i\M_{q\to q\bar q, \, {\rm tot.}}^{{\rm b2b}}=i\M_{q\to q\bar q,\,  1}^{{\rm b2b}}+i\M_{q\to q\bar q, \, 2}^{{\rm b2b}} .
\end{align}
Let's first consider the back-to-back limit of the scattering amplitude in the first mechanism. For that purpose, we perform the change of variables given in Eqs.~\eqref{eq:PK} and \eqref{eq:br} and rewrite the production amplitude of the first mechanism in the general kinematics given in Eq.~\eqref{eq:M_1_q_to_qbarq} in terms of the relative dijet momentum $\P$ and dijet momentum imbalance $\k$ and their conjugate variables $\r$ and $\b$. 
As a reminder, the back-to-back  limit, corresponding to $|\P|\gg|\k|$ or equivalently to $|\r|\ll |\b|$, allows a Taylor expansion around $\r=0$, pushing the $\r$ dependence, at zeroth order, to the phase only. For the first mechanism, this leads to 
\begin{align}
\label{eq:M_1_q_to_qbarq_b2b}
i \M_{q\to q\bar q, \, 1}^{\rm b2b}  &= 
-\frac{g^{2}}{(2p_{2}^{+})(2q^{+})^2}  
\int_{-\frac{L^+}{2}}^{\frac{L^+}{2}} dz^+
\int_{-\infty}^{-\frac{L^+}{2}} dw^+
\int_{\bb} \ e^{-i\bb \cdot (\bk- \bq)} 
\ e^{i\frac{w^{+}}{2q^{+}}\big(\frac{\P^{2}}{z(1-z)} + \frac{(1-z)m^{2}}{z}\big)}
\nn \\
&
\times
\overline{\Psi}_{\beta_{2}}(z^{+};\bb) 
\frac{\gamma^{-}\gamma^{+}}{2}\gamma^{i}
\ v(\check{p}_{2},h_{2})
\ \overline{u}(\check{p}_{1},h_{1}) \ \gamma^{+}  
\nn  \\
&
\times
\Big[ 
\{\gamma^{i},\gamma^{l}\} \bq^{l} + \frac{\gamma^{l}\gamma^{i} \bP^{l}}{z} -\gamma^{i} \Big(\frac{1-z}{z}\Big)m -\frac{2 \bP^{i}}{1-z} + 2 \bq^{i}
\Big]
\nn \\
&
\times
u(\check{q},h) 
\big[ t^{a}\ \UFd (\infty,z^{+};\bb) \ \big]_{\beta_{2}\alpha_{2}}
\ \big[ \UF(\infty,w^{+};\bb)\ t^{b}\big]_{\alpha_{1} \alpha} 
\ \UA(z^{+},w^{+};\bb)_{ab} .
\end{align}
The longitudinal coordinate $w^+$ is before the medium hence strictly outside, one can then simply take $w^+\to - \infty$ in the Wilson lines in Eq.~\eqref{eq:M_1_q_to_qbarq_b2b} since the gauge fields vanish outside the medium. Then $w^+$ dependence remains only in the phase and the integration over $w^+$ can be performed trivially. Performing this integration and taking the massless quark limit just for the simplicity of the expressions, one gets the back-to-back limit of the scattering amplitude for the first mechanism as 
\begin{align}
\label{eq:M_1_q_to_qbarq_b2b_m0}
i\M^{{\rm b2b}, \, m=0}_{q\to q\bar q, \, 1} &= 
i g^2\ \frac{z}{(2q^+)^2} \frac{1}{\P^2}   
\int_{-\frac{L^+}{2}}^{\frac{L^+}{2}}  dz^+
\int_{\bb} \ e^{-i \bb\cdot (\bk - \bq)} 
\ \overline{\Psi}_{\beta_{2}}(z^{+};\bb) 
\frac{\gamma^{-}\gamma^{+}}{2}\gamma^{i} 
\nn \\
&
\times
\ v(\check{p}_{2},h_{2}) \ \overline{u}(\check{p}_{1},h_{1}) \gamma^{+} \ 
\Big[ \frac{\gamma^{l}\gamma^{i}\P^l}{z} -\frac{2\P^i}{1-z} \Big]
u(\check{q},h) 
\nn \\
& 
\times
\big[ t^{a}\ \UFd(\infty,z^{+};\bb) \big]_{\beta_{2}\alpha_{2}}
\big[ \UF(\infty,-\infty;\bb)\ t^{b}\big]_{\alpha_{1} \alpha} 
\ \UA(z^{+},-\infty;\bb)_{ab} .
\end{align}
The color structure appearing in Eq.~\eqref{eq:M_1_q_to_qbarq_b2b_m0} can be simplified by using the Fierz identity given in Eq.~\eqref{eq:sun_Fierz} and using the properties of the Wilson lines discussed in App.~\ref{app:WL} and can be written as 
\begin{align}
\label{eq:q_to_qbarq_Simp_CS_1}
& 
\overline{\Psi}_{\beta_{2}}(z^{+};\bb)  
\big[ t^{a}\ \UFd(\infty,z^{+};\bb)\big]_{\beta_{2}\alpha_{2}}  
\big[ \UF(\infty,-\infty;\bb)\ t^{b}\big]_{\alpha_{1} \alpha} 
\ \UA(z^{+},-\infty;\bb)_{ab}
\nn \\
& 
= 
\overline{\Psi}_{\beta_{2}}(z^{+};\bb)  
\big[ t^{a}\ \UFd(\infty,z^{+};\bb) \big]_{\beta_{2}\alpha_{2}}  
\big[ \UF(\infty,z^{+};\bb)\ t^{a}\
\UF(z^{+},-\infty;\bb)\big]_{\alpha_{1} \alpha}  
\nn\\        
&
= 
\frac{1}{2}\overline{\Psi}_{\beta_2}(z^{+};\bb)  \ \UF(z^{+},-\infty;\bb)_{\beta_2\alpha} \ \delta_{\alpha_1 \alpha_2} 
\nn \\
& \hspace{4cm} 
-\frac{1}{2N_{c}}\overline{\Psi}_{\beta_2}(z^{+};\bb) 
\ \UFd(\infty,z^{+};\bb)_{\beta_2\alpha_2} 
\ \UF(\infty,-\infty;\bb)_{\alpha_{1}\alpha} .
\end{align}

Similar procedure can be performed for the second mechanism starting from the amplitude in general kinematics given in Eq.~\eqref{eq:M_2_q_to_qbarq} and one obtains the back-to-back limit of the second mechanism as 
\begin{align}
\label{eq:M_2_q_to_qbarq_b2b_m0}
 i \M_{q\to q\bar q, \, 2}^{{\rm b2b}, \, m=0} & = 
ig^2\frac{z(1-z)}{(2q^{+})^2 } \frac{1}{\P^2}  
\int_{-\frac{L^+}{2}}^{\frac{L^+}{2}} dz^+ 
\int_{\b} \ e^{-i \b \cdot (\bk-\bq)} 
\ \overline{\Psi}_{\beta}(z^{+};\bb)
\frac{\gamma^{-}\gamma^{+}}{2}\gamma^{i} 
\nn \\
& \times 
u(\check{q},h) \ \overline{u}(\check{p}_{1},h_{1}) \ \gamma^{+}
 \Big[ - \gamma^i\gamma^l \frac{\P^l}{1-z}  + \gamma^{l}\gamma^{i}  \frac{\P^l}{z} \Big] 
\nn \\
& 
\times
v(\check{p}_{2},h_{2})
\big[ t^{a} \UF(z^{+},-\infty;\bb) \ \big]_{\beta\alpha}
\big(t^{b}\big)_{\alpha_{1}\alpha_{2}} 
         \UA(\infty,z^{+};\bb)_{ba} .
    \end{align}
Similar to the case that was observed for the back-to-back limit of the first mechanism, the color structure that appears in Eq.~\eqref{eq:M_2_q_to_qbarq_b2b_m0} can be simplified as 
\begin{align}
\label{q_to_qbarq_Simp_CS_2}
& 
\overline{\Psi}_{\beta}(z^{+};\bb)  
\big[ t^{a} \ \UF(z^{+},-\infty;\bb) \big]_{\beta\alpha}
\ \big(t^{b}\big)_{\alpha_{1}\alpha_{2}} 
\UA(\infty,z^{+};\bb)_{ba}
\\
&
=  
\overline{\Psi}_{\beta}(z^{+};\bb)  
\big[ t^{a} \ \UF(z^{+},-\infty;\bb) \big]_{\beta\alpha}
\ 
\big[\UF(\infty,z^{+};\bb)\
t^{a}\ \UFd(\infty,z^{+};\bb)
\big]_{\alpha_{1}\alpha_{2}}
\nn \\
&
= 
\frac{1}{2} \overline{\Psi}_{\beta}(z^{+};\bb) 
\ \UFd(\infty,z^{+};\bb)_{\beta\alpha_{2}} 
\ \UF (\infty, -\infty;\bb)_{\alpha_{1}\alpha} 
 - \frac{1}{2N_{c}} \overline{\Psi}_{\beta}(z^{+};\bb)     \UF(z^{+},-\infty;\bb)_{\beta\alpha} \delta_{\alpha_{1}\alpha_{2}} . \nn
\end{align}
Substituting the back-to-back scattering amplitudes given in Eqs.~\eqref{eq:M_1_q_to_qbarq_b2b_m0}  and \eqref{eq:M_2_q_to_qbarq_b2b_m0} together with the simplified color structures given in Eqs.~\eqref{eq:q_to_qbarq_Simp_CS_1} and \eqref{q_to_qbarq_Simp_CS_2} into Eq.~\eqref{eq:M_tot_schm_q_to_qbarq_b2b}, one obtains the total scattering amplitude in the $q\to q\bar q$ channel in the back-to-back and massless quark limits as 
\begin{align}
\label{eq:M_tot_q_to_qbarq_b2b_m0}
&
i \M_{q\to q\bar q, \, {\rm tot.}}^{{\rm b2b}, \, m=0} = 
- i\frac{g^{2}}{(2q^{+})^2} \frac{1}{\P^2}
 \int_{-\frac{L^+}{2}}^{\frac{L^+}{2}} dz^{+}  
\int_{\b} \ e^{-i \b \cdot (\bk-\bq)}
\ \overline{\Psi}_{\beta}(z^{+};\bb)
\\
&
\times
\Big\{
\UF(z^{+},-\infty;\bb)_{\beta\alpha}  \ \delta_{\alpha_{1}\alpha_{2}} \ \mathfrak{h}_{q\to q\bar q}^{(1)}
 +
\UF^{\dagger}(\infty,z^{+};\bb)_{\beta\alpha_{2}} \ \UF(\infty, -\infty;\bb)_{\alpha_{1}\alpha} \ \mathfrak{h}_{q\to q\bar q}^{(2)}
\Big\} , \nn
\end{align}
with the coefficients defined as 
\begin{align}
\label{eq:HF_q_to_qbarq_1}
\mathfrak{h}_{q\to q\bar q}^{(1)} =&
    -\mathfrak{h'}^{(1)}_{q\to q\bar q} 
    + \frac{1}{N_c}\mathfrak{h'}^{(2)}_{q\to q\bar q} ,  \\ 
\label{eq:HF_q_to_qbarq_2}
\mathfrak{h}_{q\to q\bar q}^{(2)} =&
    -\mathfrak{h'}^{(2)}_{q\to q\bar q} 
    + \frac{1}{N_c}\mathfrak{h'}^{(1)}_{q\to q\bar q} ,
\end{align}
and 
\begin{align}
\label{eq:HFp_q_to_qbarq_1}
\mathfrak{h'}^{(1)}_{q\to q\bar q} =&
    \frac{z}{2}\frac{\gamma^{-}\gamma^{+}}{2}\gamma^{i}v(\check{p}_{2},h_{2})\overline{u}(\check{p}_{1},h_{1})\gamma^{+}
    \bigg( \frac{\gamma^{l}\gamma^{i}\P^l}{z}  - \frac{2\P^i}{1-z}\bigg)u(\check{q},h), \\
\label{eq:HFp_q_to_qbarq_2}
\mathfrak{h'}^{(2)}_{q\to q\bar q} =&
    \frac{z(1-z)}{2} \frac{\gamma^{-}\gamma^{+}}{2}\gamma^{i} u(\check{q}, h) \overline{u}(\check{p}_{1}, h_1) \gamma^{+} 
    \bigg(-\frac{\gamma^{i}\gamma^{l}\P^l}{1-z} + \frac{\gamma^{l}\gamma^{i}\P^l}{z}\bigg) v(\check{p}_{2}, h_{2}) .
\end{align}
%
%
%
%

\subsection{$q\to q\Bar{q}$ production cross section in the back-to-back limit}
\label{subsec:xs_q-qbq}

\begin{figure}[H]
\centering
\begin{subfigure}{0.49\textwidth}
\centering
\includegraphics[height=5cm]{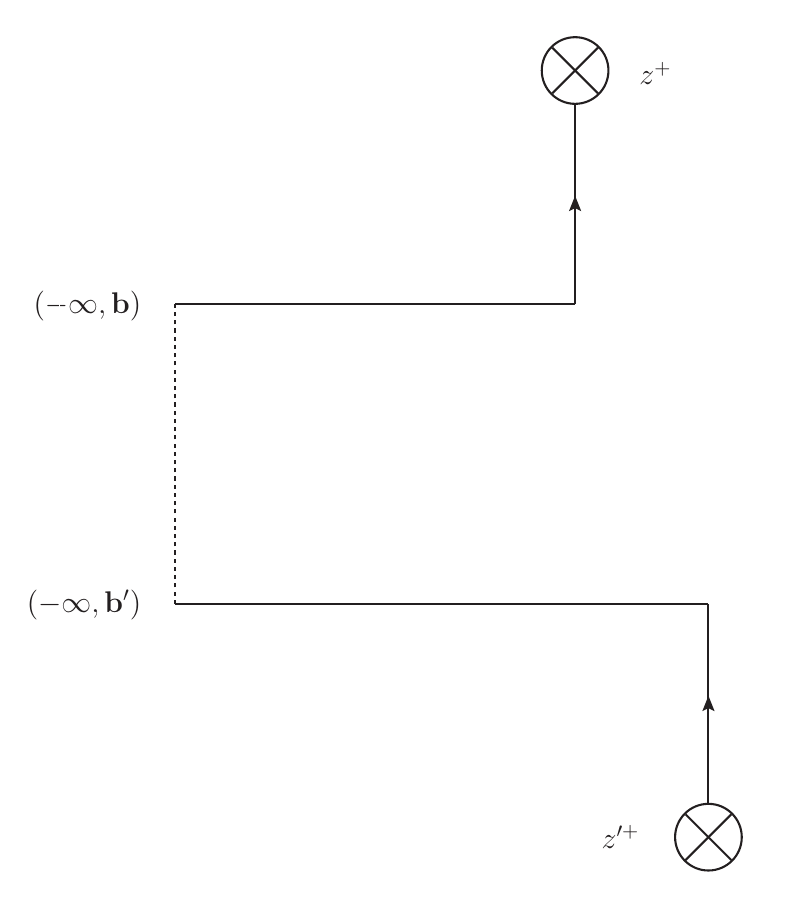}
\caption{$\overline{\mathcal{C}}^{-}$}
\label{fig:q-qbq:bC-}
\end{subfigure}
\begin{subfigure}{0.49\textwidth}
\centering
\includegraphics[height=5cm]{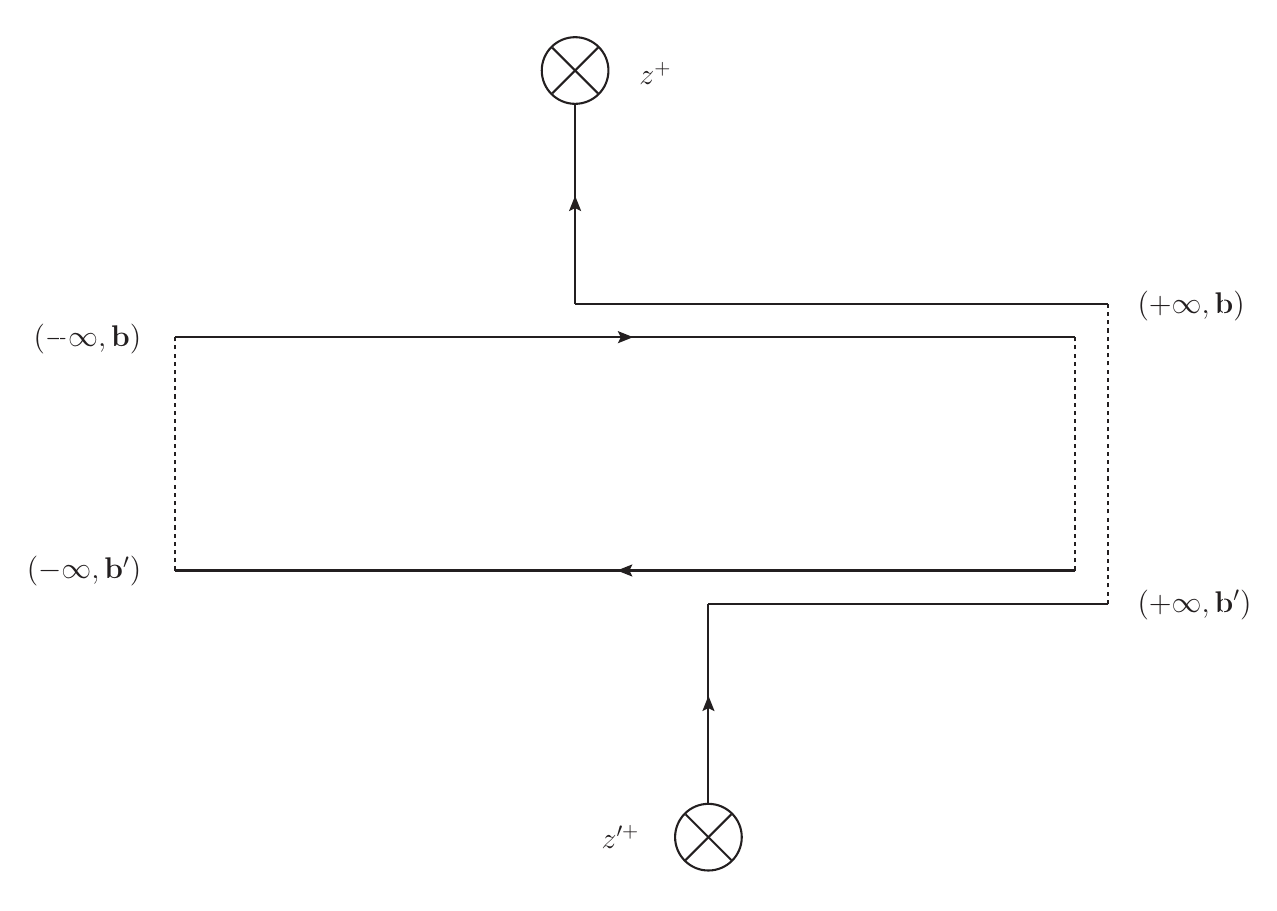}
\caption{$\overline{\mathcal{C}}^{+\square}$}
\label{fig:q-qbq:bC+l}
\end{subfigure}
\caption{Color structures appearing in the cross section of $q \to q\bar{q}$.}
\label{fig:Color_q-qbq}
\end{figure}

The partonic cross section for the production of quark-antiquark dijet in quark initiated channel in the back-to-back and massless quark limits can be written as 
\begin{equation}
 \frac{d\sigma^{{\rm b2b},\, m=0}_{q\to q\bar q}}{d{\rm P.S.}} =
 (2q^+) \ 2\pi\delta\left(p_1^+ + p_2^+ - q^+\right) 
    \frac{1}{2N_c}\sum_{h, h_1, h_2}\sum_{\alpha, \alpha_1, \alpha_2}
   \left\langle \left|i\mathcal{M}^{{\rm b2b}, \, m=0}_{q\to q\bar q, \, {\rm tot.}}\right|^2 \right\rangle,
\end{equation}
with the total amplitude given in Eq.~\eqref{eq:M_tot_q_to_qbarq_b2b_m0} together with the coefficients given in Eqs.~\eqref{eq:HF_q_to_qbarq_1} and \eqref{eq:HF_q_to_qbarq_2} in this limit. The phase space is defined in Eq.~\eqref{def:PS}. The renormalization factors of $2$ and $N_c$ comes from the averaging of the helicities and the colors, and $\langle \cdots\rangle$ stands for the target averaging within the CGC framework. Using the explicit expression of the total amplitude in the back-to-back limit, the partonic cross section can be written as 
\begin{align}
\label{eq:sigma_q_to_qbarq}
\frac{d\sigma^{{\rm b2b}, \, m=0}_{q\to q\bar q}}{d{\rm P.S.}} &=
\frac{g^4}{(2q^{+})^{3}\bP^4(2N_c)}(2\pi)\delta \left(k^+ - q^+\right)\sum_{h, h_1, h_2}
\int_{-\frac{L^+}{2}}^{\frac{L^+}{2}}dz^+
\int_{-\frac{L^+}{2}}^{\frac{L^+}{2}}dz'^+
\int_{\bb, \bb'}
e^{-i(\bb-\bb')\cdot\left(\bk - \bq\right)} 
\nn \\
&
\times \Big\{
    N_c \bigg\langle
    \Tr \bigg\{\UFd(z'^{+} ,-\infty;\bb')
    \gamma^0 \Psi(z'^{+};\bb')\overline{\Psi}(z^{+};\bb)
    \UF(z^{+},-\infty;\bb)
    \Big|\mathfrak{h}^{(1)}_{q\to q\bar q}\Big|^2\bigg\} \bigg\rangle
\nn \\
& \hspace{0.3cm}
    + \bigg\langle 
    \Tr \bigg\{\UF(\infty, z'^{+};\bb')
    \gamma^0\Psi(z'^{+};\bb')
    \overline{\Psi}(z^{+};\bb)
    \UFd(\infty, z^{+};\bb)
    \Big|\mathfrak{h}^{(2)}_{q\to q\bar q}\Big|^2 \bigg\}
\nn\\
&\hspace{6.3cm}
\times
    \Tr\left[\UFd(\infty ,-\infty;\bb')\UF(\infty ,-\infty;\bb)\right]
    \bigg\rangle
\nn \\
& \hspace{0.3cm}
    + \bigg\langle
    \Tr \bigg\{\UFd(z'^{+} ,-\infty;\bb')
    \gamma^0\Psi(z'^{+};\bb')
    \overline{\Psi}(z^{+};\bb)
    \UFd(\infty, z^{+};\bb)
\nn \\
& \hspace{6.8cm} 
\times
    \UF(\infty ,-\infty;\bb)
    \mathfrak{h}^{(2)}_{q\to q\bar q}
    \Big(\mathfrak{h}^{(1)}_{q\to q\bar q}\Big)^\dagger
    \bigg\} \bigg\rangle
\nn \\
& \hspace{0.3cm}
    + \bigg\langle
    \Tr \bigg\{\UF(\infty ,-\infty;\bb')\UFd(\infty, z'^{+} ;\bb')
    \gamma^0\Psi(z'^{+};\bb')
    \overline{\Psi}(z^{+};\bb)
\nn \\
&\hspace{6.3cm} 
\times 
    \UF( z^{+},-\infty;\bb)
    \mathfrak{h}^{(1)}_{q\to q\bar q}
    \Big(\mathfrak{h}^{(2)}_{q\to q\bar q}\Big)^\dagger
    \bigg\} \bigg\rangle
\Big\}.
\end{align}
which coefficient are given in Eqs.~\eqref{eq:HF_q_to_qbarq_11}, \eqref{eq:HF_q_to_qbarq_22} and \eqref{eq:HF_q_to_qbarq_12}.
Moreover, the color structures that appear at the cross section level can be expressed defining the following reduced one 
\begin{align}
\label{eq:bC-} 
\overline{\mathcal{C}}^- &\equiv 
\Big\langle
\Tr \bigg\{\UFd(z'^{+},-\infty;\bb')\Psi(z'^{+};\bb')
\overline{\Psi}(z^{+};\bb) \gamma^{-}\UF(z^{+},-\infty;\bb)
\bigg\} \Big\rangle, \\
\label{eq:bC+square} 
\overline{\mathcal{C}}^{+\square} &\equiv
\Big\langle 
\Tr \bigg\{\UF(\infty, z'^{+};\bb')\Psi(z'^{+};\bb')
\overline{\Psi}(z^{+};\bb) \gamma^{-}\UFd(\infty, z^{+};\bb)
\bigg\} \nn \\
& \quad
\times\Tr\left[\UF(\infty ,-\infty;\bb')\UFd(\infty ,-\infty;\bb)\right]
\Big\rangle
 .
\end{align}
The illustration of these two new color structures are provided in Fig.~\ref{fig:Color_q-qbq}. Substituting the new color structures given in Eqs.~\eqref{eq:bC-} and \eqref{eq:bC+square} into to the back-to-back cross section in the $q\to q\bar q$ channel given in Eq.~\eqref{eq:sigma_q_to_qbarq}, one arrives at the following final factorized expression
\begin{align}
\frac{d\sigma^{{\rm b2b}, \, m=0}_{q\to q\bar q}}{d{\rm P.S.}} &=
g^4 \ (2\pi)\delta\left(p_1^+ + p_2^+ - q^+\right)
\int_{-\frac{L^+}{2}}^{\frac{L^+}{2}}dz^+
\int_{-\frac{L^+}{2}}^{\frac{L^+}{2}}dz'^+
\int_{\bb, \bb'}
e^{-i(\bb-\bb')\cdot\left(\bk - \bq\right)} 
\nn\\
&
\hspace{5cm}\
\times
\Big[ \mathcal{H}^{-}_{q\to q\bar q} \  \overline{\mathcal{C}}^{-}  +  \mathcal{H}^{+\square}_{q\to q\bar q} \ \overline{\mathcal{C}}^{+\square}\Big] ,
\end{align}
with hard factors given as
\begin{align}
\label{eq:H_q_to_qbarq_-}
    \mathcal{H}^{-}_{q\to q\bar q} &= 
    \frac{z(1-z)}{4N_c \ \bP^2} 
    \left[\frac{1+z^2}{(1-z)^2}\left(N_c-\frac{2}{N_c}\right)
    -\frac{1}{N_c}\left(z^2+(1-z)^2\right)
    - \frac{2}{N_c^2}\frac{z^2}{1-z}\right], \\
\label{eq:H_q_to_qbarq_+square}
    \mathcal{H}^{+\square}_{q\to q\bar q} &= 
    \frac{z(1-z)}{4N_c \ \bP^2} 
    \left[z^2+(1-z)^2 + \frac{2}{N_c}\frac{z^2}{1-z}
    + \frac{1}{N_c^2}\frac{1+z^2}{(1-z)^2}\right] . \nn\\& 
\end{align}

%
%
%
%

\subsection{$q\to q\Bar{q}$ production cross section for different quark flavors}
\label{subsec:xs_q-qbq_f}

The results obtained in the previous section only apply to processes where all quark have the same flavor. We now present results for processes involving different quark flavors, beginning with $q_f\to q_{f'}\bar{q}_{f'}$. Only the second diagram (see \Figure{fig:q-qbq2}) contributes to this process, which directly implies that the corresponding cross-section reads
\begin{equation}
 \frac{d\sigma^{{\rm b2b},\, m=0}_{q_f\to q_{f'}\bar q_{f'}}}{d{\rm P.S.}} =
 (2q^+) \ 2\pi\delta\left(p_1^+ + p_2^+ - q^+\right) 
    \frac{1}{2N_c}\sum_{h, h_1, h_2}\sum_{\beta_1, \beta_3, \alpha_3}
   \left\langle \left|i\mathcal{M}^{{\rm b2b}, \, m=0}_{q\to q\bar q, \, 2}\right|^2 \right\rangle.
\end{equation}
Based on our previous calculation, we can provide this cross-section in a factorized form
\begin{align}
\frac{d\sigma^{{\rm b2b}, \, m=0}_{q_f\to q_{f'}\bar q_{f'}}}{d{\rm P.S.}} &=
g^4 \ (2\pi)\delta\left(p_1^+ + p_2^+ - q^+\right)
\int_{-\frac{L^+}{2}}^{\frac{L^+}{2}}dz^+
\int_{-\frac{L^+}{2}}^{\frac{L^+}{2}}dz'^+
\int_{\bb, \bb'}
e^{-i(\bb-\bb')\cdot\left(\bk - \bq\right)} 
\nn\\
&
\hspace{5cm}\
\times
\Big[ \mathcal{H}^{-}_{q_f\to q_{f'}\bar q_{f'}} \  \mathcal{C}^{-}  +  \mathcal{H}^{+\square}_{q_f\to q_{f'}\bar q_{f'}} \ \mathcal{C}^{+\square}\Big] ,
\end{align}
with hard factors given as
\begin{align}
\label{eq:H_q_to_q'barq'_-}
    \mathcal{H}^{-}_{q_f\to q_{f'}\bar q_{f'}} &= 
    -\frac{z(1-z)}{4N_c^2 \ \bP^2}\left(z^2+(1-z)^2\right), \\
\label{eq:H_q_to_q'barq'_+square}
    \mathcal{H}^{+\square}_{q_f\to q_{f'}\bar q_{f'}} &= 
    \frac{z(1-z)}{4N_c \ \bP^2}\left(z^2+(1-z)^2\right) . \nn\\& 
\end{align}
For $q_f\to q_f\bar{q}_{f'}$, only the first diagram contributes this time (see \Figure{fig:q-qbq1}), leading to
\begin{align}
    \frac{d\sigma^{{\rm b2b},\, m=0}_{q_f\to q_f\bar q_{f'}}}{d{\rm P.S.}} &=
    (2q^+) \ 2\pi\delta\left(p_1^+ + p_2^+ - q^+\right) 
    \frac{1}{2N_c}\sum_{h, h_1, h_2}\sum_{\beta_1, \beta_3, \alpha_3}\sum_{f,f'\neq f}
    \left\langle \left|i\mathcal{M}^{{\rm b2b}, \, m=0}_{q\to q\bar q, \, 1}\right|^2 \right\rangle
    \nn \\
    &= 
    g^4 \ (2\pi)\delta\left(p_1^+ + p_2^+ - q^+\right) \int_{-\frac{L^+}{2}}^{\frac{L^+}{2}}dz^+
    \int_{-\frac{L^+}{2}}^{\frac{L^+}{2}}dz'^+
    \int_{\bb, \bb'}
    e^{-i(\bb-\bb')\cdot\left(\bk - \bq\right)} 
    \nn\\
    &
    \hspace{5cm}\
    \times
    \Big[ \mathcal{H}^{-}_{q_f\to q_f\bar q_{f'}} \  \mathcal{C}^{-}  +  \mathcal{H}^{+\square}_{q_f\to q_f\bar q_{f'}} \ \mathcal{C}^{+\square}\Big] ,
\end{align}
with hard factors given as
\begin{align}
\label{eq:H_q_to_qbarq'_-}
    \mathcal{H}^{-}_{q_f\to q_f\bar q_{f'}} &= 
    \frac{z}{4N_c \ \bP^2}\frac{1+z^2}{1-z}
    \left(N_c-\frac{2}{N_c}\right), \\
\label{eq:H_q_to_qbarq'_+square}
    \mathcal{H}^{+\square}_{q_f\to q_f\bar q_{f'}} &= 
    \frac{z}{4N_c^3 \ \bP^2}\frac{1+z^2}{1-z} . \nn\\& 
\end{align}

%
%
%
%


\section{$q\to gg$ channel} 
\label{sec:qgg}                   
\begin{figure}[H]
\centering
\begin{subfigure}{0.49\textwidth}
\centering
\includegraphics[width=\textwidth]{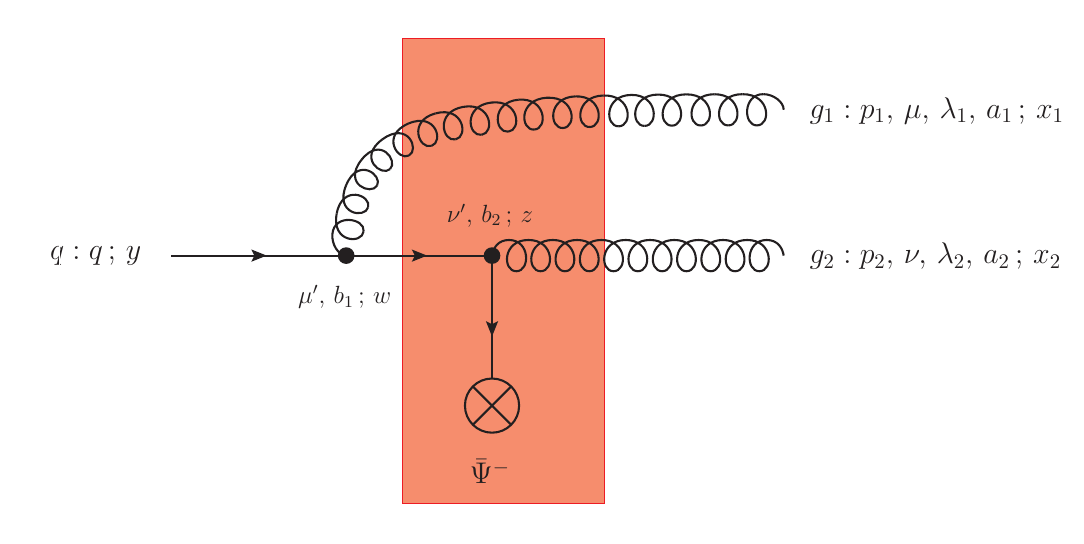}
\caption{Diagram 1}
\label{fig:q-gg1}
\end{subfigure}
\begin{subfigure}{0.49\textwidth}
\centering
\includegraphics[width=\textwidth]{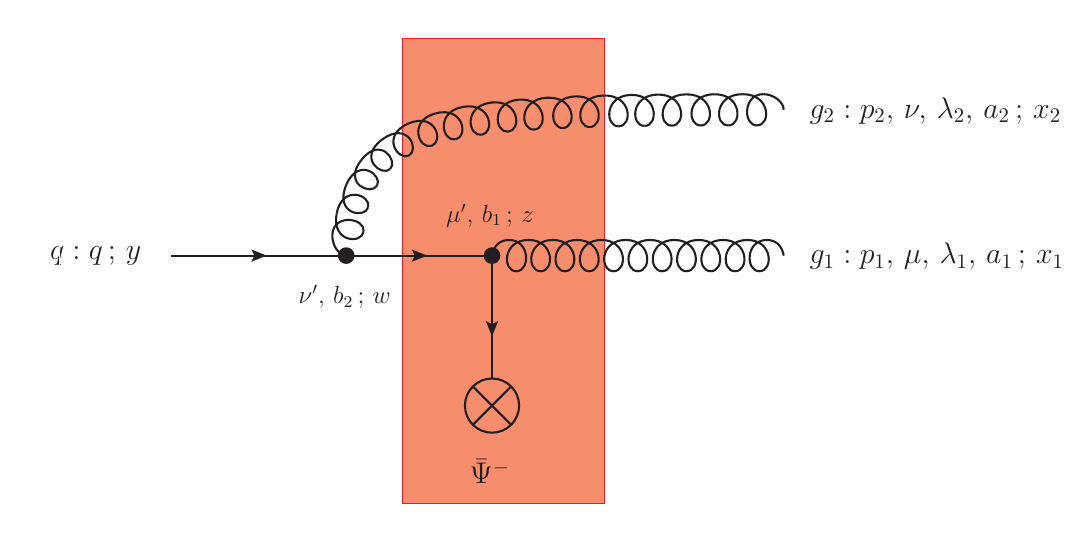}
\caption{Diagram 2}
\label{fig:q-gg2}
\end{subfigure}
\begin{subfigure}{0.5\textwidth}
\centering
\includegraphics[width=\textwidth]{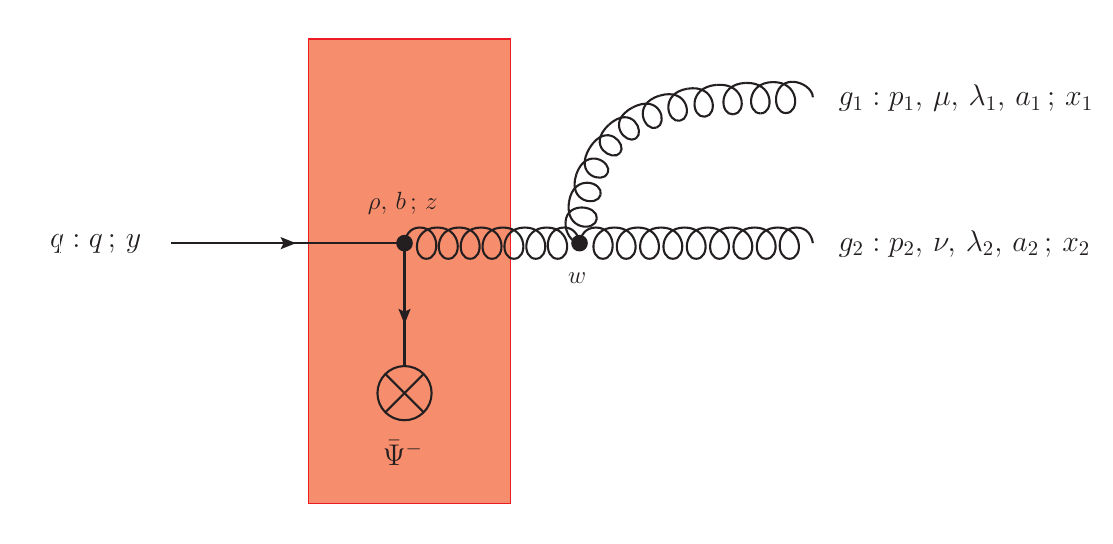}
\caption{Diagram 3}
\label{fig:q-gg3}
\end{subfigure}
\caption{Diagrams contributing to channel $q\to gg$.}
\label{fig:q-gg}
\end{figure}

The next channel we would like to discuss is gluon dijet production in quark initiated channel, i.e. $q\to gg$ channel. At NEik accuracy there are three mechanisms that contribute to this process. In the first mechanism, incoming quark splits into a quark-gluon pair before the medium which then scatters on the target. While the gluon scatters eikonally on the target, the quark scatters via a $t$-channel quark exchange and converts into a gluon. Therefore, in the final state one obtains a gluon dijet (see Fig.~\ref{fig:q-gg1}). The second mechanism corresponds to a similar mechanism except the gluons in the final state are interchanged (see Fig.~\ref{fig:q-gg2}). Finally, the third mechanism corresponds to the case where the incoming quark scatters on the target via a $t$-channel quark exchange and converts into a gluon. This gluon splits into two gluons in the final sate after the medium (see Fig.~\ref{fig:q-gg3}).  

In the rest of this section, we present the results for the scattering amplitude in general kinematics, scattering amplitude in back-to-back limit and the production cross section in the back-to-back limit for each channel. Since the computation of the amplitudes both in general kinematics and in the back-to-back limits  follow the same steps as in the previous channels, we skip the details and just provide the results to avoid repetition. On the other hand, we give a detailed discussion for the production cross section. 
\subsection{$q\to gg$ amplitude in general kinematics}
The total scattering amplitude in $q\to gg$ channel receives three contribution that corresponds to each contributing mechanism and it can be written as 
\begin{align}
\label{eq:M_tot_schm_q_to_gg}
\M_{q\to gg, \, {\rm tot.}}=\M_{q\to gg, \, 1}+\M_{q\to gg, \, 2}+\M_{q\to gg, \, 3}\, . 
\end{align}

The first mechanism is illustrated in Fig.~\ref{fig:q-gg1} and the $S$-matrix element for this mechanism can be obtained via the following LSZ-type reduction formula
\begin{align}
\S_{q\to g_{1}g_{2}, \, 1} &= 
\lim_{y^+\to-\infty}\lim_{x_1^+,x_2^+\to\infty}
\int_{\by,\bx[1],\bx[2]}\int_{y^-,x^-_1,x^-_2}
\int_{\bw, \bz}\int_{w^-, z^-}
\int_{-\frac{L^+}{2}}^{\frac{L^+}{2}}d z^+
\int_{-\infty}^{-\frac{L^+}{2}}d w^+
\nn \\
&
\times
 \ e^{ix_{1} \cdot \check{p}_{1}} \ e^{ix_{2} \cdot \check{p}_{2}} \ e^{-iy \cdot \check{q}}
 \ (-2p_{1}^{+}) {\epsilon^{\lambda_{1}}_{\mu}(p_{1})}^{*} 
 \ (-2p_{2}^{+}) {\epsilon^{\lambda_2}_{\nu}(p_{2})}^{*}
\nn \\ 
& 
\times
\Big[G^{\mu\mu'}_{F}(x_{1},w)\BA \Big]_{a_{1}b_{1}} 
\  \Big[G^{\nu\nu'}_{F}(x_{2},z) \IA\Big]_{a_{2}b_{2}}
\ \overline{\Psi}^{-}_{\beta}(z) 
\nn \\
 & 
 \times
 \Big[(-igt^{b_{2}}\gamma_{\nu'})S_{F}(z,w) \BIq 
 (-igt^{b_{1}}\gamma_{\mu'})S_{0,F}(w,y)\Big]_{\beta\alpha} 
\gamma^{+}u(\check{q},h) ,
\end{align}
where the before-to-after gluon propagator is give in Eq.~\eqref{gluon_prop_BA}, the inside-to-after gluon propagator is given in Eq.~\eqref{gluon_prop_IA}, the before-to-inside quark propagator is given in Eq.~\eqref{quark_prop_BI} and the vacuum quark propagator is given in Eq.~\eqref{eq:quark_vacuum_prop}. Using these explicit expressions and simplifying the Dirac matrices similar to the $g\to gq$ channel, one obtains 
\begin{align}
\S_{q\to g_{1}g_{2}, \, 1} &= 
\frac{g^2 \ {\varepsilon_{\lambda_1}^{j}}^{*} {\varepsilon_{\lambda_2}^{i}}^{*} }{(2q^{+})(2p_{2}^{+})} 
\ 2\pi\delta(p_{1}^{+}+p_{2}^{+}-q^{+})
\int_{-\frac{L^+}{2}}^{\frac{L^+}{2}} dz^+ 
\int_{-\infty}^{-\frac{L^+}{2}} dw^+
\\
&
\times
\int\! \frac{d^{2}\bk_{1}}{(2\pi)^{2}}\ 
e^{iw^{+}\Big(\frac{\bk_{1}^{2}}{2p_{1}^{+}} + \frac{(\bq - \bk_{1})^{2} + m^2}
{2p_{2}^{+}} - \frac{\bq^{2} + m^2}{2q^{+}}\Big)}
\int_{\bz,\bz_{1} } e^{i(\bq - \bk_1 -\p_2) \cdot \bz}
\ e^{i(\bk_1 - \bp_{1}) \cdot \bz_{1}}
\nn \\
&
\times
\Big[t^{b_{2}}\UF(z^{+},w^{+};\bz)t^{b_{1}}\Big]_{\beta\alpha} 
\UA(\infty,w^{+}; \bz_{1})_{a_{1}b_{1}} \ \UA(\infty,z^{+}; \bz)_{a_{2}b_{2}}
\overline{\Psi}_{\beta}(\underline{z})
\nn \\
&
\times\gamma^{-}\gamma^{+}
\gamma^{i}
\Big[ \big(\gamma^{j}\gamma^{l}\bq^{l} p_{2}^{+} + \gamma^{l}\gamma^{j}(\bq^{l} - \bk_{1}^{l})q^{+} + \gamma^{j}(p_{2}^{+}-q^{+})m + 2\frac{q^{+}p_{2}^{+} \bk_{1}^{j}}{p_{1}^{+}} \Big]
u(\check{q},h) , \nn
\end{align}
from which we can get the scattering amplitude by using Eq.~\eqref{eq:amplitude} and which reads in general kinematics 
\begin{align}
\label{eq:M_1_q_to_gg}
&
i\M_{q\to g_{1}g_{2}, \, 1} = 
\frac{g^2 \ {\varepsilon_{\lambda_1}^{j}}^{*} {\varepsilon_{\lambda_2}^{i}}^{*}}{(2q^{+})^2(2p_{2}^{+})} 
\int_{-\frac{L^+}{2}}^{\frac{L^+}{2}} dz^+ 
\int_{-\infty}^{-\frac{L^+}{2}} dw^+
\int\! \frac{d^{2}\bk_{1}}{(2\pi)^{2}}\ 
e^{iw^{+}\Big(\frac{\bk_{1}^{2}}{2p_{1}^{+}} + \frac{(\bq - \bk_{1})^{2} + m^2}{2p_{2}^{+}} - \frac{\bq^{2} + m^2}{2q^{+}}\Big)}
\nn \\
&
\times
\int_{\bz,\bz_{1} } e^{i(\bq - \bk_1 -\p_2) \cdot \bz}
\ e^{i(\bk_1 - \bp_{1}) \cdot \bz_{1}}
\Big[t^{b_{2}}\UF(z^{+},w^{+};\bz)t^{b_{1}}\Big]_{\beta\alpha} 
\UA(\infty,w^{+}; \bz_{1})_{a_{1}b_{1}} \ \UA(\infty,z^{+}; \bz)_{a_{2}b_{2}}
\nn \\
&
\times
\overline{\Psi}_{\beta}(\underline{z})\gamma^{-}\gamma^{+}\gamma^{i}
\Big[ \big(\gamma^{j}\gamma^{l}\bq^{l} p_{2}^{+} + \gamma^{l}\gamma^{j}(\bq^{l} - \bk_{1}^{l})q^{+} + \gamma^{j}(p_{2}^{+}-q^{+})m {+}\frac{2q^{+}p_{2}^{+} \bk_{1}^{j}}{p_{1}^{+}} \Big] 
u(\check{q},h) .
\end{align}
The second mechanism is almost the same as the first one except the gluons in the final state are exchanged (see Fig.~\ref{fig:q-gg2}). The $S$-matrix element is also very similar and it is written as 
\begin{align}
\label{eq:S_2_q_to_gg}
\S_{q\to g_{2}g_{1}, \, 2} &= 
\lim_{y^+\to-\infty}\lim_{x_1^+,x_2^+\to\infty}
(-2p_{1}^{+}) (-2p_{2}^{+})
\int_{\by,\bx[1],\bx[2]}\int_{y^-,x^-_1,x^-_2}
\int_{\bw, \bz}\int_{w^-, z^-}
\int_{-\frac{L^+}{2}}^{\frac{L^+}{2}}d z^+
\int_{-\infty}^{-\frac{L^+}{2}}d w^+
\nn \\
&
\times   
 e^{ix_{2} \cdot \check{p}_{2}} \ 
 e^{ix_{1} \cdot \check{p}_{1}} \ 
 e^{-iy \cdot \check{q}}
\ {\epsilon^{\lambda_{1}}_{\mu}(p_{1})}^{*} 
\ {\epsilon^{\lambda_2}_{\nu}(p_{2})}^{*}
\ \Big[G^{\nu\nu'}_{F}(x_{2},w)\BA\Big]_{a_{2}b_{2}}
\  \Big[G^{\mu\mu'}_{F}(x_{1},z)\IA\Big]_{a_{1}b_{1}} 
\nn \\
&
\times 
\overline{\Psi}^{-}_{\beta}(z)
\Big[(-igt^{b_{1}}\gamma_{\mu'})S_{F}(z,w) \BIq (-igt^{b_{2}}\gamma_{\nu'})S_{0,F}(w,y)\Big]_{\beta\alpha}
\gamma^{+}u(\check{q},h) ,
\end{align}
where the before-to-after gluon propagator is given in Eq.~\eqref{gluon_prop_BA}, the inside-to-after gluon propagator is given in Eq.~\eqref{gluon_prop_IA}, the before-to-inside quark propagator is given in Eq.~\eqref{quark_prop_BI} and the vacuum quark propagator is given in Eq.~\eqref{eq:quark_vacuum_prop}. After substituting these expressions into Eq.~\eqref{eq:S_2_q_to_gg} and simplifying the Dirac matrices, one obtains
\begin{align}
\S_{q\to g_{2}g_{1}, \, 2} &= 
\frac{g^2 \ {\varepsilon_{\lambda_1}^{i}}^{*} {\varepsilon_{\lambda_2}^{j}}^{*} }{(2q^{+})(2p_{1}^{+})} 
\ 2\pi\delta(p_{1}^{+}+p_{2}^{+}-q^{+})
\int_{-\frac{L^+}{2}}^{\frac{L^+}{2}} dz^+ 
\int_{-\infty}^{-\frac{L^+}{2}} dw^+
\nn \\
&
\times
\int\! \frac{d^{2}\bk_{2}}{(2\pi)^{2}}\ 
e^{iw^{+}\Big(\frac{(\bq - \bk_{2})^{2} + m^2}{2p_{1}^{+}} + \frac{\bk_{2}^{2}}{2p_{2}^{+}} - \frac{\bq^{2} + m^2}{2q^{+}}\Big)}
\int_{\bz,\bz_{1} } e^{i(\bq -\p_1 - \bk_2) \cdot \bz}
\ e^{i(\bk_2 - \bp_{2}) \cdot \bz_{1}}
\nn \\
&
\times
\overline{\Psi}_{\beta}(\underline{z})\gamma^{-}\gamma^{+}
\gamma^{i}
\Big[ \big(\gamma^{j}\gamma^{l}\bq^{l} p_{1}^{+} + \gamma^{l}\gamma^{j}(\bq^{l} - \bk_{2}^{l})q^{+} + \gamma^{j}(p_{1}^{+}-q^{+})m + 2\frac{q^{+}p_{1}^{+} \bk_{2}^{j}}{p_{2}^{+}} \Big] 
\nn \\
&
\times 
\Big[t^{b_{1}}\UF(z^{+},w^{+};\bz)t^{b_{2}}\Big]_{\beta\alpha} 
\UA(\infty,z^{+}; \bz)_{a_{1}b_{1}} \ \UA(\infty,w^{+}; \bz_{1})_{a_{2}b_{2}}
u(\check{q},h) ,
\end{align}
from which we can obtain the amplitude by using Eq.~\eqref{eq:amplitude}, and in general kinematics for the second mechanism it can be written as 
\begin{align}
\label{eq:M_2_q_to_gg}
&
i\M_{q\to g_{2}g_{1}, \, 2} = 
\frac{g^2 \ {\varepsilon_{\lambda_1}^{i}}^{*} {\varepsilon_{\lambda_2}^{j}}^{*} }{(2q^{+})^2(2p_{1}^{+})} 
\int_{-\frac{L^+}{2}}^{\frac{L^+}{2}} dz^+ 
\int_{-\infty}^{-\frac{L^+}{2}} dw^+
\int\! \frac{d^{2}\bk_{2}}{(2\pi)^{2}}\ 
e^{iw^{+}\Big(\frac{(\bq - \bk_{2})^{2} + m^2}{2p_{1}^{+}} + \frac{\bk_{2}^{2}}{2p_{2}^{+}} - \frac{\bq^{2} + m^2}{2q^{+}}\Big)}
\nn \\
&
\times
\int_{\bz,\bz_{1} } e^{i(\bq -\p_1 - \bk_2) \cdot \bz}
\ e^{i(\bk_2 - \bp_{2}) \cdot \bz_{1}}
\Big[t^{b_{1}}\UF(z^{+},w^{+};\bz)t^{b_{2}}\Big]_{\beta\alpha} 
\UA(\infty,z^{+}; \bz)_{a_{1}b_{1}} \ \UA(\infty,w^{+}; \bz_{1})_{a_{2}b_{2}}
\nn \\
&
\times
\overline{\Psi}_{\beta}(\underline{z})\gamma^{-}\gamma^{+}\gamma^{i}
\Big[ \big(\gamma^{j}\gamma^{l}\bq^{l} p_{1}^{+} + \gamma^{l}\gamma^{j}(\bq^{l} - \bk_{2}^{l})q^{+} + \gamma^{j}(p_{1}^{+}-q^{+})m {+}\frac{2q^{+}p_{1}^{+} \bk_{2}^{j}}{p_{2}^{+}} \Big] 
u(\check{q},h) .
\end{align}
Finally, in the third mechanism, that is illustrated in Fig.~\ref{fig:q-gg3}, the $S$-matrix element can be obtained as 
\begin{align}
\S_{q\to g_{1}g_{2}, \, 3} &=
\lim_{y^+\to-\infty}\lim_{x_1^+,x_2^+\to\infty}
(-2p_{1}^{+}) (-2p_{2}^{+})
\int_{\by,\bx[1],\bx[2]}\int_{y^-,x^-_1,x^-_2}
\int_{\bw, \bz}\int_{w^-, z^-}
\int_{-\frac{L^+}{2}}^{\frac{L^+}{2}}d z^+
\int_{-\infty}^{-\frac{L^+}{2}}d w^+
\nn \\
&
\times
\ e^{ix_{1} \cdot \check{p}_{1}} \ e^{ix_{2} \cdot \check{p}_{2}} \ e^{-iy \cdot \check{q}}
\ {\epsilon^{\lambda_{1}}_{\mu}(p_{1})}^{*} 
\ {\epsilon^{\lambda_2}_{\nu}(p_{2})}^{*}\ 
\Big[ G^{\mu\mu'}_{0,F}(x_{1},w)\Big]_{a_{1}b_{1}}  
\Big[G^{\nu\nu'}_{0,F}(x_{2},w) \Big]_{a_{2}b_{2}}
\nn \\
&
\times        
V^{b_{1}b_{2}c}_{\mu'\nu'\rho'}
\ \Big[G_{A}^{\rho'\rho}(w,z) \IA \Big]_{cb}
\overline{\Psi}^{-}_{\beta}(z) 
\Big[(-igt^{b}\gamma_{\rho})S_{F}(z,y) \BIq\Big]_{\beta\alpha}
\gamma^{+}u(\check{q},h) ,
\end{align}
where the vacuum gluon propagator is defined in Eq.~\eqref{eq:vacuum_gluon_prop}, the inside-to-after gluon propagator is defined in Eq.~\eqref{gluon_prop_IA} and the before-to-inside quark propagator is defined in Eq.~\eqref{quark_prop_BI}. Moreover, $V^{b_{1}b_{2}c}_{\mu'\nu'\rho'}$ is the triple gluon vertex already defined in Eq.~\eqref{eq:triple_gluon_vertex}.  Using all these expressions, one can obtain the $S$-matrix for the third mechanism as 
\begin{align}
&
 \S_{q\to g_{1}g_{2}, \, 3} =  
 -\frac{ig^{2}}{(2q^{+})} \ f^{a_{1}a_{2}c}
 \ {\varepsilon^{j}_{{\lambda_2}}}^{*} \ {\varepsilon^{i}_{{\lambda_1}}}^{*}\ 
 (2\pi) \ \delta(p_{1}^{+}+p_{2}^{+}-q^{+})
\int_{-\frac{L^+}{2}}^{\frac{L^+}{2}} dz^+
\int_{\frac{L^+}{2}}^{\infty} dw^+
\\
& 
\times
e^{iw^{+}\Big(\frac{\bp_{1}^{2}}{2p_{1}^{+}} + \frac{\bp_{2}^{2}}{2p_{2}^{+}} - \frac{(\bp_{1} + \bp_{2})^{2}}{2q^{+}}\Big)}
\int_{\bz} \  e^{-i \bz \cdot (\bp_{1}+ \bp_{2}- \bq)}\
\overline{\Psi}_{\beta}(\underline{z}) \gamma^{-}\gamma^{+} \gamma^{l} u(\check{q},h)
\Big[t^{b} \UF(z^{+},-\infty;\bz)\Big]_{\beta\alpha}
\nn \\
&
\Big\{g^{il} \Big(-\bp_{1}^{j} + \frac{p_{1}^{+}}{p_{2}^{+}} \bp_{2}^{j}\Big) 
+ g^{lj}\Big(\bp_{2}^{i} - \frac{p_{2}^{+}}{p_{1}^{+}} \bp_{1}^{i}\Big) 
+ g^{ji}\Big(\bp_{1}^{l} - \frac{p_{1}^{+}}{q^{+}}(\bp^{l}_{1}+ \bp_{2}^{l})\Big)      
\Big\}
\UA(w^{+},z^{+};\bz)_{cb} , \nn 
\end{align}
from which we can deduce the amplitude by using Eq.~\eqref{eq:amplitude} as
\begin{align}
\label{eq:M_3_q_to_gg}
&
 i\M_{q\to g_{1}g_{2}, \, 3} =  
 -\frac{ig^{2}}{(2q^{+})^2} \ f^{a_{1}a_{2}c}
 \ {\varepsilon^{j}_{{\lambda_2}}}^{*} \ {\varepsilon^{i}_{{\lambda_1}}}^{*}
\int_{-\frac{L^+}{2}}^{\frac{L^+}{2}} dz^+
\int_{\frac{L^+}{2}}^{\infty} dw^+
e^{iw^{+}\Big(\frac{\bp_{1}^{2}}{2p_{1}^{+}} + \frac{\bp_{2}^{2}}{2p_{2}^{+}} - \frac{(\bp_{1} + \bp_{2})^{2}}{2q^{+}}\Big)}
\nn \\
& 
\times
\int_{\bz} \  e^{-i \bz \cdot (\bp_{1}+ \bp_{2}- \bq)}\
\overline{\Psi}_{\beta}(\underline{z}) \gamma^{-}\gamma^{+} \gamma^{l} u(\check{q},h)
\Big[t^{b} \UF(z^{+},-\infty;\bz)\Big]_{\beta\alpha}
\UA(w^{+},z^{+};\bz)_{cb}
\nn \\
&
\Big\{g^{il} \Big(-\bp_{1}^{j} + \frac{p_{1}^{+}}{p_{2}^{+}} \bp_{2}^{j}\Big) 
+ g^{lj}\Big(\bp_{2}^{i} - \frac{p_{2}^{+}}{p_{1}^{+}} \bp_{1}^{i}\Big) 
+ g^{ji}\Big(\bp_{1}^{l} - \frac{p_{1}^{+}}{q^{+}}(\bp^{l}_{1}+ \bp_{2}^{l})\Big)      
\Big\} .
\end{align}
As mentioned earlier, the total scattering amplitude in the $q\to gg$ channel in the general kinematics is given by Eq.~\eqref{eq:M_tot_schm_q_to_gg} with each contribution, corresponding to the three mechanisms, given in Eqs.~\eqref{eq:M_1_q_to_gg}, \eqref{eq:M_2_q_to_gg} and \eqref{eq:M_3_q_to_gg}. 

\subsection{$q\to gg$ amplitude in the back-to-back limit}
The back-to-back limit in this channel can be obtained as discussed in previous channels. Namely, starting from the scattering amplitude in general kinematics for each mechanism, one first performs the change of variables given in Eqs.~\eqref{eq:PK} and \eqref{eq:br} and rewrite these amplitudes in terms of the relative dijet momenta $\P$  and dijet momentum imbalance $\k$, as well as their conjugate variables $\r$ and $\b$. Since the back-to-back limit corresponds to $|\P|\gg |\k|$, or equivalently $|\r|\ll |\b|$, one can perform a Taylor expansion around $\r=0$ and keeping only the zeroth order term in the expansion, the $\r$ and $\b$ dependence in the resulting expression are factorized and the $\r$ integration can be performed. After all said and done, the back-to-back limit of the amplitude in the $q\to gg$ channel for the first mechanism can be written as  
\begin{align}
\label{eq:M_1_q_to_gg_b2b_m0}
&
i\M_{q\to g_{1}g_{2}, \, 1}^{{\rm b2b}, \, m=0}= 
ig^{2}\frac{z(1-z)}{(2q^+)}\frac{1}{\bP^2}
\ {\varepsilon^{j}_{\lambda_1}}^{*} {\varepsilon^{i}_{\lambda_2}}^{*}
 \int_{-\frac{L^+}{2}}^{\frac{L^+}{2}} dz^{+}\ \int_{\bb}  e^{-i(\bk - \bq) \cdot \bb}
\ \overline{\Psi}_{\beta}(z^{+};\bb) 
\frac{\gamma^{-}\gamma^{+}}{2} 
\\
& 
\times
\gamma^{i}\Big[
\frac{\gamma^{l}\gamma^{j} \bP^{l}}{1-z}-2\frac{\bP^{j}}{z}
\Big] 
 \ \UA(\infty,-\infty;\bb)_{a_{1}b_{1}} 
 \ \UA(\infty,z^{+};\bb)_{a_{2}b_{2}} 
 \Big[t^{b_{2}} \UF(z^{+},-\infty;\bb)t^{b_{1}}\Big]_{\beta\alpha}u(\check{q},h) , \nn
\end{align}
where we have also considered the massless quark limit for simplicity. Before we proceed further, we would like to mention that the color structure in Eq.~\eqref{eq:M_1_q_to_gg_b2b_m0} can be simplified by using the properties of the Wilson lines discussed in \Appendix{app:sun}. The final result reads 
\begin{align}
\label{eq:q_to_gg_Simp_Color_1}
&
 \UA(\infty,-\infty;\bb)_{a_{1}b_{1}} \ \UA(\infty,z^{+};\bb)_{a_{2}b_{2}} 
\Big[  \overline{\Psi}(z^{+};\bb)  t^{b_{2}} \UF(z^{+},-\infty;\bb)t^{b_{1}}\Big]
\nn \\
&
= \ \UA(\infty, z^{+};\bb)_{a_{1}c_{1}} \UA(z^{+},-\infty;\bb)_{c_{1}b_{1}} \ \UA(\infty,z^{+};\bb)_{a_{2}b_{2}} 
\Big[ \overline{\Psi}(z^{+};\bb) t^{b_{2}} \UF(z^{+},-\infty;\bb)t^{b_{1}}\Big] 
\nn \\
&
= \ \UA(\infty, z^{+};\bb)_{a_{1}c_{1}} \ \UA(\infty,z^{+};\bb)_{a_{2}b_{2}} 
         \Big[ \overline{\Psi}(z^{+};\bb) t^{b_{2}} t^{c_{1}} \UF(z^{+},-\infty;\bb)\Big] 
\nn \\
&
= \ \UA(\infty, z^{+};\bb)_{a_{1}b_{1}} \ \UA(\infty,z^{+};\bb)_{a_{2}b_{2}} 
         \Big[ \overline{\Psi}(z^{+};\bb) t^{b_{2}} t^{b_{1}} \UF(z^{+},-\infty;\bb)\Big] ,        
\end{align}
where we renamed $c_1\to b_1$ for convenience of the computations. 

The back-to-back limit of the second mechanism can be calculated in the same way. 
After simplification, one can perform the integration over $\bk_{1}$, and obtains the back-to-back and massless quark limits of the scattering amplitude for the second mechanism as 
\begin{align}
\label{eq:M_2_q_to_gg_b2b_m0}
&
i\M_{q\to g_{2}g_{1}, \, 2}^{{\rm b2b}, \, m=0} = 
-ig^{2}\frac{z(1-z)}{(2q^+)}\frac{1}{\bP^2}
\ {\varepsilon^{i}_{\lambda_1}}^{*} {\varepsilon^{j}_{\lambda_2}}^{*}
 \int_{-\frac{L^+}{2}}^{\frac{L^+}{2}} dz^{+}\ \int_{\bb}  e^{-i(\bk - \bq) \cdot \bb}
\ \overline{\Psi}_{\beta}(z^{+};\bb) 
\frac{\gamma^{-}\gamma^{+}}{2} 
\\
& 
\times
\gamma^{i}\Big[
\frac{\gamma^{l}\gamma^{j} \bP^{l}}{z} - 2\frac{\bP^{j}}{1-z}
\Big] 
 \ \UA(\infty,z^{+};\bb)_{a_{1}b_{1}} 
 \ \UA(\infty,-\infty;\bb)_{a_{2}b_{2}} 
 \Big[t^{b_{1}} \UF(z^{+},-\infty;\bb)t^{b_{2}}\Big]_{\beta\alpha}u(\check{q},h) . \nn
\end{align}
Note that the color structure appearing in Eq.~\eqref{eq:M_1_q_to_gg_b2b_m0} is very similar to the color structure given in the last line of Eq.~\eqref{eq:q_to_gg_Simp_Color_1} and it reads 
\begin{align}
&  \hspace{-1.3cm}
 \UA(\infty,-\infty;\bb)_{a_{2}b_{2}}\  \ \UA(\infty,z^{+};\bb)_{a_{1}b_{1}} \bigg[\overline{\Psi}(z^{+};\bb) t^{b_{1}} \UF(z^{+},-\infty;\bb)t^{b_{2}}\bigg] \nn
    \\&
    =
    \ \UA(\infty,z^{+};\bb)_{a_{2}b_{2}}\ \UA(\infty,z^{+};\bb)_{a_{1}b_{1}} \bigg[\overline{\Psi}(z^{+};\bb) t^{b_{1}}t^{b_{2}} \UF(z^{+},-\infty;\bb)\bigg].
\end{align}
Finally, to get the back-to-back limit of the amplitude for the third mechanism, we start from the amplitude in general kinematics given in Eq.~\eqref{eq:M_3_q_to_gg} and follow the same procedure as in the first mechanism which yields to 
\begin{align}
\label{eq:Amp_3_b2b_q_to_gg}
&
i\M_{q\to g_{1}g_{2},\, 3}^{{\rm b2b}, \, m=0} =
g^{2}f^{a_{1}a_{2}c}
{\varepsilon^{j}_{{\lambda_2}}}^{*} 
{\varepsilon^{i}_{{\lambda_1}}}^{*}
\ \frac{z(1-z)}{(2q^+)}\frac{1}{\bP^2}
\int_{-\frac{L^+}{2}}^{\frac{L^+}{2}} dz^+ 
\int_{\bb} \ e^{-i\bb \cdot (\bk - \bq)} 
\ \UA(\infty,z^{+};\bb)_{cb}
\nn \\
& \hspace{0.5cm}
\times        
\Big[g^{il}\frac{\bP^{j}}{1-z} + g^{lj}\frac{\bP^{i}}{z} - g^{ji}\bP^{l} \Big]
\overline{\Psi}_{\beta}(z^{+};\bb) \gamma^{-}\gamma^{+}\gamma^{l} 
\ u(\check{q},h)
\Big[t^{b} \, \UF(z^{+},-\infty;\bb)\Big]_{\beta\alpha} ,
\end{align}
where we have also considered the massless quark limit for simplicity. Note that, by using the relations introduced in  \Appendix{app:sun},  the color structure that appears in Eq.~\eqref{eq:Amp_3_b2b_q_to_gg} can be simplified and it can be written as  
\begin{align}
&
f^{a_{1}a_{2}c} \ \UA(\infty,z^{+};\bb)_{cb}  \Big[ \overline{\Psi}(z^{+};\bb)\ t^{b} \ \UF(z^{+},-\infty;\bb)\Big]
\nn \\
& \hspace{1cm}
  =  \UA(\infty,z^{+};\bb)_{a_{1}b_{1}} \ \UA(\infty,z^{+};\bb)_{a_{2}b_{2}} \ f^{b_{1}b_{2}b} \   \Big[ \overline{\Psi}(z^{+};\bb)\ t^{b} \ \UF(z^{+},-\infty;\bb)\Big]
\nn \\
& \hspace{1cm}
= \UA(\infty,z^{+};\bb)_{a_{1}b_{1}} \ \UA(\infty,z^{+};\bb)_{a_{2}b_{2}} \  (-i) \ 
\Big[ \overline{\Psi}(z^{+};\bb)\ [t^{b_{1}}, t^{b_{2}}] \ \UF(z^{+},-\infty;\bb)\Big] \, . 
\end{align}
The total scattering amplitude in the $g\to qq$ channel in the back-to-back and massless quark limits is then given as 
\begin{align}
\label{eq:M_tot_schm_q_to_gg_b2b_m0}
i\M_{q\to g_{1}g_{2},\, {\rm tot.}}^{{\rm b2b}, \, m=0} =i\M_{q\to g_{1}g_{2},\, 1}^{{\rm b2b}, \, m=0}+ i\M_{q\to g_{1}g_{2},\, 2}^{{\rm b2b}, \, m=0} + i\M_{q\to g_{1}g_{2},\, 3}^{{\rm b2b}, \, m=0} ,
\end{align}
with the amplitude for each mechanism in the back-to-back limit given in Eqs.~\eqref{eq:M_1_q_to_gg_b2b_m0}, \eqref{eq:M_2_q_to_gg_b2b_m0} and \eqref{eq:Amp_3_b2b_q_to_gg}. Thus, the total amplitude can explicitly  be written as 
\begin{align}
\label{eq:M_tot_q_to_gg_b2b_m0_1}
&
i\M_{q\to g_{1}g_{2},\, {\rm tot.}}^{{\rm b2b}, \, m=0}  = 
ig^{2} \ \frac{ z(1-z)}{(2q^+)} \frac{1}{2\bP^{2}} 
{\varepsilon_{\lambda_1}^{i}}^{*}{\varepsilon_{\lambda_2}^{j}}^{*}
\int_{-\frac{L^+}{2}}^{\frac{L^+}{2}} dz^{+}
  \int_{\bb} \ e^{-i\b \cdot (\bk- \bq)} \  \UA (\infty,z^{+};\bb)_{a_{1}b_{1}}    
 \\
&
\times \overline{\Psi}
(z^{+};\bb) 
\frac{\gamma^{-}\gamma^{+}}{2}\gamma^l
\Big[
\{t^{b_{1}},t^{b_{2}}\}
\Big(\frac{\gamma^i\gamma^jP^l}{z} - \frac{\gamma^j\gamma^iP^l}{1-z}\Big)
+[t^{b_1},t^{b_2}] 
\Big(\frac{\gamma^i\gamma^jP^l}{z} + \frac{\gamma^j\gamma^iP^l}{1-z}
- 4zg^{ij}P^l\Big)
\Big] 
\nn \\
&\times \UF(z^{+},-\infty;\bb) u(\check{q},h) 
\UA (\infty,z^{+};\bb)_{a_{2}b_{2}} . \nn  
\end{align}
The color structures can be further simplified. The term with the anticommutator of two color generators can be written as 
\begin{align}
\label{eq:q_to_gg_Simp_CS_2}
&
\UA (\infty,z^{+};\bb)_{a_{1}b_{1}} \  \UA (\infty,z^{+};\bb)_{a_{2}b_{2}}
\ \overline{\Psi}(z^{+};\bb) 
\  \{t^{b_{1}},t^{b_{2}}\} 
\ \UF(z^{+},-\infty;\bb)  
\nn \\
&  \hspace{2cm}
= \UA (\infty,z^{+};\bb)_{a_{1}b_{1}} \  \UA (\infty,z^{+};\bb)_{a_{2}b_{2}} 
\  \overline{\Psi}(z^{+};\bb) 
\Big(\frac{\delta^{b_1 b_2}}{N_{c}} + d^{b_1 b_2 c}\ t^{c}\Big)
\  \UF(z^{+},-\infty;\bb)
\nn \\
&  \hspace{2cm}
= \overline{\Psi}(z^{+};\bb) \Big( \frac{\delta^{a_1 a_2}}{N_{c}} + d^{a_1 a_2 d} \ t^{c} \ \UA (\infty, z^{+};\bb)_{dc}\Big)\  \UF(z^{+},-\infty;\bb) .
\end{align}      
Similarly, the term with the commutator of two color generators can be written as 
\begin{align}
\label{eq:q_to_gg_Simp_CS_3}
&
\UA (\infty,z^{+};\bb)_{a_{1}b_{1}}
\ \UA (\infty,z^{+};\bb)_{a_{2}b_{2}}
\ \overline{\Psi}(z^{+};\bb)
\ \{t^{b_{1}},t^{b_{2}}\} 
\ \UF(z^{+},-\infty;\bb)  
\nn  \\ 
& \hspace{2cm}
= \UA(\infty,z^{+};\bb)_{a_{1}b_{1}}
\  \UA (\infty,z^{+};\bb)_{a_{2}b_{2}}
\  \overline{\Psi}(z^{+};\bb) 
\big( i f^{b_{1}b_{2}c} \ t^{c}\big) 
\ \UF(z^{+},-\infty;\bb)
\nn \\
& \hspace{2cm}
= i\overline{\Psi}(z^{+};\bb) \ f^{a_1 a_2 d} \ t^{c} 
\ \UA (\infty, z^{+};\bb)_{dc} \ \UF(z^{+},-\infty;\bb) .
 \end{align}
Substituting Eqs.~\eqref{eq:q_to_gg_Simp_CS_2} and \eqref{eq:q_to_gg_Simp_CS_3} into the total amplitude given in Eq.~\eqref{eq:M_tot_q_to_gg_b2b_m0_1}, one obtains  
\begin{align}
\label{eq:M_tot_q_to_gg_b2b_m0_2}
i\M_{q\to g_{1}g_{2},\, {\rm tot.}}^{{\rm b2b}, \, m=0}   & = 
\frac{ig^{2}}{(2q^+)} \frac{1}{2\bP^{2}}
\int_{-\frac{L^+}{2}}^{\frac{L^+}{2}} dz^{+}\
\int_{\bb} \ e^{-i\b \cdot (\bk- \bq)}
\ \overline{\Psi}
(z^{+};\bb) 
\frac{\gamma^{-}\gamma^{+}}{2} 
\nn \\
& \hspace{-1cm}
\times 
\Big[
\mathfrak{h}^{(1)}_{q\to gg}
\Big(\frac{\delta^{a_1 a_2}}{N_{c}} + d^{a_1 a_2 d}  \ \UA (\infty, z^{+};\bb)_{dc}\ t^{c}\Big)  
+ 
\mathfrak{h}^{(2)}_{q\to gg} 
\Big( if^{a_1 a_2 d}  \ \UA (\infty, z^{+};\bb)_{dc}\ t^{c} \Big)
\Big] 
\nn \\
&  \hspace{-1cm}
\times
\UF(z^{+},-\infty;\bb)  \  u(\check{q},h) ,     
\end{align}
with the two coefficients being 
\begin{align}
\label{eq:HF_q_to_gg_1}
\mathfrak{h}^{(1)}_{q\to gg} &= 
{\varepsilon_{\lambda_1}^{i}}^{*}{\varepsilon_{\lambda_2}^{j}}^{*}\gamma^l\bP^l
\left[(1-z)\gamma^i\gamma^j - z\gamma^j\gamma^i\right]
, \\
\label{eq:HF_q_to_gg_2}
\mathfrak{h}^{(2)}_{q\to gg} &= 
{\varepsilon_{\lambda_1}^{i}}^{*}{\varepsilon_{\lambda_2}^{j}}^{*}\gamma^l\bP^l
\left[(1-z)\gamma^i\gamma^j + z\gamma^j\gamma^i - 4z(1-z)g^{ij}\right] \nn \\
&= (1-2z)\mathfrak{h}^{(1)}_{q\to gg}.
\end{align}
%

\subsection{$q\to gg$ production cross section in the back-to-back limit}
\label{subsec:xsection_q-gg}
In the $q\to gg$ channel, partonic cross section in the back-to-back and massless quark limits can be written as 
\begin{equation}
\frac{d\sigma^{{\rm b2b}, \, m=0}_{q\to gg}}{d{\rm P.S.}} =
(2q^+) \ 2\pi \ \delta\left(p_1^+ + p_2^+ - q^+\right)
\ \frac{1}{2N_c}\sum_{\lambda_1,\lambda_2}\sum_{h}\sum_{a_1,a_2}
\left\langle \left|i\mathcal{M}^{{\rm b2b}, \, m=0}_{q\to gg, {\rm tot.}}\right|^2 \right\rangle ,
\end{equation}
where the phase space is defined in Eq.~\eqref{def:PS}, the back-to-back production amplitude is given in Eq.~\eqref{eq:M_tot_q_to_gg_b2b_m0_2}, and the $\langle \cdots\rangle$ stands for target averaging in the CGC formalism. Following the same procedure as in the previously discussed channels, we can write the partonic cross section in the back-to-back limit as 
\begin{align}
& 
\frac{d\sigma^{{\rm b2b}, \, m=0}_{q\to gg}}{d{\rm P.S.}}
= -\frac{g^{4} }{16 N_{c} \bP^4} \ (2\pi)\ \delta\left(k^+ - q^+\right)  \ 
\int_{z^+,{z'}^+}
\int_{\bb, \bb'} e^{-i(\bb-\bb')\cdot(\bk - \bq)} 
\\
& 
\times\,
\bigg\langle \Tr \bigg\{\Psi(z'^{+};\bb')
\overline{\Psi}(z^{+};\bb) \gamma^{-} 
\bigg(\sum_{\lambda_1 \lambda_2} 
\big|\mathfrak{h}^{(1)}_{q\to gg} \big|^{2}
\bigg)
\nn\\
& 
\times
\left[\frac{\delta^{a_1 a_2}}{N_{c}}
+\left(d^{a_1 a_2 d} + (1\!-\!2z)if^{a_1 a_2 d}
\right)
\UA (\infty, z^{+};\bb)_{dc} \ t^{c}
\right]
\, \UF(z^{+},-\infty;\bb)\,
\UF^{\dagger}(z'^{+},-\infty;\bb')
\nn\\
& 
\times
\left[\frac{\delta^{a_1 a_2}}{N_{c}}
+
\left(d^{a_1 a_2 d'} + (1\!-\!2z)(-i)f^{a_1 a_2 d'}
\right)
\ \UA (\infty, z'^{+};\bb')_{d'c'}\, t^{c'}
\right]
\bigg\}
\bigg\rangle 
\, ,
\label{eq:sigma_q_to_gg_1}
\end{align}
where the trace $\Tr$ acts both on the fundamental color indices and on the Dirac spinor indices.
The explicit computation for the coefficients $\mathfrak{h}^{1}_{q\to gg}$ is presented in \Appendix{app:dirac} and the final result is given as 
\begin{align}    
\sum_{\lambda_1,\lambda_2}\left|\mathfrak{h}^{(1)}_{q\to gg}\right|^2 &= 
-4\bP^2\left(z^{2} + (1-z)^{2}\right) 
.
\end{align}
\begin{figure}[H]
\centering
\begin{subfigure}{0.49\textwidth}
\centering
\includegraphics[height=5cm]{Color_diagrams/pdf/bC-.pdf}
\caption{$\overline{\mathcal{C}}^{-}$}
\label{fig:q-gg:bC-}
\end{subfigure}
\begin{subfigure}{0.49\textwidth}
\centering
\includegraphics[height=5cm]{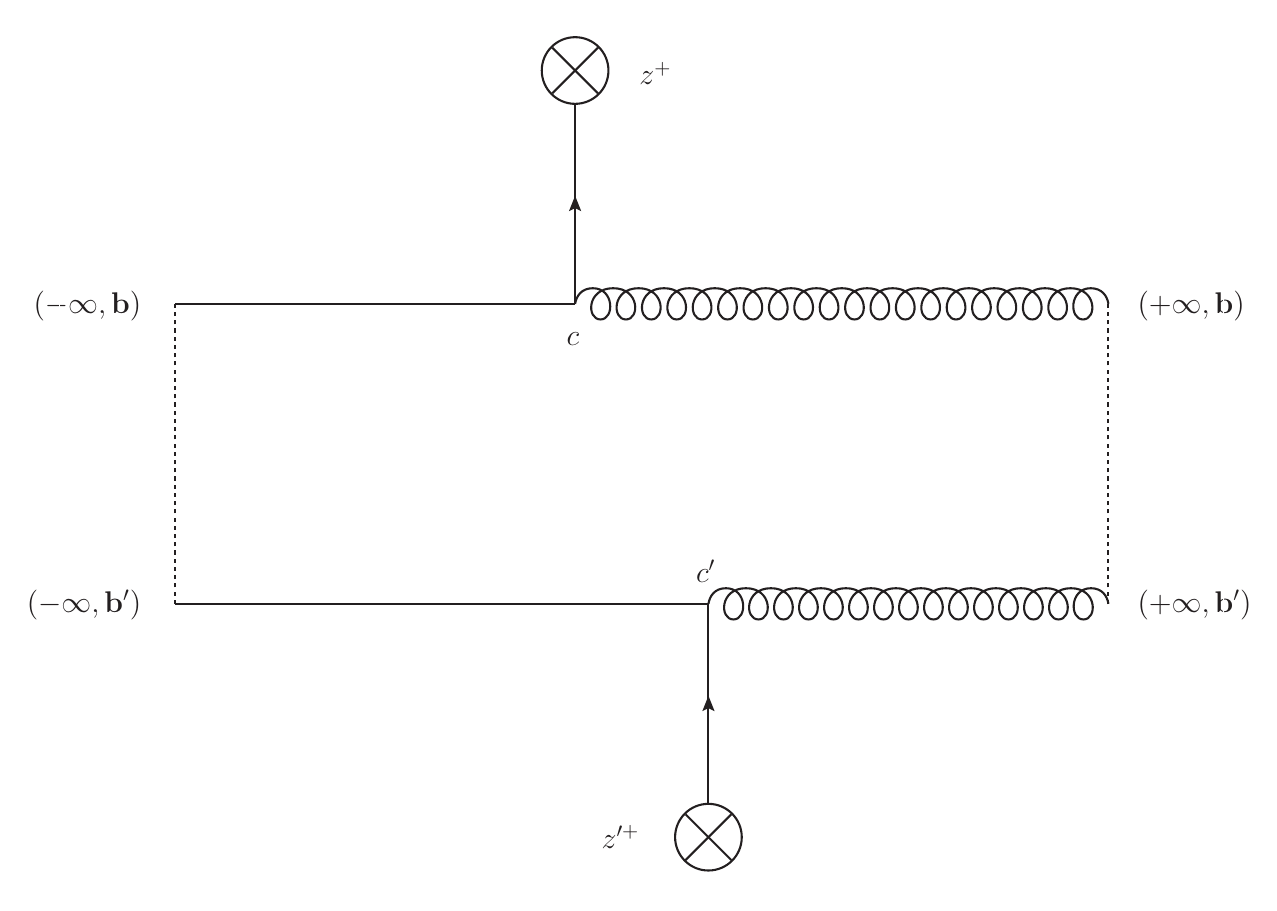}
\caption{$\overline{\mathcal{C}}^{-g}$}
\label{fig:q-gg:bC-g}
\end{subfigure}
\caption{Color structures appearing in the cross section of $q\to gg$.}
\label{fig:Color_q-gg}
\end{figure}
%
%
%
%
%
%
%
%
%
%
On the other hand, the gauge link structures that are appearing in Eq.~\eqref{eq:sigma_q_to_gg_1} can be simplified as 
\begin{align}
&
\left[\frac{\delta^{a_1 a_2}}{N_{c}}
+\left(d^{a_1 a_2 d} + (1\!-\!2z)if^{a_1 a_2 d}
\right)
\UA (\infty, z^{+};\bb)_{dc} \ t^{c}
\right]
\, \UF(z^{+},-\infty;\bb)\,
\UF^{\dagger}(z'^{+},-\infty;\bb')
\nn\\
& 
\times
\left[\frac{\delta^{a_1 a_2}}{N_{c}}
+
\left(d^{a_1 a_2 d'} + (1\!-\!2z)(-i)f^{a_1 a_2 d'}
\right)
\ \UA (\infty, z'^{+};\bb')_{d'c'}\, t^{c'}
\right]
\nn\\
=&\, 
\frac{(N_c^2\!-\!1)}{N_c^2}\ 
\UF(z^{+},-\infty;\bb)\,
\UF^{\dagger}(z'^{+},-\infty;\bb')
\nn\\
&
+ \left[\frac{(N_c^2\!-\!4)}{N_c}
+(1\!-\!2z)^2 N_c\right]\,
\UA (\infty, z^{+};\bb)_{dc}\, 
\UA (\infty, z'^{+};\bb')_{dc'}\
\nn\\
&\ \ \
 \times\ 
t^{c}\,
\UF(z^{+},-\infty;\bb)\,
\UF^{\dagger}(z'^{+},-\infty;\bb')\,t^{c'}
\, ,
\end{align}
so that the cross section for this channel can be written in terms of the reduced color structure $\overline{\mathcal{C}}^{-}$ defined in \Equation{eq:bC-} and a new one defined as
\begin{align}
\label{eq:bC-g}
\overline{\C}^{-g} &\equiv \Big\langle 
\UA(\infty,z'^{+};\bb')_{dc'}
\ \UA(\infty,z^{+};\bb)_{dc} 
\nn \\
&
\times\, \Tr \Big\{\UFd(z'^{+},-\infty;\bb') \  t^{c'} \ \Psi(z'^{+};\bb')
\overline{\Psi}(z^{+};\bb) \ \gamma^{-} \  t^{c} \ \UF(z^{+},-\infty;\bb) \Big\}  \Big\rangle ,
\end{align}
which are illustrated in Fig.~\ref{fig:Color_q-gg}. Finally, we can write the partonic cross section in the $q\to qg$ channel in the following factorized form
\begin{align}
\label{eq:sigma_q_to_gg_2}
& 
\frac{d\sigma^{{\rm b2b}, \, m=0}_{q\to gg}}{d{\rm P.S.}} = 
g^{4} \ (2\pi) \delta\left(k^+ \!-\! q^+\right) 
\int_{-\frac{L^+}{2}}^{\frac{L^+}{2}} dz^+ 
\int_{-\frac{L^+}{2}}^{\frac{L^+}{2}} d{z'}^+ 
\int_{\bb, \bb'} e^{-i(\bb-\bb')\cdot(\bk - \bq)}  
\nn \\
& \hspace{8cm}
\times 
\Big[ 
\mathcal{H}^{-}_{q\to gg} \ \overline{\mathcal{C}}^{-}
 +   
 \mathcal{H}^{-g}_{q\to gg} \  \overline{\mathcal{C}}^{-g} 
 \Big] ,
\end{align}
with the hard factors defined as 
\begin{align}
\label{eq:H_q_to_gg_-g}
\mathcal{H}^{-g}_{q\to gg} &=\frac{\left[z^{2} + (1\!-\!z)^{2}\right]}{2\P^2} 
\left[ z^{2} + (1-z)^{2} 
-\frac{2}{N_{c}^2} \right] 
, \\
\label{eq:H_q_to_gg_-}
\mathcal{H}^{-}_{q\to gg} &=\frac{(N_{c}^{2}\!-\!1)}{4 \, N_{c}^3 \P^2} \left[z^{2} + (1\!-\!z)^{2}\right] .
\end{align}
%
%


\section{$q\to qq$ channel} 
\label{sec:qq}                    

\begin{figure}[H]
\centering
\begin{subfigure}{0.49\textwidth}
\centering
\includegraphics[width=\textwidth]{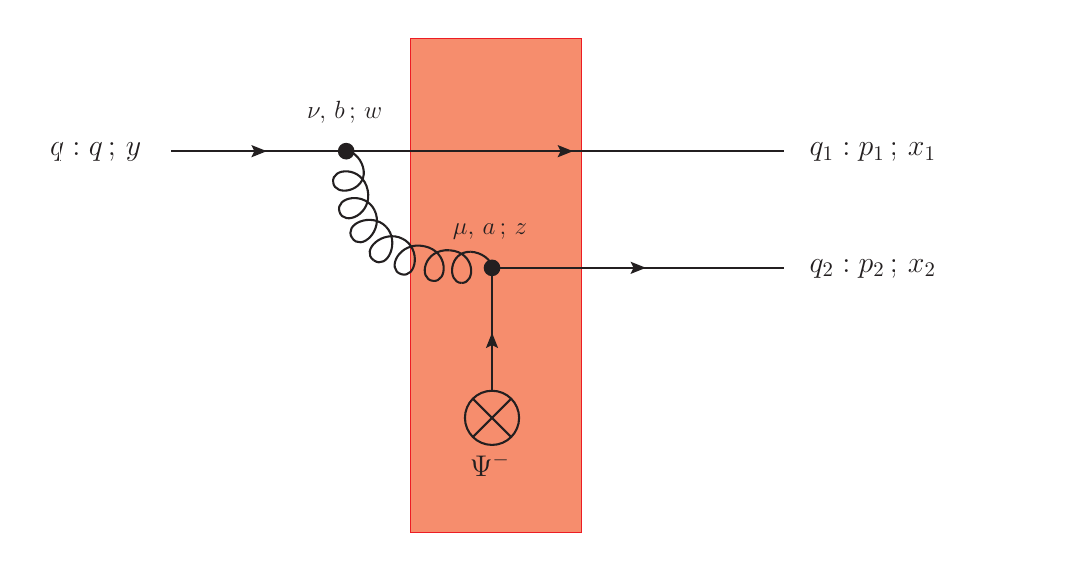}
\caption{Diagram 1}
\label{fig:q-qq1}
\end{subfigure}
\begin{subfigure}{0.49\textwidth}
\centering
\includegraphics[width=\textwidth]{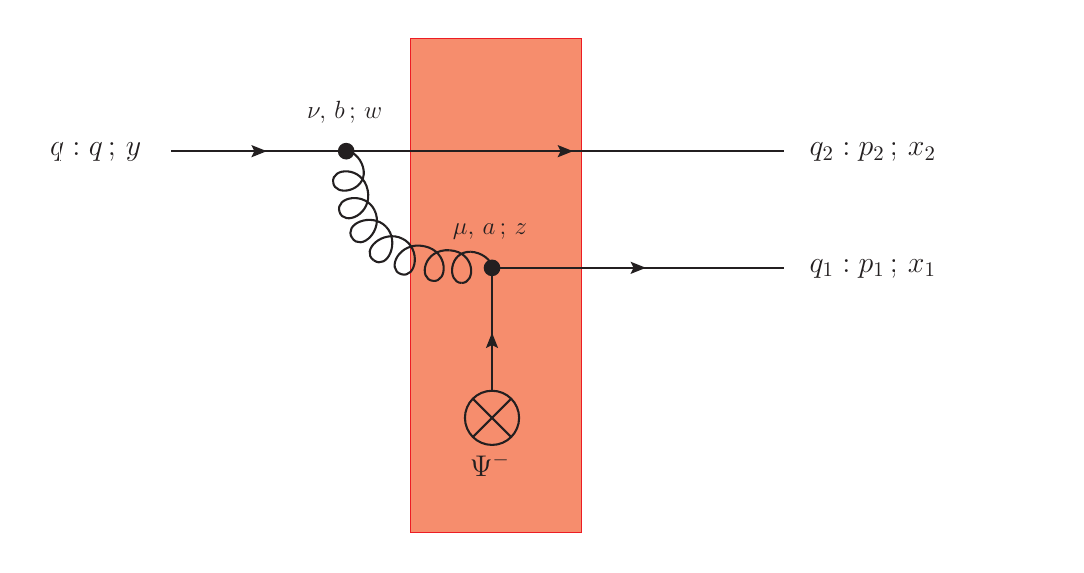}
\caption{Diagram 2}
\label{fig:q-qq2}
\end{subfigure}
\caption{Diagrams contributing to channel $q \to q q$.}
\label{fig:q-qq}
\end{figure}

The next process we would like to consider is quark-quark jet production in quark initiated channel, i.e. 
$q\to qq$ channel. 
We will first consider the case where all quarks have the same flavor.
In this case, two mechanisms contribute to this channel at NEik accuracy. One where the incoming quark splits into a quark-gluon pair before the medium, and the pair scatters on the target. While the quark scatters eikonally, the gluon scatters via a $t$-channel quark exchange and converts into a quark, producing quark dijet in the final state (see Fig.~\ref{fig:q-qq1}). The second one is similar except the quarks in the final state are interchanged (see \Figure{fig:q-qq2}).
At the end of this section, we will also consider the case where the final quarks have different flavors (more explicitly, $q_f\to q_fq_{f'}$ with $f\neq f'$), denoted $q\to qq'$. In this instance, only the first discussed diagram contributes (see Fig.~\ref{fig:q-qq1}).
%
%
%
%

%
%
\subsection{$q\to qq$ amplitude in general kinematics}
The total scattering amplitude in $q\to qq$ channel receives two contributions from the aforementioned two mechanism and can be written as 
\begin{align}
\label{eq:M_tot_schm_q_to_q}
\M_{q\to qq, \, {\rm tot.}}= \M_{q\to qq, \, 1} + \M_{q\to qq, \, 2} .
\end{align}
The $S$-matrix element for the first mechanism, illustrated in \Figure{fig:q-qq1} can be written as
\begin{align}
\S_{q\to qq,1} &= 
\lim_{y^+\to-\infty}\lim_{x_1^+,x_2^+\to\infty}
\int_{\by,\bx[1],\bx[2]}\int_{y^-,x^-_1,x^-_2}
\int_{\bw, \bz}\int_{w^-, z^-}
\int_{-\frac{L^+}{2}}^{\frac{L^+}{2}}d z^+
\int_{-\infty}^{-\frac{L^+}{2}}d w^+
\nn \\
&
\times 
e^{ix_{1} \cdot \check{p}_{1}} \ e^{ix_{2} \cdot \check{p}_{2}} \ e^{-iy \cdot \check{q}} \
\overline{u}(\check{p}_{1},h_{1}) \ \gamma^{+}  
\ \Big[  S_{F}(x_{1},w) \BAq (-igt^{b}\gamma_{\nu}) \ S_{0,F}(w,y)\Big]_{\alpha_{1}\alpha} 
\gamma^{+} u(\check{q},h)
\nn \\
&
\times  
\Big[ G^{\mu\nu}_{F}(z,w) \BI\Big]_{ab} 
\overline{u}(\check{p}_{2},h_{2}) \gamma^{+} 
\Big[ S_{F}(x_{2}, z) \IAq   (-igt^{a}\gamma_{\mu})\Big]_{\alpha_{2}\beta} 
\Psi^{-}_{\beta}(z) ,
\end{align}
where the before-to-after quark propagator is given in Eq.~\eqref{quark_prop_BA}, the before-to-inside gluon propagator is given in Eq.~\eqref{gluon_prop_BI}, the inside-to-after quark propagator is given in Eq.~\eqref{quark_prop_IA} and the vacuum quark propagator is given in Eq.~\eqref{eq:quark_vacuum_prop}. 
Using the explicit expressions for the propagators and simplifying the Dirac structure, one obtains 
\begin{align}
&
\S_{q\to qq,1} = 
\frac{g^{2}}{2p_{2}^{+}}  \ (2\pi) \delta(p_{1}^{+}+p_{2}^{+}-q^{+})  \int\! \frac{d^{2} \bk_{1}}{(2\pi)^{2}} \int_{\bz, \z_1}  e^{-i \bz \cdot (\bp_{2}+ \bk_{1}-\bq)  } \ e^{-i \bz_{1} \cdot (\bp_{1}-\bk_{1})} 
\nn \\
&
\times
 \int_{-\frac{L^+}{2}}^{\frac{L^+}{2}}dz^+
\int_{-\infty}^{-\frac{L^+}{2}} dw^+ 
e^{iw^{+}\Big(\frac{(\bk_{1}^{2}+m^2)}{2p_{1}^{+}} 
+ \frac{(\bk_{1}- \bq)^{2}}{2p_{2}^{+}} -\frac{(\bq^{2}+m^2)}{2q^{+}}  \Big)} 
\UA(z^{+},w^{+};\bz)_{ab} 
 \Big[ \UF(\infty,w^{+};\bz_{1})\ t^{b}\Big]_{\alpha_{1}\alpha}
\nn \\
&
\times
\Big[ \UF(\infty,z^{+};\bz) \ t^{a}\Big]_{\alpha_{2}\beta}  \overline{u}(\check{p}_{1},h_{1}) \gamma^{+} 
\Big[ 
\frac{\gamma^{i}\gamma^{l} \bq^{l}}{2q^{+}} 
+ \frac{\gamma^{l}\gamma^{i} \bk_{1}^{l}}{2p_{1}^{+}} 
+ \gamma^{i}\Big(\frac{1}{2q^{+}}-\frac{1}{2p_{1}^{+}}\Big)m 
+ \frac{\bq^{i}- \bk_{1}^{i}}{p_{2}^{+}}
\Big] 
\nn \\
&
\times
u(\check{q},h) \ \overline{u}(\check{p}_{2},h_{2})
\ \gamma^{i} \frac{\gamma^{+}\gamma^{-}}{2}
\ \Psi_{\beta}(\underline{z}) , 
\end{align}
from which we can extract the amplitude as 
\begin{align}
\label{eq:M_q_to_qq_1}
&
i \M_{q\to qq,1} = 
\frac{g^{2}}{(2q^+)(2p_{2}^{+})}   \int\! \frac{d^{2} \bk_{1}}{(2\pi)^{2}} \int_{\bz, \z_1}  e^{-i \bz \cdot (\bp_{2}+ \bk_{1}-\bq)  } \ e^{-i \bz_{1} \cdot (\bp_{1}-\bk_{1})} 
\nn \\
&
\times
 \int_{-\frac{L^+}{2}}^{\frac{L^+}{2}}dz^+
\int_{-\infty}^{-\frac{L^+}{2}} dw^+ 
e^{iw^{+}\Big(\frac{(\bk_{1}^{2}+m^2)}{2p_{1}^{+}} 
+ \frac{(\bk_{1}- \bq)^{2}}{2p_{2}^{+}} -\frac{(\bq^{2}+m^2)}{2q^{+}} \Big)} 
\UA(z^{+},w^{+};\bz)_{ab} 
 \Big[ \UF(\infty,w^{+};\bz_{1})\ t^{b}\Big]_{\alpha_{1}\alpha}
\nn \\
&
\times
\Big[ \UF(\infty,z^{+};\bz) \ t^{a}\Big]_{\alpha_{2}\beta}  \overline{u}(\check{p}_{1},h_{1}) \gamma^{+} 
\Big[ 
\frac{\gamma^{i}\gamma^{l} \bq^{l}}{2q^{+}} 
+ \frac{\gamma^{l}\gamma^{i} \bk_{1}^{l}}{2p_{1}^{+}} 
+ \gamma^{i}\Big(\frac{1}{2q^{+}}-\frac{1}{2p_{1}^{+}}\Big)m 
+ \frac{\bq^{i}- \bk_{1}^{i}}{p_{2}^{+}}
\Big] 
\nn \\
&
\times
u(\check{q},h) \ \overline{u}(\check{p}_{2},h_{2})
\ \gamma^{i} \frac{\gamma^{+}\gamma^{-}}{2}
\ \Psi_{\beta}(\underline{z}) . 
\end{align}
The second mechanism is almost the same as the first one except that the quarks in the final state are exchanged (see \Figure{fig:q-qq2}).
In the calculation of the first diagram, from \Figure{fig:q-qq1}, the asymptotic final state was implicitly defined as $\langle 0 | b_2 b_1 = -\langle 0 | b_1 b_2$. In order to be able to add the two diagrams, they must be calculated using the same definition of the final state. Hence, the $S$-matrix for the diagram of \Figure{fig:q-qq2} differs from the one of \Figure{fig:q-qq1} not only by exchange of the momenta and labels associated to each of the two final state quarks, but also by an overall $-$ sign. This sign difference ensures that the total amplitude, obtained from the sum of the diagrams from  \Figure{fig:q-qq1} and \Figure{fig:q-qq2} is antisymmetric by permutation of the two quarks in the final state, as expected for fermions. 
Then, we get
\begin{align}
\S_{q\to qq,2} &= -
\lim_{y^+\to-\infty}\lim_{x_1^+,x_2^+\to\infty}
\int_{\by,\bx[1],\bx[2]}\int_{y^-,x^-_1,x^-_2}
\int_{\bw, \bz}\int_{w^-, z^-}
\int_{-\frac{L^+}{2}}^{\frac{L^+}{2}}d z^+
\int_{-\infty}^{-\frac{L^+}{2}}d w^+
\nn \\
&
\times 
e^{ix_{2} \cdot \check{p}_{2}} \ e^{ix_{1} \cdot \check{p}_{1}} \ e^{-iy \cdot \check{q}} \ 
\overline{u}(\check{p}_{2},h_{2}) \ \gamma^{+}  
\ \Big[  S_{F}(x_{2},w) \BAq (-igt^{b}\gamma_{\nu}) \ S_{0,F}(w,y)\Big]_{\alpha_{2}\alpha} 
\gamma^{+} u(\check{q},h)
\nn \\
&
\times  
\Big[ G^{\mu\nu}_{F}(z,w) \BI\Big]_{ab} 
\overline{u}(\check{p}_{1},h_{1}) \gamma^{+} 
\Big[ S_{F}(x_{1}, z) \IAq   (-igt^{a}\gamma_{\mu})\Big]_{\alpha_{1}\beta} 
\Psi^{-}_{\beta}(z) ,
\end{align}
where the before-to-after quark propagator is given in Eq.~\eqref{quark_prop_BA}, the before-to-inside gluon propagator is given in Eq.~\eqref{gluon_prop_BI}, the inside-to-after quark propagator is given in Eq.~\eqref{quark_prop_IA} and the vacuum quark propagator is given in Eq.~\eqref{eq:quark_vacuum_prop}. 
Using the explicit expressions for the propagators and simplifying the Dirac structure, one obtains 
\begin{align}
&
\S_{q\to qq,2} = 
-\frac{g^{2} }{2p_{1}^{+}}  \ (2\pi) \delta(p_{1}^{+}+p_{2}^{+}-q^{+})  \int\! \frac{d^{2} \bk_{2}}{(2\pi)^{2}} \int_{\bz, \z_1}  e^{-i \bz \cdot (\bp_{1}+ \bk_{2}-\bq)  } \ e^{-i \bz_{1} \cdot (\bp_{2}-\bk_{2})} 
\nn \\
&
\times
 \int_{-\frac{L^+}{2}}^{\frac{L^+}{2}}dz^+
\int_{-\infty}^{-\frac{L^+}{2}} dw^+ 
e^{iw^{+}\Big(
 \frac{(\bk_{2}- \bq)^{2}}{2p_{1}^{+}}
+\frac{(\bk_{2}^{2}+m^2)}{2p_{2}^{+}} 
 -\frac{(\bq^{2}+m^2)}{2q^{+}} 
\Big)} 
\UA(z^{+},w^{+};\bz)_{ab} 
 \Big[ \UF(\infty,w^{+};\bz_{1})\ t^{b}\Big]_{\alpha_{2}\alpha}
\nn \\
&
\times
\Big[ \UF(\infty,z^{+};\bz) \ t^{a}\Big]_{\alpha_{1}\beta}  \overline{u}(\check{p}_{2},h_{2}) \gamma^{+} 
\Big[ 
\frac{\gamma^{i}\gamma^{l} \bq^{l}}{2q^{+}} 
+ \frac{\gamma^{l}\gamma^{i} \bk_{2}^{l}}{2p_{2}^{+}} 
+ \gamma^{i}\Big(\frac{1}{2q^{+}}-\frac{1}{2p_{2}^{+}}\Big)m 
+ \frac{\bq^{i}- \bk_{2}^{i}}{p_{1}^{+}}
\Big] 
\nn \\
&
\times
u(\check{q},h) \ \overline{u}(\check{p}_{1},h_{1})
\ \gamma^{i} \frac{\gamma^{+}\gamma^{-}}{2}
\ \Psi_{\beta}(\underline{z}) , 
\end{align}
from which we can extract the amplitude as 
\begin{align}
\label{eq:M_q_to_qq_2}
&
i \M_{q\to qq,2} = -
  \ \frac{g^{2}}{(2q^+)(2p_{1}^{+})}   \int\! \frac{d^{2} \bk_{2}}{(2\pi)^{2}} \int_{\bz, \z_1}  e^{-i \bz \cdot (\bp_{1}+ \bk_{2}-\bq)  } \ e^{-i \bz_{1} \cdot (\bp_{2}-\bk_{2})} 
\nn \\
&
\times
 \int_{-\frac{L^+}{2}}^{\frac{L^+}{2}}dz^+
\int_{-\infty}^{-\frac{L^+}{2}} dw^+ 
e^{iw^{+}\Big(
 \frac{(\bk_{2}- \bq)^{2}}{2p_{1}^{+}}
+\frac{(\bk_{2}^{2}+m^2)}{2p_{2}^{+}} 
 -\frac{(\bq^{2}+m^2)}{2q^{+}} 
\Big)} 
\UA(z^{+},w^{+};\bz)_{ab} 
 \Big[ \UF(\infty,w^{+};\bz_{1})\ t^{b}\Big]_{\alpha_{2}\alpha}
\nn \\
&
\times
\Big[ \UF(\infty,z^{+};\bz) \ t^{a}\Big]_{\alpha_{1}\beta}  \overline{u}(\check{p}_{2},h_{2}) \gamma^{+} 
\Big[ 
\frac{\gamma^{i}\gamma^{l} \bq^{l}}{2q^{+}} 
+ \frac{\gamma^{l}\gamma^{i} \bk_{2}^{l}}{2p_{2}^{+}} 
+ \gamma^{i}\Big(\frac{1}{2q^{+}}-\frac{1}{2p_{2}^{+}}\Big)m 
+ \frac{\bq^{i}- \bk_{2}^{i}}{p_{1}^{+}}
\Big] 
\nn \\
&
\times
u(\check{q},h) \ \overline{u}(\check{p}_{1},h_{1})
\ \gamma^{i} \frac{\gamma^{+}\gamma^{-}}{2}
\ \Psi_{\beta}(\underline{z}) . 
\end{align}
\subsection{$q\to qq$ amplitude in the back-to-back limit}
In order to get the back-to-back limit of the amplitude we follow the same procedure as in the previous channels. We first write the amplitudes for each mechanism in terms of the relative dijet momentum and dijet momentum imbalance together with their conjugate variables. Then one considers the back-to-back limit which corresponds to small dipole size. One can perform a Taylor expansion for the small parameter and take the zeroth order term in the expansion. All in all, the back-to-back and massless quark limits of the scattering amplitude in the $q\to qq$ channel for the first mechanism can be written as 
\begin{align}
\label{eq:M_q_to_qq_b2b_1}
&
i\M^{{\rm b2b}, \, m=0}_{q\to qq,1}= 
-ig^{2} \ \frac{z}{(2q^{+})^2 \ \bP^{2}}  
\int_{-\frac{L^+}{2}}^{\frac{L^+}{2}} dz^+ 
\int_\bb \ e^{-i \bb \cdot (\bk- \bq)  } 
\Big[ \UF(\infty, -\infty;\bb)\ t^{b}\Big]_{\alpha_{1}\alpha}  
\UA(z^{+}, -\infty;\bb)_{ab}
\nn \\
&
\times
\Big[ \UF(\infty,z^{+};\bb) \ t^{a}\Big]_{\alpha_{2}\beta}        
\overline{u}(\check{p}_{1},h_{1}) \gamma^{+}
\bigg[ \frac{\gamma^{l}\gamma^{i} \bP^{l}}{z}-\frac{2 \bP^{i}}{1-z} \bigg] 
u(\check{q},h) \  \overline{u}(\check{p}_{2},h_{2})\ 
\gamma^{i} \frac{\gamma^{+}\gamma^{-}}{2}
\ \Psi_{\beta}(z^{+};\bb) .
\end{align}
By using the properties of the Wilson lines discussed in \Appendix{app:sun}, the color structure appearing in Eq.~\eqref{eq:M_q_to_qq_b2b_1} can be simplified as 
\begin{align}
\label{eq:CS_simplified_q_to_qq}
& \hspace{-0.3cm}
\Big[ \UF(\infty, -\infty;\bb)\ t^{b}\Big]_{\alpha_{1}\alpha} 
\UA(z^{+}, -\infty;\bb)_{ab}
\Big[ \UF(\infty,z^{+};\bb) t^{a}\Big]_{\alpha_{2}\beta} 
\Psi_{\beta}(z^{+};\bb) 
\nn \\
&
=  \Big[ \UF(\infty, z^{+};\bb)\ t^{a}\  
\UF(z^{+},-\infty;\bb)
\Big]_{\alpha_{1}\alpha}  
\
\Big[ \UF(\infty,z^{+};\bb) t^{a}\Big]_{\alpha_{2}\beta} 
\Psi_{\beta}(z^{+};\bb) 
\nn \\
&
= \frac{1}{2} \
 \UF(\infty, -\infty;\bb)_{\alpha_{2}\alpha} 
 \ \UF(\infty, z^{+};\bb)_{\alpha_{1}\beta} 
 \ \Psi_{\beta}(z^{+};\bb) 
\nn \\
& \hspace{4cm}
-\frac{1}{2 N_{c}} 
\ \UF(\infty, -\infty;\bb)_{\alpha_{1}\alpha} 
\ \UF(\infty, z^{+};\bb)_{\alpha_{2}\beta}
\ \Psi_{\beta}(z^{+};\bb) .
\end{align}
We proceed the same way for the second mechanism and obtain the back-to-back and massless quark limits of the scattering amplitude in \Equation{eq:M_q_to_qq_2} as
\begin{align}
\label{eq:M_q_to_qq_b2b_2}
&
i\M^{{\rm b2b}, \, m=0}_{q\to qq,2}
= 
-ig^{2} \ \frac{(1\!-\!z)}{(2q^{+})^2 \ \bP^{2}}  
\int_{-\frac{L^+}{2}}^{\frac{L^+}{2}} dz^+ 
\int_\bb \ e^{-i \bb \cdot (\bk- \bq)  } 
\Big[ \UF(\infty, -\infty;\bb)\ t^{b}\Big]_{\alpha_{2}\alpha}  
\UA(z^{+}, -\infty;\bb)_{ab}
\nn \\
&
\times
\Big[ \UF(\infty,z^{+};\bb) \ t^{a}\Big]_{\alpha_{1}\beta}        
\overline{u}(\check{p}_{2},h_{2}) \gamma^{+}
\bigg[ \frac{\gamma^{l}\gamma^{i} \bP^{l}}{1-z}-\frac{2 \bP^{i}}{z} \bigg] 
u(\check{q},h) \  \overline{u}(\check{p}_{1},h_{1})\ 
\gamma^{i} \frac{\gamma^{+}\gamma^{-}}{2}
\ \Psi_{\beta}(z^{+};\bb) 
\end{align}
Then, the color structure can be simplified using the identity \eqref{eq:CS_simplified_q_to_qq} with the fundamental color indices $\alpha_1$ and $\alpha_2$ interchanged. 

The total scattering amplitude in the $q \to qq$ channel in the back-to-back and massless quarks limits is then given by
\begin{align}
&
\label{eq:M_tot_q_to_qq_b2b_m0}
i \M_{q\to qq, \, {\rm tot.}}^{{\rm b2b}, \, m=0}   =
i \M_{q\to qq, \, 1}^{{\rm b2b}, \, m=0}
+ i \M_{q\to qq, \, 2}^{{\rm b2b}, \, m=0}
\\
&
=- \frac{i}{2}\frac{g^{2}}{(2q^{+})^2} \frac{1}{\P^2}
\int_{-\frac{L^+}{2}}^{\frac{L^+}{2}} dz^{+}  
\int\! d^{2}\bb\ e^{-i \bb \cdot {(\bk- \bq)}}
\Bigg\{
    \UF(\infty,-\infty;\bb)_{\alpha_2\alpha}
    \UF(\infty,z^{+};\bb)_{\alpha_{1}\beta}
    \mathfrak{h}_{q\to q q}^{(1)}
    \nn \\ & \hspace{6cm}
    +\UF(\infty,-\infty;\bb)_{\alpha_1\alpha}
    \UF(\infty,z^{+};\bb)_{\alpha_{2}\beta}
    \mathfrak{h}_{q\to q q}^{(2)}
\Bigg\}
\Psi_{\beta}(z^{+};\bb), \nn
\end{align}
with the Dirac coefficients defined as 
\begin{align}
\label{eq:HF_q_to_qq_1}
\mathfrak{h}_{q\to qq}^{(1)} =&
    \mathfrak{h'}^{(1)}_{q\to qq} 
    - \frac{1}{N_c}\mathfrak{h'}^{(2)}_{q\to qq} ,  \\ 
\label{eq:HF_q_to_qq_2}
\mathfrak{h}_{q\to qq}^{(2)} =&
    \mathfrak{h'}^{(2)}_{q\to qq} 
    - \frac{1}{N_c}\mathfrak{h'}^{(1)}_{q\to qq} ,
\end{align}
and 
\begin{align}
\label{eq:HFp_q_to_qq_1}
\mathfrak{h'}^{(1)}_{q\to qq} =&
    \overline{u}(\check{p}_{1}, h_1)\gamma^{+}
    \left[\gamma^{l}\gamma^{i} \bP^{l} - 2\frac{z}{1-z}\bP^{i}\right]u(\check{q}, h)
    \overline{u}(\check{p}_{2}, h_2)
    \gamma^{i}\frac{\gamma^{+}\gamma^{-}}{2}, \\
\label{eq:HFp_q_to_qq_2}
\mathfrak{h'}^{(2)}_{q\to qq} =&
    \overline{u}(\check{p}_{2}, h_2)\gamma^{+}
    \left[\gamma^{l}\gamma^{i} \bP^{l} - 2\frac{1-z}{z}\bP^{i}\right]u(\check{q}, h)
    \overline{u}(\check{p}_{1}, h_1)
    \gamma^{i}\frac{\gamma^{+}\gamma^{-}}{2} .
\end{align}
\subsection{$q\to qq$ production cross section in the back-to-back limit}
Partonic cross section in the back-to-back and massless quark limits for the $q\to qq$ channel can be written as 
\begin{equation} 
\label{eq:sigma_q_to_qq}
\frac{d\sigma^{{\rm b2b}, \, m=0}_{q\to qq}}{d{\rm P.S}} =
(2q^+) \ (2\pi)\delta\left(p_1^+ + p_2^+ - q^+\right)
\frac{1}{2N_c}\sum_{h_{1},h_2, h}\sum_{\alpha_1, \alpha_2, \alpha}
    \left\langle \left|i\mathcal{M}^{{\rm b2b}, \, m=0}_{q\to qq, \, {\rm tot.}}\right|^2 \right\rangle ,
\end{equation}
where the back-to-back amplitude is given in Eq.~\eqref{eq:M_tot_q_to_qq_b2b_m0}. Moreover, as usual differential phase space $d{\rm P.S.}$ is defined in Eq.~\eqref{def:PS} and $\langle\cdots\rangle$ stands for the target averaging in the CGC formalism. Using the explicit expression for the amplitude, it leads to
\begin{align}
\frac{d\sigma^{{\rm b2b}, \, m=0}_{q\to qq}}{d{\rm P.S.}} &=
\frac{g^4}{4(2q^{+})^{3}\bP^4(2N_c)}(2\pi)\delta \left(p_1^+ + p_2^+ - q^+\right)
\sum_{h, h_1, h_2}
\int_{z^+,z'^+}
\int_{\bb, \bb'}
e^{-i(\bb-\bb')\cdot\left(\bk - \bq\right)} 
\\
&\times 
\Big\{
    \Big\langle
    \Psi^{\dag}(z'^+;\bb') 
    \UFd(\infty, {z'}^{+};\bb')
    \Big|\mathfrak{h}^{(1)}_{q\to qq}\Big|^2
    \UF(\infty,z^{+};\bb)\Psi(z^+;\bb)
\nn \\
& \hspace{6.5cm} 
\times
    \Tr\left[\UFd(\infty ,-\infty;\bb')\UF(\infty ,-\infty;\bb)\right]
    \Big\rangle
\nn \\
& \hspace{0.3cm}
    + \Big\langle
    {\Psi}^{\dag}(z'^+;\bb')
    \UFd(\infty, {z'}^{+};\bb')
    \Big|\mathfrak{h}^{(2)}_{q\to qq}\Big|^2
    \UF(\infty,z^{+};\bb)\Psi(z^+;\bb)
\nn \\
& \hspace{6.5cm} 
\times
    \Tr\left[\UFd(\infty ,-\infty;\bb')\UF(\infty ,-\infty;\bb)\right]
    \Big\rangle
\nn \\
& \hspace{0.3cm}
    + \Big\langle
    {\Psi}^{\dag}(z'^+;\bb')
    \UFd(\infty, {z'}^{+};\bb')
    \Big(\mathfrak{h}^{(1)}_{q\to qq}\Big)^\dagger\mathfrak{h}^{(2)}_{q\to qq} \
    \UF(\infty ,-\infty;\bb)
\nn \\
& \hspace{6cm} 
\times
    \UFd(\infty ,-\infty;\bb')\UF(\infty,z^{+};\bb)\Psi(z^+;\bb)
    \Big\rangle
\nn \\
& \hspace{0.3cm}
    + \Big\langle
    \Psi^{\dag}(z'^+;\bb')
    \UFd(\infty, {z'}^{+};\bb')
    \Big(\mathfrak{h}^{(2)}_{q\to qq}\Big)^\dagger\mathfrak{h}^{(1)}_{q\to qq} \
    \UF(\infty ,-\infty;\bb)
\nn \\
& \hspace{6cm} 
\times
    \UFd(\infty ,-\infty;\bb')\UF(\infty,z^{+};\bb)\Psi(z^+;\bb)
    \Big\rangle
\Big\} , \nn
\end{align}
which coefficient are given in Eqs.~\eqref{eq:HF_q_to_qq_11}, \eqref{eq:HF_q_to_qq_22} and \eqref{eq:HF_q_to_qq_12}.

\begin{figure}[H]
\centering
\begin{subfigure}{0.49\textwidth}
\centering
\includegraphics[height=5cm]{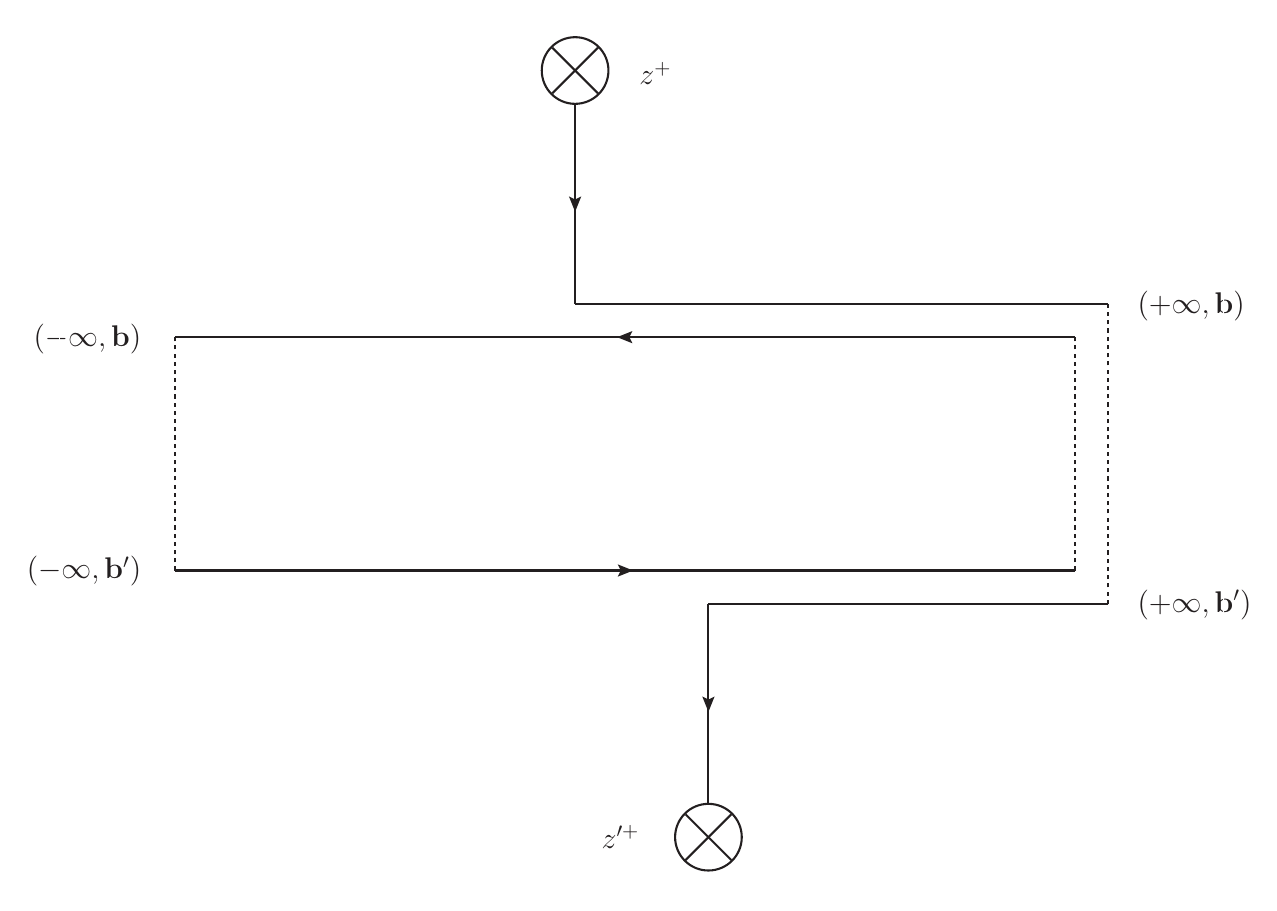}
\caption{$\mathcal{C}^{+\square}$}
\label{fig:q-qq:C+l}
\end{subfigure}
\begin{subfigure}{0.49\textwidth}
\centering
\includegraphics[height=5cm]{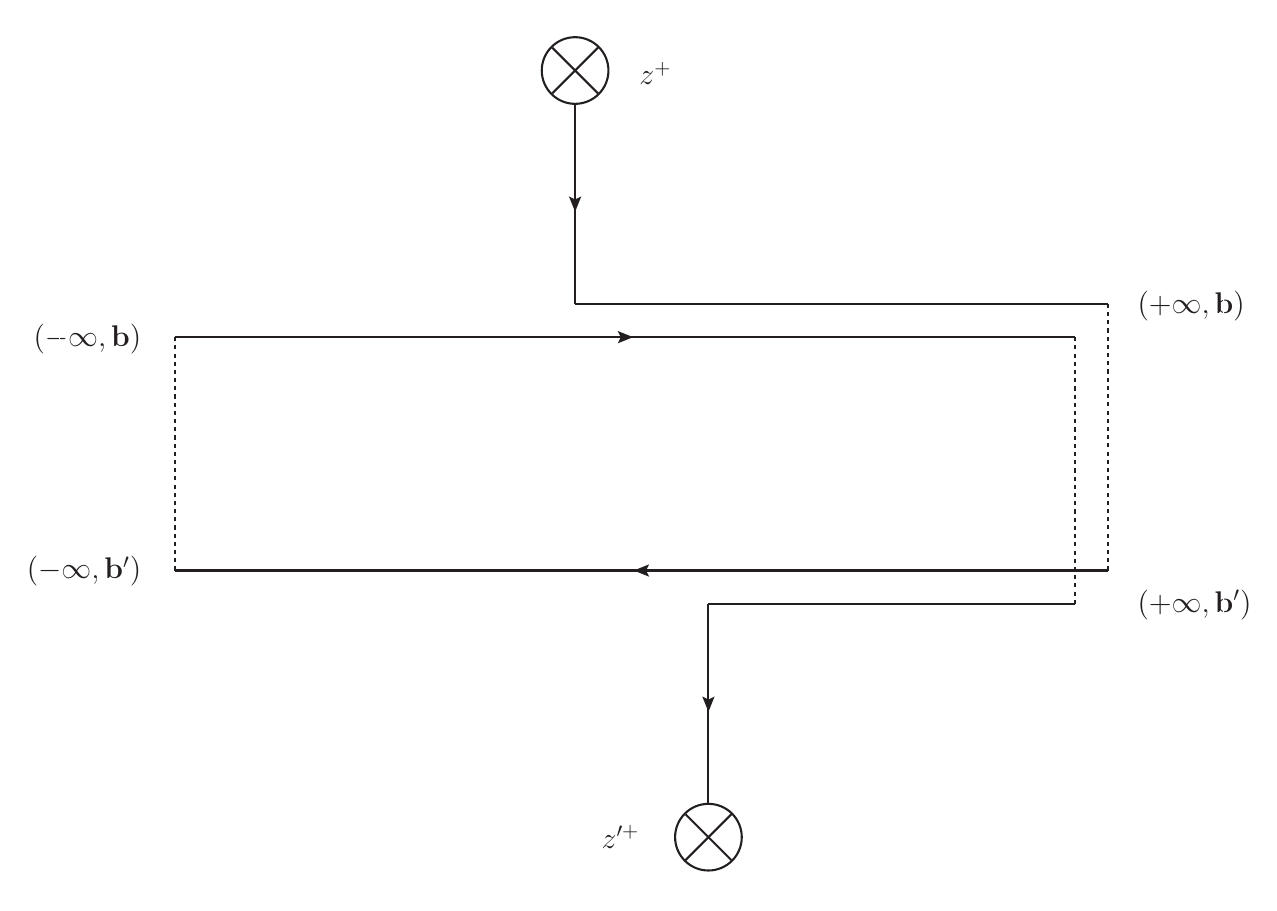}
\caption{$\mathcal{C}^{+-+}$}
\label{fig:q-qq:C+-+}
\end{subfigure}
\caption{Color structures appearing in the cross section of $q\to qq$.}
\label{fig:Color_q-qq}
\end{figure}

After performing some simplifications, the production cross section in the back-to-back limit for the $q\to qq$ channel takes the following factorized form: 
\begin{align}
\frac{d\sigma^{{\rm b2b}, \, m=0}_{q\to qq}}{d{\rm P.S.}} & = 
g^{4} \ (2\pi) \delta(p_{1}^{+} + p_{2}^{+} - q^{+})  \int_{-\frac{L^+}{2}}^{\frac{L^+}{2}} dz^+
\int_{-\frac{L^+}{2}}^{\frac{L^+}{2}} dz'^+
\int_{\bb, \bb'} e^{-i(\bb - \bb')\cdot (\bk - \bq)} 
\nn \\ 
&
\hspace{5cm}\
\times
\left[\mathcal{H}^{+\square}_{q\to qq} \  \mathcal{C}^{+\square}
+ \mathcal{H}^{+-+}_{q\to qq} \ \mathcal{C}^{+-+}\right] ,  
\end{align}
with the hard factors given as 
\begin{align}
\label{eq:H_q_to_qq_+square}
    \mathcal{H}^{+\square}_{q\to qq} &= 
    \frac{1}{4N_c \ \bP^2} 
    \left[2\left(1+\frac{1}{N_c^2}\right)\left(z(1-z) - 3 + \frac{1}{z(1-z)}\right)
    -\frac{4}{N_c}\right] , \\
    \label{eq:H_q_to_qq_+-+}
    \mathcal{H}^{+-+}_{q\to qq} &=
    \frac{1}{4N_c \ \bP^2} 
    \left[-\frac{4}{N_c}\left(z(1-z) - 3 + \frac{1}{z(1-z)}\right)
    +2\left(1+\frac{1}{N_c^2}\right)\right] ,
\end{align}
and the reduced color structures which read
\begin{align}
    \label{eq:C+square}
    \mathcal{C}^{+\square} & \equiv 
    \Big\langle 
    \Tr\left[\UFd(\infty ,-\infty;\bb')\UF(\infty ,-\infty;\bb)\right]
    \overline{\Psi}({z'}^+;\bb')\gamma^{-}\UFd(\infty, z'^{+};\bb')
    \UF(\infty, z^{+};\bb)\Psi(z^+;\bb)\Big\rangle , \\
    \label{eq:C+-+}
    \mathcal{C}^{+-+} & \equiv 
    \Big\langle \overline{\Psi}(z'^{+};\bb) \  \gamma^{-}  \    
    \UF^{\dagger}(\infty, z'^{+};\bb')  \ \UF(\infty, -\infty;\bb)
    \nn \\
    & \hspace{5.8cm}
    \times 
    \ \UF^{\dagger}(\infty, -\infty;\bb')  \ \UF(\infty, z^{+};\bb) \ \Psi(z^{+};\bb)
    \Big\rangle ,
\end{align}
that are illustrated in Fig.~\ref{fig:Color_q-qq}.  
\subsection{$q\to qq$ production cross section for different quark flavors}
\label{subsec:xs_q-qq_f}

%
%
%

As discussed in the beginning of \ref{sec:qq}, the two final quarks will have difference flavors, hence only the first mechanism (see \Figure{fig:q-qq1}) will contribute. The Partonic cross section in the back-to-back and massless quark limits for the $q\to qq$ channel with different flavor quarks in the final state can be written as 
\begin{equation} 
\label{eq:sigma_q_to_qq'}
\frac{d\sigma^{{\rm b2b}, \, m=0}_{q_{f}\to q_{f}q_{f'}}}{d{\rm P.S}} =
(2q^+) \ (2\pi)\delta\left(p_1^+ + p_2^+ - q^+\right)
\frac{1}{2N_c}\sum_{h_{1},h_2, h}\sum_{\beta_1, \beta_2, \alpha}
    \left\langle \left|i\mathcal{M}^{{\rm b2b}, \, m=0}_{q_{f}\to q_{f}q_{f'}, 1}\right|^2 \right\rangle ,
\end{equation}
where the back-to-back amplitude is given in Eq.~\eqref{eq:M_q_to_qq_b2b_1} together with Eq.~\eqref{eq:CS_simplified_q_to_qq} for the simplified color structure. As final state quarks have different flavors, sum over them is shown explicitly. Moreover, as usual differential phase space $d{\rm P.S.}$ is defined in Eq.~\eqref{def:PS} and $\langle\cdots\rangle$ stands for the target averaging in the CGC formalism. Using the explicit expression for the amplitude and after performing some simplifications, the production cross section in the back-to-back limit for the $q\to qq$ channel takes the following factorized form: 
\begin{align}
\frac{d\sigma^{{\rm b2b}, \, m=0}_{q_{f}\to q_{f}q_{f'}}}{d{\rm P.S}} & = 
g^{4} \ (2\pi) \delta(p_{1}^{+} + p_{2}^{+} - q^{+}) 
\int_{-\frac{L^+}{2}}^{\frac{L^+}{2}} dz^+
\int_{-\frac{L^+}{2}}^{\frac{L^+}{2}} dz'^+
\int_{\bb, \bb'} e^{-i(\bb - \bb')\cdot (\bk - \bq)} 
\nn \\ 
& \hspace{5cm}
\times 
\left[\mathcal{H}_{q_{f}\to q_{f}q_{f'}}^{+\square}\mathcal{C}^{+\square} 
+ \mathcal{H}_{q_{f}\to q_{f}q_{f'}}^{+-+}\mathcal{C}^{+-+} \right] ,  
\end{align}
with the hard factor given as 
\begin{align}
\label{eq:H_q_to_qq'_+square}
    \mathcal{H}^{+\square}_{q_{f}\to q_{f}q_{f'}} &= 
    \frac{1}{4 \ \bP^2}\frac{N_c^2+1}{N_c^3}
    z\frac{1+z^2}{1-z} , \\
    \label{eq:H_q_to_qq'_+-+}
    \mathcal{H}^{+-+}_{q_{f}\to q_{f}q_{f'}} &=
    -\frac{1}{2 \ \bP^2}\frac{1}{N_c^2}
    z\frac{1+z^2}{1-z} ,
\end{align}
and the reduced color structures are given in \eqref{eq:C+square} and \eqref{eq:C+-+}.



\section{Relation with the quark TMDs } 
\label{sec:TMD}                     
The color structures that appear in the back-to-back cross section in all the channels have the form of 
\begin{equation}
    \mathcal{T}^{\, (\cdots)}(\bk) \equiv
    \int_{\bb, \b'}e^{-i\bk\cdot(\bb-\bb')}
    \int\limits_{z^+, {z'}^+}
    \left\langle
    \C^{(\cdots)}\right\rangle ,
\label{eq:T_gen}
\end{equation}
where $\C^{(\cdots)}$ are the various color structures computed for various channels and $\langle \cdots\rangle$ stands for the target averaging in the spirit of the CGC formalism. The target averaging in CGC is performed with a probability distribution which can be obtained from the JIMWLK evolution with initial conditions given by the MV model. However, in the eikonal CGC only $\A^-$ fields are considered as the background field of the target. 

In this paper, where the calculations are performed beyond the eikonal approximation, and therefore beyond the scope of the standard CGC formalism, the quark background field of the target is accounted for in the scattering processes. For the moment, no generalization of the MV model is available to include the quark background field in the target averaging. Instead, we can go back to the quantum field theory expectation value that the CGC target average is supposed to model. 

As discussed in \cite{Altinoluk:2023qfr,}, 
the quantum field theory expectation of value  $\langle P_{tar}|\hat{\mathcal{O}}|P_{tar}\rangle$
of an operator $\hat{\mathcal{O}}$ in the state $| P_{tar}\rangle$ of the target with momentum $P_{tar}^{\mu}$ and the corresponding semi-classical target average $\langle {\mathcal O}\rangle$ from the CGC can be related as
\cite{Dominguez:2011wm,Marquet:2016cgx,Altinoluk:2019wyu,Belitsky:2002sm,} 
\begin{equation}
\label{eq:generic_O_QE}
    \langle \mathcal{O}\rangle =
    \lim_{P'_{tar}\to P_{tar}}
    \frac{\langle P'_{tar}|\hat{\mathcal{O}}|P_{tar}\rangle}
    {\langle P'_{tar}|P_{tar}\rangle} ,
\end{equation}
with the target states normalized as 
\begin{equation}
\label{eq:Normalization_of_TS}
    \langle P'_{tar}|P_{tar}\rangle =
    2P^-_{tar}\, (2\pi)^3 \, \delta(P'^-_{tar}-P^-_{tar}) \, 
    \delta^{(2)}(\bP'_{tar}-\bP_{tar}) .
\end{equation}
In general, in quantum field theory, the action of the momentum operator $\hat{P}_\mu$ on a local operator $\hat{\mathcal{O}}(x)$ generates  spacetime translations as 
%
%
\begin{equation}
    \hat{\mathcal{O}}(x) =
    \exp{ia^\mu\hat{P}_\mu}\hat{\mathcal{O}}(x-a)\exp{-ia^\mu\hat{P}_\mu} .
\end{equation}
%
%
For matrix elements of non-local operators, it implies 
\begin{align}
\label{eq:Matrix_nonLocal}
  \!\! \!\!\!\!  \langle P'_{tar}|\hat{\mathcal{O}}_1(x_1)
    \dots \hat{\mathcal{O}}_n(x_n)|P_{tar}\rangle &
    =
    \langle P'_{tar}|\exp{ia^\mu\hat{P}_\mu}
    \hat{\mathcal{O}}_1(x_1-a)
    \exp{-ia^\mu\hat{P}_\mu}
    \dots 
    \exp{ia^\mu\hat{P}_\mu}
    \hat{\mathcal{O}}_n(x_n-a)\exp{-ia^\mu\hat{P}_\mu}|P_{tar}\rangle \nn \\
    &= \exp{ia^\mu\left[(P'_{tar})_\mu-(P_{tar})_\mu\right]}\langle P'_{tar}|\hat{\mathcal{O}}_1(x_1-a)
    \dots \hat{\mathcal{O}}_n(x_n-a)|P_{tar}\rangle .
\end{align}
For the sake of simplicity of the discussion, let us focus on a specific case and consider the color structure $\C^+$ which reads 
\begin{align}
\C^+\equiv \left\langle\overline{\Psi}(z'^+,\b') \gamma^- \UF^{\dagger}(\infty, z'^+;\b')\UF(\infty,z^+;\b)\Psi(z^+,\b)\right\rangle .
\end{align}
Using Eq.~\eqref{eq:T_gen} together with relations given in Eqs.~\eqref{eq:generic_O_QE}, \eqref{eq:Normalization_of_TS} and \eqref{eq:Matrix_nonLocal} one obtains 
\begin{align}
    \mathcal{T}^{+}(\bk) =& 
    \lim_{P'_{tar}\to P_{tar}}
    \int_{\bb,\bb'}\frac{e^{-i\bk\cdot(\bb-\bb')}}{\langle P'_{tar}|P_{tar}\rangle} \nn \\
    &\hspace{0.6cm}\times
    \int_{z^+, {z'}^+}
    \left\langle P'_{tar}\left|\overline{\Psi}({z'}^+;\bb')\gamma^-
    \UFd(\infty,{z'}^+;\bb')\UF(\infty,z^+;\bb)\Psi({z}^+;\bb)\right|P_{tar}\right\rangle \nn \\
    &=\lim_{P'_{tar}\to P_{tar}}
    \int_{\bb,\bb'}\frac{e^{-i\bk\cdot(\bb-\bb')}}{\langle P'_{tar}|P_{tar}\rangle}
    \int_{z^+, {z'}^+}
    \exp{iz^+(P'^-_{tar}-P^-_{tar})}
    \exp{-i\bb\cdot(\bP'_{tar}-\bP_{tar})}\nn \\
    &
    \times\Big\langle P'_{tar}\left|\overline{\Psi}({z'}^+-z^+;\bb'-\bb)\gamma^-
    \UFd(\infty,{z'}^+-z^+;\bb'-\bb) 
    \UF(\infty,0;\bzer)\Psi(0;\bzer)\right|P_{tar}\Big\rangle
    \nn \\
    &=\lim_{P'_{tar}\to P_{tar}}
\int_{\Delta\bb}\frac{\exp{i\bk\cdot\Delta\bb}}{2P^-_{tar}}
    \int_{\Delta z^+}\nn \\
    &\times
    \left\langle P'_{tar}\left|\overline{\Psi}(\Delta z^+;\Delta\bb)\gamma^-
    \UFd(\infty,\Delta z^+;\Delta\bb)\UF(\infty,0;\bzer)\Psi(0;\bzer)\right|P_{tar}\right\rangle ,
\label{eq:T+ev}
\end{align}
where the limit $P'_{tar}\to P_{tar}$ can be taken safely in the last step.
We can compare this result to the unpolarized quark distribution (with future pointing staple gauge link), which reads (up to UV and rapidity renormalization) 
\begin{align}
    f_q^+({\text x},\bk) &=
    \frac{1}{(2\pi)^3}\int_{\bb}\exp{i\bk\cdot\bb}
    \int\limits_{z^+}\exp{-i{\text x}P^-_{tar}z^+}
    \Big\langle P_{tar}\Big|\overline{\Psi}(z^+;\bb)\frac{\gamma^-}{2}
    \UFd(\infty,z^+;\bb) 
    \UF(\infty,0;\bzer)\Psi(0;\bzer)\Big|P_{tar}\Big\rangle ,
\end{align}
where the transverse gauge link at infinity  is neglected. Comparing it to \Equation{eq:T+ev}, we observe the relation
\begin{equation}
    \mathcal{T}^{+}(\bk) = \frac{(2\pi)^3}{P^-_{tar}}f_q^+({\text x}=0,\bk) .
\end{equation}
%
%
%
Following the same procedure for the other color structures, one can obtain a generic relation which reads 
\begin{align}
 \int_{\bb,\b'}e^{-i\bk\cdot(\bb-\bb')}\int_{z^+, {z'}^+}
    \frac{\mathcal{C}^{(\cdots)}}{{\cal N}^{(\cdots)}}
    &= \frac{(2\pi)^3}{P^-_{tar}}f_q^{(\cdots)}({\text x}=0, \bk)
\end{align}
where $\C^{(\cdots)}$ are the color structures computed for various channels and $f_q^{(\cdots)}({\text x}=0, \bk)$ are the associated unpolarized quark TMDs. The ${\cal N}^{(\cdots)}$ are color factors required for the correct normalization of the quark TMDs. They can be obtained by replacing all Wilson lines by identity matrices in $\C^{(\cdots)}$, and then calculating the color factor appearing in that limit in front of the operator $\overline{\Psi}(z^+;\bb) \gamma^- \Psi(0;\bzer)$.
In such a way, one obtains all the required unpolarized quark TMDs as  
%
%
%
%
%
%
\begin{align}
\label{eq:quark_TMD_defs}
    f^{-}_q({\text x},\bk) =&
    \frac{1}{(2\pi)^3}\int_{\bb}\exp{i\bk\cdot\bb}
    \int\limits_{z^+}\exp{-i{\text x}P^-_{tar}z^+} \nn \\
    & \hspace{1cm} \times
    \left\langle P_{tar}\left|\overline{\Psi}({z}^+;\bb)\frac{\gamma^-}{2}
    \UF({z}^+,-\infty;\bb)\UFd(0,-\infty;\bzer)\Psi(0;\bzer)\right|P_{tar}\right\rangle , \nn
    \displaybreak[0] \\
    f^{-g}_q({\text x},\bk) =&
    \frac{1}{(2\pi)^3}\int_{\bb}\exp{i\bk\cdot\bb}
    \int\limits_{z^+}\exp{-i{\text x}P^-_{tar}z^+}
    \bigg\langle P_{tar}\bigg|\frac{1}{C_F}\, \UA(\infty,{z}^+;\bb)_{bc'}\UA(\infty,0;\bzer)_{bc} \nn \\
    & \hspace{1cm} \times
    \overline{\Psi}({z}^+;\bb)\frac{\gamma^-}{2}
    t^{c'}\UF({z}^+,-\infty;\bb)\UFd(0,-\infty;\bzer)t^c\Psi(0;\bzer)\bigg|P_{tar}\bigg\rangle , \nn
    \displaybreak[0] \\
    f^{+g}_q({\text x},\bk) =&
    \frac{1}{(2\pi)^3}\int_{\bb}\exp{i\bk\cdot\bb}
    \int\limits_{z^+}\exp{-i{\text x}P^-_{tar}z^+}    
    \bigg\langle P_{tar}\bigg|
    \frac{1}{C_F}\,
    \UA({z}^+,-\infty;\bb)_{c'a}\UA(0,-\infty;\bzer)_{ca} \nn \\
    & \hspace{1cm} \times\overline{\Psi}({z}^+;\bb)\frac{\gamma^-}{2}
    t^{c'}\UFd(\infty,{z}^+;\bb)\UF(\infty,0;\bzer)t^c\Psi(0;\bzer)\bigg|P_{tar}\bigg\rangle , \nn
    \displaybreak[0] \\
    f^{-\square}_q({\text x},\bk) =&
    \frac{1}{(2\pi)^3}\int_{\bb}\exp{i\bk\cdot\bb}
    \int\limits_{z^+}\exp{-i{\text x}P^-_{tar}z^+}
    \bigg\langle P_{tar}\bigg|\frac{1}{N_c}
    \Tr \left[ \UFd(\infty,-\infty;\bb)\UF(\infty,-\infty;\bzer)\right] \nn \\
    & \hspace{1cm} \times  
    \overline{\Psi}({z}^+;\bb)\frac{\gamma^-}{2}
    \UF({z}^+,-\infty;\bb)\UFd(0,-\infty;\bzer)\Psi(0;\bzer)\bigg|P_{tar}\bigg\rangle , \nn
    \displaybreak[0] \\
    f^{+\square}_q({\text x},\bk) =&
    \frac{1}{(2\pi)^3}\int_{\bb}\exp{i\bk\cdot\bb}
    \int\limits_{z^+}\exp{-i{\text x}P^-_{tar}z^+}
    \bigg\langle P_{tar}\bigg|\frac{1}{N_c}
    \Tr\left[\UF(\infty,-\infty;\bb)\UFd(\infty,-\infty;\bzer)\right] \nn \\
    & \hspace{1cm} \times 
    \overline{\Psi}({z}^+;\bb)\frac{\gamma^-}{2}
    \UFd(\infty,{z}^+;\bb)\UF(\infty,0;\bzer)\Psi(0;\bzer)\bigg|P_{tar}\bigg\rangle , \nn
    \displaybreak[0] \\
    f^{+\square_g}_q({\text x},\bk) =&
    \frac{1}{(2\pi)^3}\int_{\bb}\exp{i\bk\cdot\bb}
    \int\limits_{z^+}\exp{-i{\text x}P^-_{tar}z^+}
    \bigg\langle P_{tar}\bigg|
    \frac{1}{N_c^2\!-\!1}\,
    \UA(\infty,-\infty;\bb)_{ba}\UA(\infty,-\infty;\bzer)_{ba} \nn \\
    & \hspace{1cm} \times 
    \overline{\Psi}({z}^+;\bb)\frac{\gamma^-}{2}
    \UFd(\infty,{z}^+;\bb)\UF(\infty,0;\bzer)\Psi(0;\bzer)\bigg|P_{tar}\bigg\rangle , \nn
    \displaybreak[0] \\
    f^{+\square^2}_q({\text x},\bk) =&
    \frac{1}{(2\pi)^3}\int_{\bb}\exp{i\bk\cdot\bb}
    \int\limits_{z^+}\exp{-i{\text x}P^-_{tar}z^+} \nn \\
    &\times
    \bigg\langle P_{tar}\bigg|
    \frac{1}{N_c^2}
    \Tr \left[\UF(\infty,-\infty;\bb)\UFd(\infty,-\infty;\bzer)\right]
    \Tr \left[\UFd(\infty,-\infty;\bb)\UF(\infty,-\infty;\bzer)\right]
     \nn \\
    &\times 
   \overline{\Psi}({z}^+;\bb)\frac{\gamma^-}{2}
    \UFd(\infty,{z}^+;\bb)\UF(\infty,0;\bzer)\Psi(0;\bzer)\bigg|P_{tar}\bigg\rangle
\nn
    \displaybreak[0] \\
    f_q^{+-+}({\text x},\bk) &=
    \frac{1}{(2\pi)^3}\int_{\bb}\exp{i\bk\cdot\bb}
    \int\limits_{z^+}\exp{-i{\text x}P^-_{tar}z^+}
    \Big\langle P_{tar}\Big|\overline{\Psi}(z^+;\bb)\frac{\gamma^-}{2}
    \UFd(\infty,z^+;\bb)
    \UF(\infty,-\infty;\bzer)
    \nn \\
    &\times  
    \UFd(\infty,-\infty;\bb)
    \UF(\infty,0;\bzer)
    \Psi(0;\bzer)\Big|P_{tar}\Big\rangle    
    ,
\end{align}
with the associated color factors 
\begin{align}
{\cal N}^{+} =&\, {\cal N}^{-} ={\cal N}^{+-+} = 1
\nn \\
{\cal N}^{+g} =&\, {\cal N}^{-g} =C_F = \frac{(N_c^2\!-\!1)}{2N_c}
\nn \\
{\cal N}^{+\square} =&\, {\cal N}^{-\square} =d_F = N_c
\nn \\
{\cal N}^{+\square g} =&\, d_A = N_c^2-1
\nn \\
{\cal N}^{+\square^2} =&\, d_F^2 = N_c^2
\, .
\label{eq:color_factors}
\end{align}
The antiquark TMDs are defined in a similar way but from the color structures $\overline{\cal C}^{(\cdots)}$, in which the field $\overline{\Psi}$ comes from the amplitude and is on the right, whereas the field $\Psi$ comes from the complex conjugate amplitude and is on the left, as  
\begin{equation}
    \int_{\bb,\b'}e^{-i\bk\cdot(\bb-\bb')}\int_{z^+, {z'}^+}
    \frac{{\mathcal{\bar C}}^{(\cdots)}}{{\cal N}^{(\cdots)}}
    = \frac{(2\pi)^3}{P^-_{tar}}f_{\bar q}^{(\cdots)}({\text x}=0, \bk)
    \, .
\end{equation}
Thus, the unpolarized antiquark TMDs with various gauge links are defined as
\begin{align}
\label{eq:antiquark_TMD_defs}
    f_{\bar q}^+({\text x},\bk) =&\,
    \frac{1}{(2\pi)^3}\int_{\bb}\exp{i\bk\cdot\bb}
    \int\limits_{z^+}\exp{-i{\text x}P^-_{tar}z^+}
    \nn \\
    & \hspace{1cm} \times
    \Big\langle P_{tar}\Big|
    \Tr\Big\{
    \UF(\infty,z^+;\bb)\Psi(z^+;\bb)
    \overline{\Psi}(0;\bzer)\frac{\gamma^-}{2}
    \UFd(\infty,0;\bzer)     
    \Big\}
    \Big|P_{tar}\Big\rangle
    \nn \\
    f^{-}_{\bar q}({\text x},\bk) =&
    \frac{1}{(2\pi)^3}\int_{\bb}\exp{i\bk\cdot\bb}
    \int\limits_{z^+}\exp{-i{\text x}P^-_{tar}z^+} \nn \\
    & \hspace{1cm} \times
    \left\langle P_{tar}\left|
    \Tr\Big\{
    \UFd({z}^+,-\infty;\bb)\Psi({z}^+;\bb)
    \overline{\Psi}(0;\bzer)\frac{\gamma^-}{2}
    \UF(0,-\infty;\bzer)
    \Big\}
    \right|P_{tar}\right\rangle , \nn
    \displaybreak[0] \\
    f^{-g}_{\bar q}({\text x},\bk) =&
    \frac{1}{(2\pi)^3}\int_{\bb}\exp{i\bk\cdot\bb}
    \int\limits_{z^+}\exp{-i{\text x}P^-_{tar}z^+}
    \bigg\langle P_{tar}\bigg|\frac{1}{C_F}\, \UA(\infty,{z}^+;\bb)_{bc'}\UA(\infty,0;\bzer)_{bc} 
    \nn \\
    & \hspace{1cm} \times
    \Tr\Big\{
    \UFd({z}^+,-\infty;\bb)t^{c'}\Psi({z}^+;\bb)
    \overline{\Psi}(0;\bzer)\frac{\gamma^-}{2}
    t^{c}\UF(0,-\infty;\bzer)
    \Big\}
    \bigg|P_{tar}\bigg\rangle , 
    \nn
    \displaybreak[0] \\
    f^{+g}_{\bar q}({\text x},\bk) =&
    \frac{1}{(2\pi)^3}\int_{\bb}\exp{i\bk\cdot\bb}
    \int\limits_{z^+}\exp{-i{\text x}P^-_{tar}z^+}    
    \bigg\langle P_{tar}\bigg|
    \frac{1}{C_F}\,
    \UA({z}^+,-\infty;\bb)_{c'a}\UA(0,-\infty;\bzer)_{ca} 
    \nn \\
    & \hspace{1cm} \times
    \Tr\Big\{
    \UF(\infty,{z}^+;\bb)t^{c'}\Psi({z}^+;\bb)
    \overline{\Psi}(0;\bzer)\frac{\gamma^-}{2}
    t^{c}\UFd(\infty,0;\bzer)
    \Big\}
    \bigg|P_{tar}\bigg\rangle , \nn
    \displaybreak[0] \\
    f^{-\square}_{\bar q}({\text x},\bk) 
    =&
    \frac{1}{(2\pi)^3}\int_{\bb}\exp{i\bk\cdot\bb}
    \int\limits_{z^+}\exp{-i{\text x}P^-_{tar}z^+}
    \bigg\langle P_{tar}\bigg|\frac{1}{N_c}
    \Tr \left[ 
    \UF(\infty,-\infty;\bb)
    \UFd(\infty,-\infty;\bzer)
    \right] 
    \nn \\
    & \hspace{1cm} \times
    \Tr\Big\{
    \UFd({z}^+,-\infty;\bb)\Psi({z}^+;\bb)
    \overline{\Psi}(0;\bzer)\frac{\gamma^-}{2}
    \UF(0,-\infty;\bzer)
    \Big\}
    \bigg|P_{tar}\bigg\rangle , 
    \nn
    \displaybreak[0] \\
    f^{+\square}_{\bar q}({\text x},\bk) =&
    \frac{1}{(2\pi)^3}\int_{\bb}\exp{i\bk\cdot\bb}
    \int\limits_{z^+}\exp{-i{\text x}P^-_{tar}z^+}
    \bigg\langle P_{tar}\bigg|\frac{1}{N_c}
    \Tr\left[
    \UFd(\infty,-\infty;\bb)
    \UF(\infty,-\infty;\bzer)
    \right] 
    \nn \\
    & \hspace{1cm} \times
    \Tr\Big\{
    \UF(\infty,{z}^+;\bb)\Psi({z}^+;\bb)
    \overline{\Psi}(0;\bzer)\frac{\gamma^-}{2}
    \UFd(\infty,0;\bzer)
    \Big\}
    \bigg|P_{tar}\bigg\rangle , \nn
    \displaybreak[0] \\
    f^{+\square_g}_{\bar q}({\text x},\bk) =&
    \frac{1}{(2\pi)^3}\int_{\bb}\exp{i\bk\cdot\bb}
    \int\limits_{z^+}\exp{-i{\text x}P^-_{tar}z^+}
    \bigg\langle P_{tar}\bigg|
    \frac{1}{N_c^2\!-\!1}\,
    \UA(\infty,-\infty;\bb)_{ba}
    \UA(\infty,-\infty;\bzer)_{ba} 
    \nn \\
    & \hspace{1cm} \times 
    \Tr\Big\{
    \UF(\infty,{z}^+;\bb)\Psi({z}^+;\bb)
    \overline{\Psi}(0;\bzer)\frac{\gamma^-}{2}
    \UFd(\infty,0;\bzer)
    \Big\}
    \bigg|P_{tar}\bigg\rangle , 
    \nn
    \displaybreak[0] \\
    f^{+\square^2}_{\bar q}({\text x},\bk) =&
    \frac{1}{(2\pi)^3}\int_{\bb}\exp{i\bk\cdot\bb}
    \int\limits_{z^+}\exp{-i{\text x}P^-_{tar}z^+} \nn \\
    &\times
    \bigg\langle P_{tar}\bigg|
    \frac{1}{N_c^2}
    \Tr \left[\UF(\infty,-\infty;\bb)\UFd(\infty,-\infty;\bzer)\right]
    \Tr \left[\UFd(\infty,-\infty;\bb)\UF(\infty,-\infty;\bzer)\right]
     \nn \\
    &\times 
    \Tr\Big\{
    \UF(\infty,{z}^+;\bb)\Psi({z}^+;\bb)
   \overline{\Psi}(0;\bzer)\frac{\gamma^-}{2}
    \UFd(\infty,0;\bzer)
    \Big\}
    \bigg|P_{tar}\bigg\rangle
\nn
    \displaybreak[0] \\
    f_{\bar q}^{+-+}({\text x},\bk) &=
    \frac{1}{(2\pi)^3}\int_{\bb}\exp{i\bk\cdot\bb}
    \int\limits_{z^+}\exp{-i{\text x}P^-_{tar}z^+}
    \Big\langle P_{tar}\Big|
    \Tr\Big\{
    \UF(\infty,-\infty;\bb)
    \UFd(\infty,-\infty;\bzer)
    \nn \\
    &\times  
    \UF(\infty,z^+;\bb)\Psi(z^+;\bb)
    \overline{\Psi}(0;\bzer)\frac{\gamma^-}{2}
    \UFd(\infty,0;\bzer)
    \Big\}
    \Big|P_{tar}\Big\rangle    
    ,
\end{align}
with the trace acting both on the Dirac spinor indices and on the fundamental color indices.
Therefore, using the definitions of various quark and antiquark TMDs given in Eqs.~\eqref{eq:quark_TMD_defs} and \eqref{eq:antiquark_TMD_defs}, the back-to-back cross sections of each channel  can be written 
as 
\begin{align}
k^+\,\frac{d\sigma^{{\rm b2b},\, m=0}_{g\to gq}}{dk^+\, d^2\k \, d^2\P\,  dz } =&
\frac{q^+ \delta(k^+\!-\!q^+)\;  2 \alpha_s^2}{z(1\!-\!z)(2q^+P^-_{tar})} 
    \bigg[C_F\, {\mathcal{H}}^{+g}_{g\to gq} \ f^{+g}_q({\text x}=0, \bk-\bq)
\nn\\
&\hspace{4cm}
+ (N_c^2\!-\!1)\,{\mathcal{H}}^{+\square_g}_{g\to gq}\ f^{+\square_g}_q({\text x}=0,\bk-\bq)\bigg] , \nn \\
k^+\, \frac{d\sigma^{{\rm b2b},\, m=0}_{q_{f}\to q_{f_1}\bar q_{f_2}}}{dk^+\, d^2\k \, d^2\P\,  dz } =&
\frac{q^+ \delta(k^+\!-\!q^+)\;  2 \alpha_s^2}{z(1\!-\!z)(2q^+P^-_{tar})}  
\bigg[{\mathcal{H}}^{-}_{q_{f}\to q_{f_1}\bar q_{f_2}}\, f^{-}_{\bar{q}}({\text x}=0, \bk-\bq)
\nn\\
&\hspace{4cm}
+N_c\, {\mathcal{H}}^{+\square}_{q_{f}\to q_{f_1}\bar q_{f_2}}\ f^{+\square}_{\bar{q}}({\text x}=0, \bk-\bq)\bigg] , \nn \\
k^+\, \frac{d\sigma^{{\rm b2b},\, m=0}_{q\to gg}}{dk^+\, d^2\k \, d^2\P\,  dz } =&
\frac{q^+ \delta(k^+\!-\!q^+)\;  2 \alpha_s^2}{z(1\!-\!z)(2q^+P^-_{tar})}     
\bigg[{\mathcal{H}}^{-}_{q\to gg}\ f^{-}_{\bar{q}}({\text x}=0,\bk-\bq)
\nn\\
&\hspace{4cm}
+ C_F\, {\mathcal{H}}^{-g}_{q\to gg} \ 
    f^{-g}_{\bar{q}}({\text x}=0,\bk-\bq)\bigg] , \nn \\
k^+\, \frac{d\sigma^{{\rm b2b},\, m=0}_{q_{f}\to q_{f_1}q_{f_2}}}{dk^+\, d^2\k \, d^2\P\,  dz } =&
\frac{q^+ \delta(k^+\!-\!q^+)\;  2 \alpha_s^2}{z(1\!-\!z)(2q^+P^-_{tar})}
    \bigg[N_c\,{\mathcal{H}}^{+\square}_{q_{f}\to q_{f_1}q_{f_2}}\ 
    f^{+\square}_q({\text x}=0,\bk-\bq)
\nn\\
&\hspace{4cm}
+{\mathcal{H}}^{+-+}_{q_{f}\to q_{f_1}q_{f_2}} \ f^{+-+}_q({\text x}=0,\bk-\bq)\bigg] , 
\label{eq:factorization}
\end{align}
where the hard factors $\mathcal{H}^{+\square_g}_{g\to gq}$, $\mathcal{H}^{+g}_{g\to gq}$, 
$\mathcal{H}^{-g}_{q\to gg}$ and $\mathcal{H}^{-}_{q\to gg}$
are given in Eqs.~\eqref{eq:H_g_to_gq_+square_g}, \eqref{eq:H_g_to_gq_+g},
 \eqref{eq:H_q_to_gg_-g} and \eqref{eq:H_q_to_gg_-} respectively
 and the different factors represented by $\mathcal{H}^{-}_{q_{f}\to q_{f_1}\bar q_{f_2}}$, $\mathcal{H}^{+\square}_{q_{f}\to q_{f_1}\bar q_{f_2}}$, $\mathcal{H}_{q_{f}\to q_{f_1}q_{f_2}}$ and $\mathcal{H}^{+-+}_{q_{f}\to q_{f_1}q_{f_2}}$ are given in Eqs.~\eqref{eq:H_q_to_qbarq_-}, \eqref{eq:H_q_to_q'barq'_-}, \eqref{eq:H_q_to_qbarq'_-}; \eqref{eq:H_q_to_qbarq_+square}, \eqref{eq:H_q_to_q'barq'_+square}, \eqref{eq:H_q_to_qbarq'_+square}; \eqref{eq:H_q_to_qq_+square}, \eqref{eq:H_q_to_qq'_+square} and \eqref{eq:H_q_to_qq_+-+}, \eqref{eq:H_q_to_qq'_+-+} respectively.
 Eq.~\eqref{eq:factorization} is the main result of the paper where each channel is written in a factorized form of a quark TMD times the associated hard factor. 
 Moreover, since $2q^+P^-_{tar}$ corresponds to the squared center of mass energy of the parton-target collision at high energy, the suppression in $1/(2q^+P^-_{tar})$ at high energy is a characteristic of NEik contributions. 
Note that, for the $q \to q \bar q$ channel, the possibilities in term of quark flavor are limited to $f=f_1=f_2$, $f \neq f_1=f_2$ and $f=f_1 \neq f_2$ while for the $q \to qq$ channel, they are limited to $f=f_1=f_2$ and $f=f_1 \neq f_2$. 
The other partonic channels, charge conjugate of the ones presented in Eq.~\eqref{eq:factorization} have been studied in Appendix \ref{sec:antiquarks}.
The resulting cross sections can be written as
\begin{align}
k^+\,\frac{d\sigma^{{\rm b2b},\, m=0}_{g\to g \bar q}}{dk^+\, d^2\k \, d^2\P\,  dz } =&
\frac{q^+ \delta(k^+\!-\!q^+)\;  2 \alpha_s^2}{z(1\!-\!z)(2q^+P^-_{tar})} 
    \bigg[C_F\, {\mathcal{H}}^{+g}_{g\to gq} \ f^{+g}_{\bar q}({\text x}=0, \bk-\bq)
\nn\\
&\hspace{4cm}
+ (N_c^2\!-\!1)\,{\mathcal{H}}^{+\square_g}_{g\to gq}\ f^{+\square_g}_{\bar q}({\text x}=0,\bk-\bq)\bigg] , \nn \\
k^+\, \frac{d\sigma^{{\rm b2b},\, m=0}_{\bar q_{f}\to \bar q_{f_1} q_{f_2}}}{dk^+\, d^2\k \, d^2\P\,  dz } =&
\frac{q^+ \delta(k^+\!-\!q^+)\;  2 \alpha_s^2}{z(1\!-\!z)(2q^+P^-_{tar})}  
\bigg[{\mathcal{H}}^{-}_{q_{f}\to q_{f_1}\bar q_{f_2}}\, f^{-}_{{q}}({\text x}=0, \bk-\bq)
\nn\\
&\hspace{4cm}
+N_c\, {\mathcal{H}}^{+\square}_{q_{f}\to q_{f_1}\bar q_{f_2}}\ f^{+\square}_{{q}}({\text x}=0, \bk-\bq)\bigg] , \nn \\
k^+\, \frac{d\sigma^{{\rm b2b},\, m=0}_{\bar q\to gg}}{dk^+\, d^2\k \, d^2\P\,  dz } =&
\frac{q^+ \delta(k^+\!-\!q^+)\;  2 \alpha_s^2}{z(1\!-\!z)(2q^+P^-_{tar})}     
\bigg[{\mathcal{H}}^{-}_{q\to gg}\ f^{-}_{{q}}({\text x}=0,\bk-\bq)
\nn\\
&\hspace{4cm}
+ C_F\, {\mathcal{H}}^{-g}_{q\to gg} \ 
    f^{-g}_{{q}}({\text x}=0,\bk-\bq)\bigg] , \nn \\
k^+\, \frac{d\sigma^{{\rm b2b},\, m=0}_{\bar q_{f}\to \bar q_{f_1}\bar q_{f_2}}}{dk^+\, d^2\k \, d^2\P\,  dz } =&
\frac{q^+ \delta(k^+\!-\!q^+)\;  2 \alpha_s^2}{z(1\!-\!z)(2q^+P^-_{tar})}
    \bigg[N_c\,{\mathcal{H}}^{+\square}_{q_{f}\to q_{f_1}q_{f_2}}\ 
    f^{+\square}_{\bar q}({\text x}=0,\bk-\bq)
\nn\\
&\hspace{4cm}
+{\mathcal{H}}^{+-+}_{q_{f}\to q_{f_1}q_{f_2}} \ f^{+-+}_{\bar q}({\text x}=0,\bk-\bq)\bigg] , 
\label{eq:factorization_charge_conj}
\end{align}
with the same hard factors as in the channels from Eqs.~\eqref{eq:factorization}, as derived in Appendix \ref{sec:antiquarks}, but of course quark and antiquark TMDs interchanged.

Let us recall that the invariant mass of the dijet system is obtained from the jets kinematics as
\begin{align}
M_{jj}^2 =&\, \frac{\bP^2}{z(1\!-\!z)}\, .
\end{align}
On the other hand, in the back-to-back jet limit, it is given as 
\begin{align}
M_{jj}^2 =&\, 2 {\text x} P^-_{tar}\, q^+  \,,
\end{align}
where ${\text x} P^-_{tar}$ is the $-$ light-cone momentum  extracted from the target.
Thus, the corresponding momentum fraction ${\text x}$ can be obtained as
\begin{align}
{\text x} =&\, \frac{\bP^2}{(2P^-_{tar}\, q^+)z(1\!-\!z)} \, ,
\label{x_value}
\end{align}
and is parametrically small in the high-energy limit. 

Hence, our results at NEik accuracy \eqref{eq:factorization} are consistent with the expressions
\begin{align}
k^+\, \frac{d\sigma^{{\rm b2b},\, m=0}_{g\to gq}}{dk^+\, d^2\k \, d^2\P\,  dz } 
=&
q^+ \delta(k^+\!-\!q^+)\:
\frac{2 \alpha_s^2}{\bP^2} 
    \left[C_F\, {\mathcal{H}}^{+g}_{g\to gq} \ {\text x} f^{+g}_q({\text x}, \bk-\bq)
    + (N_c^2\!-\!1)\,{\mathcal{H}}^{+\square_g}_{g\to gq}\ 
    {\text x}f^{+\square_g}_q({\text x},\bk-\bq)\right] ,
\nn \\
k^+\, \frac{d\sigma^{{\rm b2b},\, m=0}_{q_{f}\to q_{f_1}\bar q_{f
{\text x}_2}}}{dk^+\, d^2\k \, d^2\P\,  dz } 
=&
q^+ \delta(k^+\!-\!q^+)\:
\frac{2 \alpha_s^2}{\bP^2}  
\left[{\mathcal{H}}^{-}_{q_{f}\to q_{f_1}\bar q_{f_2}}\, {\text x} f^{-}_{\bar{q}}({\text x}, \bk-\bq)
    +N_c\, {\mathcal{H}}^{+\square}_{q_{f}\to q_{f_1}\bar q_{f_2}}\ {\text x} f^{+\square}_{\bar{q}}({\text x}, \bk-\bq)\right] , 
\nn \\
k^+\, \frac{d\sigma^{{\rm b2b},\, m=0}_{q\to gg}}{dk^+\, d^2\k \, d^2\P\,  dz } 
=&
q^+ \delta(k^+\!-\!q^+)\:
\frac{2 \alpha_s^2}{\bP^2}     
\left[{\mathcal{H}}^{-}_{q\to gg}\ {\text x} f^{-}_{\bar{q}}({\text x},\bk-\bq)
    + C_F\, {\mathcal{H}}^{-g}_{q\to gg} \ 
    {\text x} f^{-g}_{\bar{q}}({\text x},\bk-\bq)\right] , 
\nn \\
k^+\, \frac{d\sigma^{{\rm b2b},\, m=0}_{q_{f}\to q_{f_1}q_{f_2}}}{dk^+\, d^2\k \, d^2\P\,  dz } 
=&
q^+ \delta(k^+\!-\!q^+)\:
\frac{2 \alpha_s^2}{\bP^2}
    \left[N_c\,{\mathcal{H}}^{+\square}_{q_{f}\to q_{f_1}q_{f_2}}\ 
    {\text x} f^{+\square}_q({\text x},\bk-\bq)
    +{\mathcal{H}}^{+-+}_{q_{f}\to q_{f_1}q_{f_2}} \ 
    {\text x} f^{+-+}_q({\text x},\bk-\bq)\right] , 
\nn \\ &
\label{eq:factorization_x}
\end{align}
with ${\text x}$ determined by Eq.~\eqref{x_value}, and similar expressions can be obtained for the charge conjugate channels. Nevertheless, a calculation at NNEik accuracy would be necessary to confirm from the high-energy formalism the value of ${\text x}$ taken in the TMDs, by analogy to the study of gluon TMD  in DIS dijet production  in Ref.
~\cite{Altinoluk:2024zom}.
In our results in Eqs.~\eqref{eq:factorization_x}, 
the hard factors (including their color prefactor) are in agreement with the ones derived in Ref.~\cite{Qiu:2007ey}, and the gauge link structure of the quark TMDs in the target are consistent with the results of Ref.~\cite{Bomhof:2006dp}.

Finally, let us remind that we have studied parton-target scattering cross-sections. Hence, in order to apply our results to realistic processes, like proton-nucleus collisions, we would need to convolute our parton-target cross sections with the appropriate distributions on the projectile side. In general, for back-to-back dijet production, one uses TMDs on the projectile side as well. Note that our calculation, which is LO in $\alpha_s$ in the CGC (but NEik), was enough to determine the appropriate gauge link structure for the TMDs on the target side. By contrast, our LO calculations would not determine the gauge links for the TMDs on the projectile side. For that purpose, one would have to use instead the standard techniques from the TMD formalism \cite{Bomhof:2006dp}. 
Alternatively, in the case of proton scattering on a large nucleus, one can argue that the intrinsic transverse momentum extracted from the target, of the order of the saturation momentum $Q_s$ of the target, is parametrically larger than the intrinsic transverse momentum $\bq$ of the parton from the projectile, which is a non-perturbative QCD scale. In that case, one can neglect $\bq$ compared to $\bk$, and introduce ordinary PDFs for the projectile instead.

\section{Summary and outlook}
\label{sec:outlook}

In this paper, we presented a comprehensive study of dijet production in pA collisions at forward rapidities at NEik accuracy. We restricted ourselves to the NEik contributions where the incoming parton scatters on the target via a $t$-channel quark exchange. The processes that involves interaction with the quark background field are absent in the eikonal limit and they only start at NEik accuracy. In this paper we present the first study of dijet production in pA collisions at forward rapidities that includes quark background field of the target. In \cite{Altinoluk:2023qfr,}, a similar work was performed for the production of quark-gluon dijets in DIS at NEik accuracy by including the quark background field.   

After computing the scattering amplitudes both in general kinematics and in the back-to-back limit, we computed back-to-back production cross section for all available channels, namely $g\to gq$, $q\to q\bar q$, $q\to gg$ and $q\to qq$ channels. The analogous channels, with quark and antiquark  exchanged, are also studied and the results are presented in the App.~\ref{sec:antiquarks}. Since we included the quark background field of the target as source of NEik corrections in the computation of the dijet production cross section, in the back-to-back limit we probed various different quark TMDs. For all the channels, the back-to-back cross sections are obtained in a factorized form: a quark TMD times an associated hard factor.    

Recently, in \cite{Altinoluk:2024zom}, back-to-back dijet production in DIS at NEik accuracy has been computed. In that study, NEik corrections are considered in a pure gluon background. Among others, a remarkable observation in this study is related with the leading twist gluon TMDs. For the first time, non-zero value of momentum fraction ${\text x}$ in the twist 2 gluon TMDs are recovered from NEik corrections. Therefore, as an immediate continuation of this work, we plan to study  back-to-back dijet production in pA collisions at forward rapidities at NEik accuracy  in a pure gluon background and focus on a subset of NEik corrections to obtain non-zero value of momentum fraction ${\text x}$ for leading twist gluon TMDs in various channels. 

Another interesting observables to study at NEik accuracy is photon+jet production in pA collisions at forward rapidities. As shown in \cite{Dominguez:2011wm},  at eikonal accuracy, the cross section can be factorized into a hard part and dipole gluon TMD without the need of considering the back-to-back limit. This obsrevable should be studied at NEik order, by first considering a pure gluon background field, to see whether the factorization can be obtained without the back-to-back limit or not. Moreover, one should also consider the NEik contribution arising from the quark background of the target to probe quark TMDs in the back-to-back limit.

\section*{Acknowledgements}
We would like to thank Paul Caucal, Edmond Iancu, Farid Salazar and Feng Yuan for pointing out some issues with the earlier version of the draft, providing missing references, and constructive discussions.
TA is supported in part by the National Science Centre (Poland) under the research Grant No. 2023/50/E/ST2/00133 (SONATA BIS 13). GB, EB and SM are supported in part by the National Science Centre (Poland) under the research Grant No. 2020/38/E/ST2/00122 (SONATA BIS 10). 
 
\appendix 
\section{Technical details and useful relations}
\label{app:Tech}

\subsection{$\sun$ Algebra} 
\label{app:sun}          

\begin{figure}[!h]
    \centering
    \includegraphics[width=.8\textwidth]{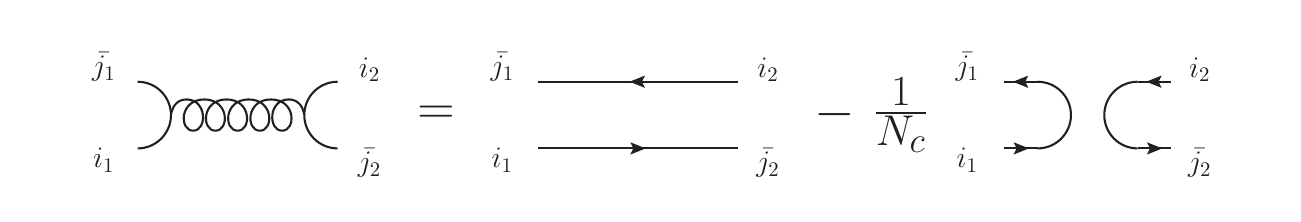}
    \caption{Diagrammatic representation of the Fierz identity (see \Equation{eq:sun_Fierz}).}
    \label{fig:Fierz}
\end{figure}

Generators of the $\sun$ Lie algebra in the fundamental representation are traceless hermitian $N_c\times N_c$ matrices that satisfy the commutation relation~:
\begin{equation}
    [t^a,t^b] = if^{abc}t^c \text{ with }
    a,b,c \in \llbracket 1,N_c^2-1 \rrbracket,
\label{eq:sun_commutator}
\end{equation}
where $f$ are totally symmetric structure constants.
We will consider the following normalization for those generators~:
\begin{equation}
    \Tr (t^at^b)=\frac{1}{2}\delta^{ab} .
\label{eq:sun_norm}
\end{equation}
The Casimir operator $C_F$ is then given by~:
\begin{equation}
    t^at^a = C_F \mathds{1}_F 
    = \frac{N_c^2-1}{2N_c}\mathds{1}_F .
\label{eq:sun_CF}
\end{equation}
One important relation for the calculations provided in this article is the Fierz identity~:
\begin{equation}
    (t^a)_{i_1j_1}(t^a)_{i_2j_2} = \frac{1}{2}\delta_{i_1j_2}\delta_{i_2j_1} - \frac{1}{2N_c}\delta_{i_1j_1}\delta_{i_2j_2}
    \text{ for } i_1, i_2, j_1, j_2 \in \llbracket 1,N_c \rrbracket ,
\label{eq:sun_Fierz}
\end{equation}
this relation is especially useful for rewriting color structure of quarks channels ($q\to qq$, $q\to q\Bar{q}$, $\bar q\to \bar q\bar q$ and $\bar q\to \bar q q$) and to write color structure involving adjoint Wilson lines with only fundamental Wilson lines.
A diagrammatic representation of this identity is shown in \Figure{fig:Fierz}.

Some other important relations implied when calculating the presented cross sections can be obtained using the Fierz identity, such as~:
\begin{align}
    t^at^bt^a &= -\frac{1}{2N_c}t^b , \nn \\
    \Tr(t^at^bt^at^c) &= -\frac{1}{4N_c}\delta^{bc} .
\end{align}

The commutation relation \Equation{eq:sun_commutator} can be generalized as~:
\begin{equation}
    t^at^b = \frac{1}{2}\left[\frac{1}{N_c}\delta^{ab}\mathds{1}_F
    +(d_{abc}+if_{abc})t^c\right], 
\label{eq:sun_tt}
\end{equation}
with the introduction of totally symmetric $d$-coefficients which, similarly to $f$, can be introduced through the anti-commutation relation~:
\begin{equation}
    \{t^a,t^b\} = \frac{1}{N_c}\delta^{ab}\mathds{1}_F + d_{abc}t^c .
\label{eq:sun_anti-commutator}
\end{equation}

Note that, from \Equation{eq:sun_tt}, one can define $f$ and $d$ as~:
\begin{align}
    f_{abc} = -2i\Tr\left[{[t^a,t^b]t^c}\right] , \nn \\
    d_{abc} = 2\Tr\left[\{t^a,t^b\}t^c\right] .
\end{align}
This implies, in particular, that $d_{aab}=d_{aba}=d_{baa}=0$. 
Since $d_{abc}$ is symmetric, while $f_{abc}$ is antisymmetric, we have $f_{abc}d_{abd} = 0$. Using the previous relation, one can show that~:
\begin{align}
    f_{abc}f_{abd} &= N_c\delta_{cd} , \nn \\
    d_{abc}d_{abd} &= \frac{N_c^2-4}{N_c}\delta_{cd} .
\end{align}

\subsection{Wilson lines properties} 
\label{app:WL}                    

Straight Wilson lines, between coordinates along the light-cone $y^+$ and $x^+$, at constant transverse coordinate $\bz$ are defined in the standard way as~:
\begin{equation}
    \UR(x^+,y^+; \bz) \equiv 
    \mathcal{P}_+e^{\left\{-ig{\textstyle\int}_{y^+}^{x^+}\rmd z^+ 
    T_\rmR\cdot\mathcal{A}^-(z^+;\bz)\right\}} ,
\label{eq:WL}
\end{equation}
where $\mathcal{P}_+$ denotes path-ordering of color matrices along the $+$ direction and R stands for the $\sun$ generators representation, being either A or F (for adjoint or fundamental respectively).

\begin{figure}[!h]
    \centering
    \includegraphics[width=.8\textwidth]{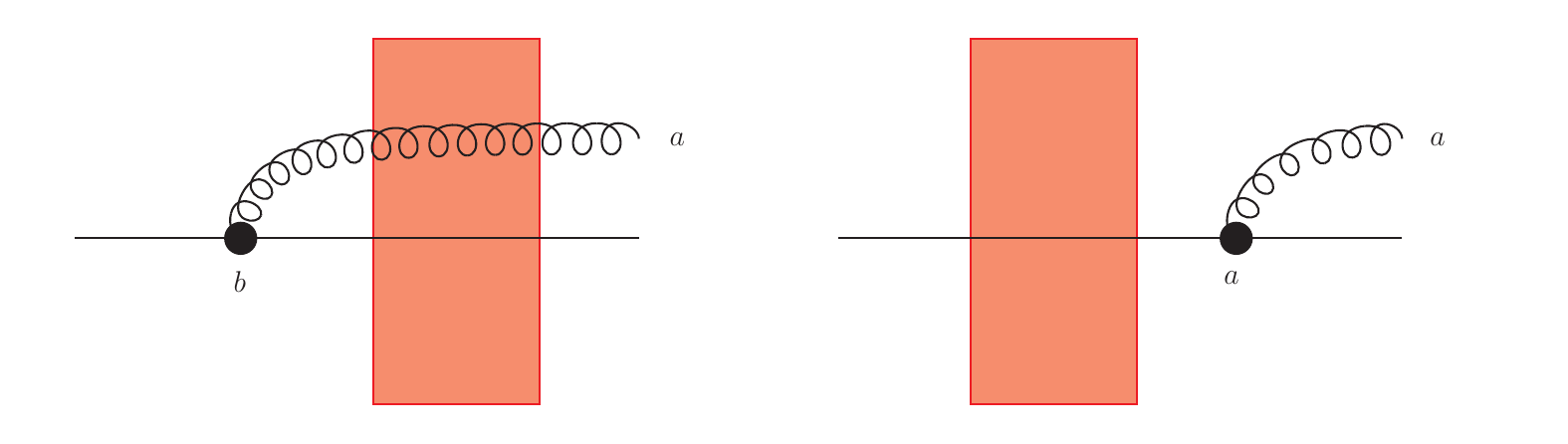}
    \includegraphics[width=.8\textwidth]{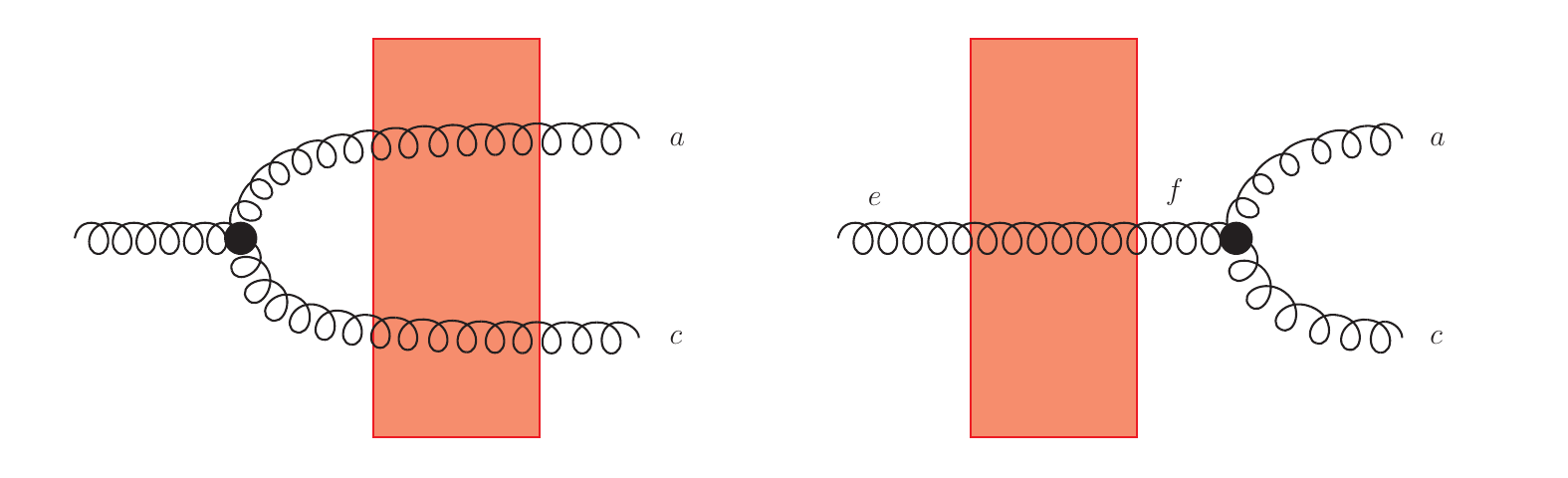}
    \caption{
    Diagrammatic representation of equation \Equation{eq:WL-relations}. The diagrams on the left and on the right have respectively an equivalent color structure in the correlation limit.}
    \label{fig:WL-relations}
\end{figure}

In the correlation limit, in which we provide most of our results, we need the following relations among Wilson lines~:
\begin{align}
    \UA^{ab}\left[\UR T_R^b\right] = \left[T_R^a\UR\right] , \nn \\
    \UA^{ab}\UA^{cd}f^{bde} = f^{acf}\UA^{fe} , \nn \\
    \UA^{ab}\UA^{cd}d^{bde} = d^{acf}\UA^{fe} ,
\label{eq:WL-relations}
\end{align}
where all the Wilson lines are along the same implicit path.
A diagrammatic representation of these relations is provided in \Figure{fig:WL-relations}.

Based on the previous relations, one can show that an adjoint Wilson line can be expressed in terms of fundamental ones via~:
\begin{equation}
    \UA^{ab} = 2\Tr(\UFd t^a\UF t^b) .
\label{eq:WL_atof}
\end{equation}

\subsection{Dirac Algebra for the cross section} 
\label{app:dirac}                             

The generators of the Dirac algebra satisfy the anti-commutation relation~:
\begin{equation}
    \{\gamma^\mu,\; \gamma^\nu\} = 2g^{\mu\nu}\mathds{1}_4 ,
\label{eq:Dirac_anticommutator}
\end{equation}
and which hermitian conjugate reads~:
\begin{equation}
    (\gamma^\mu)^\dagger = \gamma^0\gamma^\mu\gamma^0 .
\label{eq:Dirac_dagger}
\end{equation}

In particular, in the light-cone coordinates, the following relations involving $\gamma^\pm$ arise~:
\begin{align}
    &\{\gamma^\pm,\; \gamma^i\} = 0, \quad
    \{\gamma^+,\; \gamma^-\} = 2\mathds{1}_4, \nn \\
    &(\gamma^\pm)^2 = 0, \quad
    \gamma^\pm\gamma^\mp\gamma^\pm = 2\gamma^\pm\mathds{1}_4 .
\label{eq:Dirac_gamma+}
\end{align}

For the cross sections given in this article, mainly transverse gamma matrices are involved. Those satisfy the following relations~:
\begin{align}
    &\{\gamma^i,\; \gamma^j\} = 2g^{ij}\mathds{1}_4, \nn \\ 
    &\gamma^i\gamma^i = \frac{1}{2}\{\gamma^i,\; \gamma^i\} = g^{ii}\mathds{1}_4 = -2\mathds{1}_4, \nn \\
    &\gamma^i\gamma^j\gamma^i = \left(\{\gamma^i,\; \gamma^j\} - \gamma^j\gamma^i\right)\gamma^i = 2g^{ij}\gamma^i - \gamma^j\gamma^i\gamma^i = -2\gamma^j + 2\gamma^j = 0, \nn \\
    &\gamma^i\gamma^j\gamma^k\gamma^i = \left(\{\gamma^i,\; \gamma^j\} - \gamma^j\gamma^i\right)\gamma^k\gamma^i = -2\gamma^k\gamma^j .
\label{eq:Dirac_gammai}
\end{align}


The completeness relations for Dirac spinors reads~:
\begin{align}
    \sum_{h_1=\pm\frac{1}{2}}u(\check{p}_1,h_1)\overline{u}(\Check{p}_1,h_1) =
    \slashed{\check{p}}_1 + m, \nn \\
    \sum_{h_1=\pm\frac{1}{2}}v(\check{p}_1,h_1)\overline{v}(\check{p}_1,h_1) =
    \slashed{\check{p}}_1 - m ,
\label{eq:Dirac_ubu}
\end{align}
while for the transverse polarization vectors in the light-cone gauge, it writes~:
\begin{equation}
    \sum_{\lambda_1}\varepsilon_{\lambda_1}^{i}\varepsilon_{\lambda_1}^{j*} =
    \delta^{ij} .
\label{eq:Dirac_ees}
\end{equation}
Based on Eqs.~\eqref{eq:Dirac_gammai} and \eqref{eq:Dirac_ees}, we can write the hard factors for channel $g\to gq$ as squares of the coefficients $\mathfrak{h}^{(1)}_{g\to gq}$ and $\mathfrak{h}^{(2)}_{g\to gq}$ defined in Eqs.~\eqref{eq:HF_g_to_gq_1} and \eqref{eq:HF_g_to_gq_2}, as follows
\begin{align}
\label{eq:h_1_g_gq}
\sum_{\lambda_1,\lambda_2}
\left|\mathfrak{h}^{(1)}_{g\to gq}\right|^2 &=
    z^2\gamma^{l'}\bP^{l'}\left(\gamma^j\gamma^i + 2\frac{z}{1-z}g^{ij}\right)
    \left(\gamma^i\gamma^j + 2\frac{z}{1-z}g^{ij}\right) \gamma^{l}\bP^{l}\nn \\
    &= -4\bP^2z^2\left(1 + 2\frac{z}{1-z} + 2\frac{z^2}{(1-z)^2}\right)
    = -4\bP^2z^2 \frac{1+z^2}{(1-z)^2} , \\
\label{eq:h_2_g_gq}
    \sum_{\lambda_1,\lambda_2}
    \left|\mathfrak{h}^{(2)}_{g\to gq}\right|^2 &=
    \frac{1}{z^2}\sum_{\lambda_1,\lambda_2}\left|\mathfrak{h}^{(1)}_{g\to gq}\right|^2
    = -4\bP^2 \frac{1+z^2}{(1-z)^2} , \\
\label{eq:h_12_g_gq}
    \sum_{\lambda_1,\lambda_2}
    \left(\mathfrak{h}^{(1)}_{g\to gq}\right)^\dagger\mathfrak{h}^{(2)}_{g\to gq} &=
    \sum_{\lambda_1,\lambda_2}
    \left(\mathfrak{h}^{(2)}_{g\to gq}\right)^\dagger\mathfrak{h}^{(1)}_{g\to gq}
    = -\frac{1}{z}\sum_{\lambda_1,\lambda_2}\left|\mathfrak{h}^{(1)}_{g\to gq}\right|^2
    = 4\bP^2z \frac{1+z^2}{(1-z)^2}
\end{align}
Similarly, the hard factors of $q\to gg$ are obtained from the coefficients $\mathfrak{h}^{(1)}_{q\to gg}$ and $\mathfrak{h}^{(2)}_{q\to gg}$, defined in Eqs.~\eqref{eq:HF_q_to_gg_1} and \eqref{eq:HF_q_to_gg_2}, via~:
\begin{align}
\sum_{\lambda_1,\lambda_2}\Big|\mathfrak{h}^{(1)}_{q\to gg}\Big|^2
    &= \big[(1-z)\gamma^j\gamma^i - z\gamma^i\gamma^j\big]
    \bP^{l'}\gamma^{l'}\bP^l\gamma^l
    \big[(1-z)\gamma^i\gamma^j - z\gamma^j\gamma^i\big] \nn \\
    &= -4\bP^2\big[(1-z)^2 + z^2\big] 
    , \nn \\
    \sum_{\lambda_1,\lambda_2}\Big|\mathfrak{h}^{(2)}_{q\to gg}\Big|^2
    &= (1-2z)^2\sum_{\lambda_1,\lambda_2}\Big|\mathfrak{h}^{(1)}_{q\to gg}\Big|^2
    =  -4\bP^2(1-2z)^2\big[z^2 + (1-z)^2\big] \nn \\
    &= -4\bP^2\left[2\left(z^{2} + (1-z)^{2}\right)^2
    - \left(z^{2} + (1-z)^{2}\right)\right] .
\end{align}
In order to calculate the square of the coefficients $\mathfrak{h}_{q\to q\bar q}^{(1)}$ and $\mathfrak{h}_{q\to q\bar q}^{(1)}$, defined in Eqs~\eqref{eq:HF_q_to_qbarq_1} and \eqref{eq:HF_q_to_qbarq_2}, appearing in the $q\to q\bar q$ channel, let's first calculate the following term
\begin{align}
\sum_{h, h_1, h_2}\left|\mathfrak{h'}^{(1)}_{q\to q\bar q}\right|^2 
&=
\frac{1}{16}\sum_{h, h_1, h_2}
\gamma^{-}\gamma^{+}\gamma^{i}v(\check{p}_{2},h_{2})\overline{u}(\check{p}_{1},h_{1})\gamma^{+}
\bigg(\gamma^{l}\gamma^{i}\P^l - 2\frac{z}{1-z}\P^i\bigg)u(\check{q},h)
\nn \\
& \times
\overline{u}(\check{q},h)
\bigg(\gamma^{j}\gamma^{l'}\P^{l'} - 2\frac{z}{1-z}\P^j\bigg)
\gamma^{+}u(\check{p}_{1},h_{1})
\overline{v}(\check{p}_{2},h_{2})\gamma^{j}\gamma^{+}\gamma^{-}\gamma^0
\nn \\
&=
-\frac{1}{8}(2q^+)(2p_2^+)\gamma^{-}\gamma^0\gamma^{i}\gamma^{j}
\Tr\Bigg[\slashed{p}_1\gamma^+\bigg\{\gamma^l\gamma^i\gamma^j\gamma^{l'}P^lP^{l'}
\nn \\
& \hspace{3.5cm}
- 2\frac{z}{1-z}\bigg(\gamma^j\gamma^{l'}\bP^i\bP^{l'} + \gamma^l\gamma^i\bP^j\bP^l\bigg)
+ 4\frac{z^2}{(1-z)^2}\P^i\P^j\bigg\}\Bigg] \nn \\
&=
-2q^+p_1^+p_2^+\gamma^{-}\gamma^0\gamma^{i}\gamma^{j}
\bigg(-\bP^2g^{ij} + 4\frac{z}{1-z}\P^i\P^j + 4\frac{z^2}{(1-z)^2}\P^i\P^j\bigg)
\nn \\
&= \frac{\bP^2}{2}(2q^+)^3z(1-z)\gamma^-\gamma^0\bigg(1 + 2\frac{z}{(1-z)^2}\bigg)
= \frac{\bP^2}{2}(2q^+)^3\gamma^-\gamma^0z\frac{1+z^2}{1-z} ,
\end{align}
where we have used the cyclicity of the trace and in particular $\Tr[\gamma^{i'}\gamma^{i}] = \Tr\Big[\frac{1}{2}\{\gamma^{i'},\gamma^{i}\} \Big] = 4g^{i'i}$.
The same way, we calculate
\begin{align}
\sum_{h, h_1, h_2}\left|\mathfrak{h'}^{(2)}_{q\to q\bar q}\right|^2 &=
\frac{\bP^2}{2}(2q^+)^3\gamma^-\gamma^0 z(1-z)\left(z^2 + (1-z)^2\right)
\nn \\
\sum_{h, h_1, h_2}
\left(\mathfrak{h'}^{(1)}_{q\to q\bar q}\right)
\left(\mathfrak{h'}^{(2)}_{q\to q\bar q}\right)^\dagger 
&=
\sum_{h, h_1, h_2}\left(\mathfrak{h'}^{(2)}_{q\to q\bar q}\right)
\left(\mathfrak{h'}^{(1)}_{q\to q\bar q}\right)^\dagger
= -\frac{\bP^2}{2}(2q^+)^3\gamma^-\gamma^0
z^3
\end{align}
From these results, we can write
\begin{align}
\label{eq:HF_q_to_qbarq_11}
\sum_{h, h_1, h_2}\left|\mathfrak{h}_{q\to q\bar q}^{(1)}\right|^2 =&
    \sum_{h, h_1, h_2}\left(-\mathfrak{h'}^{(1)}_{q\to q\bar q} 
    + \frac{1}{N_c}\mathfrak{h'}^{(2)}_{q\to q\bar q}\right)^\dagger
    \left(-\mathfrak{h'}^{(1)}_{q\to q\bar q} 
    + \frac{1}{N_c}\mathfrak{h'}^{(2)}_{q\to q\bar q}\right)
\nn \\
    &= \sum_{h, h_1, h_2}\left[\left|\mathfrak{h'}^{(1)}_{q\to q\bar q}\right|^2
    - \frac{2}{N_c}\left(\mathfrak{h'}^{(2)}_{q\to q\bar q}\right)^\dagger\left(\mathfrak{h'}^{(1)}_{q\to q\bar q}\right)
    + \frac{1}{N_c^2}\left|\mathfrak{h'}^{(2)}_{q\to q\bar q}\right|^2\right]
\nn \\
& = \frac{\bP^2}{2}(2q^+)^3\gamma^-\gamma^0 z(1-z)
    \left[\frac{1+z^2}{(1-z)^2} + \frac{2}{N_c}\frac{z^2}{1-z}
    + \frac{1}{N_c^2}\left(z^2 + (1-z)^2\right)\right] ,
\\
\label{eq:HF_q_to_qbarq_22}
\sum_{h, h_1, h_2}\left|\mathfrak{h}_{q\to q\bar q}^{(2)}\right|^2 
=&
    \sum_{h, h_1, h_2}\left(-\mathfrak{h'}^{(2)}_{q\to q\bar q} 
    + \frac{1}{N_c}\mathfrak{h'}^{(1)}_{q\to q\bar q}\right)^\dagger
    \left(-\mathfrak{h'}^{(2)}_{q\to q\bar q} 
    + \frac{1}{N_c}\mathfrak{h'}^{(1)}_{q\to q\bar q}\right)
\nn \\
& =
    \sum_{h, h_1, h_2}\left[\left|\mathfrak{h'}^{(1)}_{q\to q\bar q}\right|^2
    - \frac{2}{N_c}\left(\mathfrak{h'}^{(1)}_{q\to q\bar q}\right)^\dagger\left(\mathfrak{h'}^{(2)}_{q\to q\bar q}\right)
    + \frac{1}{N_c^2}\left|\mathfrak{h'}^{(1)}_{q\to q\bar q}\right|^2\right]
\nn \\
& = \frac{\bP^2}{2}(2q^+)^3\gamma^-\gamma^0 z(1-z)
    \left[z^2 + (1-z)^2 + \frac{2}{N_c}\frac{z^2}{1-z}
    + \frac{1}{N_c^2}\frac{1+z^2}{(1-z)^2}\right] ,
\end{align}
and
\begin{align}
\label{eq:HF_q_to_qbarq_12}
&
\sum_{h, h_1, h_2}\left(\mathfrak{h}^{(1)}_{q\to q\bar q}\right)\left(\mathfrak{h}^{(2)}_{q\to q\bar q}\right)^\dagger
=
    \sum_{h, h_1, h_2}\left(-\mathfrak{h'}^{(1)}_{q\to q\bar q} 
    + \frac{1}{N_c}\mathfrak{h'}^{(2)}_{q\to q\bar q}\right)
    \left(-\mathfrak{h'}^{(2)}_{q\to q\bar q} 
    + \frac{1}{N_c}\mathfrak{h'}^{(1)}_{q\to q\bar q}\right)^\dagger
\nn \\
& = \sum_{h, h_1, h_2}\left[
    \left(1+\frac{1}{N_c^2}\right)\left(\mathfrak{h'}^{(1)}_{q\to q\bar q}\right)
    \left(\mathfrak{h'}^{(2)}_{q\to q\bar q}\right)^\dagger
    -\frac{1}{N_c}\left(\left|\mathfrak{h'}^{(1)}_{q\to q\bar q}\right|^2
    + \left|\mathfrak{h'}^{(2)}_{q\to q\bar q}\right|^2\right)\right]
\nn \\
& = -\frac{\bP^2}{2}(2q^+)^3\gamma^-\gamma^0 z(1-z)\left[
    \left(1+\frac{1}{N_c^2}\right)\frac{z^2}{1-z}
    + \frac{1}{N_c}\left(z^2 + (1-z)^2 + \frac{1+z^2}{(1-z)^2}\right)\right]
    \nn \\
& =
\sum_{h, h_1, h_2}\left(\mathfrak{h}^{(2)}_{q\to q\bar q}\right)\left(\mathfrak{h}^{(1)}_{q\to q\bar q}\right)^\dagger
.
\end{align}
The same procedure can be applied to calculate the hard factors for $q\to qq$ channel, starting with
\begin{align}
\sum_{h, h_1, h_2}\left|\mathfrak{h'}^{(1)}_{q\to qq}\right|^2 
&=
2\bP^2(2q^+)^3\gamma^0\gamma^-\frac{z}{1-z}\left(1+z^2\right)
\nn \\
\sum_{h, h_1, h_2}\left|\mathfrak{h'}^{(2)}_{q\to qq}\right|^2 
&=
2\bP^2(2q^+)^3\gamma^0\gamma^-\frac{1-z}{z}\left(1+(1-z)^2\right)
\nn \\
\sum_{h, h_1, h_2}\left(\mathfrak{h'}^{(1)}_{q\to qq}\right)^\dagger\left(\mathfrak{h'}^{(2)}_{q\to qq}\right) 
&=
\sum_{h, h_1, h_2}\left(\mathfrak{h'}^{(2)}_{q\to qq}\right)^\dagger\left(\mathfrak{h'}^{(1)}_{q\to qq}\right)
= 2\bP^2(2q^+)^3\gamma^0\gamma^- ,
\end{align}
which leads to
\begin{align}
\label{eq:HF_q_to_qq_11}
\sum_{h, h_1, h_2}\left|\mathfrak{h}_{q\to qq}^{(1)}\right|^2 =&
    \sum_{h, h_1, h_2}
\left(\mathfrak{h'}^{(1)}_{q\to qq} 
    - \frac{1}{N_c}\mathfrak{h'}^{(2)}_{q\to qq}\right)^\dagger
    \left(\mathfrak{h'}^{(1)}_{q\to qq} 
    - \frac{1}{N_c}\mathfrak{h'}^{(2)}_{q\to qq}\right)
\nn \\
    &= \sum_{h, h_1, h_2}\left[\left|\mathfrak{h'}^{(1)}_{q\to qq}\right|^2
    - \frac{2}{N_c}\left(\mathfrak{h'}^{(2)}_{q\to qq}\right)^\dagger\left(\mathfrak{h'}^{(1)}_{q\to qq}\right)
    + \frac{1}{N_c^2}\left|\mathfrak{h'}^{(2)}_{q\to qq}\right|^2\right]
\nn \\
& = 2\bP^2(2q^+)^3\gamma^0\gamma^-
    \left[\frac{z}{1-z}\left(1+z^2\right) - \frac{2}{N_c}
    +\frac{1}{N_c^2}\frac{1-z}{z}\left(1+(1-z)^2\right)\right] ,
\\
\label{eq:HF_q_to_qq_22}
\sum_{h, h_1, h_2}\left|\mathfrak{h}_{q\to qq}^{(2)}\right|^2 =&
    \sum_{h, h_1, h_2}\left(\mathfrak{h'}^{(2)}_{q\to qq} 
    - \frac{1}{N_c}\mathfrak{h'}^{(1)}_{q\to qq}\right)^\dagger
    \left(\mathfrak{h'}^{(2)}_{q\to qq} 
    - \frac{1}{N_c}\mathfrak{h'}^{(1)}_{q\to qq}\right)
\nn \\
& =
    \sum_{h, h_1, h_2}\left[\left|\mathfrak{h'}^{(2)}_{q\to qq}\right|^2
    - \frac{2}{N_c}\left(\mathfrak{h'}^{(1)}_{q\to qq}\right)^\dagger\left(\mathfrak{h'}^{(2)}_{q\to qq}\right)
    + \frac{1}{N_c^2}\left|\mathfrak{h'}^{(1)}_{q\to qq}\right|^2\right]
\nn \\
& = 2\bP^2(2q^+)^3\gamma^0\gamma^-
    \left[ \frac{1-z}{z}\left(1+(1-z)^2\right)
    - \frac{2}{N_c}
    +\frac{1}{N_c^2}\frac{z}{1-z}\left(1+z^2\right)\right] ,
\end{align}
and
\begin{align}
\label{eq:HF_q_to_qq_12}
&
\sum_{h, h_1, h_2}\left(\mathfrak{h}^{(1)}_{q\to qq}\right)^\dagger\left(\mathfrak{h}^{(2)}_{q\to qq}\right) 
=\sum_{h, h_1, h_2}\left(\mathfrak{h}^{(2)}_{q\to qq}\right)^\dagger\left(\mathfrak{h}^{(1)}_{q\to qq}\right) 
\nn \\
& =    \sum_{h, h_1, h_2}\left(\mathfrak{h'}^{(1)}_{q\to qq} 
    - \frac{1}{N_c}\mathfrak{h'}^{(2)}_{q\to qq}\right)^\dagger
    \left(\mathfrak{h'}^{(2)}_{q\to qq} 
    - \frac{1}{N_c}\mathfrak{h'}^{(1)}_{q\to qq}\right)
\nn \\
& = \sum_{h, h_1, h_2}\left[
    \left(1+\frac{1}{N_c^2}\right)\left(\mathfrak{h'}^{(1)}_{q\to qq}\right)^\dagger
    \left(\mathfrak{h'}^{(2)}_{q\to qq}\right)
    -\frac{1}{N_c}\left(\left|\mathfrak{h'}^{(1)}_{q\to qq}\right|^2
    + \left|\mathfrak{h'}^{(2)}_{q\to qq}\right|^2\right)\right]
\nn \\
& = 2\bP^2(2q^+)^3\gamma^0\gamma^-
    \left[\left(1+\frac{1}{N_c^2}\right)
    -\frac{2}{N_c}\left(z(1-z) - 3 + \frac{1}{1-z}\right)\right] .
\end{align}
%
%
%
%
%

\subsection{Fundamental Wilson line only color structures}
\label{app:fonlycolor}                                    

\begin{figure}[H]
\centering
\includegraphics[width=0.75 \textwidth]{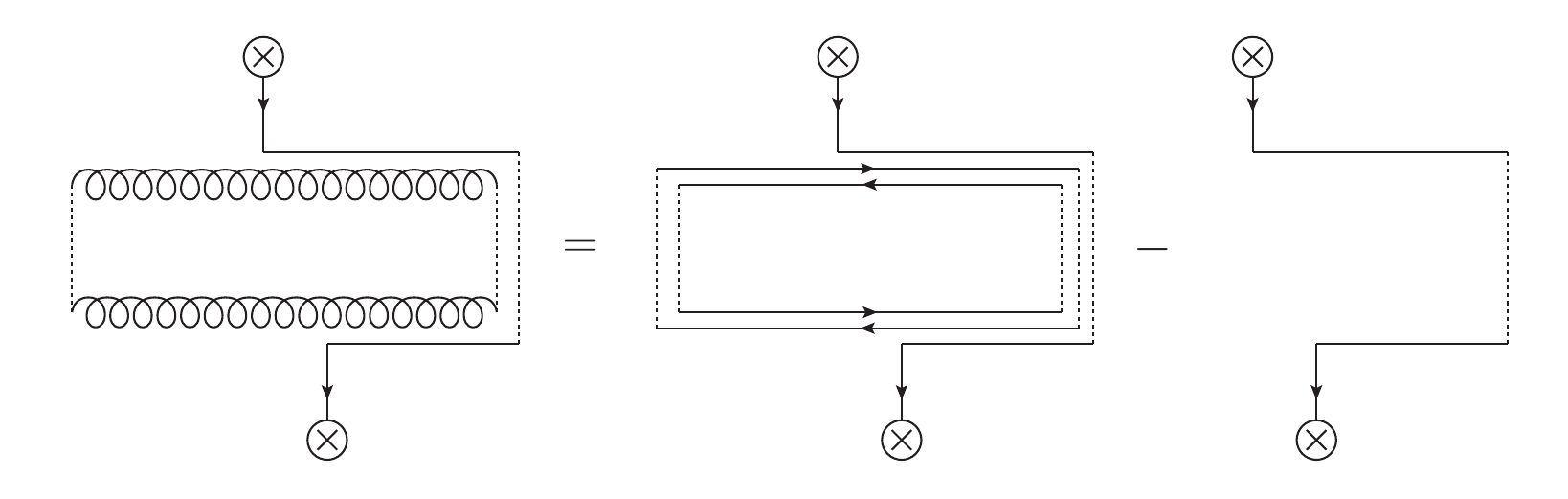}
\caption{Decomposition of $\mathcal{C}^{+\square_g}$ in terms of color structures involving fundamental Wilson lines only.}
\label{fig:C+lgtoF}
\end{figure}

In the results presented in this paper, we have shown the color structure encountered in dijet cross section coming from exotic channels. We have kept them as simple as possible which, in particular, implied keeping adjoint Wilson lines.
It is possible to write those color structure, to be more precise, to rewrite $\mathcal{C}^{-g}$, $\mathcal{C}^{+g}$ and $\mathcal{C}^{+\square_g}$ in terms of colors structures involving only fundamental Wilson lines.
This can be done in several ways, using different relations as shown in \Appendix{app:WL}. For coherence between calculation, we will use here mainly \Equation{eq:WL_atof} which directly writes adjoint Wilson lines in term of fundamental ones. Also, one may notice that, in the color structure presented, these adjoint Wilson lines come by pair.
Hence, let's start by an intermediate calculation~:
\begin{align}
    \UA(x^+,&y^+;\bb)_{ba}\UA({x'}^+,{y'}^+;\bb')_{ba} \nn \\
    =& 4\Tr\left[\UFd(x^+,y^+;\bb)t^b\UF(x^+,y^+;\bb)t^a\right]
    \Tr\left[\UFd({x'}^+,{y'}^+;\bb')t^b\UF({x'}^+,{y'}^+;\bb')t^a\right] \nn \\
    =& 2\Tr\left[\UFd(x^+,y^+;\bb)\UF({x'}^+,{y'}^+;\bb')t^a
    \UFd({x'}^+,{y'}^+;\bb')\UF(x^+,y^+;\bb)t^a\right] \nn \\
    & - \frac{2}{N_c}\cancel{\Tr\left[t^a\right]}
    \cancel{\Tr\left[t^a\right]} \nn \\
    =& \Tr\left[\UFd(x^+,y^+;\bb)\UF({x'}^+,{y'}^+;\bb')\right]
    \Tr\left[\UFd({x'}^+,{y'}^+;\bb')\UF(x^+,y^+;\bb)\right] \nn \\
    & - \frac{1}{N_c}\Tr\left[\mathds{1}_F\right] \nn \\
    =& \Tr\left[\UFd(x^+,y^+;\bb)\UF({x'}^+,{y'}^+;\bb)\right]
    \Tr\left[\UFd({x'}^+,{y'}^+;\bb)\UF(x^+,y^+;\bb)\right]
    - 1 ,
\label{eq:UAUA}
\end{align}
where, in the first step, we applied the Fierz identity (see \Equation{eq:sun_Fierz}) on $t^b$ then on $t^a$ in the second step.
We can directly apply this result to $\mathcal{C}^{+\square_g}$ with $x^+={x'}^+=\infty$ and $y^+={y'}^+=-\infty$ and conclude~:
\begin{equation}
    \mathcal{C}^{+\square_g} = \mathcal{C}^{+\square^2}
    -\mathcal{C}^{+} ,
\label{eq:C+lgtoF}
\end{equation}

which makes appear two color structures not present in other channels~:
\begin{align}
    \mathcal{C}^{+\square^2} \equiv &
    \bigg\langle 
    \Tr\left[\UF(\infty,-\infty;\bb')\UFd(\infty,-\infty;\bb)\right]
    \Tr\left[\UFd(\infty,-\infty;\bb')\UF(\infty,-\infty;\bb)\right] 
    \nn \\
    &\times 
    \overline{\Psi}({z'}^+;\bb')\gamma^-
    \UFd(\infty,{z'}^+;\bb')\UF(\infty,z^+;\bb)\Psi({z}^+;\bb)\bigg\rangle
    ,
\label{eq:C+ll} \\
    \mathcal{C}^{+} \equiv &
    \left\langle\overline{\Psi}({z'}^+;\bb')\gamma^-
    \UFd(\infty,{z'}^+;\bb')\UF(\infty,z^+;\bb)\Psi({z}^+;\bb)\right\rangle .
\label{eq:C+}
\end{align}
The relation in \Equation{eq:C+lgtoF} can be visualized in \Figure{fig:C+lgtoF}.

\begin{figure}[H]
\centering
\includegraphics[width=0.75 \textwidth]{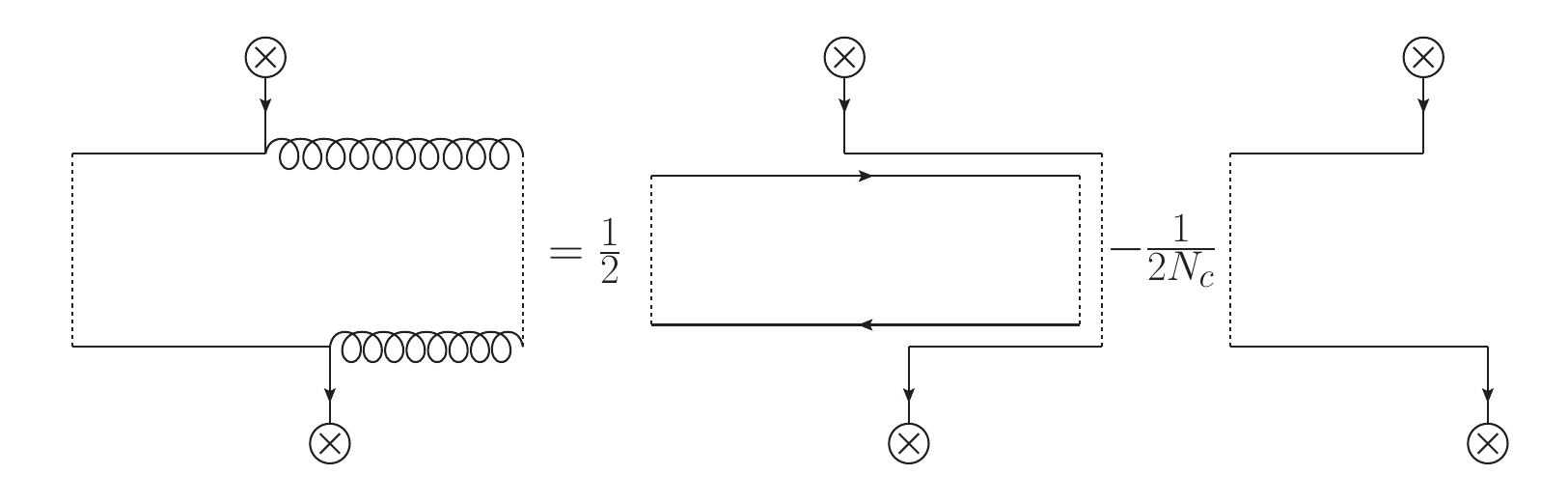}
\caption{Decomposition of $\mathcal{C}^{-g}$ in terms of color structures involving fundamental Wilson lines only.}
\label{fig:C-gtoF}
\end{figure}

We can generalize \Equation{eq:UAUA} for two adjoint Wilson lines with only one color index in common, for which we can apply the Fierz identity~:
\begin{align}
    \UA(x^+,&y^+;\bb)_{ba}\UA({x'}^+,{y'}^+;\bb')_{ba'} \nn \\
    =& 4\Tr\left[\UFd(x^+,y^+;\bb)t^b\UF(x^+,y^+;\bb)t^a\right]
    \Tr\left[\UFd({x'}^+,{y'}^+;\bb')t^b\UF({x'}^+,{y'}^+;\bb')t^{a'}\right] \nn \\
    =& 2\Tr\left[\UFd(x^+,y^+;\bb)\UF({x'}^+,{y'}^+;\bb')t^{a'}
    \UFd({x'}^+,{y'}^+;\bb')\UF(x^+,y^+;\bb)t^a\right] \nn \\
    & - \frac{2}{N_c}\cancel{\Tr\left[t^a\right]}
    \cancel{\Tr\left[t^{a'}\right]} .
\label{eq:UAUA2}
\end{align}
This is the first step to rewrite both $\mathcal{C}^{-g}$ and $\mathcal{C}^{+g}$. Let's begin with $\mathcal{C}^{-g}$~:
\begin{align}
    \mathcal{C}^{-g} =&
    \Big\langle\overline{\Psi}({z'}^+;\bb')\gamma^-t^{c'}
    \UF({z'}^+,-\infty;\bb')\UFd(z^+,-\infty;\bb)t^c\Psi({z}^+;\bb) 
    \nn \\
    &\hspace{5cm}\times
    \UA(\infty,{z'}^+;\bb')_{bc'}\UA(\infty,z^+;\bb)_{bc}\Big\rangle 
    \nn \\    
    =& 2\bigg\langle\overline{\Psi}({z'}^+;\bb')\gamma^-t^{c'}
    \UF({z'}^+,-\infty;\bb')\UFd(z^+,-\infty;\bb)t^c\Psi({z}^+;\bb)
    \nn \\
    &\times \Tr\left[\UFd(\infty,{z'}^+;\bb')\UF(\infty,z^+;\bb)t^{c}
    \UFd(\infty,z^+;\bb)\UF(\infty,{z'}^+;\bb')t^{c'}\right]
    \bigg\rangle
    \nn \\
    =& \bigg\langle\overline{\Psi}({z'}^+;\bb')\gamma^-t^{c'}
    \UF({z'}^+,-\infty;\bb')
    \UFd(\infty,-\infty;\bb) 
    \nn \\
    &\hspace{2cm}\times
    \UF(\infty,{z'}^+;\bb')t^{c'}
    \UFd(\infty,{z'}^+;\bb')\UF(\infty,z^+;\bb)\Psi({z}^+;\bb)\bigg\rangle 
    \nn \\
    &-\frac{1}{N_c}\left\langle\overline{\Psi}({z'}^+;\bb')\gamma^-t^{c'}
    \UF({z'}^+,-\infty;\bb')\UFd(z^+,-\infty;\bb)\Psi({z}^+;\bb)\right\rangle\cancel{\Tr\left[ t^{c'}\right]} 
    \nn \\
    =& \frac{1}{2}\bigg\langle\overline{\Psi}({z'}^+;\bb')\gamma^-
    \UFd(\infty,{z'}^+;\bb')\UF(\infty,z^+;\bb)\Psi({z}^+;\bb)
    \nn \\
    &\hspace{2cm}\times
    \Tr\left[\UF(\infty,-\infty;\bb')\UFd(\infty,-\infty;\bb)\right]
    \bigg\rangle 
    \nn \\
    & - \frac{1}{2N_c}\left\langle\overline{\Psi}({z'}^+;\bb')\gamma^-
    \UF({z'}^+,-\infty;\bb')\UFd(\infty,-\infty;\bb)
    \mathds{1}_F\UF(\infty,z^+;\bb)\Psi({z}^+;\bb)\right\rangle 
    \nn \\
    =& \frac{1}{2}\mathcal{C}^{+\square} 
    - \frac{1}{2N_c}\mathcal{C}^{-} ,
\label{eq:C-gtoF}
\end{align}

where, after applying the result of \Equation{eq:UAUA2}, we have applied 2 Fierz identity, first on $t^c$ then on $t^{c'}$.
The obtained relation is diagrammatically pictured in \Figure{fig:C-gtoF}.

\begin{figure}[H]
\centering
\includegraphics[width=0.75 \textwidth]{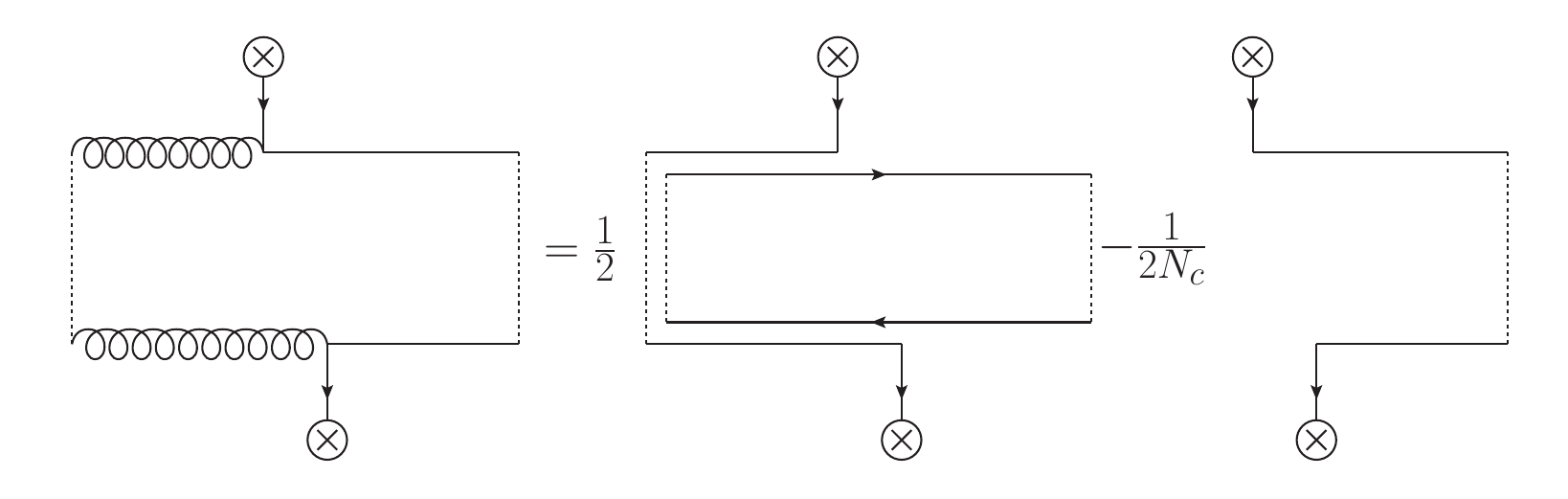}
\caption{Decomposition of $\mathcal{C}^{+g}$ in terms of color structures involving fundamental Wilson lines only.}
\label{fig:C+gtoF}
\end{figure}

Similarly for $\mathcal{C}^{+g}$, we have~:
\begin{align}
    \mathcal{C}^{+g} =&
    \bigg\langle\overline{\Psi}({z'}^+;\bb')\gamma^-t^{c'}
    \UFd(\infty,{z'}^+;\bb')\UF(\infty,z^+;\bb)t^c\Psi({z}^+;\bb)
    \nn \\
    &\times 
    \UA({z'}^+,-\infty;\bb')_{c'a}\UA(z^+,-\infty;\bb)_{ca}
    \bigg\rangle 
    \nn \\
    =& 2\bigg\langle\overline{\Psi}({z'}^+;\bb')\gamma^-t^{c'}
    \UFd(\infty,{z'}^+;\bb')\UF(\infty,z^+;\bb)t^c\Psi({z}^+;\bb) \nn \\
    &\times \Tr\left[\UFd({z'}^+,-\infty;\bb')t^{c'}\UF({z'}^+,-\infty;\bb')
    \UFd(z^+,-\infty;\bb)t^{c}\UF(z^+,-\infty;\bb)\right]
    \bigg\rangle
    \nn \\
    =& \frac{1}{2}\bigg\langle\overline{\Psi}({z'}^+;\bb')\gamma^-
    \UF({z'}^+,-\infty;\bb')\UFd(z^+,-\infty;\bb)\Psi({z}^+;\bb)
    \nn \\
    &\times
    \Tr\left[\UFd(\infty,-\infty;\bb')\UF(\infty,-\infty;\bb)\right]
    \bigg\rangle
    \nn \\
    & - \frac{1}{2N_c}\left\langle\overline{\Psi}({z'}^+;\bb')\gamma^-
    \UFd(\infty,{z'}^+;\bb')\UF(\infty,-\infty;\bb)
    \mathds{1}_F\UFd(z^+,-\infty;\bb)\Psi({z}^+;\bb)\right\rangle \nn \\
    =& \frac{1}{2}\mathcal{C}^{-\square} 
    - \frac{1}{2N_c}\mathcal{C}^{+} ,
\label{eq:C+gtoF}
\end{align}
with another color structures that wasn't appearing in previous results~:
\begin{align}
    \mathcal{C}^{-\square} 
    \equiv &
\bigg\langle\overline{\Psi}({z'}^+;\bb')\gamma^-
    \UF({z'}^+,-\infty;\bb')\UFd(z^+,-\infty;\bb)\Psi({z}^+;\bb)
    \nn \\
    &\times
    \Tr\left[\UFd(\infty,-\infty;\bb')\UF(\infty,-\infty;\bb)\right]
    \bigg\rangle
, 
\label{eq:C-l}
\end{align}
This result is diagrammatically shown in \Figure{fig:C+gtoF}.

Following the method from Sec.~\ref{sec:TMD}, one can relate the color structures ${\cal C}^{(\cdots)}$ to the corresponding unpolarized quark TMDs $f^{(\cdots)}_q$. Then at the TMD level, the relations \eqref{eq:C+lgtoF}, \eqref{eq:C-gtoF} and \eqref{eq:C+gtoF} become
\begin{align}
(N_c^2\!-\!1)\, f^{+\square_g}_q({\text x},\bk) 
=&\, 
N_c^2\, f^{+\square^2}_q({\text x},\bk) 
-f^{+}_q({\text x},\bk) 
\nn\\
(N_c^2\!-\!1)\, f^{-g}_q({\text x},\bk) 
=&\, 
N_c^2\, f^{+\square}_q({\text x},\bk) 
-f^{-}_q({\text x},\bk) 
\nn\\
(N_c^2\!-\!1)\, f^{+g}_q({\text x},\bk) 
=&\, 
N_c^2\, f^{-\square}_q({\text x},\bk) 
-f^{+}_q({\text x},\bk) 
\, ,
\end{align}
using the color factors listed in Eq.~\eqref{eq:color_factors}.

Note that the relations Eqs.~\eqref{eq:C+lgtoF}, \eqref{eq:C-gtoF} and \eqref{eq:C+gtoF} hold for the $\overline{\mathcal{C}}$ color structures which then leads to the same relations between antiquark TMDs.

\section{Eikonal parton propagators} 
\label{app:eikprop}           
This appendix is devoted to recalling various parton propagators at eikonal accuracy that are needed to perform the computations in this paper. In recent work \cite{Altinoluk:2024dba}, the explicit expressions are provided.  
For the gluon propagator we have three types. The before-to-after gluon propagator: 
\begin{align}
\label{gluon_prop_BA}
&
{G_\rmF^{\mu\nu}(x,y)\BA} = 
\int\frac{d^3\underline{k_1}}{(2\pi)^3}\, \theta(k_1^+) \, \exp{-ix\cdot\Check{k}_1}
\int\frac{d^3\underline{k_2}}{(2\pi)^3} \, \frac{\theta(k_2^+)}{k_1^++k_2^+} \, \exp{iy\cdot\Check{k}_2} 
\\
&
\times
\left[-g^{\mu\nu} + \frac{\Check{k}_2^\mu\eta^\nu}{k_2^+} + \frac{\eta^\mu\Check{k}_1^\nu}{k_1^+} - \frac{\eta^\mu\eta^\nu}{k_1^+k_2^+}\Check{k}_1\cdot\Check{k}_2\right] 
\int_{\z}\exp{-i \z\cdot(\k_1-\k_2)}
\int_{z^-}\exp{iz^-(k_1^+-k_2^+)} \ \UA(x^+,y^+;\z) . \nn 
 \end{align}
 The before-to-inside gluon propagator (starting from before the medium and ending inside the medium)
 \begin{align}
 \label{gluon_prop_BI}
 &
 {G_\rmF^{\mu\nu}(x,y)\BI} = 
 \int\frac{d^3\underline{k}}{(2\pi)^3}\frac{\theta(k^+)}{2k^+}
\exp{iy\cdot\Check{k}}\exp{-ix\cdot\underline{k}} \nn \\
&
\times
\left[-g^\mu_{\:\:  j}\, g^{j\nu} + g^\mu_{\:\:j} \eta^\nu\frac{\k^j}{k^+} + i\left(\frac{\eta^\mu g^\nu_{\:\: j}}{k^+} - \eta^\mu\eta^\nu\frac{\k^j}{(k^+)^2}\right)
\left(\RPartial_{\x^j}^A+i\k^j\right)\right]    
    \UA(x^+,y^+;\x) .
\end{align}
 The inside-to-after gluon propagator (starting from inside the medium and ending after the medium)
\begin{align}
\label{gluon_prop_IA}
& \hspace{-1.7cm}
{G_\rmF^{\mu\nu}(x,y)\IA}= 
\int\frac{d^3\underline{k}}{(2\pi)^3} \, \frac{\theta(k^+)}{2k^+}
\, \exp{-ix\cdot\Check{k}} \, 
\UA(x^+,y^+;\y) \nn \\
& \hspace{-1.7cm}
\times\left[-g^\mu_{\:\: j}g^{j\nu} + \eta^\mu g^\nu_{\:\: j}\frac{\k^j}{k^+}
+ i\left(\frac{g^\mu_{\:\: j}\eta^\nu}{k^+} + \eta^\mu\eta^\nu\frac{\k^j}{(k^+)^2}\right)
\left(\LPartial_{\y^j}^A-i\k^j\right)\right]\exp{iy\cdot\underline{k}} .
\end{align}
The before-to-after quark propagator: 
\begin{align}
\label{quark_prop_BA}
& \hspace{-2cm}
S_\rmF(x,y)\BAq =
    \int\frac{d^3\underline{k_1}}{(2\pi)^3}\frac{\theta(k_1^+)}{2k_1^+}\exp{-ix\cdot\Check{k}_1}
    \int\frac{d^3\underline{k_2}}{(2\pi)^3}\frac{\theta(k_2^+)}{2k_2^+}\exp{iy\cdot\Check{k}_2}
    \int_{\z}\exp{-i\z\cdot(\k_1-\k_2)} 
\nn \\
& \hspace{-2cm}
\times 
\int_{z^-}\exp{iz^-(k_1^+-k_2^+)}
 (\Check{\slashed{k_1}}+m) \gamma^+ (\Check{\slashed{k_2}}+m)
\ \UF(x^+,y^+;\z) .
 \end{align}
The before-to-inside quark propagator: 
\begin{align}
\label{quark_prop_BI}
&     
S_\rmF(x,y)\BIq =
\int\frac{d^3\underline{k}}{(2\pi)^3}\frac{\theta(k^+)}{2k^+} \
\exp{-ix\cdot\underline{k}} \ \exp{iy\cdot\Check{k}}
\left[1-i\frac{\gamma^+\gamma^j}{2k^+}\RPartial_{\x^j}^F\right] \ (\Check{\slashed{k}}+m)
\ \UF(x^+,y^+;\x) .
 \end{align}
 The inside-to-after quark propagator  
 \begin{align}
\label{quark_prop_IA}
& 
S_\rmF(x,y)\IAq =
\int\frac{d^3\underline{k}}{(2\pi)^3}\frac{\theta(k^+)}{2k^+}
\  \UF(x^+,y^+;\y) \ (\Check{\slashed{k}}+m)
\  \left[1-i\frac{\gamma^+\gamma^j}{2k^+}\LPartial_{\y^j}^F\right]
\ \exp{-ix\cdot\Check{k}} \ \exp{iy\cdot\underline{k}} .
\end{align}
And finally the antiquarks propagators read, starting with before-to-after antiquark propagator
\begin{align}
\label{antiquark_prop_BA}
 &    
S_\rmF(x,y)\BAbq = (-1)
\int\frac{d^3\underline{k_1}}{(2\pi)^3}\frac{\theta(-k_1^+)}{2k_1^+}\exp{-ix\cdot\Check{k}_1}
\int\frac{d^3\underline{k_2}}{(2\pi)^3}\frac{\theta(-k_2^+)}{2k_2^+}\exp{iy\cdot\Check{k}_2}
\int_{\z}\exp{-i\z\cdot(\k_1-\k_2)} 
\nn \\
&
\times \int\limits_{z^-}\exp{iz^-(k_1^+-k_2^+)}
 \  (\Check{\slashed{k_1}}+m)\gamma^+(\Check{\slashed{k_2}}+m)
 \ \UF^\dagger(y^+,x^+;\z) .
\end{align}
The before-to-inside antiquark propagator 
\begin{align}
\label{antiquark_prop_BI}
& \hspace{-2cm}   
S_\rmF(x,y)\BIbq =
\int\frac{d^3\underline{k}}{(2\pi)^3}(-1)\frac{\theta(-k^+)}{2k^+}
\  \exp{-ix\cdot\Check{k}}
 \ \UF^\dagger(y^+,x^+;\y)
 \nn \\
 & \hspace{5cm}
 \times
 (\Check{\slashed{k}}+m)
    \left[1-i\frac{\gamma^+\gamma^j}{2k^+}\LPartial_{\y^j}^F\right]
    \exp{iy\cdot\underline{k}} .
\end{align}
The inside-to-after antiquark propagator  
\begin{align}     
\label{antiquark_prop_IA}
& \hspace{-1cm}
S_\rmF(x,y)\IAbq =
\int\frac{d^3\underline{k}}{(2\pi)^3}(-1)\frac{\theta(-k^+)}{2k^+}
\ \exp{iy\cdot\Check{k}} \ \exp{-ix\cdot\underline{k}}
\nn \\
& \hspace{5cm}
\times    
\left[1-i\frac{\gamma^+\gamma^j}{2k^+}\RPartial_{\x^j}^F\right]
(\Check{\slashed{k}}+m) \ \UF^\dagger(y^+,x^+;\x) ,
\end{align}
with the covariant derivatives defined as
\begin{align}
        \RPartial^\rmR_{z^\mu} & \equiv \Rpartial_{z^\mu} + igT_\rmR\cdot\mathcal{A}_\mu(z)\\
        \LPartial^\rmR_{z^\mu} & \equiv \Lpartial_{z^\mu} - igT_\rmR\cdot\mathcal{A}_\mu(z) ,
\end{align}
where R denotes the $\sun$ generators representation (as in \Equation{eq:WL}). 

\section{Antiquark channels}
\label{sec:antiquarks}

\subsection{$g\to g\bar q$ channel } 
\label{sec:ggbq}                        
We continue our analysis with the $g\to g\bar q$ channel. This channel is very similar to the $g\to gq$ channel except the quark jet in the final state is replaced by an antiquark jet. Since the computation of the scattering amplitude (both in general kinematics and in the back-to-back limit) and the cross section follow the same steps as in the computation presented in Sec.~\ref{sec:ggq}, here we only present the results and omit the details in order to avoid repetition. 

\begin{figure}[H]
\centering
\begin{subfigure}{0.49\textwidth}
\centering
\includegraphics[width=\textwidth]{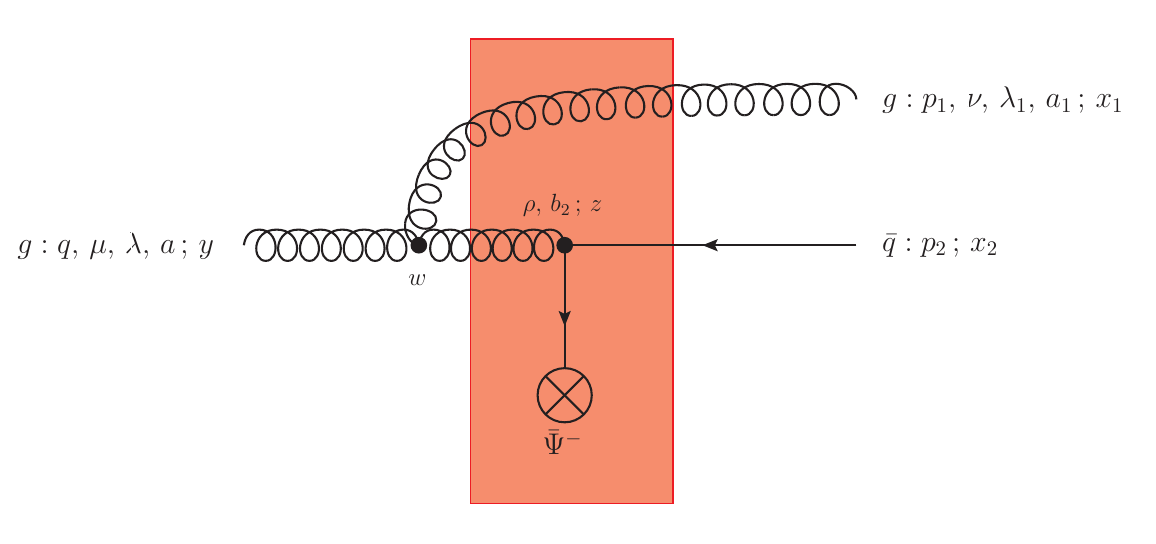}
\caption{Diagram 1}
\label{fig:g-gbq1}
\end{subfigure}
\begin{subfigure}{0.49\textwidth}
\centering
\includegraphics[width=\textwidth]{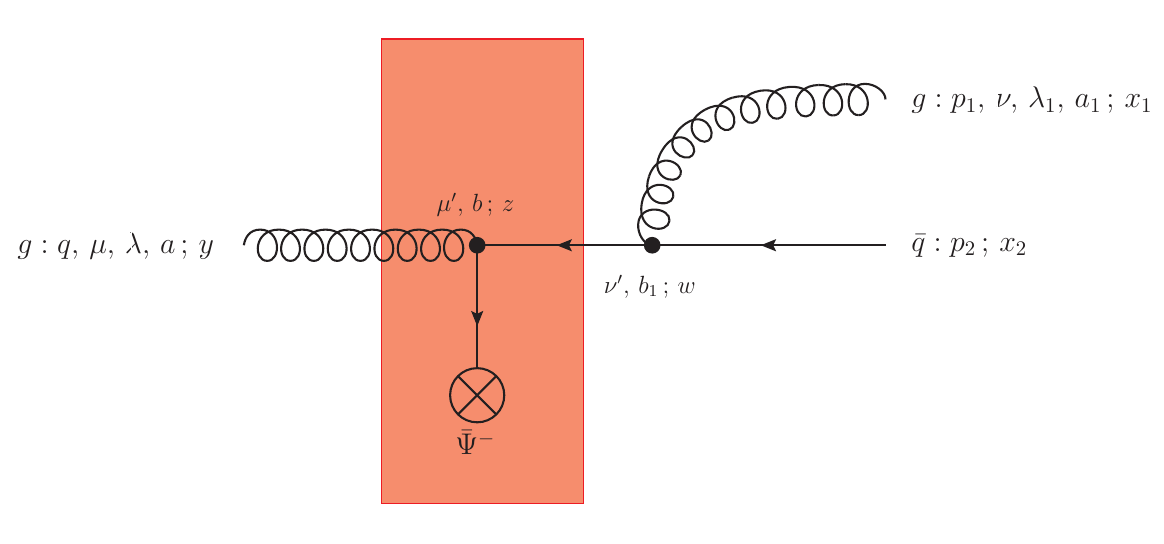}
\caption{Diagram 2}
\label{fig:g-gbq2}
\end{subfigure}
\begin{subfigure}{0.5\textwidth}
\centering
\includegraphics[width=\textwidth]{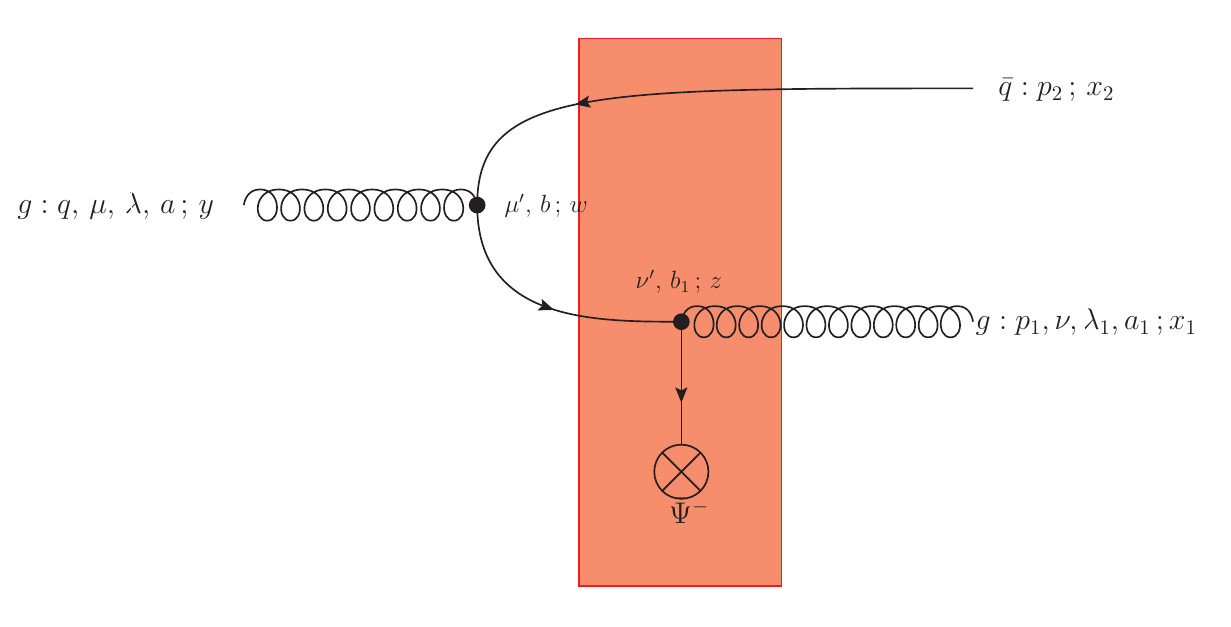}
\caption{Diagram 3}
\label{fig:g-gbq3}
\end{subfigure}
\caption{Diagrams contributing to channel $g\to g\bar q$.}
\label{fig:g-gbq}
\end{figure}

Similar to the $g\to gq$ channel, there are three mechanisms contributing to the production of antiquark-gluon dijet in gluon initiated channel at NEik accuracy. The first mechanism corresponds to the case where the incoming gluon splits into two gluons before the medium. The gluon pair enters the medium, one interacts eikonally while the second one interacts via a $t$-channel quark exchange and converts into an antiquark (see Fig.~\ref{fig:g-gbq1}). In the second mechanism, the incoming gluon scatters on the target via a $t$-channel quark exchange converting into an antiquark and this antiquark splits into an antiquark-gluon pair in the final state (see Fig.~\ref{fig:g-gbq2}). The third mechanism corresponds to the case where the incoming gluon splits into a quark-antiquark pair before entering the medium. Then, the antiquark scatters on the target eikonally while the quark scatters via a $t$-channel quark exchange and converts into a gluon, and thus one obtains an antiquark-gluon pair in the final state (see Fig.~\ref{fig:g-gbq3}). In the rest of the section we present the scattering amplitude (both in general kinematics and in the back-to-back limit) for each of these three mechanisms and we also present the total production cross section in this channel in the back-to-back limit.

%
%
The total scattering amplitude for the antiquark-gluon dijet in gluon initiated channel writes as a sum of the scattering amplitudes of the aforementioned three mechanisms
\begin{align}
\label{M_tot_schm_g_to_gbarq}
\M_{g\to g\bar q, \, {\rm tot.}} = \M_{g\to g\bar q, \, {\rm 1}}+\M_{g\to g\bar q, \, {\rm 2}}+ \M_{g\to g\bar q, \, {\rm 3}} .
\end{align}
In order to compute the amplitude for the first mechanism (see Fig.~\ref{fig:g-gbq1}) we start with the $S$-matrix element which can be obtained via the following LSZ-type reduction formula
\begin{align}
\mathcal{S}_{g\to g\bar q, \, 1}  & = 
\lim_{y^+\to-\infty} \ \lim_{x_1^+,x_2^+\to\infty}
\int_{\by,\bx[1],\bx[2]}\int_{y^-,x^-_1,x^-_2}
\int_{\bw, \bz}
\int_{w^-, z^-}
\int_{-\frac{L^+}{2}}^{\frac{L^+}{2}}d z^+
\int_{-\infty}^{-\frac{L^+}{2}}d w^+ 
\nn \\
&
\times 
\exp{ix_1\cdot\Check{p}_1} \ \exp{ix_2\cdot\Check{p}_2} \ \exp{-iy\cdot\Check{q}}
\ (-2q^+) \ \eps{0}{\mu}(-2p_1^+) \ \eps{1}{\nu} 
\nn \\
&
\times 
\left[G_{0,\rmF}^{\mu'\mu}(w,y)\right]_{a'a}
\left[G_\rmF^{\nu\nu'}(x_1,w)\BA\right]_{a_1b_1}
\left[G_\rmF^{\rho\rho'}(z,w)\BI\right]_{b_2b}V_{\mu'\nu'\rho'}^{a'b_1b} 
\nn \\
&
\times 
\overline{\Psi}_\beta^-(z)\gamma_{\rho}
\left[(-ig)t^{b_2}S_\rmF(z,x_2)\IAbq\right]_{\beta\alpha_2}
\gamma^+u(\Check{p}_2,h) ,
\end{align}
where the vacuum gluon propagator is given in Eq.~\eqref{eq:vacuum_gluon_prop}, before-to-after and before-to-inside gluon propagators are given in Eqs.~\eqref{gluon_prop_BA} and \eqref{gluon_prop_IA} respectively, the inside-to-after antiquark propagator is given in Eq.~\eqref{quark_prop_IA} and the triple gluon vertex is defined in Eq.~\eqref{eq:triple_gluon_vertex}. By using these explicit expressions and following the same steps as in the computation of $S$-matrix element for the first mechanism in the $g\to gq$ channel one obtains 
\begin{align}
\mathcal{S}_{g\to g\bar q, \, 1}  & =
\frac{-ig^2}{2p_2^+}f^{ab_1b} \ (2\pi)\delta(q^+-p_1^+-p_2^+)
\int_{\bz,\bz[1]}
\int_{-\frac{L^+}{2}}^{\frac{L^+}{2}}d z^+
\int_{-\infty}^{-\frac{L^+}{2}}d w^+ 
\int\frac{d^2 \bk[1]}{(2\pi)^2} 
\nn \\
&
\times 
\exp{-iw^+\left(\frac{\bq^2}{2q^+}-\frac{\bk[1]^2}{2p_1^+}-\frac{(\bk[1]-\bq)^2}{2p_2^+}\right)}
\exp{-i\bz\cdot(\bp[2]+\bk[1]-\bq)}
\ \exp{-i\bz[1]\cdot(\bp[1]-\bk[1])}
\nn \\
&
\times 
\overline{\Psi}(\underline{z}) \ \gamma^-\gamma^+\gamma^l
\ t^{b_2} \ \UFd(\infty,z^+;\bz)
\ \UA(z^+,w^+;\bz[1])_{b_2b}
\ \UA(\infty,w^+;\bz)_{a_1b_1}
\ u(\Check{p}_2,h) 
\nn \\
&
\times 
\varepsilon_\lambda^i \ \varepsilon_{\lambda_1}^{j*}
\left[g^{ij}\left(\frac{p_1^+}{p_2^+}\bq^l 
    - \frac{q^+}{p_2^+}\bk[1]^l\right)
    +g^{il}\left(\bq^j - \frac{q^+}{p_1^+}\bk[1]^j\right)
    -g^{jl}\left(\frac{p_1^+}{q^+}\bq^i - \bk[1]^i\right)\right] .
\end{align}
The relation between the $S$-matrix and the scattering amplitude given in Eq.~\eqref{eq:amplitude} holds for any channel. Thus, using this relation we can write the scattering amplitude for the first mechanism in the general kinematics as 
\begin{align}
i \mathcal{M}_{g\to g\bar q, \, 1}  & =
\frac{-ig^2}{(2p_2^+)(2q^+)}f^{ab_1b} 
\int_{\bz,\bz[1]}
\int_{-\frac{L^+}{2}}^{\frac{L^+}{2}}d z^+
\int_{-\infty}^{-\frac{L^+}{2}}d w^+ 
\int\frac{d^2 \bk[1]}{(2\pi)^2} 
\nn \\
&
\times 
\exp{-iw^+\left(\frac{\bq^2}{2q^+}-\frac{\bk[1]^2}{2p_1^+}-\frac{(\bk[1]-\bq)^2+m^2}{2p_2^+}\right)}
\exp{-i\bz\cdot(\bp[2]+\bk[1]-\bq)}
\ \exp{-i\bz[1]\cdot(\bp[1]-\bk[1])}
\nn \\
&
\times 
\overline{\Psi}(\underline{z}) \ \gamma^-\gamma^+\gamma^l
\ t^{b_2} \ \UFd(\infty,z^+;\bz)
\ \UA(z^+,w^+;\bz[1])_{b_2b}
\ \UA(\infty,w^+;\bz)_{a_1b_1}
\ u(\Check{p}_2,h) 
\nn \\
&
\times 
\varepsilon_\lambda^i \ \varepsilon_{\lambda_1}^{j*}
\left[g^{ij}\left(\frac{p_1^+}{p_2^+}\bq^l 
    - \frac{q^+}{p_2^+}\bk[1]^l\right)
    +g^{il}\left(\bq^j - \frac{q^+}{p_1^+}\bk[1]^j\right)
    -g^{jl}\left(\frac{p_1^+}{q^+}\bq^i - \bk[1]^i\right)\right] .
\label{eq:M_1_g_to_gbarq}
\end{align}

The $S$-matrix element for the second mechanism (see Fig.~\ref{fig:g-gbq2}) can be obtained as 
\begin{align}
\label{eq:S_2_g_to_gbarq_1}
&
\mathcal{S}_{g\to g\bar q, \, 2}   =
-\lim_{y^+\to-\infty} \ \lim_{x_1^+,x_2^+\to\infty}
\ \int_{\by,\bx[1],\bx[2]}\int_{y^-,x^-_1,x^-_2}
\int_{\bw, \bz}
\int_{w^-, z^-}
\int_{-\frac{L^+}{2}}^{\frac{L^+}{2}}d z^+
\int_{\frac{L^+}{2}}^{\infty}d w^+  
\nn \\
& 
\times 
\exp{ix_1\cdot\Check{p}_1} \ \exp{ix_2\cdot\Check{p}_2} \ \exp{-iy\cdot\Check{q}}
\ (-2q^+)\ \epsilon_\mu^{\lambda}(q) \ (-2p_1^+) \ \epsilon_\nu^{\lambda_1}(p_1)^* 
\left[G_\rmF^{\mu'\mu}(z,y)\BI\right]_{ba}
\\
&
\times
\left[G_{0,\rmF}^{\nu\nu'}(x_1,w)\right]_{a_1b_1}
\overline{\Psi}_\beta^-(z)\gamma_{\mu'} 
\left[(-ig)t^{b}S_\rmF(z,w)\IAbq \ \gamma_{\nu'}
(-ig)t^{b_1}S_{0,\rmF}(w,x_2)\right]_{\beta\alpha_2}
 \gamma^+u(\Check{p}_2, h) , \nn
\end{align}
where the before-to-inside gluon propagator is given in Eq.~\eqref{gluon_prop_BI}, the inside-to-after antiquark propagator is given in Eq.~\eqref{antiquark_prop_IA}, vacuum gluon and vacuum quark propagators are given in Eqs.~\eqref{eq:vacuum_gluon_prop} and \eqref{eq:quark_vacuum_prop} respectively. Substituting these expressions in Eq.~\eqref{eq:S_2_g_to_gbarq_1}  and following the same steps, one obtains 
\begin{align}
\mathcal{S}_{g\to g\bar q, \, 2} & =
\frac{g^2}{p_2^+(2q^+)}
(2\pi)\delta(q^+-p_1^+-p_2^+)
\int_{\bz}
\int_{-\frac{L^+}{2}}^{\frac{L^+}{2}}d z^+
\int_{\frac{L^+}{2}}^{\infty}d w^+  
\int\frac{d^2 \bk[0]}{(2\pi)^2} 
\nn \\
&
\times
\exp{-iw^+(\check{k}_0^--\check{p}_1^--\check{p}_2^-)}
\ \exp{-i\bz\cdot(\bk[0]-\bq)}
\ (2\pi)^2\delta^{(2)}(\bk[0]-\bp[1]-\bp[2])
\nn \\
&
\times 
\varepsilon^i_{\lambda} \ \varepsilon_{\lambda_1}^{j*}
\overline{\Psi}(\underline{z}) \frac{\gamma^{-}\gamma^{+}}{2}\gamma^i
\left[p_2^+(\bp[1]^l+\bp[2]^l)\gamma^l\gamma^j
+q^+\bp[2]^l\gamma^j\gamma^l - mp_1^+\gamma^j
+ 2\frac{q^+p_2^+}{p_1^+}\bp[1]^j\right]
\nn \\
&
\times
\ t^{b} \ \UFd(w^+,z^+;\bz) \ t^{a_1}
\ \UA(z^+,-\infty;\bz)_{ba}
\ u(\Check{p}_2, h) .
\label{eq:S_2_g_to_gbarq_2}
\end{align}
Using Eqs.~\eqref{eq:amplitude} and \eqref{eq:M_1_g_to_gbarq}, we obtain the scattering amplitude for the second mechanism in general kinematics as 
\begin{align}
i\mathcal{M}_{g\to g\bar q, \, 2} & =
\frac{g^2}{p_2^+(2q^+)^2}
\int_{\bz}
\int_{-\frac{L^+}{2}}^{\frac{L^+}{2}}d z^+
\int_{\frac{L^+}{2}}^{\infty}d w^+  
\int\frac{d^2 \bk[0]}{(2\pi)^2} 
\nn \\
&
\times
\exp{-iw^+(\check{k}_0^--\check{p}_1^--\check{p}_2^-)}
\ \exp{-i\bz\cdot(\bk[0]-\bq)}
\ (2\pi)^2\delta^{(2)}(\bk[0]-\bp[1]-\bp[2])
\nn \\
&
\times 
\varepsilon^i_{\lambda} \ \varepsilon_{\lambda_1}^{j*}
\overline{\Psi}(\underline{z}) \frac{\gamma^{-}\gamma^{+}}{2}\gamma^i
\left[p_2^+(\bp[1]^l+\bp[2]^l)\gamma^l\gamma^j
+q^+\bp[2]^l\gamma^j\gamma^l - mp_1^+\gamma^j
+ 2\frac{q^+p_2^+}{p_1^+}\bp[1]^j\right]
\nn \\
&
\times
\ t^{b} \ \UFd(w^+,z^+;\bz) \ t^{a_1}
\ \UA(z^+,-\infty;\bz)_{ba}
\ u(\Check{p}_2, h) .
\label{eq:M_2_g_to_gbarq}
\end{align}

The $S$-matrix element for the third mechanism (see Fig.~\ref{fig:g-gbq3}) is expressed as 
\begin{align}
&
\mathcal{S}_{g\to g\bar q,\, 3} =
-\lim_{y^+\to-\infty} \ \lim_{x_1^+,x_2^+\to\infty}
\int_{\by,\bx[1],\bx[2]}\int_{y^-,x^-_1,x^-_2}
\int_{\bw, \bz}
\int_{w^-, z^-}
\int_{-\frac{L^+}{2}}^{\frac{L^+}{2}}d z^+
\int_{-\infty}^{-\frac{L^+}{2}}d w^+  
\nn \\
&
\times
\exp{ix_1\cdot\Check{p}_1} \ \exp{ix_2\cdot\Check{p}_2} \ \exp{-iy\cdot\Check{q}}
(-2q^+) \ \epsilon_\mu^{\lambda}(q)^* \ (-2p_1^+) \ \epsilon_\nu^{\lambda_1}(p_1)^* 
 \left[G_{0,\rmF}^{\mu'\mu}(w,y)\right]_{ba}
  \left[G_\rmF^{\nu\nu'}(x_1,z)\IA\right]_{a_1b_1}
\nn \\
&
\times
\overline{\Psi}_\beta^-(z)\gamma_{\nu'}
\left[(-ig)t^{b_1}S_\rmF(z,w)\BIq\gamma_{\mu'}
(-ig)t^{b}S_\rmF(w,x_2)\BAbq\right]_{\beta\alpha_2}
\gamma^+ u(\Check{p}_2, h) ,
\end{align}
where the inside-to-after gluon propagator is given in Eq.~\eqref{gluon_prop_IA}, the before-to-inside quark propagator is given in Eq.~\eqref{quark_prop_BI}, the before-to-after antiquark propagator is given in Eq.~\eqref{antiquark_prop_BA} and the vacuum gluon propagator is given in Eq.~\eqref{eq:vacuum_gluon_prop}. Using these expressions, one can get 
\begin{align}
\mathcal{S}_{g\to g\bar q,\, 3}    & =
\frac{g^2}{(2p_1^+)(2p_2^+)}(2\pi)\delta\left(p_1^+ + p_2^+ - q^+\right)
\int_{-\frac{L^+}{2}}^{\frac{L^+}{2}}d z^+
\int_{-\infty}^{-\frac{L^+}{2}}d w^+  
\int_{\bz,\bz[1]}\int\frac{d^2\bk[1]}{(2\pi)^2}
\nn \\
&
\times 
e^{iw^+\left[\frac{(\bk[1]+\bq)^2+m^2}{2p_2^+} + \frac{\bk[1]^2+m^2}{2p_1^+} - \frac{\bq^2}{2q^+}\right]}e^{-i\bz\cdot(\bp[1] + \bk[1])}
e^{-i\bz[1]\cdot(\bp[2]-\bk[1]-\bq)}
\nn \\
&
\times 
\varepsilon^i_{\lambda} \ \varepsilon_{\lambda_1}^{j*}
\ \overline{\Psi}(\underline{z}) \ \gamma^-\gamma^+\gamma^j
\left[p_2^+\bk[1]^l\gamma^l\gamma^i
- p_1^+(\bk[1]^l+\bq^l)\gamma^i\gamma^l + mq^+\gamma^i
- 2\frac{p_1^+p_2^+}{q^+}\bq^i\right]u(\Check{p}_2) 
\nn \\
&
\times 
\UA(\infty,z^+;\bz)_{a_1b_1}
\ t^{b_1} \ \UF(z^+,w^+;\bz) \ t^{a} \ \UFd(\infty,w^+;\bz[1]) ,
\end{align}
from which we can obtain the scattering amplitude for the third mechanism in general kinematics as 
\begin{align}
i \mathcal{M}_{g\to g\bar q,\, 3}    & =
\frac{g^2}{(2p_1^+)(2p_2^+)(2q^+)}(2\pi)\delta\left(p_1^+ + p_2^+ - q^+\right)
\int_{-\frac{L^+}{2}}^{\frac{L^+}{2}}d z^+
\int_{-\infty}^{-\frac{L^+}{2}}d w^+  
\int_{\bz,\bz[1]}\int\frac{d^2\bk[1]}{(2\pi)^2}
\nn \\
&
\times 
e^{iw^+\left[\frac{(\bk[1]+\bq)^2+m^2}{2p_2^+} + \frac{\bk[1]^2+m^2}{2p_1^+} - \frac{\bq^2}{2q^+}\right]}e^{-i\bz\cdot(\bp[1] + \bk[1])}
e^{-i\bz[1]\cdot(\bp[2]-\bk[1]-\bq)}
\nn \\
&
\times 
\varepsilon^i_{\lambda} \ \varepsilon_{\lambda_1}^{j*}
\ \overline{\Psi}(\underline{z}) \ \gamma^-\gamma^+\gamma^j
\left[p_2^+\bk[1]^l\gamma^l\gamma^i
- p_1^+(\bk[1]^l+\bq^l)\gamma^i\gamma^l + mq^+\gamma^i
+ 2\frac{p_1^+p_2^+}{q^+}\bq^i\right]u(\Check{p}_2) 
\nn \\
&
\times 
\UA(\infty,z^+;\bz)_{a_1b_1}
\ t^{b_1} \ \UF(z^+,w^+;\bz) \ t^{a} \ \UFd(\infty,w^+;\bz[1]) .
\label{eq:M_3_g_to_gbarq}
\end{align}
The total scattering amplitude in the general kinematics in the $g\to g\bar q$ channel is then given by Eq.~\eqref{M_tot_schm_g_to_gbarq} with the contributions from each mechanism given in Eqs.~\eqref{eq:M_1_g_to_gbarq}, \eqref{eq:M_2_g_to_gbarq} and \eqref{eq:M_3_g_to_gbarq}. 
%
%

We can now consider the back-to-back limit of the scattering amplitudes in the $g\to g\bar q$ channel. We follow the same strategy as in Sec.~\ref{subsec:b2b_g_to_gq_Amp}. Namely, we start with the amplitudes given in Eqs.~\eqref{eq:M_1_g_to_gbarq}, \eqref{eq:M_2_g_to_gbarq} and \eqref{eq:M_3_g_to_gbarq} written in general kinematics, and perform the change of variables given in Eqs.~\eqref{eq:PK} and \eqref{eq:br} in order to write the amplitudes in terms of relative dijet momentum $\P$, dijet momentum imbalance $\k$ and their conjugate variables $\r$ and $\b$. Since in the back-to-back limit one has $|\P|\gg |\k|$ or equivalently $|\r|\ll|\b|$, one can perform a small $|\r|$ expansion of the Wilson lines. Keeping only the zeroth order term in this expansion, one gets to the following expressions for the scattering amplitudes in the $g \to g \bar q$ channel in the back-to-back and massless quark limits
\begin{align}
i\mathcal{M}_{g\to g\bar q, \, 1}^{{\rm b2b}, \, m=0} & =
-\frac{i g^2}{(2q^+)}\frac{1}{\P^2}
\int_{-\frac{L^+}{2}}^{\frac{L^+}{2}}d z^+
\int_{\bb}e^{-i\bb\cdot(\bk-\bq)}
\overline{\Psi}(z^+;\bb) \ \gamma^-\gamma^+\gamma^l
\ \UFd(\infty,z^+;\bb) \ \left[t^{a_1},t^{c}\right]
\nn \\
&
\times 
\UA(\infty,-\infty;\bb)_{ca} 
\ \varepsilon_{\lambda}^i \ \varepsilon_{\lambda_1}^{j*}\left[
\frac{z}{1-z}g^{ij}{\bP^l} + g^{il}{\bP^j} - z g^{jl}{\bP^i}
\right]u(\Check{p}_2, h) ,
\end{align}
\begin{align}
i \mathcal{M}_{g\to g\bar q, \, 2}^{{\rm b2b}, \, m=0} & =
-\frac{ig^2}{(2q^+)} \frac{1}{\P^2}
\int_{-\frac{L^+}{2}}^{\frac{L^+}{2}}d z^+
\int_{\b}
\exp{-i\b\cdot(\bk - \bq)}
\varepsilon_{\lambda}^i \ \varepsilon_{\lambda_1}^{j*}
\overline{\Psi}(z^+;\bb) \frac{\gamma^{-}\gamma^{+}}{2}\gamma^i
\left[z\gamma^j\gamma^l{\bP^l} - 2(1-z)\bP^j\right] 
\nn \\
&
\times 
\UFd(\infty,z^+;\bb) \ t^{c} \ t^{a_1} \
\ \UA(+\infty,-\infty;\bb)_{ca} \ u(\Check{p}_2, h)  ,
\end{align}
and 
\begin{align}
i \mathcal{M}_{g\to g\bar q, \, 3}^{{\rm b2b}, \, m=0}  & = 
\frac{ig^2}{(2q^+)}\frac{1}{\P^2}
\int_{-\frac{L^+}{2}}^{\frac{L^+}{2}}d z^+
\int_{\bb}\exp{-i\bb\cdot(\bk -\bq)}
\varepsilon_{\lambda}^i \ \varepsilon_{\lambda_1}^{j*}
\ \overline{\Psi}(z^+;\bb) \frac{\gamma^{-}\gamma^{+}}{2}\gamma^j
\nn \\
& \hspace{-1.5cm}
\times 
\left[-z {\bP^l}\gamma^i\gamma^l
    + (1-z){\bP^l}\gamma^l\gamma^i\right]
u(\check{q}_2, h)
\ \UFd(\infty,z^+;\bb)\ t^{a_1} \ t^{c}\ 
\UA(\infty,-\infty;\bb)_{ca} . 
\end{align}
Summing these three contributions we obtain the total scattering amplitude for the $g\to g\bar q$ channel in the back-to-back and massless quarks limit as 
\begin{align}
\label{eq:M_tot_g_to_gbarq_b2b_m0_2}
i\mathcal{M}_{g\to g\bar q, \, {\rm tot.}}^{{\rm b2b}, \, m=0} & = 
\frac{-ig^2}{(2q^+)} \frac{1}{\P^2}
\int_{-\frac{L^+}{2}}^{\frac{L^+}{2}}d z^+
\int_{\bb}e^{-i\bb\cdot\left(\bk - \bq\right)}
\ \UA(\infty,-\infty;\bb)_{ba}
\nn \\    
&
\times 
\overline{\Psi}(z^+;\bb) \frac{\gamma^{-}\gamma^{+}}{2} \
\UFd(\infty,z^+;\bb)
\left[
    t^{a_1}t^b\mathfrak{h}^{(1)}_{g\to g\bar q} 
    +t^bt^{a_1}\mathfrak{h}^{(2)}_{g\to g\bar q}
    \right]u(\check{q}_2,h) ,
\end{align}
with the associated coefficients defined as 
\begin{align}
\label{eq:HF_g_to_gbarq_1}
\mathfrak{h}^{(1)}_{g\to g\bar q} &= 
\varepsilon_{\lambda}^i\ \varepsilon_{\lambda_1}^{j*}
\left(
- z{\bP^l} \ \gamma^j\gamma^i\gamma^l
+ (1-z){\bP^l} \ \gamma^j\gamma^l\gamma^i 
+ 2{\bP^j} \gamma^i
- 2z{\bP^i} \gamma^j
- 2\frac{z}{1-z}{\bP^l}g^{ij}\gamma^l\right)
, \\
\label{eq:HF_g_to_gbarq_2} 
\mathfrak{h}^{(2)}_{g\to g\bar q} &=
\varepsilon_{\lambda}^i \ \varepsilon_{\lambda_1}^{j*}
\left(
 - z{\bP^l} \ \gamma^i\gamma^j\gamma^l
 - 2z{\bP^j} \ \gamma^i 
 + 2z{\bP^i} \gamma^j
 + 2\frac{z}{1-z}{\bP^l} \ g^{ij}\gamma^l\right)
\end{align}
We would like to emphasize that the associated hard coefficients in the $g\to g\bar q$ channel given in Eqs.~\eqref{eq:HF_g_to_gbarq_1} and \eqref{eq:HF_g_to_gbarq_2} are conjugate to the ones in the $g\to gq$ channel given in Eqs.~\eqref{eq:HF_g_to_gq_1} and \eqref{eq:HF_g_to_gq_2}, such that they satisfy 
\begin{align}
\mathfrak{h}^{(1)}_{g\to g\bar q} & = (\mathfrak{h}^{(2)}_{g\to gq})^\dagger , \nn \\
\mathfrak{h}^{(2)}_{g\to g\bar q} & = (\mathfrak{h}^{(1)}_{g\to gq})^\dagger ,
\end{align}
due to charge conjugation. 

\label{sec:ggbqXS} 
The partonic cross section for the  $g\to g\bar q$ channel in the back-to-back limit can be written as 
\begin{equation}
\frac{d\sigma^{{\rm b2b}, \, m=0}_{g\to g\bar q}}{d{\rm P.S}} =
    (2q^+)2\pi\delta\left(p_1^+ + p_2^+ - q^+\right)
    \frac{1}{2(N_c^2-1)}\sum_{\lambda,\lambda_1}\sum_{h_2}\sum_{a,a_1}
    \left\langle \left|i\mathcal{M}^{{\rm b2b}, \, m=0}_{g\to g\bar q}\right|^2 \right\rangle ,
\end{equation}
where the differential phase space is defined in Eq.~\eqref{def:PS} and the amplitude is given in Eq.~\eqref{eq:M_tot_g_to_gbarq_b2b_m0_2}. 
One can simply repeat the same procedure that was adopted to compute the back-to-back production cross section in $g\to gq$ channel and show that 
\begin{align}
\frac{d\sigma_{g\to g\bar q}^{{\rm b2b}, \, m=0}}{d{\rm P.S.}} & =
g^4 \ (2\pi)\delta\left(p_1^+ + p_2^+ - q^+\right)
\int_{-\frac{L^+}{2}}^{\frac{L^+}{2}} dz^+
\int_{-\frac{L^+}{2}}^{\frac{L^+}{2}} d{z'}^+ 
 \nn \\
& \hspace{2cm}
\times 
\int_{\bb, \bb'}
    e^{-i(\bb-\bb')\cdot\left(\bk - \bq\right)}
\ \left[\mathcal{H}^{+g}_{g\to g\bar q} \ \overline{\mathcal{C}}^{+g}
    + \mathcal{H}^{+\square_g}_{g\to g\bar q} \ \overline{\mathcal{C}}^{+\square_g}\right] ,
\end{align}
where the color structures $\overline{\mathcal{C}}^{+g}$ and $\overline{\mathcal{C}}^{+\square_g}$ are similar to $\mathcal{C}^{+g}$ and $\mathcal{C}^{+\square_g}$ defined in Eqs.~\eqref{eq:C+g} and \eqref{eq:C+square_g} and read
\begin{align}
\label{eq:bC+g}
\overline{\mathcal{C}}^{+g} & \equiv 
\Big\langle \Tr \Big\{\UF(\infty,{z'}^+;\bb')\ t^{c'} \ \Psi(z'^{+};\bb')
\overline{\Psi}(z^{+};\bb) \ \gamma^{-}
\ t^c \ \UFd(\infty,z^+;\bb) \Big\}
\nn \\
& \hspace{6.2cm}
\times \UA({z'}^+,-\infty;\bb')_{c'a} \ \UA(z^+,-\infty;\bb)_{ca} \Big\rangle
\\
\label{eq:bC+square_g} 
\overline{\mathcal{C}}^{+\square_g} & \equiv 
\Big\langle \Tr \Big\{\UF(\infty,{z'}^+;\bb') \ \Psi(z'^{+};\bb')
\overline{\Psi}(z^{+};\bb) \ \gamma^{-} \ \UFd(\infty,z^+;\bb) \Big\}
\nn \\
& \hspace{6.2cm}
\times \UA(\infty,-\infty;\bb')_{ba} \ \UA(\infty,-\infty;\bb)_{ba} \Big\rangle ,
\end{align}
One can show that the hard factors computed in the $ g\to gq$ and $ g\to g\bar q$ channels are indeed the same 
 \begin{equation}
    \mathcal{H}^{+\square_g}_{g\to g\bar q} =
    \mathcal{H}^{+\square_g}_{g\to gq}
    \text{ and }
    \mathcal{H}^{+g}_{g\to g\bar q} =
    \mathcal{H}^{+g}_{g\to gq}.
\end{equation}

\subsection{$\bar q\to \bar q q$ channel} 
\label{sec:bqbqq}                             

\begin{figure}[H]
\centering
\begin{subfigure}{0.49\textwidth}
\centering
\includegraphics[width=\textwidth]{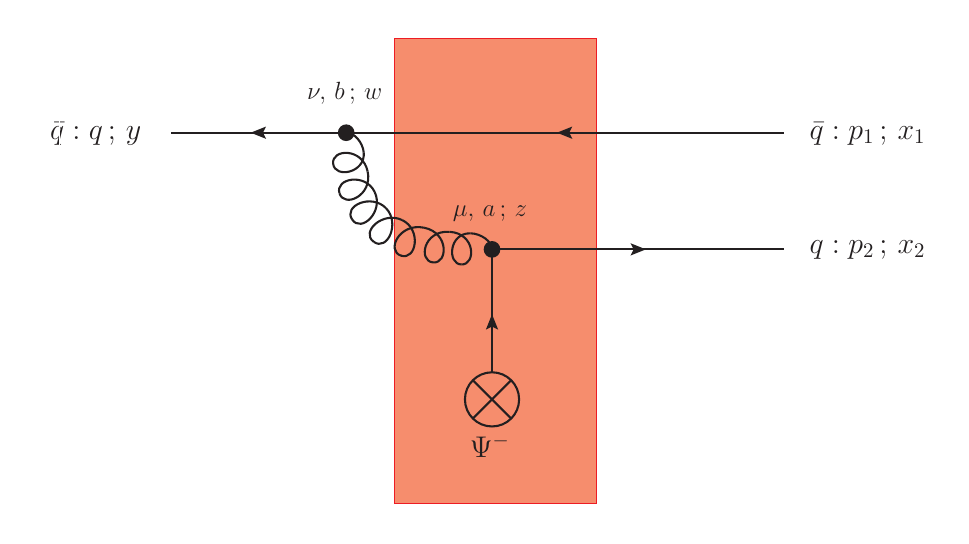}
\caption{Diagram 1}
\label{fig:bq-qbq1}
\end{subfigure}
\begin{subfigure}{0.49\textwidth}
\centering
\includegraphics[width=\textwidth]{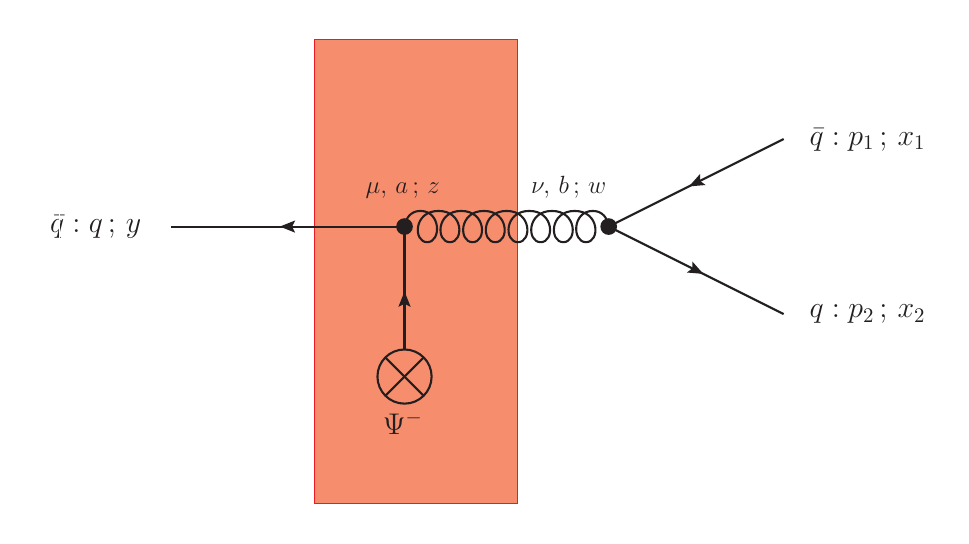}
\caption{Diagram 2}
\label{fig:bq-qbq2}
\end{subfigure}
\caption{Diagrams contributing to channel $\bar{q} \to \bar{q}q$.}
\label{fig:bq-bqq}
\end{figure}

This channel is very similar to the $q\to q\bar q$ with the incoming quark replaced with an antiquark in the initial state. The computation follows very closely the one performed in Sec.~\ref{sec:qqbq}. For the case where all quarks have the same flavor, there are two mechanisms that contribute to the quark-antiquark dijet production in antiquark initiated channel at NEik accuracy. The first one corresponds to the case where the antiquark in the initial state splits into an antiquark-gluon pair before the medium which then scatters on the target. In this mechanism, the antiquark scatters eikonally while the gluon scatters via a $t$-channel quark exchange and converts the gluon to a quark. Thus, one gets a quark-antiquark pair produced in the final state (see Fig.~\ref{fig:bq-qbq1}). In the second mechanism, the antiquark scatters on the target via a $t$-channel quark exchange and converts into a gluon. Then, the gluon splits into a quark-antiquark pair in the final state (see Fig.~\ref{fig:bq-qbq2}).\\ 
Furthermore, two other cases exist where all quarks do not share the same flavor.  If the final antiquarks-quark pair has the same flavor, but is different
than the initial antiquark one (more explicitly, $\bar q_{f} \to  \bar q_{f'} q_{f'}$, with $f \neq f'$), only the second mechanism
contributes (\ref{fig:bq-qbq2}). When the final quark has a different flavor than the initial and final antiquarks (more explicitly, $\bar q_{f} \to \bar{q}_{f} q_{f'}$  ), only the first mechanism contributes (see Fig.\ref{fig:bq-qbq1}).

The $S$-matrix element for the first mechanism can be obtained from the following LSZ-type reduction formula
\begin{align}
\S_{\bar q\to \bar q q, \, 1} &= 
\lim_{y^+\to-\infty}\lim_{x_1^+,x_2^+\to\infty}
\int_{\by,\bx[1],\bx[2]}\int_{y^-,x^-_1,x^-_2}
\int_{\bw, \bz}\int_{w^-, z^-}
\int_{-\frac{L^+}{2}}^{\frac{L^+}{2}}d z^+
\int_{-\infty}^{-\frac{L^+}{2}}d w^+
\nn \\
&
\times 
 e^{ix_{1} \cdot \check{p}_{1}} \ e^{ix_{2} \cdot \check{p}_{2}} 
 \ e^{-iy \cdot \check{q}} \
\overline{v}(\check{q},h)
\gamma^{+} \Big[ S_{0,F}(y,w)  (-igt^{b}\gamma_{\nu}) \  S_{F}(w,x_{1}) \BAbq\Big]_{\alpha \alpha_{1}}
 \gamma^{+} v(\check{p}_{1},h_{1})  
\nn \\
&
\times
\Big[ G^{\mu\nu}_{F}(z,w) \BI \Big]_{ab} 
\overline{u}(\check{p}_{2},h_{2}) \gamma^{+}  
\Big[ S_{F}(x_{2},z) \IAq (-igt^{a}\gamma_{\mu})   \Big]_{\alpha_{2}\beta}
\Psi^{-}_{\beta}(z) ,
\end{align}
where the before-to-after antiquark propagator is given in Eq.~\eqref{antiquark_prop_BA}, the before-to-inside gluon propagator is given in Eq.~\eqref{gluon_prop_BI}, the inside-to-after quark propagator is given in Eq.~\eqref{quark_prop_IA}, and the vacuum quark propagator is given in Eq.~\eqref{eq:quark_vacuum_prop}. Using these explicit expressions and following the same steps as in Sec.~\ref{sec:qqbq} one obtains the $S$-matrix for the first mechanism as 
\begin{align}
\S_{\bar q\to \bar q q, \, 1} &= 
-\frac{g^{2}}{2p_{2}^{+}}  \ (2\pi) \delta(p_{1}^{+}+p_{2}^{+}-q^{+}) 
\int\! \frac{d^{2} \bk_{1}}{(2\pi)^{2}} \int_{\bz} \ e^{-i \bz \cdot (\bp_{2}+ \bk_{1}- \bq)} \int_{\bz_{1}}  \ e^{-i\bz_{1} \cdot (\p_{1}-\bk_{1})} 
\nn \\
&
\times
\int_{-\frac{L^+}{2}}^{\frac{L^+}{2}} dz^{+} 
\int_{-\infty}^{-\frac{L^+}{2}} dw^{+} 
e^{iw^{+}\Big(\frac{\bk_{1}^{2}}{2p_{1}^{+}} + \frac{(\bq- \bk_{1})^{2}}{2p_{2}^{+}} -\frac{\bq^{2}}{2q^{+}}+ \frac{m^{2}}{2p^{+}_{1}} -\frac{m^{2}}{2q^{+}}\Big)} 
\nn  \\
&
\times
       \Big[ \UF(\infty,z^{+}; \bz) \ t^{a}\Big]_{\alpha_{2}\beta} \
       \UA(z^{+},w^{+}; \bz)_{ab}  \Big[ t^{b}\  \UF^{\dagger}(\infty,w^{+}; \bz_{1})\ \Big]_{\alpha\alpha_{1}}  
\nn \\& \times
\overline{v}(\check{q},h)  \gamma^{+} \nn
\bigg[ \frac{\gamma^{l}\gamma^{i} \bq^{l}}{2q^{+}} + \frac{\gamma^{i}\gamma^{l}\bk_{1}^{l}}{2p_{1}^{+}} - \gamma^{i} \Big(\frac{1}{2q^{+}}-\frac{1}{2p_{1}^{+}}\Big)m +\frac{\bq^{i}-\bk_{1}^{i}}{p_{2}^{+}}\bigg]  \nn
\\&\times      
        v(\check{p}_{1},h_{1}) \overline{u}(\check{p}_{2},h_{2})  \gamma^{i} \frac{\gamma^{+}\gamma^{-}}{2}     \Psi_{\beta}(\underline{z}) ,
\end{align}
which leads to the following scattering amplitude
\begin{align}
\label{eq:M_1_barq_to_barqq}   
i \M_{\bar q\to \bar q q, \, 1} &= 
-\frac{g^{2}}{(2p_{2}^{+})(2q^+)}  
\int\! \frac{d^{2} \bk_{1}}{(2\pi)^{2}} \int_{\bz} \ e^{-i \bz \cdot (\bp_{2}+ \bk_{1}- \bq)} \int_{\bz_{1}}  \ e^{-i\bz_{1} \cdot (\p_{1}-\bk_{1})} 
\nn \\
&
\times
\int_{-\frac{L^+}{2}}^{\frac{L^+}{2}} dz^{+} 
\int_{-\infty}^{-\frac{L^+}{2}} dw^{+} 
e^{iw^{+}\Big(\frac{\bk_{1}^{2}}{2p_{1}^{+}} + \frac{(\bq- \bk_{1})^{2}}{2p_{2}^{+}} -\frac{\bq^{2}}{2q^{+}}+ \frac{m^{2}}{2p^{+}_{1}} -\frac{m^{2}}{2q^{+}}\Big)} 
\nn  \\
&
\times
       \Big[ \UF(\infty,z^{+}; \bz) \ t^{a}\Big]_{\alpha_{2}\beta} \
       \UA(z^{+},w^{+}; \bz)_{ab}  \Big[ t^{b}\  \UF^{\dagger}(\infty,w^{+}; \bz_{1})\ \Big]_{\alpha\alpha_{1}}  
\nn \\& \times
\overline{v}(\check{q},h)  \gamma^{+} \nn
\bigg[ \frac{\gamma^{l}\gamma^{i} \bq^{l}}{2q^{+}} + \frac{\gamma^{i}\gamma^{l}\bk_{1}^{l}}{2p_{1}^{+}} - \gamma^{i} \Big(\frac{1}{2q^{+}}-\frac{1}{2p_{1}^{+}}\Big)m +\frac{\bq^{i}-\bk_{1}^{i}}{p_{2}^{+}}\bigg]  \nn
\\&\times      
        v(\check{p}_{1},h_{1}) \overline{u}(\check{p}_{2},h_{2})  \gamma^{i} \frac{\gamma^{+}\gamma^{-}}{2}     \Psi_{\beta}(\underline{z})  .   
\end{align}
Similarly, the $S$-matrix element for the second mechanism can be obtained via 
%
%
%
%
%
%
%
%
\begin{align}
 \S_{\bar q\to \bar q q, \, 2} &=
-\lim_{y^+\to-\infty} \ \lim_{x_1^+,x_2^+\to\infty}
\ \int_{\by,\bx[1],\bx[2]} \ \int_{y^-,x^-_1,x^-_2}
\ \int_{\bw, \bz}\int_{w^-, z^-}
\int_{-\frac{L^+}{2}}^{\frac{L^+}{2}}d z^+
\int_{\frac{L^+}{2}}^{\infty}d w^+
        \nn\\&\times   
        e^{ix_{1}\cdot \check{p}_{1}} \ e^{ix_{2} \cdot \check{p}_{2}} \ e^{-iy \cdot \check{q}} \  
       \overline{u}(\check{p}_{2},h_{2}) \gamma^{+}     \Big[  S_{0,F}(x_{2},w) \ (-igt^{b}\gamma_{\nu}) \ S_{0,F}(w,x_{1})\Big]_{\alpha_{2}\alpha_{1}}
       \gamma^{+} v(\check{p}_{1},h_{1})     \nn
       \\&\times                    
        \Big[ G^{\nu\mu}_{F}(w,z) \IA \Big]_{ba}
        \overline{v}(\check{q},h) \gamma^{+} \Big[ S_{F}(y,z)\BIbq  (-igt^{a}\gamma_{\mu})\Big]_{\alpha\beta} \Psi_{\beta}^{-}(z) .
 \end{align}
Using Eq.~\eqref{gluon_prop_IA} for the inside-to-after gluon propagator, Eq.~\eqref{antiquark_prop_BI} for the before-to-inside antiquark propagator  and Eq.~\eqref{eq:quark_vacuum_prop} for the vacuum quark propagator, one obtains
\begin{align}       
&
 \S_{\bar q\to \bar q q, \, 2} =       
 \frac{g^{2}}{2q^{+}}  \ (2\pi) \delta(p_{1}^{+}+p_{2}^{+}-q^{+}) 
 \int_{-\frac{L^+}{2}}^{\frac{L^+}{2}} dz^{+} 
 \int_{\frac{L^+}{2}}^{\infty} dw^{+} \ e^{iw^{+}\Big(\check{p}_{1}^{-} + \check{p}_{2}^{-} - \frac{( \bp_{1}+ \bp_{2})^{2}}{2p_{1}^{+}}\Big)} 
\nn \\
&
\times
\int_{\z} \ e^{- \bz \cdot ( \bp_{1}+ \bp_{2}-\bq)}
\ \UA(w^{+},z^{+};\bz)_{ba} \big(t^{b}\big)_{\alpha\beta} 
\Big[ \UF^{\dagger}(z^{+},-\infty; \bz) \ t^{a}\Big]_{\alpha_{2}\alpha_{1}}
\overline{u}(\check{p}_{2},h_{2}) \gamma^{+} 
\nn \\
&
\times
\bigg[ \frac{\gamma^{l}\gamma^{j} \bp_{2}^{l}}{2p_{2}^{+}} + \frac{\gamma^{j}\gamma^{l} \bp_{1}^{l}}{2p_{1}^{+}} + \gamma^{j} (\frac{1}{2p_{1}^{+}}+\frac{1}{2p_{2}^{+}})m +\frac{\bp_{1}^{j}+ \bp_{2}^{j}}{q^{+}}
        \bigg] v(\check{p}_{1},h_{1})
        \overline{v}(\check{q},h)\gamma^{j}\frac{\gamma^{+}\gamma^{-}}{2} \Psi_{\beta}(\underline{z}) ,     
    \end{align}
from which the scattering amplitude can be extracted as 
\begin{align}   
\label{eq:M_2_barq_to_barqq}    
&
i \M_{\bar q\to \bar q q, \, 2} =       
 \frac{g^{2}}{(2q^{+})^2}   
 \int_{-\frac{L^+}{2}}^{\frac{L^+}{2}} dz^{+} 
 \int_{\frac{L^+}{2}}^{\infty} dw^{+} \ e^{iw^{+}\Big(\check{p}_{1}^{-} + \check{p}_{2}^{-} - \frac{( \bp_{1}+ \bp_{2})^{2}}{2p_{1}^{+}}\Big)} 
\nn \\
&
\times
\int_{\z} \ e^{- \bz \cdot ( \bp_{1}+ \bp_{2}-\bq)}
\ \UA(w^{+},z^{+};\bz)_{ba} \big(t^{b}\big)_{\alpha\beta} 
\Big[ \UF^{\dagger}(z^{+},-\infty; \bz) \ t^{a}\Big]_{\alpha_{2}\alpha_{1}}
\overline{u}(\check{p}_{2},h_{2}) \gamma^{+} 
\nn \\
&
\times
\bigg[ \frac{\gamma^{l}\gamma^{j} \bp_{2}^{l}}{2p_{2}^{+}} + \frac{\gamma^{j}\gamma^{l} \bp_{1}^{l}}{2p_{1}^{+}} + \gamma^{j} (\frac{1}{2p_{1}^{+}}+\frac{1}{2p_{2}^{+}})m +\frac{\bp_{1}^{j}+ \bp_{2}^{j}}{q^{+}}
        \bigg] v(\check{p}_{1},h_{1})
        \overline{v}(\check{q},h)\gamma^{j}\frac{\gamma^{+}\gamma^{-}}{2} \Psi_{\beta}(\underline{z}) .  
    \end{align}




In order to compute the back-to-back limit of the scattering amplitudes in the $\bar q\to \bar q q $ channel, one starts from Eqs.~\eqref{eq:M_1_barq_to_barqq} and \eqref{eq:M_2_barq_to_barqq}, performs the change of variables given in Eqs.~\eqref{eq:PK} and \eqref{eq:br}, performs the small $|\r|$ expansion in the resulting expression and keeps only the zeroth order term in the expansion. This procedure yields to    
\begin{align}
\label{eq:M_1_barq_to_barqq_b2b}
i\M^{{\rm b2b}, \, m=0 }_{\bar q\to \bar q q, \, 1} &= 
i g^2\ \frac{z}{(2q^+)^2} \frac{1}{\P^2}
\int_{-\frac{L^+}{2}}^{\frac{L^+}{2}}  dz^+
\int_{\bb} \ e^{-i \bb\cdot (\bk - \bq)}
\Big[ t^{b}\  \UF^{\dagger}(\infty, -\infty;\bb)\Big]_{\beta_{1}\alpha_{1}} 
\nn \\
&
\times 
 \Big[ t^{b}\  \UF^{\dagger}(\infty, -\infty;\bb)\Big]_{\alpha\alpha_{1}} 
\UA(z^{+}, -\infty;\bb)_{ab}
\Big[ \UF(\infty,z^{+};\bb) \ t^{a}\Big]_{\alpha_{2}\beta_{2}} 
\nn \\
&
\times
\overline{v}(\check{q},h)  \gamma^{+} 
        \bigg[  \frac{\gamma^{i}\gamma^{l} \bP^{l}}{z}   - \frac{2 \bP^{i}}{1-z}
        \bigg]
        v(\check{p}_{1},h_{1}) \overline{u}(\check{p}_{2},h_{2})  \gamma^{i} \frac{\gamma^{+}\gamma^{-}}{2}  \Psi_{\beta_{2}}(z^{+};\bb)  ,   
\end{align}
for the first mechanism and to 
\begin{align}
\label{eq:M_2_barq_to_barqq_b2b}
&
i\M^{{\rm b2b}, \, m=0}_{\bar q\to \bar q q, \, 2} = 
ig^2\frac{z(1-z)}{(2q^{+})^2 } \frac{1}{\P^2}  
\int_{-\frac{L^+}{2}}^{\frac{L^+}{2}} dz^+ 
\int_{\b} \ e^{-i \b \cdot (\bk-\bq)} 
\ \UA(\infty,z^{+};\bb)_{ba} \big(t^{b}\big)_{\alpha{2}\alpha_{1}}  \\
&
\times
\Big[ \UF^{\dagger}(z^{+},-\infty;\bb) \ t^{a}\Big]_{\alpha\beta}
\overline{u}(\check{p}_{2},h_{2}) \gamma^{+}
        \Big[ -\gamma^{l}\gamma^{i}\frac{\bP^{l}}{1-z} + \gamma^{i}\gamma^{l}\frac{\bP^{l}}{z} 
        \Big] v(\check{p}_{1},h_{1})
        \overline{v}(\check{q},h)\gamma^{i}\frac{\gamma^{+}\gamma^{-}}{2} \Psi_{\beta}(z^{+};\bb) , \nn       
\end{align}
for the second mechanism. The total amplitude in the back-to-back and massless quark limits is given by the sum of Eqs.~\eqref{eq:M_1_barq_to_barqq_b2b} and \eqref{eq:M_2_barq_to_barqq_b2b} which can be written as  
\begin{align}
\label{eq:M_tot_barq_to_barqq_b2b}
&
i\M^{{\rm b2b}, \, m=0}_{\bar q\to \bar q q, \, {\rm tot.}} = 
- i\frac{g^{2}}{(2q^{+})^2} \frac{1}{\P^2}
 \int_{-\frac{L^+}{2}}^{\frac{L^+}{2}} dz^{+}  
\int_{\b} \ e^{-i \b \cdot (\bk-\bq)} 
\\
&
\times
\Big\{
  \UFd(z^{+},-\infty;\bb)_{\alpha\beta} \ \delta_{\alpha_{1}\alpha_{2}} \
\mathfrak{h}_{\bar q\to \bar q q}^{(1)}
 +
\UFd(\infty, -\infty;\bb)_{\alpha\alpha_{1}} \ \UF(\infty,z^{+};\bb)_{\alpha_{2}\beta} \ 
\mathfrak{h}_{\bar q\to \bar q q}^{(2)}
\Big\}\Psi_{\beta}(z^{+};\bb) , \nn
\end{align}
with the coefficients defined as
\begin{equation}
\label{eq:HF_barq_to_barqq_1}
    \mathfrak{h}_{\bar q\to \bar q q}^{(1)} =
    \left(\mathfrak{h}_{q\to q\bar q}^{(1)}\right)^{\dagger} ,
    \quad \quad
    \mathfrak{h}_{\bar q\to \bar q q}^{(2)} =
    \left(\mathfrak{h}_{q\to q\bar q}^{(2)}\right)^{\dagger} .
\end{equation}

Now, the partonic cross section for the $\bar q \to \bar q q$ channel in the back-to-back limit can be written as
\begin{equation} 
\label{eq:sigma_barq_to_barq q}
\frac{d\sigma^{{\rm b2b}, \, m=0}_{\bar q\to \bar q q}}{d{\rm P.S}} =
(2q^+) \ (2\pi)\delta\left(p_1^+ + p_2^+ - q^+\right)
\frac{1}{2N_c}\sum_{h_{1},h_2, h}\sum_{\beta_1, \beta_3, \alpha_3}
    \left\langle \left|i\mathcal{M}^{{\rm b2b}, \, m=0}_{\bar q\to \bar q q, {\rm tot.}}\right|^2 \right\rangle .
\end{equation}
Comparing the total scattering amplitude in the back-to-back limit given in Eq.~\eqref{eq:M_tot_barq_to_barqq_b2b} in the $\bar q\to \bar qq$ channel and the total scattering amplitude in the back-to-back limit given in Eq.~\eqref{eq:M_tot_q_to_qbarq_b2b_m0} in the $q \to q  \bar q$ channel, one realizes that the coefficients of the color structures in $\bar q\to \bar q  q$ channel are conjugate to the ones in $q \to q \bar q$. Thus one can trivially write down the partonic cross section in the back-to-back and massless quark limits for the $\bar q \to \bar q q$ channel. This can directly be done for the different flavor cases as  

\begin{align}
\frac{d\sigma^{{\rm b2b}, \, m=0}_{\bar q_{f}\to \bar q_{f_{1}} q_{f_{2}}}}{d{\rm P.S.}} &=
g^4 \ (2\pi)\delta\left(p_1^+ + p_2^+ - q^+\right)
\int_{-\frac{L^+}{2}}^{\frac{L^+}{2}}dz^+
\int_{-\frac{L^+}{2}}^{\frac{L^+}{2}}dz'^+
\int_{\bb, \bb'}
e^{+i(\bb-\bb')\cdot\left(\bk - \bq\right)} 
\nn\\
&
\hspace{5cm}\
\times
\Big[ \mathcal{H}^{-}_{\bar q_{f}\to \bar q_{f_{1}} q_{f_{2}}} \  \mathcal{C}^{-}  +  \mathcal{H}^{+\square}_{\bar q_{f}\to \bar q_{f_{1}} q_{f_{2}}} \ \mathcal{C}^{+\square}\Big] ,
\end{align}
where $\mathcal{C}^{+\square}$ is defined in \eqref{eq:C+square} and $\mathcal{C}^-$ reads
\begin{equation}
\label{eq:C-} 
\mathcal{C}^- \equiv 
\Big\langle\overline{\Psi}(z'^+;\bb')
\gamma^-\UF(z'^{+},-\infty;\bb')\UFd(z^{+},-\infty;\bb)
\Psi({z}^+;\bb)\Big\rangle .
\end{equation}
Similar to $q \to q \bar q$ channel, the possibilities in terms of quark flavor are limited to $f=f_1=f_2$, $f \neq f_1=f_2$ and $f=f_1 \neq f_2$ and the hard factors are defined for the case $f=f_1=f_2$ as 
\begin{align}
    \mathcal{H}^{-}_{\bar q\to \bar qq}  = \mathcal{H}^{-}_{q\to q\bar q} \;\; \text{and}\;\;
    \mathcal{H}^{+\square}_{\bar q\to \bar qq}  = \mathcal{H}^{+\square}_{q\to q\bar q} ,
\end{align}
for the case $f \neq f_1=f_2$ as 
\begin{align}
    \mathcal{H}^{-}_{\bar q_{f}\to \bar q_{f'} q_{f'}}  = \mathcal{H}^{-}_{q_{f}\to q_{f'}\bar q_{f'}} \;\; \text{and}\;\;
    \mathcal{H}^{+\square}_{\bar q_{f}\to \bar q_{f'} q_{f'}}  = \mathcal{H}^{+\square}_{q_{f}\to q_{f'}\bar q_{f'}} ,
\end{align}
and for the case $f=f_1 \neq f_2$ as
\begin{align}
    \mathcal{H}^{-}_{\bar q_{f}\to \bar q_{f} q_{f'}}  = \mathcal{H}^{-}_{q_{f}\to q_{f}\bar q_{f'}} \;\; \text{and}\;\;
    \mathcal{H}^{+\square}_{\bar q_{f}\to \bar q_{f} q_{f'}}  = \mathcal{H}^{+\square}_{q_{f}\to q_{f}\bar q_{f'}}.
\end{align}

\subsection{$\bar q\to gg$ channel} 
\label{sec:bqgg}                        
\begin{figure}[H]
\centering
\begin{subfigure}{0.49\textwidth}
\centering
\includegraphics[width=\textwidth]{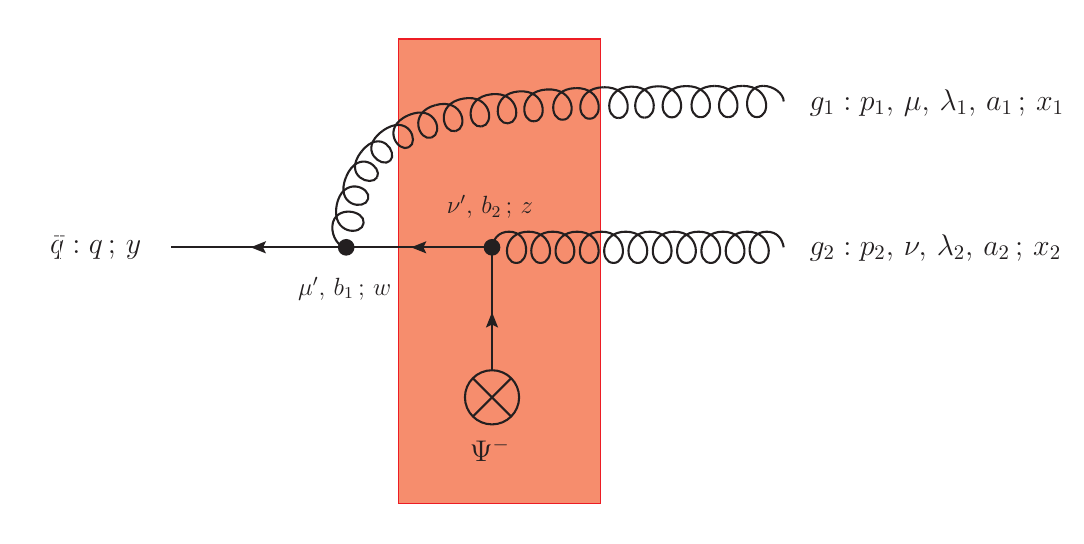}
\caption{Diagram 1}
\label{fig:bq-gg1}
\end{subfigure}
\begin{subfigure}{0.49\textwidth}
\centering
\includegraphics[width=\textwidth]{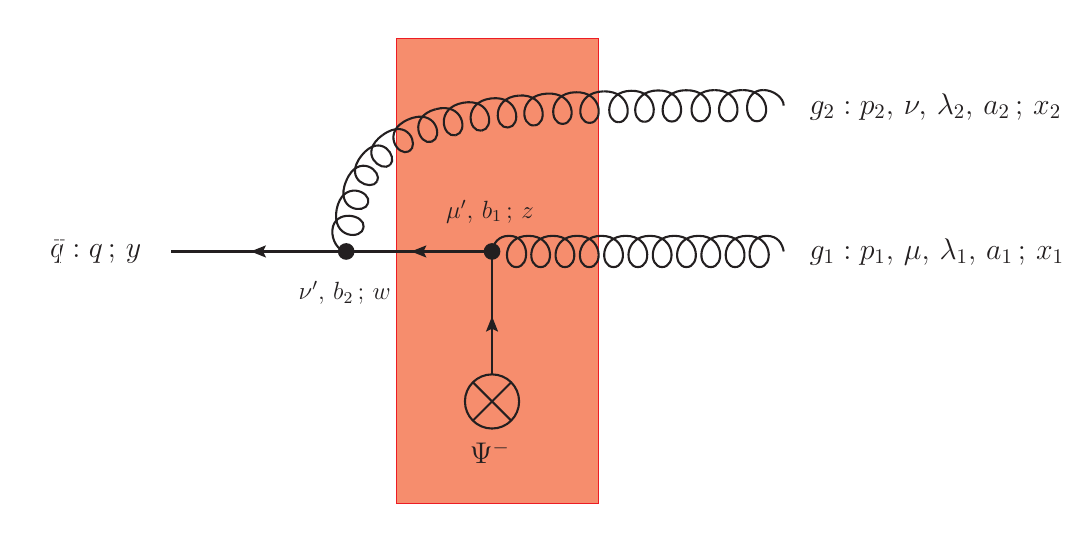}
\caption{Diagram 2}
\label{fig:bq-gg2}
\end{subfigure}
\begin{subfigure}{0.5\textwidth}
\centering
\includegraphics[width=\textwidth]{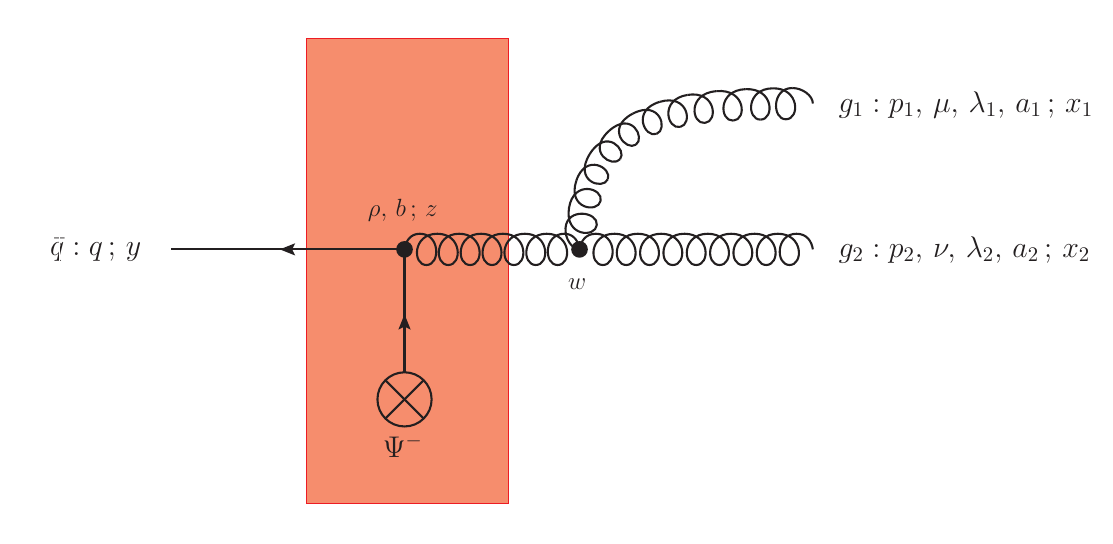}
\caption{Diagram 3}
\label{fig:bq-gg3}
\end{subfigure}
\caption{Diagrams contributing to channel $\bar{q} \to gg$.}
\label{fig:bq-gg}
\end{figure}

The computation of the scattering amplitude and the cross section in this channel is very similar to the $q\to gg$ channel. Since the computations are very similar, in this channel we briefly present the results. 
There are three mechanisms that contribute to the production of gluon-gluon dijet in antiquark initiated channel at NEik accuracy. Each of these three mechanisms are illustrated in Fig.~\ref{fig:bq-gg}.
%
%
%
The first mechanism (see Fig.~\ref{fig:bq-gg1}) corresponds to the case where the incoming antiquark splits into an antiquark-gluon pair before the medium which then scatters on the target. While the gluon scatters eikonally, the antiquark scatters via a $t$-channel quark exchange and converts into a gluon, producing a gluon dijet in the final state.
The $S$-matrix element for this mechanism can be obtained from
\begin{align}
\S_{\bar q\to g_{1}g_{2}, \, 1} &= 
-\lim_{y^+\to-\infty}\lim_{x_1^+,x_2^+\to\infty}
\int_{\by,\bx[1],\bx[2]}\int_{y^-,x^-_1,x^-_2}
\int_{\bw, \bz}\int_{w^-, z^-}
\int_{-\frac{L^+}{2}}^{\frac{L^+}{2}}d z^+
\int_{-\infty}^{-\frac{L^+}{2}}d w^+
\nn \\
&
\times        
\ e^{ix_{1} \cdot \check{p}_{1}} \ e^{ix_{2} \cdot \check{p}_{2}} \ e^{-iy \cdot \check{q}}
\ (-2p_{1}^{+}) {\epsilon^{\lambda_{1}}_{\mu}(p_{1})}^{*} 
(-2p_{2}^{+}) {\epsilon^{\lambda_2}_{\nu}(p_{2})}^{*}
\nn  \\
&
\times 
\ \Big[ G^{\mu\mu'}_{F}(x_{1},w) \BA \Big]_{a_{1}b_{1}}  
\Big[ G^{\nu\nu'}_{F}(x_{2},z) \IA \Big]_{a_{2}b_{2}} 
\nn  \\
&
\times 
 \overline{v}(\check{q},h)  \ 
 \gamma^{+} 
 \Big[ S_{0,F}(y,w) \ (-igt^{b_{1}}\gamma_{\mu'}) \ S_{F}(w,z)\BIbq  (-igt^{b_{2}}\gamma_{\nu'})\Big]_{\alpha\beta}  
\Psi^{-}_{\beta}(z) ,
\end{align}
where the before-to-after gluon propagator is given in Eq.~\eqref{gluon_prop_BA}, the inside-to-after gluon propagator is given in Eq.~\eqref{gluon_prop_IA}, the before-to-inside antiquark propagator is given in Eq.~\eqref{antiquark_prop_BI} and the vacuum quark propagator is given in Eq.~\eqref{eq:quark_vacuum_prop}. Using the explicit expressions of the propagators, one can obtain the $S$-matrix element for the first mechanism as 
\begin{align}
\S_{\bar q\to g_{1}g_{2}, \, 1} &= 
\frac{g^2 \ {\varepsilon_{\lambda_1}^{j}}^{*} {\varepsilon_{\lambda_2}^{i}}^{*} }{(2q^{+})(2p_{2}^{+})} 
\ 2\pi\delta(p_{1}^{+}+p_{2}^{+}-q^{+})
\int_{-\frac{L^+}{2}}^{\frac{L^+}{2}} dz^+ 
\int_{-\infty}^{-\frac{L^+}{2}} dw^+
\\
&
\times
\int\! \frac{d^{2}\bk_{1}}{(2\pi)^{2}}\ 
e^{iw^{+}\Big(\frac{\bk_{1}^{2}}{2p_{1}^{+}} + \frac{(\bq - \bk_{1})^{2} + m^2}{2p_{2}^{+}} - \frac{\bq^{2} + m^2}{2q^{+}}\Big)}
\int_{\bz,\bz_{1} } e^{i(\bq - \bk_1 -\p_2) \cdot \bz}
\ e^{i(\bk_1 - \bp_{1}) \cdot \bz_{1}}
\nn \\
&
\times
\Big[ t^{b_{1}}\ \UF^{\dagger}(z^{+},w^{+};\bz)\ t^{b_{2}}\Big]_{\alpha\beta}
\UA(\infty,w^{+}; \bz_{1})_{a_{1}b_{1}} \ \UA(\infty,z^{+}; \bz)_{a_{2}b_{2}}
\overline{v}(\check{q},h)  
\nn \\
&
\times
\Big[ 
\big(\gamma^{l}\gamma^{j}\bq^{l} p_{2}^{+} + \gamma^{j}\gamma^{l}(\bq^{l} - \bk_{1}^{l})q^{+} - \gamma^{j}(p_{2}^{+}-q^{+})m + 2\frac{q^{+}p_{2}^{+} \bk_{1}^{j}}{p_{1}^{+}} \Big] 
\gamma^i\gamma^{+}\gamma^{-} \Psi_{\beta}(\underline{z}) , \nn
\end{align}
from which we can extract the amplitude in general kinematics as
\begin{align}
\label{eq:M_1_barq_to_gg}
&
i\M_{\bar q\to g_{1}g_{2}, \, 1} = 
\frac{g^2 \ {\varepsilon_{\lambda_1}^{j}}^{*} {\varepsilon_{\lambda_2}^{i}}^{*} }{(2q^{+})^2(2p_{2}^{+})} 
\int_{-\frac{L^+}{2}}^{\frac{L^+}{2}} dz^+ 
\int_{-\infty}^{-\frac{L^+}{2}} dw^+
\int\! \frac{d^{2}\bk_{1}}{(2\pi)^{2}}
e^{iw^{+}\Big(\frac{\bk_{1}^{2}}{2p_{1}^{+}} + \frac{(\bq - \bk_{1})^{2} + m^2}{2p_{2}^{+}} - \frac{\bq^{2} + m^2}{2q^{+}}\Big)}
\nn \\
&
\times
\int_{\bz,\bz_{1} } e^{i(\bq - \bk_1 -\p_2) \cdot \bz}
\ e^{i(\bk_1 - \bp_{1}) \cdot \bz_{1}}
\Big[ t^{b_{1}}\ \UF^{\dagger}(z^{+},w^{+};\bz)\ t^{b_{2}}\Big]_{\alpha\beta}
\UA(\infty,w^{+}; \bz_{1})_{a_{1}b_{1}} \ \UA(\infty,z^{+}; \bz)_{a_{2}b_{2}}
\nn \\
&
\times
\overline{v}(\check{q},h) \Big[ 
\big(\gamma^{l}\gamma^{j}\bq^{l} p_{2}^{+} + \gamma^{j}\gamma^{l}(\bq^{l} - \bk_{1}^{l})q^{+} - \gamma^{j}(p_{2}^{+}-q^{+})m + 2\frac{q^{+}p_{2}^{+} \bk_{1}^{j}}{p_{1}^{+}} \Big] 
\gamma^i\gamma^{+}\gamma^{-} \Psi_{\beta}(\underline{z})  .   
\end{align}
The second mechanism is the same as the first with the gluons in the final state interchanged (see Fig.~\ref{fig:bq-gg2}). Thus, the $S$-matrix element for the second mechanism can be written as
\begin{align}
&
\S_{\bar q\to g_{1}g_{2}, \, 2} = 
-\lim_{y^+\to-\infty}\lim_{x_1^+,x_2^+\to\infty}
(-2p_{1}^{+})(-2p_{2}^{+})
\int_{\by,\bx[1],\bx[2]}\int_{y^-,x^-_1,x^-_2}
\int_{\bw, \bz}\int_{w^-, z^-}
\int_{-\frac{L^+}{2}}^{\frac{L^+}{2}}d z^+
\int_{-\infty}^{-\frac{L^+}{2}}d w^+
\nn \\
&
\times       
\ e^{ix_{1} \cdot \check{p}_{1}} \ e^{ix_{2} \cdot \check{p}_{2}} \ e^{-iy \cdot \check{q}}
\ {\epsilon^{\lambda_{1}}_{\mu}(p_{2})}^{*} 
{\epsilon^{\lambda_2}_{\nu}(p_{1})}^{*}
\ \Big[G^{\nu\nu'}_{F}(x_{2},w)\BA\Big]_{a_{2}b_{2}}
\  \Big[G^{\mu\mu'}_{F}(x_{1},z)\IA\Big]_{a_{1}b_{1}} 
\nn \\
&
\times 
\overline{v}(\check{q},h) \ \gamma^{+} 
\Big[ S_{0,F}(y,w) (-igt^{b_{2}}\gamma_{\mu'}) \ S_{F}(w,z)\BIbq  (-igt^{b_{1}}\gamma_{\nu'})\Big]_{\alpha\beta} 
\Psi^{-}_{\beta}(z) ,
\end{align}
where the propagators here are the same as in the first mechanism. Again using the explicit expressions for these propagators we get the $S$-matrix element for the second mechanism as
\begin{align}    
\S_{\bar q\to g_{1}g_{2}, \, 2} &= 
\frac{g^2 \ {\varepsilon_{\lambda_1}^{i}}^{*} {\varepsilon_{\lambda_2}^{j}}^{*} }{(2q^{+})(2p_{1}^{+})} 
\ 2\pi\delta(p_{1}^{+}+p_{2}^{+}-q^{+})
\int_{-\frac{L^+}{2}}^{\frac{L^+}{2}} dz^+ 
\int_{-\infty}^{-\frac{L^+}{2}} dw^+
\nn \\
&
\times
\int\! \frac{d^{2}\bk_{2}}{(2\pi)^{2}}\ 
e^{iw^{+}\Big(\frac{(\bq - \bk_{2})^{2} + m^2}{2p_{1}^{+}} + \frac{\bk_{2}^{2}}{2p_{2}^{+}} - \frac{\bq^{2} + m^2}{2q^{+}}\Big)}
\int_{\bz,\bz_{1} } e^{i(\bq -\p_1 - \bk_2) \cdot \bz}
\ e^{i(\bk_2 - \bp_{2}) \cdot \bz_{1}}
\nn \\
&
\times
\overline{v}(\check{q},h) 
\Big[ 
\big(\gamma^{l}\gamma^{j}\bq^{l} p_{1}^{+} + \gamma^{j}\gamma^{l}(\bq^{l} - \bk_{2}^{l})q^{+} - \gamma^{j}(p_{1}^{+}-q^{+})m + 2\frac{q^{+}p_{1}^{+} \bk_{2}^{j}}{p_{2}^{+}}
\Big] \gamma^i
\nn \\
& 
\times 
\Big[ t^{b_{2}}\ \UF^{\dagger}(z^{+},w^{+}; \bz)\ t^{b_{1}}\Big]_{\alpha\beta} 
\UA(\infty,z^{+}; \bz)_{a_{1}b_{1}} \ \UA(\infty,w^{+}; \bz_{1})_{a_{2}b_{2}}
\gamma^{+}\gamma^{-}\Psi_{\beta}(\underline{z}) , 
\end{align}
from which we can obtain the amplitude for the second mechanism as 
\begin{align}    
\label{eq:M_2_barq_to_gg}
&
i \M_{\bar q\to g_{1}g_{2}, \, 2}  = 
\frac{g^2 \ {\varepsilon_{\lambda_1}^{i}}^{*} {\varepsilon_{\lambda_2}^{j}}^{*} }{(2q^{+})^2(2p_{1}^{+})} 
\int_{-\frac{L^+}{2}}^{\frac{L^+}{2}} dz^+ 
\int_{-\infty}^{-\frac{L^+}{2}} dw^+
\int\! \frac{d^{2}\bk_{2}}{(2\pi)^{2}}\ 
e^{iw^{+}\Big(\frac{(\bq - \bk_{2})^{2} + m^2}{2p_{1}^{+}} + \frac{\bk_{2}^{2}}{2p_{2}^{+}} - \frac{\bq^{2} + m^2}{2q^{+}}\Big)}
\nn \\
&
\times
\int_{\bz,\bz_{1} } e^{i(\bq -\p_1 - \bk_2) \cdot \bz}
\ e^{i(\bk_2 - \bp_{2}) \cdot \bz_{1}}
\Big[ t^{b_{2}}\ \UF^{\dagger}(z^{+},w^{+}; \bz)\ t^{b_{1}}\Big]_{\alpha\beta} 
\UA(\infty,z^{+}; \bz)_{a_{1}b_{1}} \ \UA(\infty,w^{+}; \bz_{1})_{a_{2}b_{2}}
\nn \\
&
\times
\overline{v}(\check{q},h) 
\Big[ 
\big(\gamma^{l}\gamma^{j}\bq^{l} p_{1}^{+} + \gamma^{j}\gamma^{l}(\bq^{l} - \bk_{2}^{l})q^{+} - \gamma^{j}(p_{1}^{+}-q^{+})m + 2\frac{q^{+}p_{1}^{+} \bk_{2}^{j}}{p_{2}^{+}}
\Big] \gamma^i
\gamma^{+}\gamma^{-}\Psi_{\beta}(\underline{z})  . 
\end{align}
The third mechanism corresponds to the case where the incoming antiquark converts into a gluon upon scattering via $t$-channel quark exchange, and then splits into two gluons in the final state (see Fig.~\ref{fig:bq-gg3}). For that case, the $S$-matrix element can be written as 
\begin{align}
\S_{\bar q\to g_{1}g_{2}, \, 3} &= 
\lim_{y^+\to-\infty} \ \lim_{x_1^+,x_2^+\to\infty}
(-2p_{1}^{+})(-2p_{2}^{+})
\ \int_{\by,\bx[1],\bx[2]} \ \int_{y^-,x^-_1,x^-_2}
\ \int_{\bw, \bz}\int_{w^-, z^-}
\int_{-\frac{L^+}{2}}^{\frac{L^+}{2}}d z^+
\int_{\frac{L^+}{2}}^{\infty}d w^+
\nn \\
&
\times
\ e^{ix_{1} \cdot \check{p}_{1}} \ e^{ix_{2} \cdot \check{p}_{2}} \ e^{-iy \cdot \check{q}}
\ {\epsilon^{\lambda_{1}}_{\mu}(p_{1})}^{*} 
\ {\epsilon^{\lambda_2}_{\nu}(p_{2})}^{*} 
\Big[ G^{\mu\mu'}_{0,F}(x_{1},w)\Big]_{a_{1}b_{1}} 
\Big[G^{\nu\nu'}_{0,F}(x_{2},w) \Big]_{a_{2}b_{2}} 
\nn \\
&
\times 
V^{b_{1}b_{2}c}_{\mu'\nu'\rho'}
\Big[ G_{F}^{\rho'\rho}(w,z) \IA\Big]_{cb}
v(\check{q},h')  \gamma^{+}
\Big[S_{F}(y,z) \BIbq (-igt^{b}\gamma_{\rho})\Big]_{\alpha\beta}
\Psi^{-}_{\beta}(z) ,
\end{align}
with $V^{b_{1}b_{2}c}_{\mu'\nu'\rho'}$ being the triple gluon vertex defined in Eq.~\eqref{eq:triple_gluon_vertex}. Using the explicit expressions for the inside-to-after gluon propagator given in Eq.~\eqref{gluon_prop_IA}, the before-to-inside antiquark propagator given in Eq.~\eqref{antiquark_prop_BI} and the vacuum gluon propagator given in Eq.~\eqref{eq:vacuum_gluon_prop}, one can obtain the $S$-matrix for the third mechanism as 
\begin{align}
&
 \S_{\bar q\to g_{1}g_{2}, \, 3} =
 -\frac{ig^{2}}{(2q^{+})} \ f^{a_{1}a_{2}c}
 \ {\varepsilon^{j}_{{\lambda_2}}}^{*} \ {\varepsilon^{i}_{{\lambda_1}}}^{*}\ 
 (2\pi) \ \delta(p_{1}^{+}+p_{2}^{+}-q^{+})
\int_{-\frac{L^+}{2}}^{\frac{L^+}{2}} dz^+
\int_{\frac{L^+}{2}}^{\infty} dw^+
\\
& 
\times
e^{iw^{+}\Big(\frac{\bp_{1}^{2}}{2p_{1}^{+}} + \frac{\bp_{2}^{2}}{2p_{2}^{+}} - \frac{(\bp_{1} + \bp_{2})^{2}}{2q^{+}}\Big)}
\int_{\bz} \  e^{-i \bz \cdot (\bp_{1}+ \bp_{2}- \bq)}\
\Big[t^{b} \ \UFd (z^{+},-\infty; \bz)\Big]_{\alpha\beta}
\UA(w^{+},z^{+};\bz)_{cb}
\nn \\
&
v(\check{q},h') \ \gamma^{l}\gamma^{+}\gamma^{-}
\Psi_{\beta}(\underline{z})
\Big\{g^{il} \Big(-\bp_{1}^{j} + \frac{p_{1}^{+}}{p_{2}^{+}} \bp_{2}^{j}\Big) 
+ g^{lj}\Big(\bp_{2}^{i} - \frac{p_{2}^{+}}{p_{1}^{+}} \bp_{1}^{i}\Big) 
+ g^{ji}\Big(\bp_{1}^{l} - \frac{p_{1}^{+}}{q^{+}}(\bp^{l}_{1}+ \bp_{2}^{l})\Big)      
\Big\} , \nn
\end{align}
from which the amplitude can be extracted and reads 
\begin{align}
\label{eq:M_3_barq_to_gg}
&
i \M_{\bar q\to g_{1}g_{2}, \, 3} =
 -\frac{ig^{2}}{(2q^{+})^2} \ f^{a_{1}a_{2}c}
 \ {\varepsilon^{j}_{{\lambda_2}}}^{*} \ {\varepsilon^{i}_{{\lambda_1}}}^{*}\ 
\int_{-\frac{L^+}{2}}^{\frac{L^+}{2}} dz^+
\int_{\frac{L^+}{2}}^{\infty} dw^+
e^{iw^{+}\Big(\frac{\bp_{1}^{2}}{2p_{1}^{+}} + \frac{\bp_{2}^{2}}{2p_{2}^{+}} - \frac{(\bp_{1} + \bp_{2})^{2}}{2q^{+}}\Big)}
\nn \\
& 
\times
\int_{\bz} \  e^{-i \bz \cdot (\bp_{1}+ \bp_{2}- \bq)}\
\Big[t^{b} \ \UFd (z^{+},-\infty; \bz)\Big]_{\alpha\beta}
\UA(w^{+},z^{+};\bz)_{cb}
\ v(\check{q},h') \ \gamma^{l}\gamma^{+}\gamma^{-}
\Psi_{\beta}(\underline{z})
\nn \\
&
\Big\{g^{il} \Big(-\bp_{1}^{j} + \frac{p_{1}^{+}}{p_{2}^{+}} \bp_{2}^{j}\Big) 
+ g^{lj}\Big(\bp_{2}^{i} - \frac{p_{2}^{+}}{p_{1}^{+}} \bp_{1}^{i}\Big) 
+ g^{ji}\Big(\bp_{1}^{l} - \frac{p_{1}^{+}}{q^{+}}(\bp^{l}_{1}+ \bp_{2}^{l})\Big)      
\Big\}  .
\end{align}
The total scattering amplitude is the sum of the three contributions from each mechanism and it can be written as 
\begin{align}
i \M_{\bar q\to g_{1}g_{2}, \, {\rm tot.}} = i \M_{\bar q\to g_{1}g_{2}, \, 1} + i \M_{\bar q\to g_{1}g_{2}, \, 2}  + i \M_{\bar q\to g_{1}g_{2}, \, 3} ,
\end{align}
with the contributions from each mechanism given in Eqs.~\eqref{eq:M_1_barq_to_gg}, \eqref{eq:M_2_barq_to_gg} and \eqref{eq:M_3_barq_to_gg}.


The back-to-back limit of the scattering amplitudes can be computed as in the $q\to gg$ channel. Namely, one starts from the production amplitudes written in general kinematics, then perform the change of variables to rewrite the production amplitude in terms of the relative dijet momenta $\P$ and dijet momentum imbalance $k$ as well as their conjugate variables $\r$ and $\b$. As discussed previously, the back-to-back limit corresponds to a kinematic regime where $|\P|\gg|\k|$ which is equivalent to $|\r|\ll|\b|$. Therefore, in this regime one can perform a small $|\r|$ expansion of the Wilson lines and keep only the first non-trivial term in the expansion. After the expansion, $\r$ and $\b$ integrations factorize and one can perform $\r$ integration trivially. After all said and done, the back-to-back limit of the amplitudes for the three mechanisms can be written as 
\begin{align}
&
i\M^{{\rm b2b}, \, m=0}_{\bar q\to g_{1}g_{2}, \, 1} = 
ig^2\ \frac{z(1-z)}{(2q^+)\, \bP^2}
\ {\varepsilon_{\lambda_{1}}^{j}}^{*} 
{\varepsilon_{\lambda_2}^{i}}^{*}  
\int_{-\frac{L^+}{2}}^{\frac{L^+}{2}} dz^+
\int_{\b}e^{-i \bb \cdot (\bk-\bq)}
\overline{v}(\check{q},h) 
\Big[  
\frac{\gamma^{j}\gamma^{l}\bP^{l}}{1-z}  
-2\frac{\bP^{j}\gamma^{j}}{z} 
\Big]\gamma^{i}
\nn \\
& 
\times
\Big[ t^{b_{1}}\ \UF^{\dagger}(-\infty,z^{+};\bb) \ t^{b_{2}}\Big]_{\alpha\beta}
\frac{\gamma^{+}\gamma^{-}}{2} \Psi_{\beta}(z^{+};\bb)
\UA(\infty,-\infty;\bb)_{a_{1}b_{1}}
\ \UA(\infty,z^{+};\bb)_{a_{2}b_{2}}    ,  
\end{align}
\begin{align}
&
i\M^{{\rm b2b}, \, m=0}_{\bar q\to g_{2}g_{1}, \, 2} = 
-ig^2\ \frac{z(1-z)}{(2q^+)\, \bP^2}
\ {\varepsilon_{\lambda_{1}}^{i}}^{*} 
{\varepsilon_{\lambda_2}^{j}}^{*}  
\int_{-\frac{L^+}{2}}^{\frac{L^+}{2}} dz^+
\int_{\b}e^{-i \bb \cdot (\bk-\bq)}
\overline{v}(\check{q},h) 
\Big[  
\frac{\gamma^{j}\gamma^{l}\bP^{l}}{z}  
-2\frac{\bP^{j}\gamma^{j}}{1-z} 
\Big]\gamma^{i}
\nn \\
& 
\times
\Big[ t^{b_{2}}\ \UF^{\dagger}(-\infty,z^{+};\bb) \ t^{b_{1}}\Big]_{\alpha\beta}
\frac{\gamma^{+}\gamma^{-}}{2} \Psi_{\beta}(z^{+};\bb)
\UA(\infty,z^{+};\bb)_{a_{1}b_{1}}
\ \UA(\infty,-\infty;\bb)_{a_{2}b_{2}}   ,  \nn  
\end{align}
and 
\begin{align}
&
i\M^{{\rm b2b}, \, m=0}_{\bar q\to g_{1}g_{2}, \, 3} =  
g^{2}f^{a_{1}a_{2}c}
{\varepsilon^{j}_{{\lambda_2}}}^{*} 
{\varepsilon^{i}_{{\lambda_1}}}^{*}
\ \frac{z(1-z)}{(2q^+)}\frac{1}{\bP^2}
\int_{-\frac{L^+}{2}}^{\frac{L^+}{2}} dz^+ 
\int_{\bb} \ e^{-i\bb \cdot (\bk - \bq)} 
\ \UA(\infty,z^{+};\bb)_{cb}
\nn \\
& \hspace{0.5cm}
\times        
\Big[g^{il}\frac{\bP^{j}}{1-z} + g^{lj}\frac{\bP^{i}}{z} - g^{ji}\bP^{l} \Big]
\Big[\UFd (z^{+},-\infty;\bb) \ t^{b}\Big]_{\alpha\beta} 
v(\check{q},h') \ \gamma^{l}\gamma^{+}\gamma^{-}
\ \Psi_{\beta}(z^{+};\bb) .
\end{align}
The total amplitude for the $\bar q\to gg$ channel is given as the sum of the above three amplitudes in the back-to-back limit and it can be written as 
\begin{align}
&
i\M^{{\rm b2b}, \, m=0}_{\bar q\to gg, \, {\rm tot.} } = 
\frac{ig^{2}}{(2q^+)} \frac{1}{2\bP^{2}}
\int_{-\frac{L^+}{2}}^{\frac{L^+}{2}} dz^{+}\
\int_{\bb} \ e^{-i\b \cdot (\bk- \bq)}
\overline{v}(\check{q},h) \ \UFd(z^{+},-\infty;\bb) 
\frac{\gamma^{+}\gamma^{-}}{2} 
\\
&
\times 
\Big[
\mathfrak{h}^{(1)}_{\bar q\to gg}
\Big(\frac{\delta^{a_1 a_2}}{N_{c}} + d^{a_1 a_2 d}  \ \UA (\infty, z^{+};\bb)_{dc}\ t^{c}\Big)  
+ 
\mathfrak{h}^{(2)}_{\bar q\to gg} 
\Big(if^{a_1 a_2 d}  \ \UA (\infty, z^{+};\bb)_{dc}\ t^{c} \Big)
\Big] 
\ \Psi(z^{+};\bb)  \  ,  \nn   
\end{align}
with the two coefficients being 
\begin{equation}
\label{eq:HF_barq_to_gg}
\mathfrak{h}^{(1)}_{\bar q\to gg} = 
\left(\mathfrak{h}^{(1)}_{q\to gg}\right)^\dagger
, \quad \quad
\mathfrak{h}^{(2)}_{\bar q\to gg} = 
\left(\mathfrak{h}^{(2)}_{q\to gg}\right)^\dagger.
\end{equation}

The partonic cross section in the back-to-back limit can now be written as 
\begin{equation}
\frac{d\sigma^{{\rm b2b}, \, m=0}_{\bar q\to gg}}{d{\rm P.S.}} =
(2q^+) \ (2\pi) \ \delta\left(p_1^+ + p_2^+ - q^+\right)
\ \frac{1}{2N_c}\sum_{\lambda_1,\lambda_2}\sum_{h}\sum_{a_1,a_2}
\left\langle \left|i\mathcal{M}^{{\rm b2b}, \, m=0}_{\bar q\to gg, {\rm tot.}}\right|^2 \right\rangle .
\end{equation}
Following the same steps as in Sec.~\ref{subsec:xsection_q-gg}, the cross section can be written in the following factorized form: 
\begin{align}
\frac{d\sigma^{{\rm b2b}, \, m=0}_{\bar q\to gg}}{d{\rm P.S.}} &= 
g^{4} \, (2\pi) \ \delta\left(p_1^+ + p_2^+ - q^+\right) 
\int_{-\frac{L^+}{2}}^{\frac{L^+}{2}} dz^+ 
\int_{-\frac{L^+}{2}}^{\frac{L^+}{2}} d{z'}^+ \int_{\bb, \bb'} e^{i(\bb-\bb')\cdot(\bk - \bq)} 
\nn \\
& \hspace{4cm}
\times
\Big[ \mathcal{H}^{-}_{\bar q\to gg} \ \mathcal{C}^{-}
+   \mathcal{H}^{-g}_{\bar q\to gg}  \ \mathcal{C}^{-g} \Big] ,
\label{eq:sigma_barq_to_gg}
\end{align}
with the same hard factors defined in Eqs.~\eqref{eq:H_q_to_gg_-g} and \eqref{eq:H_q_to_gg_-},  $\mathcal{C}^{-}$ defined in Eq.~\eqref{eq:C-} and $\mathcal{C}^{-g}$ defined as 
\begin{align}
\label{eq:C-g}
\C^{-g} &\equiv \Big\langle \overline{\Psi}(z^{+};\bb) \ \gamma^{-} \ t^{c} \ \UF(z^{+},-\infty;\bb) \  \UAd(\infty,z'^{+};\bb')_{d'c'}
\ \UA(\infty,z^{+};\bb)_{dc} 
\nn \\
&\hspace{7cm} 
\times \UFd(z'^{+},-\infty;\bb') \  t^{c'} \ \Psi(z'^{+};\bb') \Big\rangle ,
\end{align}

\subsection{$\bar q\to \bar q\bar q$ channel} 
\begin{figure}[H]
\centering
\begin{subfigure}{0.49\textwidth}
\centering
\includegraphics[width=\textwidth]{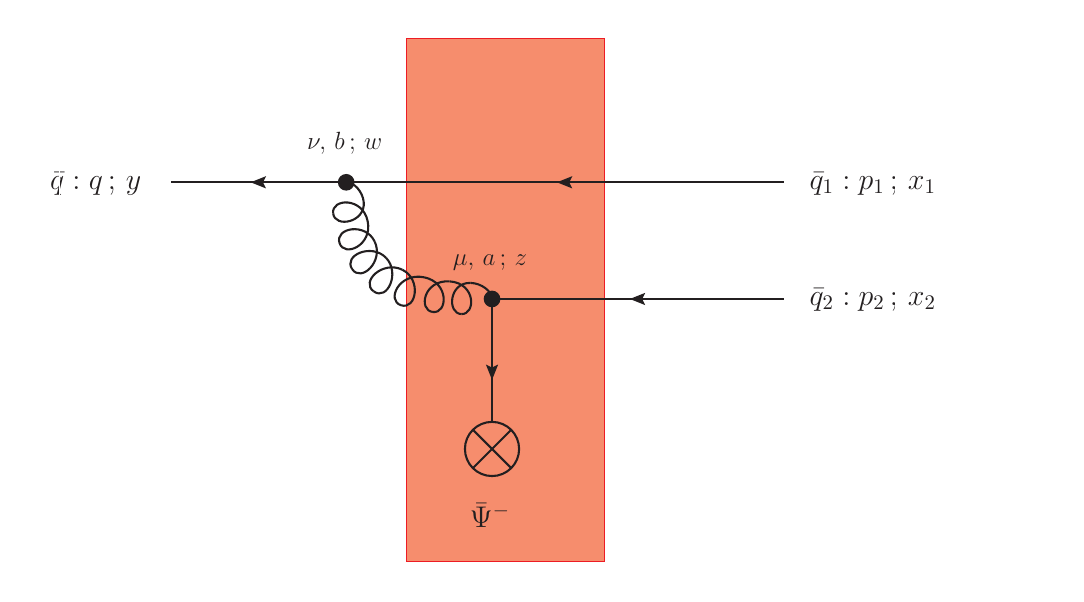}
\caption{Diagram 1}
\label{fig:bq-bqbq1}
\end{subfigure}
\begin{subfigure}{0.49\textwidth}
\centering
\includegraphics[width=\textwidth]{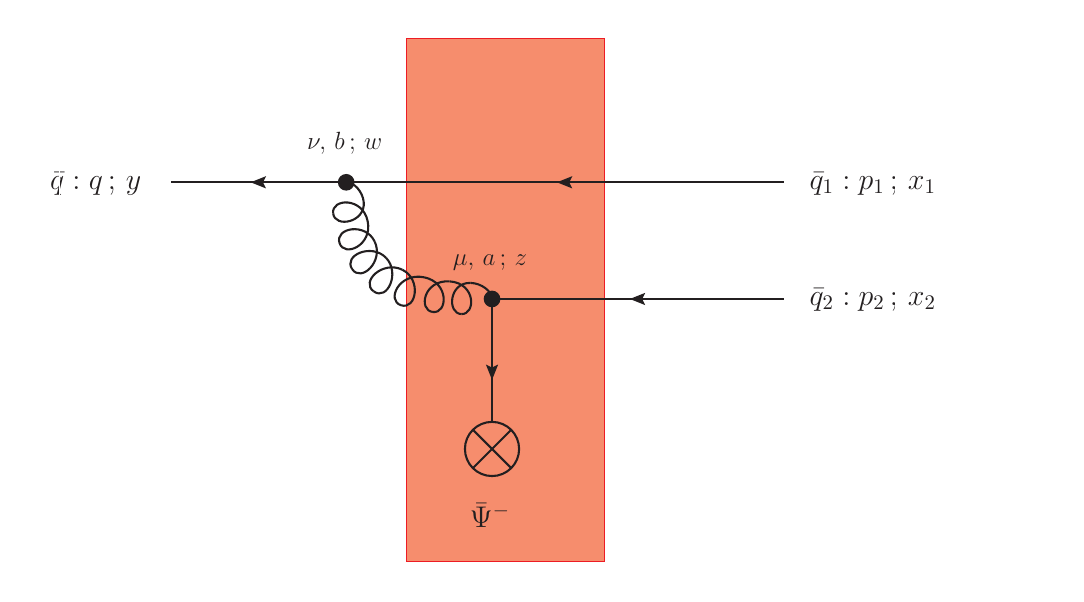}
\caption{Diagram 2}
\label{fig:bq-bqbq2}
\end{subfigure}
\caption{Diagrams contributing to channel $\bar{q} \to \bar{q}\bar{q}$.}
\label{fig:bq-bqbq}
\end{figure}

The computations of this channel are very similar to the ones performed in $q\to qq$ channel.  We will first consider the case where all antiquarks have the same flavor. In this case at NEik accuracy, the two mechanisms contribute to this channel. The one where the incoming antiquark splits into an antiquark-gluon pair before the medium. Then the pair interacts with the target. While the antiquark scatters eikonally, the gluon scatters via a $t$-channel quark exchange and converts into an antiquark, producing an antiquark-antiquark dijet in the final state (see Fig.~\ref{fig:bq-bqbq}). The second one is similar except the antiquarks in the final state are interchanged (see Fig.~\ref{fig:bq-bqbq2}). We will also consider the case where the final antiquarks have different flavors (more explicitly, $\bar q_{f} \to \bar q_{f} \bar q_{f'}$, with $f \neq f'$). In this instance, only the first mechanism contributes (see Fig.~\ref{fig:bq-bqbq1}).

The $S$-matrix element for the first mechanism can be written as 
\begin{align}
\S_{\bar q\to \bar q\bar q, \, 1} &= 
\lim_{y^+\to-\infty}\lim_{x_1^+,x_2^+\to\infty}
\int_{\by,\bx[1],\bx[2]}\int_{y^-,x^-_1,x^-_2}
\int_{\bw, \bz}\int_{w^-, z^-}
\int_{-\frac{L^+}{2}}^{\frac{L^+}{2}}d z^+
\int_{-\infty}^{-\frac{L^+}{2}}d w^+
\nn \\
&
\times 
e^{ix_{1} \cdot \check{p}_{1}} \ e^{ix_{2} \cdot \check{p}_{2}} \ e^{-iy \cdot \check{q}} \     
\overline{\Psi}^{-}_{\beta_2}(z) 
\Big[ (-igt^{a}\gamma_{\mu})  S_{F}(z,x_{2}) \IAbq \Big]_{\beta_{2}\alpha_{2}}
\gamma^{+} v(\check{p}_{2},h_{2})
\Big[ G^{\mu\nu}_{F}(z,w) \BI \Big]_{ab}  
\nn \\
&
\times
\overline{v}(\check{q},h)
\gamma^{+}     
\Big[S_{0,F}(y,w)  (-igt^{b}\gamma_{\nu})   S_{F}(w,x_{1}) \BAbq \Big]_{\alpha\alpha_{1}}
\gamma^{+} v(\check{p}_{1},h_{1}) , 
\end{align}
where the inside-to-after antiquark propagator is given in Eq.~\eqref{antiquark_prop_IA}, the before-to-inside gluon propagator is given in Eq.~\eqref{gluon_prop_BI}, the before-to-after antiquark propagator is given in Eq.~\eqref{antiquark_prop_BA}, and the vacuum quark propagator is given in Eq.~\eqref{eq:quark_vacuum_prop}. 
After substituting the corresponding propagators and simplifying, we get
\begin{align}
&
\S_{\bar q\to \bar q\bar q, \, 1} = 
\frac{g^{2}}{2p_{2}^{+}}  \ (2\pi) \delta(p_{1}^{+}+p_{2}^{+}-q^{+})  \int\! \frac{d^{2} \bk_{1}}{(2\pi)^{2}} \int_{\bz, \z_1}  e^{-i \bz \cdot (\bp_{2}+ \bk_{1}-\bq)  } \ e^{-i \bz_{1} \cdot (\bp_{1}-\bk_{1})} 
\nn \\
&
\times
 \int_{-\frac{L^+}{2}}^{\frac{L^+}{2}}dz^+
\int_{-\infty}^{-\frac{L^+}{2}} dw^+ 
e^{iw^{+}\Big(\frac{(\bk_{1}^{2}+m^2)}{2p_{1}^{+}} 
+ \frac{(\bk_{1}- \bq)^{2}}{2p_{2}^{+}} -\frac{(\bq^{2}+m^2)}{2q^{+}}  \Big)} 
\ \overline{\Psi}_{\beta_{2}}(\underline{z}) 
\frac{\gamma^{-}\gamma^{+}}{2}\gamma^{i}
\nn \\
&
\times
v(\check{p}_{2},h_{2}) \ \overline{v}(\check{q},h) \ 
\Big[ 
\frac{\gamma^{l}\gamma^{i} \bq^{l}}{2q^{+}} 
+ \frac{\gamma^{i}\gamma^{l} \bk_{1}^{l}}{2p_{1}^{+}} 
- \gamma^{i}\Big(\frac{1}{2q^{+}}-\frac{1}{2p_{1}^{+}}\Big)m 
+ \frac{\bq^{i} - \bk_{1}^{i}}{p_{2}^{+}}
\Big] 
\gamma^{+} v(\check{p}_{1},h_{1})
\nn \\
& 
\times
\Big[ t^{a}\ \UF^{\dagger}(\infty,z^{+}; \bz) \Big]_{\beta_{2}\alpha_{2}} 
\ \UA(z^{+},w^{+}; \bz)_{ab}
\Big[ t^{b}\ \UF^{\dagger}(\infty,w^{+}; \bz_{1})\Big]_{\alpha\alpha_{1}} ,     
\end{align}
from which one can extract the scattering amplitude as 
\begin{align}
\label{eq:M_barq_to_barqbarq_1}
&
i\M_{\bar q\to \bar q\bar q, \, 1} = 
\frac{g^{2}}{(2p_{2}^{+})(2q^+)}  \int\! \frac{d^{2} \bk_{1}}{(2\pi)^{2}} \int_{\bz, \z_1}  e^{-i \bz \cdot (\bp_{2}+ \bk_{1}-\bq)  } \ e^{-i \bz_{1} \cdot (\bp_{1}-\bk_{1})} 
\nn \\
&
\times
 \int_{-\frac{L^+}{2}}^{\frac{L^+}{2}}dz^+
\int_{-\infty}^{-\frac{L^+}{2}} dw^+ 
e^{iw^{+}\Big(\frac{(\bk_{1}^{2}+m^2)}{2p_{1}^{+}} 
+ \frac{(\bk_{1}- \bq)^{2}}{2p_{2}^{+}} -\frac{(\bq^{2}+m^2)}{2q^{+}}  \Big)} 
\ \overline{\Psi}_{\beta_{2}}(\underline{z}) 
\frac{\gamma^{-}\gamma^{+}}{2}\gamma^{i}
\nn \\
&
\times
v(\check{p}_{2},h_{2}) \ \overline{v}(\check{q},h) \ 
\Big[ 
\frac{\gamma^{l}\gamma^{i} \bq^{l}}{2q^{+}} 
+ \frac{\gamma^{i}\gamma^{l} \bk_{1}^{l}}{2p_{1}^{+}} 
- \gamma^{i}\Big(\frac{1}{2q^{+}}-\frac{1}{2p_{1}^{+}}\Big)m 
+ \frac{\bq^{i} - \bk_{1}^{i}}{p_{2}^{+}}
\Big] 
\gamma^{+} v(\check{p}_{1},h_{1})
\nn \\
& 
\times
\Big[ t^{a}\ \UF^{\dagger}(\infty,z^{+}; \bz) \Big]_{\beta_{2}\alpha_{2}} 
\ \UA(z^{+},w^{+}; \bz)_{ab}
\Big[ t^{b}\ \UF^{\dagger}(\infty,w^{+}; \bz_{1})\Big]_{\alpha\alpha_{1}} .     
\end{align}
Similarly, the $S$-matrix element for the second mechanism can be obtained via
\begin{align}
\S_{\bar q\to \bar q\bar q, \, 2} &= 
-\lim_{y^+\to-\infty}\lim_{x_1^+,x_2^+\to\infty}
\int_{\by,\bx[1],\bx[2]}\int_{y^-,x^-_1,x^-_2}
\int_{\bw, \bz}\int_{w^-, z^-}
\int_{-\frac{L^+}{2}}^{\frac{L^+}{2}}d z^+
\int_{-\infty}^{-\frac{L^+}{2}}d w^+
\nn \\
&
\times 
e^{ix_{1} \cdot \check{p}_{1}} \ e^{ix_{2} \cdot \check{p}_{2}} \ e^{-iy \cdot \check{q}} \     
\overline{\Psi}^{-}_{\beta_1}(z) 
\Big[ (-igt^{a}\gamma_{\mu})  S_{F}(z,x_{1}) \IAbq \Big]_{\beta_{1}\alpha_{1}}
\gamma^{+} v(\check{p}_{1},h_{1})
\Big[ G^{\mu\nu}_{F}(z,w) \BI \Big]_{ab}  
\nn \\
&
\times
\overline{v}(\check{q},h)
\gamma^{+}     
\Big[S_{0,F}(y,w)  (-igt^{b}\gamma_{\nu})   S_{F}(w,x_{2}) \BAbq \Big]_{\alpha\alpha_{2}}
\gamma^{+} v(\check{p}_{2},h_{2}) , 
\end{align}
Using the Eq.~\eqref{antiquark_prop_IA} for inside-to-after antiquark propagator, Eq.~\eqref{gluon_prop_BI} for the before-to-inside gluon propagator, for Eq.~\eqref{antiquark_prop_BA} the before-to-after antiquark propagator, and Eq.~\eqref{eq:quark_vacuum_prop} the vacuum quark propagator and simplifying, we get
\begin{align}
&
\S_{\bar q\to \bar q\bar q, \, 2} = 
-\frac{g^{2} }{2p_{1}^{+}}  \ (2\pi) \delta(p_{1}^{+}+p_{2}^{+}-q^{+})  \int\! \frac{d^{2} \bk_{2}}{(2\pi)^{2}} \int_{\bz, \z_1}  e^{-i \bz \cdot (\bp_{1}+ \bk_{2}-\bq)  } \ e^{-i \bz_{1} \cdot (\bp_{2}-\bk_{2})} 
\nn \\
&
\times
 \int_{-\frac{L^+}{2}}^{\frac{L^+}{2}}dz^+
\int_{-\infty}^{-\frac{L^+}{2}} dw^+ 
e^{iw^{+}\Big(
 \frac{(\bk_{2}- \bq)^{2}}{2p_{1}^{+}}
+\frac{(\bk_{2}^{2}+m^2)}{2p_{2}^{+}} 
 -\frac{(\bq^{2}+m^2)}{2q^{+}} 
\Big)} 
\ \overline{\Psi}_{\beta_{1}}(\underline{z}) 
\frac{\gamma^{-}\gamma^{+}}{2}\gamma^{i}
\nn \\
&
\times
v(\check{p}_{1},h_{1}) \ \overline{v}(\check{q},h) \ 
\Big[ 
\frac{\gamma^{l}\gamma^{i} \bq^{l}}{2q^{+}} 
+ \frac{\gamma^{i}\gamma^{l} \bk_{2}^{l}}{2p_{2}^{+}} 
- \gamma^{i}\Big(\frac{1}{2q^{+}}-\frac{1}{2p_{2}^{+}}\Big)m 
+ \frac{\bq^{i} - \bk_{2}^{i}}{p_{1}^{+}}
\Big] 
\gamma^{+} v(\check{p}_{2},h_{2})
\nn \\
& 
\times
\Big[ t^{a}\ \UF^{\dagger}(\infty,z^{+}; \bz) \Big]_{\beta_{1}\alpha_{1}} 
\ \UA(z^{+},w^{+}; \bz)_{ab}
\Big[ t^{b}\ \UF^{\dagger}(\infty,w^{+}; \bz_{1})\Big]_{\alpha\alpha_{2}} ,     
\end{align}
from which one can extract the scattering amplitude as 
\begin{align}
\label{eq:M_barq_to_barqbarq_2}
&
i\M_{\bar q\to \bar q\bar q, \, 2} = 
-\frac{g^{2}}{(2p_{1}^{+})(2q^+)}    \int\! \frac{d^{2} \bk_{2}}{(2\pi)^{2}} \int_{\bz, \z_1}  e^{-i \bz \cdot (\bp_{1}+ \bk_{2}-\bq)  } \ e^{-i \bz_{1} \cdot (\bp_{2}-\bk_{2})} 
\nn \\
&
\times
 \int_{-\frac{L^+}{2}}^{\frac{L^+}{2}}dz^+
\int_{-\infty}^{-\frac{L^+}{2}} dw^+ 
e^{iw^{+}\Big(
 \frac{(\bk_{2}- \bq)^{2}}{2p_{1}^{+}}
+\frac{(\bk_{2}^{2}+m^2)}{2p_{2}^{+}} 
 -\frac{(\bq^{2}+m^2)}{2q^{+}} 
\Big)} 
\ \overline{\Psi}_{\beta_{1}}(\underline{z}) 
\frac{\gamma^{-}\gamma^{+}}{2}\gamma^{i}
\nn \\
&
\times
v(\check{p}_{1},h_{1}) \ \overline{v}(\check{q},h) \ 
\Big[ 
\frac{\gamma^{l}\gamma^{i} \bq^{l}}{2q^{+}} 
+ \frac{\gamma^{i}\gamma^{l} \bk_{2}^{l}}{2p_{2}^{+}} 
- \gamma^{i}\Big(\frac{1}{2q^{+}}-\frac{1}{2p_{2}^{+}}\Big)m 
+ \frac{\bq^{i} - \bk_{2}^{i}}{p_{1}^{+}}
\Big] 
\gamma^{+} v(\check{p}_{2},h_{2})
\nn \\
& 
\times
\Big[ t^{a}\ \UF^{\dagger}(\infty,z^{+}; \bz) \Big]_{\beta_{1}\alpha_{1}} 
\ \UA(z^{+},w^{+}; \bz)_{ab}
\Big[ t^{b}\ \UF^{\dagger}(\infty,w^{+}; \bz_{1})\Big]_{\alpha\alpha_{2}} .     
\end{align}
Following the same procedure as in the previously discussed channels, starting from the scattering amplitude in general kinematics given in Eq.~\eqref{eq:M_barq_to_barqbarq_1} and ~\eqref{eq:M_barq_to_barqbarq_2}, the scattering amplitudes in the  back-to-back and massless quark limits can be obtained as 
\begin{align}
\label{eq:M_barq_to_barqbarq_b2b_1}
&
i\M^{{\rm b2b}, \, m=0}_{\bar q\to \bar q\bar q, \, 1} = 
-i g^{2} \frac{z}{(2q^+)}\frac{1}{\P^2}    
\int_{-\frac{L^+}{2}}^{\frac{L^+}{2}} dz^+
\int_\bb \ e^{-i\bb \cdot (\bk - \bq)} 
\ \overline{\Psi}_{\beta_{2}}(z^{+};\bb) 
\frac{\gamma^{-}\gamma^{+}}{2}\gamma^{i}
\ v(\check{p}_{2},h_{2}) \  \overline{v}(\check{q},h)
 \nn \\
 &
\times
\bigg[ \frac{\gamma^{i}\gamma^{l} \bP^{l}}{z} - \frac{2 \bP^{i}}{1-z} \bigg]  
\gamma^{+} v(\check{p}_{1},h_{1}) 
\Big[ t^{a}\ \UF^{\dagger}(\infty,z^{+};\bb) \Big]_{\beta_{2}\alpha_{2}}
\ \UA(z^{+},-\infty;\bb)_{ab}
\Big[ t^{b}\ \UF^{\dagger}(\infty, -\infty;\bb)\Big]_{\alpha\alpha_{1}} ,
\end{align}
and
\begin{align}
\label{eq:M_barq_to_barqbarq_b2b_2}
&
i\M^{{\rm b2b}, \, m=0}_{\bar q\to \bar q\bar q, \, 2} = 
-i g^{2} \frac{(1-z)}{(2q^+)}\frac{1}{\P^2}    
\int_{-\frac{L^+}{2}}^{\frac{L^+}{2}} dz^+
\int_\bb \ e^{-i\bb \cdot (\bk - \bq)} 
\ \overline{\Psi}_{\beta_{1}}(z^{+};\bb) 
\frac{\gamma^{-}\gamma^{+}}{2}\gamma^{i}
\ v(\check{p}_{1},h_{1}) \  \overline{v}(\check{q},h)
 \nn \\
 &
\times
\bigg[ \frac{\gamma^{i}\gamma^{l} \bP^{l}}{1-z} - \frac{2 \bP^{i}}{z} \bigg]  
\gamma^{+} v(\check{p}_{2},h_{2}) 
\Big[ t^{a}\ \UF^{\dagger}(\infty,z^{+};\bb) \Big]_{\beta_{1}\alpha_{1}}
\ \UA(z^{+},-\infty;\bb)_{ab}
\Big[ t^{b}\ \UF^{\dagger}(\infty, -\infty;\bb)\Big]_{\alpha\alpha_{2}} .
\end{align}
The total scattering amplitude in the $\bar q\to \bar q\bar q$ channel in the back-to-back and massless quarks limits is then given by
\begin{align}
\label{eq:M_tot_bq_to_bqbq_b2b_m0}
&
i \M_{\bar q\to \bar q\bar q, \, {\rm tot.}}^{{\rm b2b}, \, m=0}   =
i \M_{\bar q\to \bar q\bar q, \, 1}^{{\rm b2b}, \, m=0}
+ i \M_{\bar q\to \bar q\bar q, \, 2}^{{\rm b2b}, \, m=0}
\\
&=
\frac{i}{2}\frac{g^{2}}{(2q^{+})^2} \frac{1}{\P^2}
\int_{-\frac{L^+}{2}}^{\frac{L^+}{2}} dz^{+}  
\int\! d^{2}\bb\ e^{-i \bb \cdot {(\bk- \bq)}}
\overline{\Psi}_{\beta}(z^{+};\bb) 
\nn \\
& \times
\Bigg\{
    \mathfrak{h}_{\bar q\to \bar q\bar q}^{(1)}
    \UFd(\infty,z^{+};\bb)_{\beta\alpha_1}
    \UFd(\infty,-\infty;\bb)_{\alpha\alpha_2}
    +\mathfrak{h}_{\bar q\to \bar q\bar q}^{(2)}
    \UFd(\infty,z^{+};\bb)_{\beta\alpha{2}}
    \UFd(\infty,-\infty;\bb)_{\alpha_1\alpha}
\Bigg\} , \nn
\end{align}
with the Dirac coefficients obtained defined as
\begin{equation}
\label{eq:HF_barq_to_barqbarq_1}
    \mathfrak{h}_{\bar q\to \bar q\bar q}^{(1)} =
    \left(\mathfrak{h}_{q\to qq}^{(1)}\right)^{\dagger} ,
    \quad \quad
    \mathfrak{h}_{\bar q\to \bar q\bar q}^{(2)} =
    \left(\mathfrak{h}_{q\to qq}^{(2)}\right)^{\dagger} ,
\end{equation}
due to charge conjugation.

The partonic cross section in the back-to-back limit is then given by 
\begin{equation} 
\label{eq:sigma_barq_to_barqbarq}
\frac{d\sigma^{{\rm b2b}, \, m=0}_{\bar q\to \bar q\bar q}}{d{\rm P.S}} =
(2q^+) \ (2\pi)\delta\left(p_1^+ + p_2^+ - q^+\right)
\frac{1}{2N_c}\sum_{h_{1},h_2, h}\sum_{\beta_1, \beta_2, \alpha}
    \left\langle \left|i\mathcal{M}^{{\rm b2b}, \, m=0}_{\bar q\to \bar q\bar q, {\rm tot.}}\right|^2 \right\rangle .
\end{equation}
In the end, the partonic cross section in the back-to-back limit can be written in a general form for the different $\bar q_{f}\to \bar q_{f_1}\bar q_{f_2}$ channels as
\begin{align}
\frac{d\sigma^{{\rm b2b}, \, m=0}_{\bar q_{f}\to \bar q_{f_1}\bar q_{f_2}}}{d{\rm P.S}} &= 
g^{4}  (2\pi) \delta(p_{1}^{+} + p_{2}^{+} - q^{+}) 
\int_{-\frac{L^+}{2}}^{\frac{L^+}{2}} dz^+
\int_{-\frac{L^+}{2}}^{\frac{L^+}{2}} dz'^+
\int_{\bb, \bb'}e^{i(\bb - \bb')\cdot (\bk - \bq)}
\nn \\
& \hspace{6cm}
\times
\left[\mathcal{H}^{+\square}_{\bar q_{f}\to \bar q_{f_1}\bar q_{f_2}} \  \overline{\mathcal{C}}^{+\square}
+ \mathcal{H}^{+-+}_{\bar q_{f}\to \bar q_{f_1}\bar q_{f_2}} \ \overline{\mathcal{C}}^{+-+}\right] ,   
\end{align}
where, hard factors $\mathcal{H}^{+\square}_{\bar q_{f}\to \bar q_{f_1}\bar q_{f_2}}$ and  $\mathcal{H}^{+-+}_{\bar q_{f}\to \bar q_{f_1}\bar q_{f_2}}$  are the same as in the $q\to qq$ channel, given in Eqs.~\eqref{eq:H_q_to_qq_+square} and \eqref{eq:H_q_to_qq_+-+} (for the all same flavor antiquarks case); \eqref{eq:H_q_to_qq'_+square} and  \eqref{eq:H_q_to_qq'_+-+} (for the different flavor antiquark case) respectively.  
Moreover, the color structures $\overline{\mathcal{C}}^{+\square}$ is defined in \Equation{eq:bC+square} and $\overline{\mathcal{C}}^{+-+}$ reads
\begin{align}
\label{eq:bC+-+} 
\overline{\mathcal{C}}^{+-+} &\equiv 
\Big\langle
\Tr \bigg\{\UFd(\infty, -\infty;\bb) \ \UF(\infty, z'^{+};\bb')
\  \gamma^{-} \ \Psi(z'^{+};\bb)
\nn \\
& \hspace{5cm}
\times 
\overline{\Psi}(z^{+};\bb) \ \UFd(\infty, z^{+};\bb) \ \UF(\infty, -\infty;\bb') \bigg\} \Big\rangle . 
\end{align}

\printbibliography

\end{document}